\documentclass[12pt,a4paper,oneside]{book}
\makeglossary

\usepackage[dvips]{graphics}
\usepackage[dvips]{graphicx}
\usepackage{makeidx}
\usepackage{doublespace}
\usepackage{array}
\usepackage{multicol}
\usepackage{subfigure}
\usepackage{cite}
\usepackage{afterpage}
\usepackage{rotating}
\usepackage{bm}
\usepackage{feynarts}
\usepackage{amsmath}

\usepackage{epsfig}

\allowdisplaybreaks[3]

\newcommand{\mylinespace}{1.66} %1.66 is doublespace
\def\mytitle{{Computer Simulations of High Energy Physics}}

\newcommand{\newc}{\newcommand}

\newcommand{\prepareAbbrev}[6]{\newcounter{#5}\newcommand{#1}{\glossary{#2 @{\bf #6) & #4}}\ifnum\arabic{#5}=0 {#4 (#3)}\else#3\fi\addtocounter{#5}{1}}}
\prepareAbbrev\SM{SM}{SM}{Standard Model}{SMcounter}{SM}
\prepareAbbrev\MSSM{MSSM}{MSSM}{minimal supersymmetric standard model}
{MSSMcounter}{MSSM}
\prepareAbbrev\mSUGRA{MSUGRA}{mSUGRA}{minimal supergravity}
{MSUGRAcounter}{mSUGRA}
\prepareAbbrev\GMSB{GMSB}{GMSB}{gauge mediated \sus\ breaking}
{GMSBcounter}{GMSB}
\prepareAbbrev\AMSB{AMSB}{AMSB}{anomaly mediated \sus\ breaking}
{AMSBcounter}{AMSB}
\prepareAbbrev\LEP{LEP}{LEP}{large electron-positron}{LEPcounter}{LEP}
\prepareAbbrev\LHC{LHC}{LHC}{large hadron collider}{LHCcounter}{LHC}
\prepareAbbrev\QFT{QFT}{QF$\Theta$}{quantum field theory}
{QFTcounter}{QF$\Theta$} 
\prepareAbbrev\VEV{VEV}{VEV}{vacuum expectation value}{VEVcounter}{VEV} 
\prepareAbbrev\GUT{GUT}{GUT}{grand unified theory}{GUTcounter}{GUT}
\prepareAbbrev\RGEs{RGEs}{RGEs}{renormalization group equations}
{RGEscounter}{RGEs} 
\prepareAbbrev\RPV{RPV}{$R_P$V}{$R$-parity violating}{RPVcounter}{$R_P$V}
\prepareAbbrev\RPC{RPC}{$R_P$C}{$R$-parity conserving}{RPCcounter}{$R_P$C}
\prepareAbbrev\EWSB{EWSB}{EWSB}{electro-weak symmetry breaking}
{EWSBcounter}{EWSB}
\newc{\psibar}{\bar{\psi}}
\newc{\eps}{\epsilon}
\newc{\prt}[3]{Phys. Rept. #1 (#2) #3}
\newc{\prd}[3]{Phys. Rev. D#1 (#2) #3}
\newc{\npb}[3]{Nucl. Phys. B#1 (#2) #3}
\newc{\plb}[3]{Phys. Lett. B#1 (#2) #3}

\def\half{{\frac 1 2}}

\def\slashchar#1{\setbox0=\hbox{$#1$}           % set a box for #1
   \dimen0=\wd0                                 % and get its size
   \setbox1=\hbox{/} \dimen1=\wd1               % get size of /
   \ifdim\dimen0>\dimen1                        % #1 is bigger
      \rlap{\hbox to \dimen0{\hfil/\hfil}}#1 
   \else                                        % / is bigger
      \rlap{\hbox to \dimen1{\hfil$#1$\hfil}}/                                    \fi}          

\def\as{\alpha_{\mbox{\tiny S}}}
\def\rs{\rho_{\mbox{\tiny S}}}
\def\ee{e^+e^-}

\def\eps{\epsilon}

\def\MSbar{\overline{{\rm MS}}}

\def\0t{{\mbox{\bf 0}}}
\def\kt{{\mbox{\bf k}_\perp}}
\def\pt{{\mbox{\bf p}_\perp}}
\def\qb{{\bar q}}
\def\sb{{\bar s}}
\def\tb{{\bar t}}
\def\ub{{\bar u}}

\def\qt{{\mbox{\bf q}_\perp}}
\def\Qt{{\mbox{\bf Q}_\perp}}
\def\ptr#1{{\mbox{\bf p}_{\perp #1}}}
\def\qtr#1{{\mbox{\bf q}_{\perp #1}}}

\def\tk{\tilde{\kappa}}
\def\tq{\tilde{q}}

\def\frac#1#2{ {{#1} \over {#2} }}
\def\GeV{\mbox{\rm{GeV}}}
\def\VEV#1{\left\langle #1\right\rangle}
\def\HWP{\mbox{\textsf{Herwig++}}}
\def\ee{e^+e^-}
\def\qq{q\bar q}

\def\HW{\textsc{\small HERWIG}}
\def\pythia{\textsc{\small PYTHIA}}
\def\hpp{\mbox{\textsf{Herwig++}}}
\def\ThePEG{\textsf{ThePEG}}
\def\Pythia7{\textsf{Pythia7}}
\def\SHERPA{\textsf{SHERPA}}
\def\CLHEP{\textsf{CLHEP}}
\def\EFF{\textsf{Effective}}
\def\ycut{y_{\rm cut}}

\def\Clm{{\texttt{Cl}_{\texttt{max}}}}
\def\Clp{{\texttt{Cl}_{\texttt{pow}}}}
\def\D{\cite{DELPHIshapes}}
\def\O{\cite{OPALnjet}}
\def\DF{\cite{DELPHIfourj}}
\def\OC{\cite{Acton:1991aa}}
\def\Sx{\cite{SLD03}}
\def\A{\cite{ALEPHLambda}}
\def\SB{\cite{SLDbfrag}}
\def\AB{\cite{ALEPHbfrag}}

\newcommand{\Fs}[1]{{}\kern-.45em \not \kern-.12em #1 \hspace{.2pt}}
\newcommand{\Tr}[0]{\mathrm{Tr}}
\def\Btp{\texttt{BCl}_{\texttt{pow}}}

\begin{document}
\DeclareGraphicsExtensions{.pdf,.eps,.png,.ps,.gif,.jpg}
% This bit controls what is numbered and shown in the contents page
\setcounter{tocdepth}{3} % show down to sub subsections in contents
\setcounter{secnumdepth}{3} % number down to sub subsections

\frontmatter

\begin{singlespace}

%%%%%%%%%%%%%%%%%%%%%%%%%%% Title %%%%%%%%%%%%%%%%%%%%%%%%%%%%
\begin{titlepage}
\begin{center}
\begin{spacing}{2.2}
${}^{}$ \\
\vspace{20mm}
{\huge\bf \mytitle} \\
\end{spacing}
\vspace{40mm}
\large
Philip John Stephens\\
${}^{}$\\
Trinity College \\
\vspace{82mm} \normalsize A dissertation submitted to the University
of Cambridge \\ for the degree of Doctor of Philosophy \\ May 2004
\end{center}
\end{titlepage}

%%%%%%%%%%%%%%%%%%%%%%%%% Abstract %%%%%%%%%%%%%%%%%%%%%%%%%%
%$\ $
%\newpage
\begin{center}
\begin{spacing}{1.4}
%${}^{}$ \\
%\vspace{0.5cm}
{\Large\bf \mytitle} \\
\end{spacing}
\vspace{0.5cm}
{\large Philip John Stephens}\\
\vspace{1.0cm}
{\Large\bf Abstract} \\
\end{center}
\normalsize
\begin{spacing}{1.5} 
This thesis describes the development of two independent computer programs,
\HWP{} and \EFF{}. Both of these programs are used for phenomenological
predictions of high energy physics. The former is used to simulate events
as measured at particle colliders. The latter is used to generate the mass
spectrum of supersymmetric models.

Simulation of collider events requires the implementation of several different 
aspects of particle phenomenology. After a brief introduction on the relevant 
aspects of the Standard Model and numerical techniques, a new set of variables 
for parton shower evolution are presented. These new variables retain the 
angular ordering feature of the variables found in the original \HW{} 
software, while improving the Lorentz invariance of the shower and improving 
the coverage of the phase space. These new variables define new initial 
conditions for the shower depending on how the partons are colour connected. 
Also developed is a new model for hadronization. By changing the distribution 
of probabilities of cluster decays into hadron pairs this model is able to 
enforce desired results, such as isospin symmetry or meson-baryon ratios, more 
intuitively.

The physics of the \HWP{} software is described in detail. The improvements
to the new evolution variables and the new hadronization model provide are 
illustrated by comparing them against data for $e^+ e^-$ events. 

\EFF{} is a program that is able to provide the 1-loop effective potential
and 1-loop mass matrices for an arbitrary $N=1$ supersymmetric model. This
software also is able to solve the renormalization group equations at one-loop
for the parameters of the model and, in turn, provide the scale dependent 
values of these parameters. The program is described and some results 
indicative of its potential are also presented.
\end{spacing}
%%%%%%%%%%%%%%%%%%%%%%%%% END OF SINGLESPACE %%%%%%%%%%%%%%%%%%%%%%%%%%

\end{singlespace}

%%%%%%%%%%%%%%%%%%%%%% Declaration %%%%%%%%%%%%%%%%%%%%%%%%
\newpage
\vspace*{65mm}
\centerline{\Huge {\bf Declaration}}
\vspace{10mm}
This dissertation is the result of my own work, except where explicit
reference is made to the work of others, and has not been submitted
for another qualification to this or any other university.
\\
\\
\\
\\
\vspace{30mm}
\hspace{10cm}
Philip John Stephens

%%%%%%%%%%%%%%%%%%%%%%%%% Preface %%%%%%%%%%%%%%%%%%%%%%%%%%
%\newpage
%{\ }
\newpage
{\ }
\vspace{10mm}
\begin{center}
{\Large\bf Preface}
\end{center}
\vspace{5mm}
%%%%%%%%%%%%%%%%%%% Collaboration notes %%%%%%%%%%%%%%%%%%%%%
This thesis contains the work I have done on \HWP{} and \EFF{} during the
course of my high-energy physics PhD. The first chapter is an original summary
of theoretical and practical issues pertaining to my research. The results of
chapter 2 were derived independently by Prof. Bryan Webber and myself and
cross checked for errors. Chapter 3 is the result of my own independent work.
Chapter 4 is a complete discussion of \HWP{} which has been implemented by
myself and Dr. Stefan Gieseke, with other coding contributions from Dr. Alberto
Ribon. Theoretical contributions to \HWP{} have also been made by Prof. Bryan
Webber and Dr. Mike Seymour. Chapter 5 is the result of comparison studies of
\HWP{} to LEP data. This is the result of work done in conjunction by Dr.
Gieseke and myself. Chapter 6 is the extension of work done by Dr. James
Hetherington as part of his PhD thesis. His original idea of \EFF{} has been 
extended to include the renormalization group equations and the correct 
one-loop mass corrections have been implemented.

I would like to thank Dr. Stefan Gieseke for his continual help and guidance
throughout my time at Cambridge. I would also like to thank my supervisor
Prof. Bryan
Webber and the rest of the members of the HEP Group over the last three years, 
in particular
Dr. Agustin Sabio Vera and Dr. Jeppe Anderson for numerous fruitful discussions.

The work of chapter 6 would not have been possible if not for the time and
energy spent by Dr. James Hetherington helping me to understand the ideas
he had implemented in \EFF{}.

I would like to thank Trinity College and the University Rugby Club 
for membership and support. Also the Universities UK and the Cambridge Overseas
trust for
their financial support. And I'd like to thank CERN for allowing me to
spend a summer there while attending the Monte Carlo Workshop for LHC, during
which discussions with the SHERPA team helped me to understand the content of
chapter 3 better.

Mostly I would like to thank my family, old and new. Thanks to my parents
for encouragement, support and belief that I would achieve this goal. Thanks
to my brothers and sister for constantly supporting me and for their
confidence in my abilities. Lastly I would like to thank Jenna. This is the
first step in our long life of happiness together.

%%%%%%%%%%%%%%%%%%% lists of everything %%%%%%%%%%%%%%%%%%%%%
\begin{singlespace}
\tableofcontents
\listoffigures
\listoftables
\end{singlespace}
\mainmatter

\pagenumbering{arabic}

\chapter{Introduction}
\label{chap:Intro}

\section{Field Theory Introduction}
This section briefly introduces a few key ideas that are used throughout
high energy physics for calculating predictions of physics. I start by
introducing the Klein-Gordon and Dirac field equations. This is followed by
a discussion of the \emph{Lagrangian} for both \emph{Abelian} and 
\emph{non-Abelian} gauge groups. Lastly I explain how calculations are 
performed in \emph{perturbation theory}.

We start by introducing the Klein-Gordon field. This field obeys Bose
statistics and is used to describe all bosons: scalars and vectors. Free fields of this form obey the Klein-Gordon equation
\begin{equation}
(\partial^2 +m^2)\phi(x) = 0.
\end{equation}

\begin{figure}[htb]
\centering
\includegraphics[width=5in]{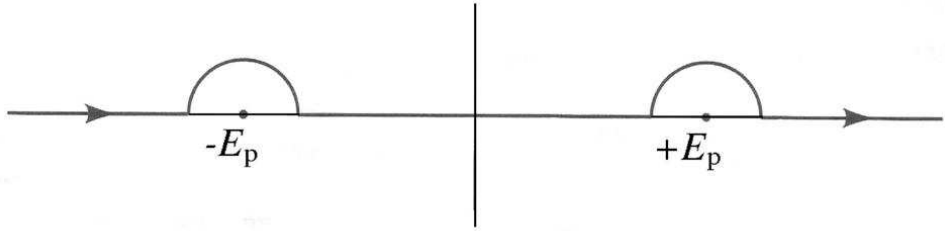}
\caption{The contour taken to find the retarded Green's function. For $x_0 > 
y_0$ we can close the contour below. For $x_0<y_0$ we can close the contour
above, giving zero.}
\label{fig:retarded_contour}
\end{figure} 
The propagator of a field is the Green's function of the field equation. In
this case it is the Green's function of the Klein-Gordon equation. In order
to find this function we must introduce a pole prescription for the integral. 
Figure~\ref{fig:retarded_contour} shows the contour used for the integral 
over $p^0$. When the contour is closed below we find the Green's function only 
over the range $x_0 > y_0$. This is called the \emph{retarded} Green's 
function. For $x^0 > y^0$ it is 
\begin{equation}
D_R(x) = \int \frac{d^4 p}{(2\pi)^4}\frac{i e^{-i p \cdot (x-y)}}{p^2 - m^2}.
\end{equation}
and zero for $x_0 < y^0$.
\begin{figure}[htb]
\centering
\includegraphics[width=5in]{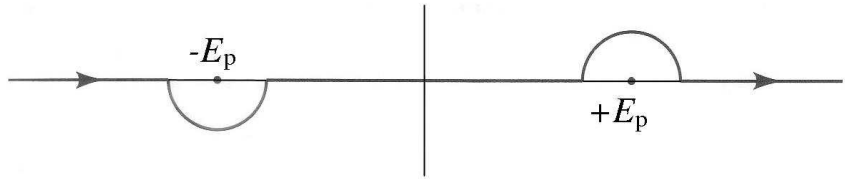}
\caption{The contour for the Feynman prescription. When $x_0>y_0$ the contour
can be closed below and when $x_0<y_0$ the contour can be closed above.}
\label{fig:feyn_contour}
\end{figure}

Using the \emph{Feynman prescription} of pole contours we find the propagator, 
known as the \emph{Feynman propagator}, for bosonic fields over all $x^0$ and
$y^0$ is
\begin{equation}
\label{eq:BosProp}
D_F(x-y) \equiv \int \frac{d^4p}{(2\pi^4)} \frac{e^{-i p \cdot (x-y)}}
{p^2 - m^2 + i \epsilon}.
\end{equation}
This can be written as
\begin{equation}
D_F(x-y) = \left\{ \begin{array}{cc} 
\left< 0 \right| \phi(x) \phi(y) \left| 0 \right> & {\rm for}~~x^0 > y^0 \\ 
\left< 0 \right| \phi(y) \phi(x) \left| 0 \right> & {\rm for}~~x^0 < y^0 
\end{array} \right. .\end{equation}

We now want to see how fields that obey Fermi statistics behave. The Dirac 
Equation is the field equation for spin 1/2 fermions. We first start by 
introducing the Dirac matrices $\gamma^\mu$. These are in four-dimensional 
Minkowski space and are given in a popular representation by
\begin{equation}
\gamma^0 = \left( \begin{array}{cc} 0 & {\bm 1} \\ {\bm 1} & 0 \end{array} 
\right);~~\gamma^i = \left( \begin{array}{cc} 0 & \sigma^i \\ -\sigma^i & 0 
\end{array} \right),
\end{equation}
where ${\bm 1}$ is the $2 \times 2$ identity matrix and $\sigma^i$ are the 
Pauli sigma matrices. Using these matrices we define the following notation: 
$\Fs{a} \equiv \gamma^\mu a_\mu$ and $\bar \psi \equiv \psi^\dagger \gamma^0$. 
The Dirac equation is then
\begin{equation}
(i\Fs{\partial} - m) \psi(x) = 0.
\end{equation}
Doing the same as for the Klein-Gordon equation, we can find that the retarded 
Green's function, in momentum space, of a Dirac field is
\begin{equation}
\tilde{S}_R = \frac{i}{\Fs{p}-m} = \frac{i(\Fs{p}+m)}{p^2-m^2}.
\end{equation}
Along with the anti-commuting nature of fermions this leads to the
Feynman propagator
\begin{equation}
\label{eq:FermiProp}
S_F(x-y) = \int \frac{d^4p}{(2\pi)^4} \frac{i (\Fs{p} + m)}{p^2 - m^2 + i 
\epsilon} e^{-i p \cdot (x-y)},
\end{equation}
which is 
\begin{equation}
S_F(x-y) = \left\{ \begin{array}{cc} 
\left< 0 \right| \psi(x) \bar \psi(y) \left| 0 \right> & {\rm for}~~x^0>y^0 \\ 
- \left< 0 \right| \bar \psi(y) \psi(x) \left| 0 \right> & {\rm for}~~x^0<y^0 
\end{array} \right. .
\end{equation}

\subsection{Lagrangian}
In classical mechanics, the fundamental quantity is the action, $S$. This is
the time integral of the Lagrangian $L$. The quantity $L$ can be written as
a spatial integration of a Lagrangian density, ${\cal L}$. It is ${\cal L}$ 
that is generally referred to as the Lagrangian in field theory. This is also
done so throughout this thesis.

The Lagrangian contains the information about all of the fields and 
interactions of a theory. The different types of interactions are mediated by 
particles which transform according to the adjoint representation of a gauge 
group. In the Standard Model (SM), these particles are gauge bosons; spin 1 
particles. The kinetic term for a gauge boson,
${\cal A}^A_\mu$, is
\begin{equation}
{\cal L}_{\rm gauge} = -\frac{1}{4} F^{A}_{\mu \nu} F^{A \mu \nu},
\end{equation}
where 
\begin{equation}
\label{eq:gauge_kinetic}
F^A_{\mu\nu} = \partial_\mu {\cal A}^A_\nu - \partial_\nu {\cal A}^A_\mu - g 
f^{ABC} {\cal A}^B_\mu {\cal A}^C_\nu.
\end{equation}
In equation (\ref{eq:gauge_kinetic}) the indices $A,B$ and $C$ are the indices 
in the adjoint representation of the group and the indices $\mu$ and $\nu$ 
are Lorentz indices. $g$ is known as a \emph{coupling constant}. These 
define the relative scale of the term in the Lagrangian when compared to other 
terms. The terms $f^{ABC}$ are the structure constants of the gauge group. Each
group has a set of generators $t^A$. The structure constants for the gauge 
group are given by the relation
\begin{equation}
\left[ t^A, t^B \right] = i t^C f^{ABC}.
\end{equation}
When $f^{ABC}$ is equal to 0 for all indices the group is called an Abelian 
group. When they
are non-zero it is called a non-Abelian group. From (\ref{eq:gauge_kinetic}) 
it can be seen that for an Abelian group the gauge fields don't interact
with themselves. For quantum electrodynamics (QED) the group is $U(1)$ which is
an Abelian group. The gauge field of QED is the photon and therefore we can
see that photons don't interact with themselves.  For quantum chromodynamics 
(QCD) this group is $SU(3)$, which is non-Abelian. It is the non-Abelian 
nature of this group that gives rise to the triplet and quartic gluon 
self-interaction terms which complicate QCD calculations. 

In order for the symmetries of a gauge group to be preserved, all the terms in 
the Lagrangian must be invariant under the local gauge transformations of 
that group. The Lagrangian is composed of gauge fields, matter fields and
derivatives of both. As the Lagrangian needs to be invariant under these
transformations, only certain combinations of these fields are possible.

The matter fields furnish the fundamental representation of the group. In the 
Standard Model these are fermions or scalars; spin $\frac{1}{2}$ or spin 0 
particles. For a set of local transformations, $\theta^A(x)$, we have the
transformation
\begin{equation}
\Psi(x)_i \rightarrow \left( {\rm e}^{i \sum_A t^A \theta^A(x)} \right)_{ij} 
\Psi(x)_j \equiv \Omega(x)_{ij} \Psi(x)_j.
\end{equation}
Here, the indices $i$ and $j$ indicate the fundamental indices and the index 
$A$ is the index of the adjoint representation of the group. In $SU(N)$ the 
adjoint 
representation runs from $1$ to $(N^2 -1)$ and $N$ is the number
of fundamental indices. In $U(N)$ the adjoint indices run from $1$ to $N^2$.
For 
example in $SU(2)$ there are 3 adjoint index values and 2 fundamental index 
values. Likewise in $SU(3)$ there are 8 adjoint index values and 3 fundamental 
index values.

We now define the \emph{covariant derivative} for a local transformation. This
is
\begin{equation}
D_\mu \equiv \partial_\mu + ig \sum_A t^A {\cal A}^A_\mu(x).
\end{equation}
Here $g$ is the coupling constant of the group. As mentioned before, these 
govern the relative strength the interactions of one group have when 
compared with the other groups. In order to use the Klein-Gordon and Dirac
field equations in the Lagrangian we must replace the partial derivate 
$\partial_\mu$ by the covariant derivative $D_\mu$. We also require the 
covariant derivative to transform in the same way as the matter fields. This
is the requirement
\begin{equation}
D_\mu \Psi(x)_i \rightarrow \Omega(x)_{ij} D_\mu \Psi(x)_j.
\end{equation}

We can use these transformations to define the transformation of the gauge 
fields. In the Abelian case this is
\begin{equation}
{\cal A}_\mu(x) \rightarrow {\cal A}_\mu(x) - \frac{1}{g} \partial_\mu 
\theta(x),
\end{equation}
while the non-Abelian case the transformation is more complicated. It is given
by
\begin{equation}
\sum_A t^A {\cal A}^A_\mu \rightarrow \Omega(x) \left( \sum_A t^A 
{\cal A}^A_\mu \right) \Omega^{-1}(x)  + \frac{i}{g} \left( \partial_\mu 
\Omega(x) \right) \Omega^{-1}(x).
\end{equation}
This requirement prevents the addition of a term like $\frac{1}{2}m^2 
{\cal A}_\mu {\cal A}^\mu$ to give the field mass. Instead this must be done
by the Higgs mechanism, which is described in section \ref{sec:Higgs}. 

The coupling constants entered into the Lagrangian are known as \emph{bare
couplings}. Due to higher order corrections, these are not the same as the 
physically observed couplings. Instead the physical coupling is a 
\emph{renormalized coupling}. This means that the higher order corrections 
introduce a shift in the coupling.

At first it may seem that there are an infinite number of possible terms in
the Lagrangian that can satisfy gauge invariance. It will be shown in Section
\ref{sec:Perturbation} that higher order terms in perturbation theory will 
involve integrals over 4-momenta of virtual particles. These are divergent 
integrals. In order to do the calculations these must have some cutoff imposed
at finite momentum, $\Lambda$. Theories where the observables, expressed in 
terms of the suitably renormalized parameters, have values which are
independent on $\Lambda$ are known as \emph{renormalizable} theories. Using 
this it can be shown that theories which contain a coupling constant of mass to
the \emph{negative} power are not renormalizable. 

The last constraint on the Lagrangian is that it must have dimension $\left(
{\rm mass}\right)^4$. Using the constraints of gauge invariance, 
renormalizability and the correct dimension of the Lagrangian the only
allowable terms in the Lagrangian, for spinors $\psi$, scalars $\phi$ and 
vectors ${\cal A}_\mu$, are
\begin{eqnarray}
\bar \psi \psi,~~\phi^\dagger \phi,~~(\phi^\dagger \phi)^2,~~\bar \psi \psi
\phi, \nonumber \\ 
\bar \psi \gamma^\mu D_\mu \psi,~~(D_\mu \phi)^\dagger (D^\mu \phi),
~~F_{\mu \nu} F^{\mu \nu}. \nonumber
\end{eqnarray}
All of these terms can have some relevant coupling that has mass dimension 
larger than or equal to 0. For a more thorough discussion of gauge invariance, 
renormalizability and Lagrangian formalism see~\cite{WeinbergI, WeinbergII,
PeskinSchroeder, Ellis:1996qj, Collins, Ryder}.

\subsection{Perturbation Theory}
\label{sec:Perturbation}
Earlier we found the Feynman propagator for a field that obeys the 
Klein-Gordon equation. This can be written as
\begin{equation}
D_F(x-y) = \int \frac{d^4p}{(2 \pi)^4}\frac{i {\rm e}^{-ip\cdot(x-y)}}
{p^2-m^2+i\epsilon} \equiv \left< 0 \left| T \left\{ \phi(x) \phi(y) \right\}
\right| 0\right>,
\end{equation}
where $T$ indicates the time-ordering property of the Feynman propagator. This
propagator describes the free field theory. Physical predictions can only be
made on theories that interact, however. To do so we work in the 
\emph{interaction picture}. We now define a unitary operator, 
$U(t,t_0)$, that takes a field at time $t_0$ to time $t$ in the presence of
an interaction. This is known as the interaction picture propagator and is
defined as
\begin{equation}
U(t,t_0) = {\rm e}^{iH_0(t-t_0)}{\rm e}^{-iH(t-t_0)},
\end{equation}
where we have divided the Hamiltonian into the free field part, $H_0$ and the 
interaction part $H_{\rm int}$. We find $U$ obeys the Schr\"odinger equation
\begin{equation}
i\frac{\partial}{\partial t} U(t,t_0) = {\rm e}^{i H_0 (t-t_0)} (H-H_0)
{\rm e}^{-iH(t-t_0)} = H_I(t) U(t,t_0).
\end{equation}
$H_I$ is the interaction Hamiltonian written in the interaction picture. This 
has the form
\begin{equation}
H_I(t) = {\rm e}^{i H_0(t-t_0)}\left(H_{\rm int}\right) {\rm e}^{-iH_0(t-t_0)}.
\end{equation}
Solving this differential equation with the initial condition $U(t_0,t_0)=1$
we can find the solution as
\begin{equation}
\label{eq:Utt0}
U(t,t_0) = T\left\{\exp\left[-i\int_{t_0}^t dt' H_I(t')\right]\right\},
\end{equation}
where the time-ordering of an exponential is the Taylor series with each term 
time ordered. It is this Taylor series that is used when doing perturbative
calculations. Before we can define these perturbative calculations we must
introduce \emph{Wick's Theorem}.

In the interaction picture we can decompose the field $\phi$ into its positive
and negative energy parts
\begin{equation}
\phi(x) = \phi^+(x) + \phi^-(x).
\end{equation}
The \emph{contraction} of two fields is defined as
\begin{equation}
\underline{\phi(x)}\underline{\phi(y)} \equiv \left\{ \begin{array}{cc} 
\left[ \phi^+(x),\phi^-(y) \right] & {\rm for}~~x^0 > y^0; \\ 
\left[ \phi^+(y), \phi^-(x) \right] & {\rm for}~~y^0 > x^0. \end{array} \right.
\end{equation}
This is exactly the Feynman propagator that was encountered before
\begin{equation}
\underline{\phi(x)}\underline{\phi(y)} = D_F(x-y).
\end{equation}
This allows us to write the time-ordering of fields as
\begin{equation}
\label{eq:Wicks}
T\left\{\phi(x_1)\phi(x_2)\dots\phi(x_m)\right\} =
N\left\{\phi(x_1)\phi(x_2)\dots\phi(x_m) + {\rm all~possible~contractions}
\right\},
\end{equation}
where $N$ indicates the \emph{normal ordering}. This is just the ordering of
having all creation operators to the right of all annihilation operators. The
identity in (\ref{eq:Wicks}) is Wick's theorem. 

When computing the vacuum expectation value of a time-ordered product 
any uncontracted operators from applying Wick's theorem give zero
($\left<0\left|N({\rm any~operator})\right|0\right> = 0$) and the 
contracted operators are simply Feynman propagators! It is this 
decomposition of the time ordered products that leads to Feynman diagrams.

When we wish to construct higher order terms we will have states that are
created and destroyed and never produce observable particles. These are known
as \emph{virtual} particles. When computing observables, the contributions
due to these particles must be integrated over their momenta. As we can
see from (\ref{eq:BosProp}) and (\ref{eq:FermiProp}), these
integrals contain $p^{-2}$ for each virtual particle and $d^4p$ for the 
integrals over these particles. Using power counting it can be shown that 
some diagrams will contain integrals like 
\begin{equation}
\int^{\Lambda} \frac{dp}{p} \sim \ln\left( \Lambda \right).
\end{equation}
These integrals need to be bound above\footnote{These also need to be bound
below, known as an infrared cutoff. This is another problem that is not related
to renormalizability.} by an ultra-violet cutoff, $\Lambda$, in order to be
computed. Observables must be independent of this cutoff and this condition 
is what leads to renormalizability, as discussed previously.

The power of perturbation theory is fully exploited when the coupling 
constants are small. This allows us to write (\ref{eq:Utt0}) 
as a Taylor series in order of the coupling constant. Using Wick's theorem,
we can then decompose the terms of the Taylor series into normal ordered 
products. This allows us to use Feynman diagrams to describe each of the 
normal ordered products. Where perturbation theory is a good approximation we 
only need to evaluate the first few terms of the series in order to 
approximately describe the physics. Unfortunately, the magnitude of the higher 
order terms cannot be predicted beforehand. Only by calculating them can one 
decide how accurate the initial calculation really is.

\section{The Standard Model}
\label{sec:SM}
The Standard Model (SM) is well established as a model that describes the 
particles and all the interactions, except gravity. The predictions of the 
model have been tested to high accuracy by the series of LEP experiments. 
Recent experimental evidence~\cite{Neutrino} shows that neutrino flavours 
oscillate which means they must have mass. This is in direct contradiction to 
the SM and is the first evidence of physics beyond the SM.
Apart from the incorrect description of the neutrino flavour oscillation,
the model also predicts the existence of the Higgs boson. This has not been
experimentally confirmed to date. This section explains the particle content of
the model, the Higgs mechanism and properties of QCD. 

There are theoretical reasons to believe that at higher energies the Standard 
Model will also break down. In section \ref{sec:susy} I discuss what 
shortcomings there are believed to be and a theoretical solution to these 
shortcomings, known as supersymmetry (SUSY).

\subsection{Particle Content}
The Standard Model is composed of the $SU(3)_c, SU(2)_L$ and the
$U(1)_Y$ gauge groups. Properties of the interaction eigenstates are governed 
by these groups. The weak and and electromagnetic interactions are not directly
governed by $SU(2)_L$ and $U(1)_Y$ groups, however. Instead the mass 
eigenstates are given by the symmetry breaking of these two groups. This will 
be discussed in section~\ref{sec:Higgs}. 

The gauge fields for these groups are ${\cal A}^a_\mu$, $W^i_\mu$ and $B_\mu$. 
The field ${\cal A}$ is known as the gluon field. As mentioned earlier, these 
gluons have self-interaction terms. That means a gluon, unlike the neutral 
photon, carries a (colour) charge. These terms make calculations with QCD much 
more complex than with QED. The $W$ and $B$ bosons mix through the Higgs 
mechanism (see section~\ref{sec:Higgs}) to form the $W^\pm$, $Z^0$ bosons and 
the photon, $A$\footnote{Note the different font between the gluon field and 
the photon field.}. It is the photon that mediates the electromagnetic 
interaction we observe. For each field we have a term in the Lagrangian given 
by (\ref{eq:gauge_kinetic}).

The matter content of the Standard Model can be summarized quite simply. There
are leptons, quarks and the Higgs boson. Leptons have $SU(2)_L$ and $U(1)_Y$
charges and the quarks have $SU(3)_c$, $SU(2)_L$ and $U(1)_Y$
charges. The $L$ subscript of the $SU(2)$ gauge means that it only couples to
left handed particles. We can see from table~\ref{table:SMfermions} that there
are right and left handed charged leptons, up-type quarks and down-type quarks
but there are no right handed neutrinos. This means that the left handed 
leptons interact with the $W$ and $B$ bosons, but not with the gluons, 
${\cal A}$, while the right handed leptons only interact with the $B$ boson. 
The left and right handed quarks interact with the gluons and the $B$ boson. 
The left handed quarks also interact with the $W$ bosons while the right handed
quarks don't. Table~\ref{table:SMfermions} shows the charges of the fields in 
each gauge. As will be discussed later, the Higgs boson has $SU(2)_L$ and 
$U(1)_Y$ charges.

The strength of the different interactions (QED, Weak, QCD) are dictated by the
relative size of their coupling constants. In QED this constant is $e$ which
is related to the $SU(2)_L$ coupling, $g_W$ and the $U(1)_Y$ coupling, $g'$. In
weak interactions the $SU(2)_L$ symmetry is broken and the coupling depends on
whether we are coupling to the $W^\pm$ bosons or the $Z^0$ boson. Either way
this coupling is smaller then $e$, thus the weak interaction is weaker than the
electromagnetic one. The coupling constant for QCD is $g_3$. This is larger 
than $e$; for fields that interact with gluons the QCD terms are most often
the dominant ones. 

There is more to this picture than the couplings being 
constant, however. In fact these terms are scale dependent. This is known as 
the running of the couplings and will be discussed in 
section~\ref{sec:runningCoup}. It is believed that at some large scale, the 
couplings of all the interactions, including gravity, will be of the same size,
thus unifying the theories.

It is known from experiments that these particles have mass. In the SM the
fermions have right and left handed components which have different charges 
under the groups. Due to gauge invariance, a term like $ m \bar \psi_L \psi_R$ 
cannot be added to the Lagrangian to give these fields mass as
they have different $SU(2)_L$ and $U(1)_Y$ charges. Instead, the Higgs 
mechanism is used again to define a Yukawa coupling. This will also be 
explained in the next section.

\begin{table}
\begin{center}
\begin{tabular}{lccc}
Field             & ${\rm U(1)}_Y$ & ${\rm SU(2)}_L$ & ${\rm SU(3)}_c$ \\ 
\hline
$e_R$             &   -1           &   $\bm{1}$      &  $\bm{1}$ \\
$\ell$            & $-\frac{1}{2}$ &   $\bm{2}$      &  $\bm{1}$ \\
$Q$               & $\frac{1}{6}$  &   $\bm{2}$      &  $\bm{3}$ \\
$u_R$             & $\frac{2}{3}$  &   $\bm{1}$      &  $\bm{3}$ \\
$d_R$             & $-\frac{1}{3}$ &   $\bm{1}$      &  $\bm{3}$ \\
\hline
\end{tabular}
\caption{The fermionic fields of the Standard Model and their charges. 
There are left handed ${\rm SU(2)}_L$ doublets, $\ell$ and $Q$, and right 
handed $SU(2)_L$ singlets, $e_R$, $u_R$ and $d_R$. There is also three 
families of each type of fermion.}
\label{table:SMfermions}
\end{center}
\end{table}

\subsection{Higgs Mechanism}
\label{sec:Higgs}
The Higgs mechanism is a mechanism for generating the masses of the Standard 
Model particles while keeping the Lagrangian gauge invariant. The idea hinges 
on a scalar field being added to the model which has a non-zero vacuum 
expectation value (VEV). The scalar can then couple to the particles in the 
Standard Model and by doing so defines their masses. Because the SM Higgs boson
has charges under $SU(2)_L$ and $U(1)_Y$ the non-zero VEV breaks the gauge 
invariance. This is why the $B$ boson of the $U(1)_Y$ group is not the mediator
of the electromagnetic interaction. Instead the photon is, which is composed of
both the $B$ boson and the $W_3$ boson.

The SM Higgs boson field, $\Phi$, is a ${\rm SU(2)}_L$ doublet and carries 
${\rm U(1)}_Y$ charge. This means that it couples to the $W$ bosons and the $B$
boson through the covariant derivative. The field can be defined as
\begin{equation}
\Phi = U(\xi) \left( \begin{array}{c} 0 \\ \frac{1}{\sqrt{2}} \left( \upsilon
+ H \right) \end{array} \right),
\end{equation}
where $U(\xi)$ is a unitary operator in $SU(2)$ with three degrees of freedom. 
These three degrees of freedom are known as the Goldstone bosons. It is these
bosons that are `eaten' in order to give the $W^\pm$ and $Z^0$ bosons mass.
Though these terms do not appear as physical particles they do play a role in 
calculations depending on the choice of gauge fixing term, $\xi$. $H$ is a real
scalar field which has a zero VEV and is interpreted as the physical Higgs 
boson field.

\begin{figure}
\centering
\includegraphics{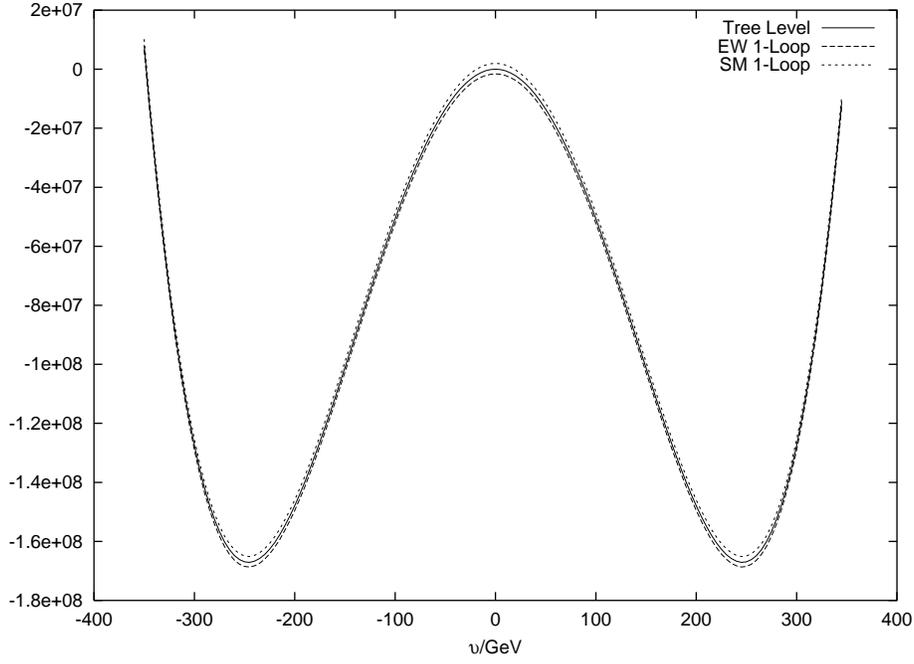}
\caption{The effective potential of the Standard Model as $\upsilon$ is varied.
The three lines are the tree level potential, the 1-Loop correction only 
with the Electroweak particles (leptons, W,Z,$\gamma$,Higgs boson) and the 
one-loop 
correction for all Standard Model particles. These are all for the combination 
$\frac{\mu}{\sqrt{\lambda}} = 246.0~{\rm GeV}$}
\label{fig:higgs_pot}
\end{figure}

The potential of the Higgs boson is given by a linear combination of 
$\Phi^\dagger 
\Phi$ term and a $(\Phi^\dagger \Phi)^2$ term. The result is added to the 
Lagrangian in the SM. The Lagrangian for the Higgs boson is
\begin{equation}
\label{eq:higgs_lag}
{\cal L}_{\rm Higgs} = (D^\mu \Phi)^\dagger D_\mu \Phi + \mu^2 \Phi^\dagger 
\Phi - \lambda \left( \Phi^\dagger \Phi \right)^2.
\end{equation}
The positive sign in front of the $\mu^2$ is different from the standard mass
term in a Lagrangian. Therefore, when the parameters $\mu^2, \lambda > 0$ this
gives the potential a minimum at $\left< \Phi \right> \neq 0$. The $\mu$ and 
$\lambda$ parameters also dictate what the VEV of the field is that minimizes 
the potential. The particular combination 
\begin{equation}
\left< \Phi \right> \equiv \frac{\upsilon}{\sqrt{2}} 
= \frac{\mu}{\sqrt{2\lambda}},
\end{equation}
defines the value of the VEV that minimizes the potential. Figure 
\ref{fig:higgs_pot} shows the potential as the VEV is varied for a choice of
$\frac{\mu}{\sqrt{\lambda}} = 246.0~{\rm GeV}$. It can be seen that at 
$\upsilon = 246.0$ GeV this potential is a minimum. Also in the figure is the 
1-loop effective potential including only the electroweak (EW) particles and 
the 1-loop effective potential when all the SM particles are included. These  
are also minimized at the same value of the VEV. This figure was generated by 
putting the model into the software \EFF{}, which will be explained in 
Chapter \ref{chap:Effective}. 

The covariant derivative term, ${\cal L}_{\rm Cov~Higgs} = (D_\mu \Phi)^* 
(D^\mu \Phi)$ from (\ref{eq:higgs_lag}), when expanded out is
\begin{equation}
{\cal L}_{\rm Cov~Higgs} = \left| \left( \begin{array}{cc} 
\partial_\mu - \frac{i}{2}(g_W W^3_\mu + g'B_\mu) & 
-\frac{i}{2}\left(W^1_\mu - iW^2_\mu\right) \\ 
-\frac{i}{2}\left(W^1_\mu + iW^2_\mu\right) &
\partial_\mu - \frac{i}{2}(-g_W W^3_\mu + g'B_\mu)
\end{array} \right) \left( \begin{array}{c} 0 \\ \frac{\upsilon + H}{\sqrt{2}}
\end{array} \right) \right|^2.
\end{equation} 
From this term we find that the $W^3$ boson and the $B$ boson mix.
This is known as \emph{spontaneous symmetry breaking} as the exact symmetries 
${\rm SU(2)}_L$ and ${\rm U(1)}_Y$ are broken down to $U(1)_{\rm em}$ and a 
broken ${SU(2)}_W$. We can see the $SU(2)_W$ is broken as the $W^\pm$ bosons
and the $Z^0$ have different masses. The mixing between $W_3$ and $B$ produces 
mass eigenstates with mass squared $\frac{1}{4} \left(g'^2 + g_W^2 \right) 
\upsilon^2$ and 0. These are then interpreted as the $Z^0$ boson and the 
photon, $A$, respectively. This means that the mass eigenstates are defined as
\begin{equation}
\left( \begin{array}{c} Z^{0\mu}\\ A^\mu \end{array} \right) = 
\left( \begin{array}{cc} \cos\theta_W & -\sin\theta_W \\ \sin\theta_W 
& \cos\theta_W \end{array} \right) \left( \begin{array}{c} W_3^\mu \\ B^\mu
\end{array} \right).
\end{equation}
where $\cos^2\theta_W = \frac{g_W^2}{g_W^2 + g'^2}$ and $\sin^2 \theta_W =
\frac{g'^2}{g_W^2+g'^2}$. Here $\theta_W$ is known as the Weinberg angle.
The mass squared of the $W^\pm$ boson is $\frac{1}{4} g_W^2 \upsilon^2$, where 
the $W^+$ bosons is the combination 
\begin{equation}
W^{\mu+} = \frac{1}{\sqrt{2}} (W_1^\mu - i W_2^\mu),
\end{equation}
and $W^{\mu-}$ is the complex conjugate.

The electric charge, $Q$, of a field is then given by $eQ=eT_3+eY$. $T_3$
is the eigenvalue of the third generator of the $SU(2)_L$ group. For a 
${\rm SU(2)}_L$ doublet this is $\frac{1}{2}$ for the first component of the 
doublet and $-\frac{1}{2}$ for the second component. $e$ is given as
$e = g_W \sin\theta_W = g' \cos\theta_W$. We can see from table 
\ref{table:SMfermions} that the charged leptons have charge $-1$, neutrinos 
have charge 0, up-type quarks have charge $+\frac{2}{3}$ and down-type quarks 
have charge $-\frac{1}{3}$. 

The value of the VEV can be determined from the Fermi constant, $G_F$. This is
measured as $1.166 \times 10^{-5}~{\rm GeV}^{-2}$. It is defined as
\begin{equation}
\frac{g_W^2}{8 M_W^2} \equiv \frac{1}{2\upsilon^2} = \frac{G_F}{\sqrt{2}}
\end{equation}
and yields the value $\upsilon=  246$ GeV. This value, along with the value
of $\theta_W$ then correctly predicts the mass of the $Z^0$ boson as well. The
fact that this mechanism predicts the masses of the $Z^0$ and $W^\pm$ bosons as
well as the fact that the photon is massless, in a gauge invariant way, has 
led to the belief that the Higgs boson must exist. Fortunately, even though the
particle hasn't been found yet its mass is given by $m_H = \sqrt{2}\mu$.  
Since $\mu$ is only restricted by loop corrections it gives bounds
of the Higgs mass~\cite{HiggsMass} between 117 GeV and 251 GeV with 95\%
confidence; experimentally 
this ceiling has not yet been 
reached. Finding the Higgs boson is one of the main goals of the upcoming Large 
Hadron Collider (LHC).

The Higgs boson also allows a gauge invariant way to introduce fermion masses 
as well. Instead of having a term $m_x \bar \psi_{xL} \psi_{xR}$ 
for each fermionic field $x$ we introduce instead a Yukawa matrix $Y^f_{ij}$ 
for each fermionic family (leptons and quarks). The Yukawa term in the 
Lagrangian is
\begin{equation}
{\cal L}_{\rm Yukawa} = Y^f_{ij} \bar \psi_{fL}\Phi \psi_{fR} + {\rm h.c.},
\end{equation}
where the subscript $L$ and $R$ denote the left and right handed fields. The 
VEV of $\Phi$ then generates the mass terms for all the fermions and the mixing
between families. Of course, we still have independent parameters to define the
mass of the fields. 

The most general gauge invariant term that can be added for the quark masses is
\begin{equation}
{\cal L}_m = -Y_{ij}^d \bar Q_{ia} \phi_a d_{Rj} - Y_{ij}^u \epsilon^{ab} \bar 
Q_{ia} \phi_b^\dagger u_{Rj} + {\rm h.c.}.
\label{eq:quarkMass}
\end{equation}
The Yukawa matrices are not necessarily symmetric or Hermitian. In fact,
there is no principle that even requires that they are real valued! However, 
if $CP$ is conserved, this would be true. We can simplify the form of 
(\ref{eq:quarkMass}) by diagonalizing the matrices obtained from squaring the
Yukawa matrices, $Y_u, Y_d$. Each one defines two unitary matrices, $U_{i}$ 
and $W_{i}$, for $i = u,d$ by
\begin{equation}
Y_{i} Y_{i}^\dagger = U_{i} D^2_{i} U_{i}^\dagger,
~~~Y_{i}^\dagger Y_{i} = W_{i} D^2_{i} W_{i}^\dagger,
\end{equation}
where $D_{i}$ is the diagonal matrix with elements which are the positive
square roots of the eigenvalues. The Yukawa matrices can then be defined as
\begin{equation}
Y_{i} = U_{i} D_{i} W_{i}^\dagger.
\end{equation}
If we now make a chiral rotation of the right handed fields by $W$ and the 
left handed fields by $U$, these transformations don't affect the couplings
of these particles to the Higgs field. In this basis, we also find that $P$, 
$C$ and $T$ are conserved.

Since all the up and down type quarks have the identical couplings in QCD the
$U$ matrices commute with the covariant derivative of QCD. This transformation,
however, mixes $u_L$ and $d_L$ and does affect the 
$SU(2) \times U(1)$ couplings. If we now neglect the QCD interactions and write
the Lagrangian in the basis of the $Z^0$, $W^\pm$ and $A$, rather than the 
$W$'s and $B$, we have
\begin{eqnarray}
{\cal L} &=& \bar \ell(i\Fs{\partial})\ell + \bar e_R (i\Fs{\partial})e_R + 
\bar Q (i\Fs{\partial}) Q + \bar u_R (i\Fs{\partial}) u_R +
d_R (i\Fs{\partial}) d_R \nonumber \\
&&~~+ g\left( W_\mu^+ J_W^{\mu+} + W_\mu^- J_W^{\mu-} + Z^0_\mu J_Z^\mu \right)
+ e A_\mu J_{\rm EM}^\mu,
\end{eqnarray}
where
\begin{eqnarray}
\label{eq:currents}
J_W^{\mu+} &=& \frac{1}{\sqrt{2}} \left( \bar \nu_L \gamma^\mu e_L + \bar u_L 
\gamma^\mu d_L \right); \nonumber \\
J_W^{\mu-} &=& \frac{1}{\sqrt{2}} \left( \bar e_L \gamma^\mu \nu_L + \bar d_L
\gamma^\mu u_L \right); \nonumber \\
J^\mu_Z &=& \frac{1}{\cos \theta_W}\left[ \bar \nu_L \frac{\gamma^\mu}{2} \nu_L
+ \bar e_L \gamma^\mu \left( -\frac{1}{2}+\sin^2\theta_W \right) e_L
+ \bar e_R \gamma^\mu \sin^2 \theta_W e_R \right. \nonumber \\
&&~~+\bar u_L\gamma^\mu\left(\frac{1}{2} - \frac{2}{3}\sin^2\theta_W\right)u_L
+ \bar u_R \gamma^\mu \left( -\frac{2}{3} \sin^2 \theta_W \right)u_R\nonumber\\
&&~~\left. +\bar d_L \gamma^\mu \left(-\frac{1}{2} + \frac{1}{3}\sin^2\theta_W
\right) d_L + \bar d_R \gamma^\mu \left(\frac{1}{2}\sin^2\theta_W \right)d_R
\right]; \nonumber \\
J^\mu_{\rm EM} &=& -\bar e \gamma^\mu e + \frac{2}{3} \bar u \gamma^\mu u
-\frac{1}{3} \bar d \gamma^\mu d.
\end{eqnarray}
These currents contain an abundance of information. For example, the first two
show that the $W^\pm$ couple up-type quarks to down-type quarks and neutrinos 
to charged leptons. From the electromagnetic current one can directly read
off the electric charges of the different particles.

We can now see how these currents change when the chiral transformations, $U$ 
and $W$, are applied. We can see that in the electromagnetic current, 
$J_{\rm EM}^\mu$, the transformation matrices cancel out
\begin{equation}
\bar d_L^i \gamma^\mu d^i_L \rightarrow \bar d_L^i \left(U^{\dagger}_d
\right)^{ij} \gamma^\mu U^{jk}_d d_L^k = \bar d^i_L \gamma^\mu d^i_L.
\end{equation}
This is also true for $J_Z^\mu$. The current that couples to the $W^\pm$, 
however, does change. This is
\begin{equation}
J_W^{\mu +} = \frac{1}{\sqrt{2}} \bar u^i_L \gamma^\mu d^i_L \rightarrow
\frac{1}{\sqrt{2}} \bar u^i_L \gamma^\mu \left(U^\dagger_u U_d\right)^{ij}
d^j_L.
\end{equation}
This defines a new matrix
\begin{equation}
V = U^\dagger_u U_d,
\end{equation}
which is known as the \emph{Cabibbo-Kobayashi-Maskawa} (CKM) mixing matrix. 
This explains why strange quarks enter into weak interactions. The $W^\pm$ 
boson is able to not only turn up-type quarks into down-type quarks but also
change the generation in the process.

The same arguments that were just given can also be applied to the
lepton families. Since the neutrinos don't interact in any way except by the
weak interactions we can by convention choose to label the mass eigenstates
of the neutrinos according to the charged lepton partner it is formed with. 
Unlike the case of the quarks, there is no way to distinguish these states
in another way. This way a $W^\pm$ boson only couples the neutrinos of one 
generation to a charged lepton of the same generation.

\subsection{Running Coupling}
\label{sec:runningCoup}
We start by defining $\alpha$ for a theory as $\alpha = \frac{g^2}{4\pi}$.
In QED $g$ is the electric charge of the positron and the $\alpha$ is the fine
structure constant and is denoted simply by
$\alpha$. In QCD $g$ is $g_3$ and the $\alpha$ is labelled $\alpha_S$. In order
to remove the ultra-violet divergences in the perturbative series a 
renormalization procedure is used. This procedure introduces a mass scale 
$\mu$. Therefore, when we want to calculate a dimensionless physical observable
at mass scale $Q$, it can only depend on the ratio $Q^2/\mu^2$, which is not 
constant. The choice of $\mu$ is arbitrary and therefore if we were to
hold the bare coupling of the Lagrangian fixed a physical quantity, $R$,
cannot depend on $\mu$. Instead it must depend only on $Q^2/\mu^2$ and the 
renormalized couplings. We will consider first the running of $\alpha_S$, as it
plays an important role in QCD. In QCD this can all be expressed as
\begin{equation}
\mu^2 \frac{d}{d\mu^2} R(Q^2/\mu^2, \alpha_S) \equiv \left[ \mu^2 
\frac{\partial}{\partial \mu^2} + \mu^2 \frac{\partial \alpha_S}{\partial \mu^2}
\frac{\partial}{\partial \alpha_S} \right] R = 0.
\end{equation}
Identifying 
\begin{equation}
t = \ln\left(\frac{Q^2}{\mu^2}\right),~~\beta\left(\alpha_S\right) = \mu^2 
\frac{\partial \alpha_S}{\partial \mu^2},
\end{equation}
the $\mu$ independence of the observable $R$ for massless particles can be 
expressed as
\begin{equation}
\label{eq:Beta}
\left[ - \frac{\partial}{\partial t} + \beta\left(\alpha_S\right) 
\frac{\partial}{\partial \alpha_S} \right] R\left({\rm e}^t, \alpha_S\right)
= 0.
\end{equation}
It is from this that we define the \emph{running coupling} $\alpha_S(Q^2)$ as
\begin{equation}
t = \int_{\alpha_S}^{\alpha_S(Q^2)} \frac{dx}{\beta(x)},~~\alpha_S(\mu^2) 
\equiv \alpha_S.
\end{equation}
We can then see that
\begin{equation}
\frac{\partial \alpha_S(Q^2)}{\partial t} = \beta(\alpha_S(Q^2)),~~
\frac{\partial \alpha_S(Q^2)}{\partial \alpha_S} = \frac{\partial \beta(
\alpha_S(Q^2))}{\partial \beta(\alpha_S)}.
\end{equation}
A solution to (\ref{eq:Beta}) is $R(1,\alpha_S(Q^2))$. Therefore, the scale 
dependence of $R$ is due solely to the running of the coupling. By the same 
means we can find the running coupling of $\alpha$ for QED interactions.

\begin{figure}[htb]
\centering
\input{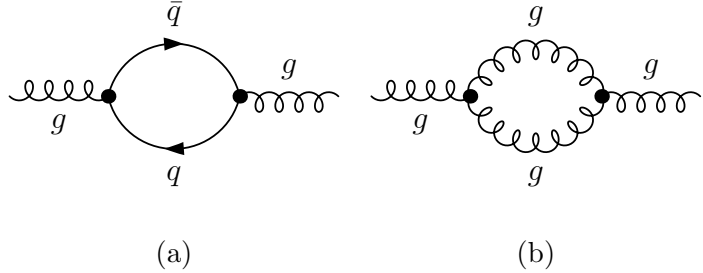}
\caption{Graphs which contribute to the QCD $\beta$ function in the one-loop
approximation (in a physical gauge).}
\label{fig:BetaCorr}
\end{figure}

The running coupling constants are then determined by the renormalization
group equation (RGE),
\begin{equation}
Q^2 \frac{\partial \alpha}{\partial Q^2} = \beta(\alpha).
\end{equation}
These always have at least $\alpha^2$ dependence as they are extracted from
higher-order loop corrections to the bare vertices of a theory. 
Fig.~\ref{fig:BetaCorr}~shows contributions to the $\beta$ function for
QCD at one-loop. Figure~\ref{fig:BetaCorr}b shows a new interaction due to the
non-Abelian nature of QCD. The $\beta$ function in QCD then has the 
perturbative expansion
\begin{equation}
\beta(\alpha) = -b \alpha_S^2 \left(1 + b' \alpha_S + b'' \alpha_S^2 
+ O(\alpha_S^3)
\right).
\end{equation}
The coefficient $b''$ depends on the renormalization scheme. $b$ and
$b'$ don't, however, and are given by
\begin{eqnarray}
b &=& \frac{11C_A - 2n_f}{12 \pi} \nonumber \\
b' &=& \frac{17 C_A^2 - 5C_A n_f - 3 C_F n_f}{2\pi(11 C_A - 2n_f)},
\end{eqnarray}
where $C_A = 3$, $C_F = \frac{4}{3}$ and $n_f$ is the number of active 
flavours. From this we see that in order for $b$ to be negative (which
corresponds to the leading term of the expansion being positive) we must
have $n_f \geq \frac{11C_A}{2}$. For $C_A=3$ this is $n_f \geq 16$. If we
look at the QED $\beta$ function
\begin{equation}
\beta_{\rm QED}(\alpha) = \frac{1}{3\pi}\alpha^2 + O(\alpha^3).
\end{equation}
The leading term of this $\beta$ is always positive. This is where we can see
the difference the non-Abelian interactions (triplet and quartic gluon 
vertices) of QCD makes. QCD is known as an \emph{asymptotically free} theory. 
This means that $\alpha_S$ becomes smaller as $Q^2$ increases, corresponding
to the opposite sign of the $\beta$ function.

A fuller discussion of RGEs is given in chapter~\ref{chap:Effective} as a part 
of the development of the software \EFF{}. 

\subsection{Asymptotic Freedom}
\label{sec:AsFree}
\begin{figure}[htb]
\centering
\includegraphics[width=3.2in]{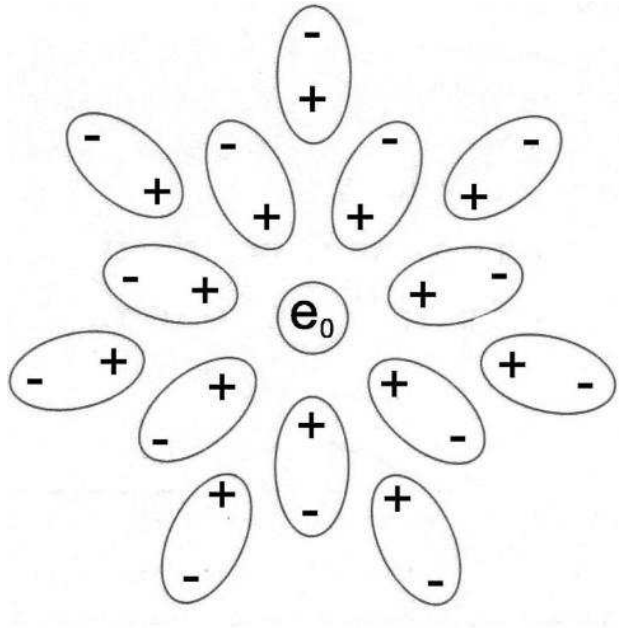}
\caption{Virtual $e^+e^-$ pairs are effectively dipoles that screen the bare
charge of the electron.}
\label{fig:electron_screening}
\end{figure}

We start first by explaining why the observed charge of the electron decreases
with distance. Referring back to our previous discussion of the running 
coupling we see that as $Q^2$ increases the QED coupling increases. This
corresponds to it growing at small distances. This can be easily explained as
follows. The larger the distance between the observer and the charge means that
more electron-positron pairs will be temporarily created out of the vacuum. 
These can be considered temporary electric dipoles which are preferentially 
aligned with the positive end towards the electron and the negative end away.  
Fig.~\ref{fig:electron_screening} shows this situation. This effectively 
screens the bare charge of the electron and it appears to have a smaller charge
the further away from the electron you are.

In QCD, the picture is not quite so clear. We now have three degrees
of colour charge, as opposed to one in QED, and non-Abelian 
interactions. There are two ways of describing the difference: either as a 
dielectric effect or as a paramagnetic effect.

We start by giving the dielectric effect argument. We can define a running
charge at scale $q^2$  in terms of the charge at an ultra-violet cutoff
$\Lambda_{\rm UV} \gg q^2$ and a scale-dependent dielectric constant, 
$\epsilon(q^2)$. This is
\begin{equation}
\alpha(q^2) = \frac{\alpha\left(\Lambda_{\rm UV}^2\right)}{\epsilon(q^2)}
\end{equation}
and from the previous discussion we find
\begin{equation}
\label{eq:dielectric}
\frac{1}{\epsilon(q^2)} = 1 - \frac{\beta(\alpha)}{\alpha} 
\ln\frac{\Lambda_{\rm UV}^2}{q^2}.
\end{equation}
This implies that the running charge satisfies the equation
\begin{equation}
\frac{d\alpha(q^2)}{d\ln q^2} = \beta(\alpha).
\end{equation}
We see that in QED $\beta$ is positive meaning the dielectric constant is 
greater than one. This corresponds to a screening of the charge. In QCD we have
the opposite case. Here the dielectric constant is smaller than one. This is
an antiscreening of the charge. 

If we now assume the vacuum of a relativistic quantum field theory can be 
treated as a polarizable medium we can define a magnetic permeability, 
$\mu(q^2)$ that due to Lorentz invariance must satisfy the equation
\begin{equation}
\epsilon(q^2)\mu(q^2) = 1
\end{equation}
for all $q$. We can now analyze the behavour of the magnetic susceptibility,
$\chi \equiv \mu - 1$. From (\ref{eq:dielectric}) we have
\begin{equation}
\label{eq:perm}
\chi(q^2) = -\frac{\beta(\alpha)}{\alpha}\ln\frac{\Lambda^2_{\rm UV}}{q^2}.
\end{equation}
This can be broken up into two parts. The first term, known as Pauli 
paramagnetism, describes how the spins interact with the magnetic field and 
the second term, known as the Landau diamagnetism, describes how the orbital
motion of the particles interacts with the magnetic field. When the Pauli
paramagnetic term is larger then the Landau diamagnetic term the system is 
considered a paramagnetic system, otherwise it is diamagnetic. In QCD the
Pauli paramagnetism term has some dependence on both spin $\frac{1}{2}$ quarks
and spin 1 gluons. The higher spin gluons make a larger contribution than the
quarks. The permeability is given by (\ref{eq:perm}) when
\begin{equation}
\beta(\alpha) = -b \alpha^2.
\end{equation}
In QCD a contribution from a particle with spin $S$ to $b$ is given by
\begin{equation}
b = \frac{(-1)^{2S}}{2\pi} \left[ (2S)^2 -\frac{1}{3} \right].
\end{equation}
This gives a total $b$ term in QCD of
\begin{equation}
b=\frac{11 C_A - 2 n_f}{12 \pi}.
\end{equation}
From this we can see that the permeability can only be negative (making it
a diamagnetic system) when $n_f > 16$. This shows that QCD is asymptotically
free due to the colour charge carrying spin 1 gluons.

QCD is an asymptotically free theory. It also has a coupling which runs.
The running decreases $\alpha_S$ as $Q^2$ increases and therefore 
high energy QCD allows the methods of perturbation theory to be applied.

This is by no means an exhaustive reference of the Standard Model and more 
complete descriptions are readily available \cite{WeinbergI, WeinbergII, 
PeskinSchroeder,Griffiths,HalzenMartin}. This section
has, however, described the basics of the particle content, the Higgs 
mechanism and two topics of importance in QCD: running couplings and asymptotic
freedom. The existence of the Higgs boson is taken as an assumption for the
development of SUSY, and will be done so for the remainder of this thesis.

\section{Event Generation}

We now move into the discussion of Monte Carlo Event Generators. Chapters 
\ref{chap:NewVars} through \ref{chap:Results} are all 
discussions on the development of the event generator \HWP{}. In this section
I start by introducing the Monte Carlo integration technique and explain why 
this is so useful for high-energy particle physics simulations. I then 
introduce all the various parts of an event: hard process, parton shower, 
hadronization and hadron decays.

\subsection{Monte Carlo Approach}
\label{sec:monte}
The idea of the Monte Carlo approach is that the value of an integral can be 
calculated using random numbers. The same ideas are also applicable to sampling
based on a distribution. Many calculations of quantum field theory 
involve matrix elements where the amplitude squared is interpreted as a 
probability; the Monte Carlo approach is an excellent fit for computer 
simulations due to these probability distributions. This is because points can 
be drawn according to a distribution, thus simulating a physical event with 
the correct probabilities. A more thorough discussion of Monte Carlo 
integration is given in \cite{NumericalRecipes}.

\subsubsection{Simple Monte Carlo Integration}
If we pick $N$ random points which are uniformly distributed in a 
multidimensional volume $V$, the basic theorem of Monte Carlo integration is 
that the integral of a function $f$ over the multidimensional volume can be 
approximated by

\begin{equation}
\label{eqn:simplemonte}
\int f dV \sim V \left( \left< f \right> \pm \sqrt{\frac{\left< f^2 \right> - 
\left< f \right>^2}{N}} \right).
\end{equation}  
The angle brackets indicate the arithmetic mean over the $N$ points in $V$ 
space,
\begin{equation}
\label{eq:averageF}
\left< f \right> \equiv \frac{1}{N}\sum_{i=1}^N f(x_i)~~~~~\left< f^2 \right> 
\equiv \frac{1}{N^2}\sum_{i=1}^N f^2(x_i).
\end{equation}
The ``plus-minus'' term in (\ref{eqn:simplemonte}) indicates an estimation of 
one standard deviation of the integral.

To calculate the value of the integral using (\ref{eqn:simplemonte}), one
simply generates a set of $x$ values in the volume $V$ which are the argument of
$f(x)$. Using the two parts of (\ref{eq:averageF}) and the set of values
generated the approximate value of the integral is given.

\begin{figure}[htb]
\centering
\includegraphics{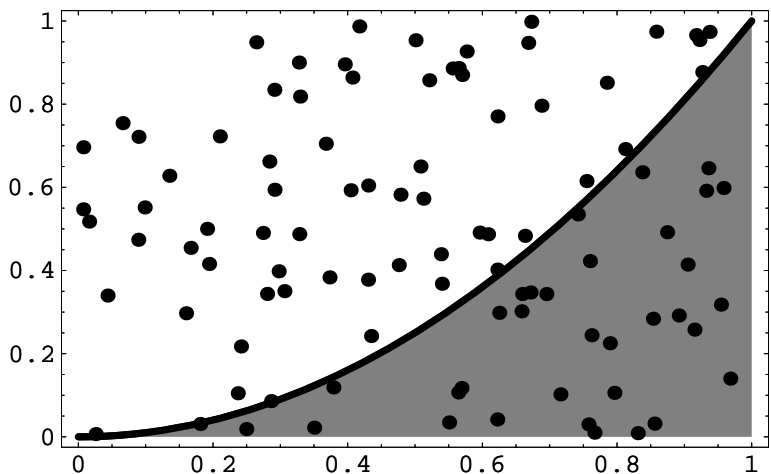}
\caption{This shows the randomly distributed points in a square of area 1. The 
value of the integral is the percentage of points under the function}
\label{monte1}
\end{figure}

There is also an algorithm known as the rejection algorithm which is useful 
when generating values according to a distribution. In this case we don't want
to calculate the integral but rather generate a point according to the 
distribution. This can be done by simply generating the points uniformly in a
volume which is $V$ plus an extra dimension bounded by the function. If the 
point generated lies inside the function we accept the point. Otherwise we 
reject it and generate a new one. If we wanted to take the integral, this would
then be the percentage of accepted points times the volume sampled over.
If we consider a function of one variable, $f(x)$, we would generate $N$ pairs
of points, $(x,y)$. We find the ratio of points where $y<f(x)$, and multiply 
this ratio by the area the points are generated in. Figure \ref{monte1} shows 
an example of this technique. This technique is called integration by 
rejection. In the example above the points are generated in an unweighted 
manner. It is also possible to make the algorithm more efficient or add some
more information by generating points in a weighted manner as well. 

The method can easily be expanded to integrate over regions with an unknown 
volume. Say we want to integrate over a strange region, $C$, of which we do not
know the volume. This can also be easily done with the Monte Carlo technique. 
Simply expand the region the points are generated in to a simpler region, $M$, 
which contains $C$, and the multi-dimensional volume of $M$ is known. When the 
point lies outside of the desired region, set the value of the function to 0.

\subsubsection{Non-Uniform Sampling for Monte Carlo Integration}
When working with probability distributions, we will be using the points that 
lie inside of the function to be a physical quantity. Therefore, in the simple 
approach, there may be many points that are generated, that aren't used in the 
simulation. In particle physics, many of this distributions have very sharp 
peaks and valleys. These peaks and valleys cause a large number of points to be
generated which aren't used. This is a large waste of CPU time and we would 
like to optimize this.

Fortunately, this can be done. If we instead generate our points according to 
some other distribution, rather than a uniform one, we can reduce the number of
unused points. Optimally, we would want to generate them according to the 
function being integrated, but this would require that we know the value of the
integral beforehand! Instead, we find another function that we can integrate, 
that is always larger than the function we would like to integrate. We want to 
choose this function so that the sharp peaks and valleys of our unknown 
integral are also, approximately, present in our known function. But, in order 
to sample according to our new function, it must be invertible.

\begin{figure}[htb]
\centering
\includegraphics{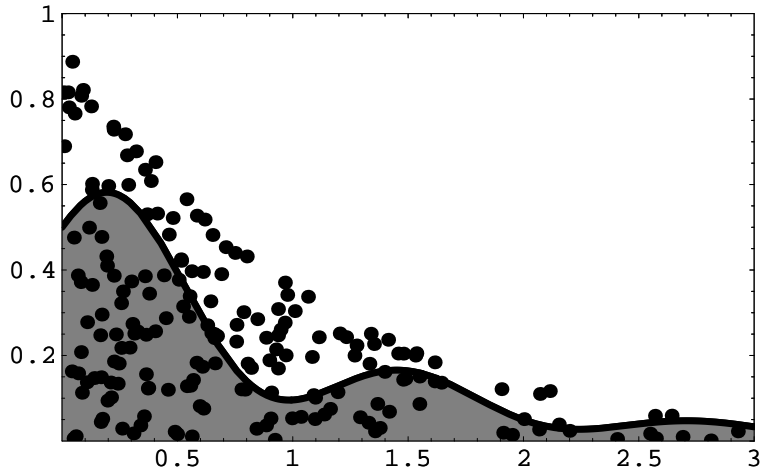}
\caption{This shows the random numbers sampled under a known function, $g(x)$. 
The area of the integral is now $G$, the integral under $g(x)$, times the 
percent of points under $f(x)$.}
\label{monte2}
\end{figure}

The evaluation of the integral with this new distribution is not much different
than for a uniform distribution. Again, we will consider a function of one 
variable, $f(x)$. Our known function is $g(x)$ with integral $G$. Now we 
randomly generate the $x$'s and evaluate both $f(x)$ and $g(x)$. A random 
number, ${\cal R}$ between 0 and 1 is also generated. If ${\cal R}$ is is less 
than the ratio $f(x)/g(x)$, then the point is under the function. Otherwise it 
is not. The value of the integral is then just $G$ times the percentage of the 
points under the function. This technique can greatly reduce the number of 
unused points that are generated. The improvement is dependent on how well the 
known function, $g$, estimates $f(x)$. An example of this is shown in Figure 
\ref{monte2}.

Luckily, there are some general features for QCD matrix elements that can be 
used to take advantage of non-uniform sampling. These are well known 
\cite{RAMBO, SARGE} and general algorithms have been developed to improve the 
efficiency of generating points for these matrix elements.

There are also some other optimizations that can be performed, such as 
stratified sampling. One program that is particularly useful for particle 
physics integrals is VEGAS \cite{VEGAS}. More detail on these optimizations is
also given in~\cite{NumericalRecipes}.

\subsection{Hard Process}
There are two separate momentum regimes that different methods of high energy 
physics can be used at. The first is the perturbative regime. This regime is 
for high momentum transfer, short distances. In this regime calculations can be
approximated by truncating the time-ordered series at any order and applying 
Wick's theorem. Calculations are usually calculated in orders of $\alpha$ 
(couplings). The \emph{hard process} fits into this regime. There is also the 
non-perturbative regime. In this regime calculations are complicated to
perform because as we saw in Section~\ref{sec:Perturbation} calculations depend
on a time-ordered exponential and each term is itself a complicated expression.
As was shown earlier, the value of $\alpha_S$ is largest in the low momentum 
transfer regime. This means the calculations would need to be performed to
many orders of $\alpha_S$. Currently calculations to next-to-next-to leading
order (leading order + 2 more orders) are state-of-the-art. In fact, in the
low momentum transfer regime one of the tools of perturbative physics, the
free field propagator is not valid anymore. Since non-perturbative calculations
are so difficult, models of physics are usually used instead of trying exact 
calculations. The process of \emph{hadronization} is an example of this.

The hard process is the underlying process that occurs when two beams
of particles are collided. The description of what happens is divided into two 
different parts, the actual hard process and the Parton Distribution Functions 
(PDFs). A similar division in this discussion is used. Here I discuss the hard 
process and in the next section explain the PDFs and how they affect the hard 
process.

The hard process of an event is what describes the interaction of high energy 
particles that `collide'. There are numerous processes for a given set of 
incoming particles. For example, in an $e^+e^-$ collision we could produce 
another $e^+e^-$ pair through a $e^+e^- \rightarrow \gamma^*\rightarrow e^+e^-$
or a $e^+e^-\rightarrow Z^0 \rightarrow e^+e^-$ process. We could also produce 
a $q\overline{q}$ pair in similar fashion. In fact, there is an infinite number
of possible final states as any number of photons or gluons can be emitted by 
an electrically or colour charged particle. The higher order terms fall off 
rapidly however and contribute only a fraction of the total cross section.  The
actual process that occurs is proportional to the fraction of the total cross 
section for a given process. 

In event generators a specific set of hard processes can be used, without
considering all possibilities. This way different properties particular to 
certain hard process can be studied. Given a set of processes, one is chosen 
based on the ratio of each cross section to the total cross sections in the 
allowable set. At the lowest order of perturbation theory, there are relatively
few processes to calculate for a given pair of incoming particles. As the next
order of perturbation will contain a new factor of the coupling, these 
processes are suppressed. It is impossible to know beforehand what the 
cross section for the next order will be; it must be calculated. In particular
regions of phase space (soft and collinear) there is a formalism for the 
splitting of a particle, $a$, into particles $b+c$. These can then be used to 
generate approximations to higher order 
diagrams. More detail on this is given in Section \ref{sec:shower}. These 
splitting functions are then used to generate the QED and QCD emissions
from a parton. This process is known as the \emph{parton shower}.

The development of matrix elements for a given process is a complete discussion
in itself\cite{HalzenMartin,Griffiths,PeskinSchroeder}. Here we only give the 
results of the main processes studied in this thesis. The first is 
$e^+e^-\rightarrow f \overline{f}$~\cite{Ellis:1996qj}. The differential cross 
section for this process with either a $\gamma$ or $Z^0$ in the $s$-channel and
$\theta$ the center-of-mass scattering angle of the outgoing fermions is
\begin{eqnarray}
\label{eq:e+e-ME}
\frac{d\sigma}{d \cos \theta} &=& \frac{\pi \alpha^2}{2 s}\left[ (1+\cos\theta)
\left\{Q_f^2 - 2 Q_f V_e V_f F_{Z^0}^{\gamma}(s) \right. \right.  \nonumber \\
&& \left. + (A_e^2 + V_e^2)(A_f^2 + V_f^2) F_{Z^0}^{Z^0}(s) \right\}\nonumber\\
&& \left. + \cos \theta \left\{ -4 Q_f A_e A_f F_{Z^0}^{\gamma}(s) + 8 A_e 
V_e A_f V_f F_{Z^0}^{Z^0}(s) \right\} \right],
\end{eqnarray}
where
\begin{eqnarray}
F_{Z^0}^{\gamma}(s) &=& \frac{\sqrt{2} G_F M_Z^2}{16 \pi \alpha} 
\frac{s(s-M_Z^2)}{(s-M_Z^2)^2 + \Gamma_Z^2 M_Z^2}, \nonumber \\
F_{Z^0}^{Z^0}(s) &=& \frac{2 G_F^2 M_Z^4}{256 \pi^2 \alpha^2} \frac{s^2}
{(s-M_Z^2)^2 + \Gamma_Z^2 M_Z^2}.
\end{eqnarray}
$s$ is the centre-of-mass energy and $A_f$ and $V_f$ are the vector and axial 
couplings of fermion $f$ to the $Z^0$ boson. These are
\begin{equation}
V_f = T_f^3 - 2 Q_f \sin^2 \theta_W,~~~A_f = T_f^3.
\end{equation}
The functions $F_{Z^0}^\gamma(s)$ and $F_{Z^0}^{Z^0}(s)$ are the contributions 
from the $Z^0$-$\gamma$ interference and the $Z^0$-exchange, respectively. This
process is of particular importance when comparing the simulations to data 
taken at LEP. For center-of-mass energies well below the $Z^0$ peak, the two
functions $F(s)$ can be ignored yielding a differential cross section of
\begin{equation}
\frac{d\sigma}{d \cos \theta} = \frac{\pi \alpha^2 Q_f^2}{2 s} (1- \cos^2
\theta).
\end{equation}
Integrating over $\theta$ gives a total cross section of
\begin{equation}
\label{eq:e+e-XSec}
\sigma_0 = \frac{4 \pi \alpha^2 Q_f^2}{3 s}.
\end{equation}
Around the $Z$ pole, the $F_{Z^0}^{Z^0}(s)$ term dominates and the cross 
section is approximately
\begin{equation}
\sigma_0 = \frac{G_F M_Z^2}{96 \pi \Gamma_Z^2}(A_e^2 + V_e^2)(A_f^2 + V_f^2).
\end{equation}

To use simulations for hadron-hadron events another set of matrix elements is
used. Below are the matrix elements squared that have 
been spin and colour averaged (summed) over the initial (final) states for 
massless partons~\cite{Ellis:1996qj}.
\begin{eqnarray}
\label{eq:hhME}
&\frac{1}{g_3^4}\bar \sum \left| {\cal M} \right|^2_{qq'\to qq'} &= 
\frac{4}{9}\frac{\hat{s}^2 + \hat{u}^2}{\hat{t}^2}; \nonumber \\
&\frac{1}{g_3^4}\bar \sum \left| {\cal M} \right|^2_{q\bar q'\to q\bar q'} &=
\frac{4}{9}\frac{\hat{s}^2 + \hat{u}^2}{\hat{t}^2}; \nonumber \\
&\frac{1}{g_3^4}\bar \sum \left| {\cal M} \right|^2_{qq\to qq} &= \frac{4}{9}
\left( \frac{\hat{s}^2 + \hat{u}^2}{\hat{t}^2} + \frac{\hat{s}^2 + \hat{t}^2}
{\hat{u}^2} \right) -\frac{8}{27} \frac{\hat{s}^2}{\hat{u}\hat{t}};\nonumber \\
&\frac{1}{g_3^4}\bar \sum \left| {\cal M} \right|^2_{q\bar q\to q' \bar q'} &= 
\frac{4}{9}\frac{\hat{t}^2 + \hat{u}^2}{\hat{s}^2}; \nonumber \\
&\frac{1}{g_3^4}\bar \sum \left| {\cal M} \right|^2_{q \bar q\to q \bar q} &= 
\frac{4}{9}\left( \frac{\hat{s}^2 + \hat{u}^2}{\hat{t}^2} + \frac{\hat{t}^2 + 
\hat{u}^2}{\hat{s}^2} \right) -\frac{8}{27} \frac{\hat{u}^2}{\hat{s}\hat{t}}; 
\nonumber \\
&\frac{1}{g_3^4}\bar \sum \left| {\cal M} \right|^2_{q \bar q\to gg} &= 
\frac{32}{27}\frac{\hat{t}^2 + \hat{u}^2}{\hat{t}\hat{u}} - \frac{8}{3}\frac{
\hat{t}^2+\hat{u}^2}{\hat{s}^2}; \nonumber \\
&\frac{1}{g_3^4}\bar \sum \left| {\cal M} \right|^2_{gg\to q \bar q} &= 
\frac{1}{6}\frac{\hat{t}^2 + \hat{u}^2}{\hat{t}\hat{u}} -\frac{3}{8} \frac{
\hat{t}^2 + \hat{u}^2}{\hat{s}^2}; \nonumber \\
&\frac{1}{g_3^4}\bar \sum \left| {\cal M} \right|^2_{gq\to gq} &= 
-\frac{4}{9}\frac{\hat{s}^2 + \hat{u}^2}{\hat{s}\hat{u}} + \frac{\hat{u}^2+
\hat{s}^2}{\hat{t}^2}; \nonumber \\
&\frac{1}{g_3^4}\bar \sum \left| {\cal M} \right|^2_{gg\to gg} &= 
\frac{9}{2}\left( 3 - \frac{\hat{t}\hat{u}}{\hat{s}^2} - \frac{\hat{s}\hat{u}}
{\hat{t}^2} - \frac{\hat{s}\hat{t}}{\hat{u}^2} \right),
\end{eqnarray}
where $\bar \sum$ is the spin averaged sum and $\hat{s}, \hat{t}, \hat{u}$ are 
the usual Mandelstam variables for the process $AB \to CD$: $\hat{s} 
\equiv (p_A+p_B)^2$, $\hat{t} \equiv (p_A-p_C)^2$ and $\hat{u} 
\equiv (p_A-p_D)^2$.

\subsection{Parton Distribution Functions}
Hadrons are composed of quarks. This means that when two hadrons are collided, 
it is some components of the hadrons that interact fundamentally. Determining
which part of the hadrons are the ones that interact is a complex issue.  This
is defined by the \emph{Parton Distribution Functions}. These are 
developed from the data of deep inelastic scattering (DIS) experiments. This 
section provides a brief discussion of these functions. For a more detailed 
discussion, see \cite{Ellis:1996qj, Pomerons, Thorne}. We also
discuss how the remaining parts of the hadrons are handled in a Monte Carlo
event generator. These are called the \emph{beam remnants}.

The \emph{factorization theorem} allows the study of the parton constituents
to be factorized into a non-perturbative part and a perturbative part. The
non-perturbative part is determined from experiments, while the perturbative
part can be calculated as a perturbation series ordered in the strong coupling
constant, $\alpha_S$. 

\begin{figure}
\centering
\input{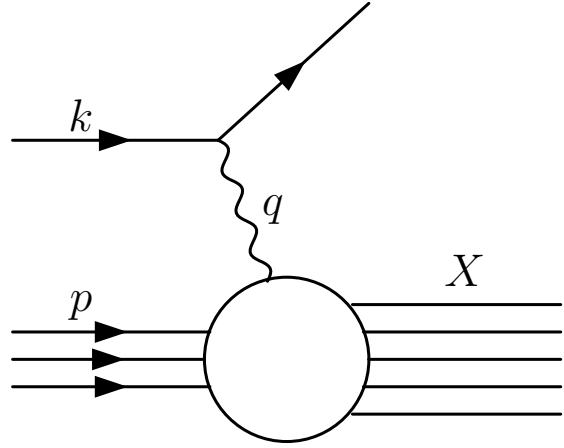}
\caption{The electron emits a virtual photon which probes the structure of the
proton.}
\label{fig:DIS}
\end{figure}

We start by considering the DIS process $e p \rightarrow e X$. Figure 
\ref{fig:DIS} shows this process with the relavant momenta $k,q$ and $p$. 
We start by introducing the variables
\begin{equation}
\label{eq:DISVars}
Q^2 = -q^2,~~~\nu = p \cdot q,~~~x = \frac{Q^2}{2\nu},~~~y = \frac{p\cdot q}
{p \cdot k}.
\end{equation}
We now define the hadronic tensor, $W^{\mu\nu}$. This defines how the hadron 
will interact with the photon. This is given by
\begin{equation}
W^{\mu\nu}(p,q) = \frac{1}{4} \sum_r \sum_X \left< p,r \left| J^\mu(0) \right|
X \right> \left<X \left| J^\nu(0) \right| p,r \right> (2\pi)^3 \delta^4
(p+q-P_X),
\end{equation}
where $J^\mu$ is the electromagnetic current and $r$ is the spin of the proton.
The variable $X$ is summed over all of the possible final products of the 
process. If we require that this tensor conserves parity and we have 
unpolarized protons, we find that we can decompose this unknown tensor into two
independent amplitudes, $F_1(x,Q^2)$ and $F_2(x,Q^2)$, called structure 
functions. We now write this tensor as
\begin{equation}
W^{\mu\nu} (p,q) = -\left(g^{\mu\nu} + \frac{q^\mu q^\nu}{Q^2}\right)F_1(x,Q^2)
+ \frac{1}{\nu}\left(p^\mu + \frac{\nu}{Q^2}q^\mu\right)\left(p^\nu + 
\frac{\nu}{Q^2}q^\nu\right) F_2(x,Q^2).
\end{equation}
The total differential cross section for this process in terms of these
structure functions is
\begin{equation}
\label{eq:CrossSecF}
\frac{d^2 \sigma}{dx dy} = \frac{4 \pi \alpha^2}{x y Q^2} \left( (1-y)F_2 + 
xy^2 F_1 - \frac{m^2}{Q^2} x^2 y^2 F_2 \right).
\end{equation}
The \emph{Bjorken limit} is defined as $Q^2,\nu \rightarrow \infty$ with $x$ 
fixed. In this limit the structure functions depend, approximately, only on
$x$. This implies the photons scatter off \emph{pointlike constituents}. If we
work in the `infinite momentum frame' we can ignore the mass of the proton.
Comparing (\ref{eq:CrossSecF}) to the spin averaged matrix element for the 
process $e^-q\rightarrow e^-q$ yields the relation
\begin{equation}
\label{eq:BjorkStruct}
\hat F_2 = x e_q^2 \delta (x - \xi) = 2x \hat F_1,
\end{equation}
where the hat indicates that the stucture function refers to a quark,
not a proton and the $\xi$ in this context is the fraction of the protons
momentum that the quark constituent carries. Measurements show that 
the momentum is not given by a delta function but by a distribution. This 
means that the quarks carry a range of momentum fractions. 

The ideas generated by studying the structure functions in the Bjorken limit 
are incorporated into the `naive parton model'. In this model the following
assumptions are made: 
\begin{itemize}
\item{$f_i(\xi)d\xi$ is the probability that quark of flavour $i$ carries a
momentum fraction between $\xi$ and $\xi + d\xi$,}
\item{the photon scatters incoherently off the quark constituents.}
\end{itemize}
(\ref{eq:BjorkStruct}) can then be written as
\begin{eqnarray}
F_2(x) = 2xF_1(x) &=& \sum_{i} \int_0^1 d\xi f_i(\xi) x e_q^2 \delta(x-\xi) 
\nonumber \\     &=& \sum_i e_q^2 f_i(x).
\end{eqnarray}

Beyond leading order the `naive' parton model is broken in QCD by logarithms 
of $Q^2$. In the Bjorken limit the transverse momentum is assumed to be small.
The higher order contributions show that is is not
the case though. A quark is able to emit a gluon with large transverse momentum
with probability $\alpha_S \frac{dk_T^2}{k_T^2}$ at large $k_T$. These 
contributions give terms proportional to $\alpha_S \log(Q^2)$. It is these
terms that break scaling. When $Q^2$ is large enough, these terms compensate 
for the small value of $\alpha_S$. The approximation where these terms are 
summed to all orders is known as the leading-log approximation (LLA). It is 
this approximation that the parton shower is developed at.

The breaking of the scaling means that the functions $f_i(x)$ gain a scale
dependence and are instead $f_i(x,Q^2)$. These can be written at some 
renormalization scale $\mu$ as
\begin{eqnarray}
f_i(x,\mu^2) &=& f_{i0}(x) + \frac{\alpha_S}{2\pi} \int_x^1 \frac{d\xi}{\xi} 
\left[ f_{i0}(\xi) \left\{ P_{qq}\left(\frac{x}{\xi}\right) \ln \frac{\mu^2}
{\kappa^2} + C_q\left(\frac{x}{\xi}\right) \right\} \right. \nonumber \\
&& + \left. f_{g0}(\xi) \left\{ P_{qg}\left( \frac{x}{\xi} \right) \ln 
\frac{\mu^2}{\kappa^2} + C_g\left(\frac{x}{\xi}\right)\right\} \right] + \dots,
\label{eqn:rendist}
\end{eqnarray}
where $\kappa$ is an infrared cutoff and $f_g$ is a new function, similar to
$f_i$ but describing a gluon rather than a quark. $P_{qq}$ is a splitting 
function in the case of a quark emitting a gluon before scattering off the 
virtual photon, whereas the $P_{qg}$ is a splitting function for the case of a 
gluon splitting into a $q \bar q$ pair and the $q$ scattering off the virtual 
photon. $f_{i0}(x)$ is an unmeasurable, bare distribution. This distribution
absorbs all of the collinear singularities at a `factorization scale' $\mu$.
$F_2(x,Q^2)$ can then be written in terms of (\ref{eqn:rendist}) by
\begin{eqnarray}
\label{eq:F2glory}
F_2(x,Q^2) &=& x \sum_{i=q,\bar q} e_q^2 \int_x^1 \frac{d\xi}{\xi} \left[
f_i(\xi,\mu^2) \left\{ \delta(1-\frac{x}{\xi}) + \frac{\alpha_S}{2\pi} 
P_{qq} \left( \frac{x}{\xi} \right) \ln \frac{Q^2}{\mu^2} + \dots \right\} 
\right. \nonumber \\
&& \left. + f_g(\xi,\mu^2) \left\{ 
\frac{\alpha_S}{2\pi} P_{qg} \left(\frac{x}{\xi} \right) \ln \frac{Q^2}{\mu^2}
 + \dots \right\} \right].
\label{eqn:f2sl}
\end{eqnarray}
This has absorbed all of the finite contribution, $C$, into the parton
distributions. It is possible to factor out an arbitrary finite term from the
distributions which leaves behind an additional finite contribution. This
depends on the `factorization scheme'. A common choice is the $\MSbar$ scheme
in which (\ref{eqn:f2sl}) is expressed as
\begin{eqnarray}
\label{eq:F2MS}
F_2(x,Q^2) &=& x \sum_{i=q,\bar q} e_q^2 \int_x^1 \frac{d\xi}{\xi} \left[
f_i(\xi,Q^2) \left\{ \delta(1-\frac{x}{\xi}) + \frac{\alpha_S}{2\pi} 
C_q^{\MSbar}\left(\frac{x}{\xi} \right) + \dots \right\} \right. 
\nonumber \\
&& \left. + f_g(\xi,Q^2) \left\{ 
\frac{\alpha_S}{2\pi} C_g^{\MSbar} \left(\frac{x}{\xi} \right) + 
\dots \right\} \right].
\end{eqnarray}
The functions $C_q(x)$ and $C_g(x)$ are called the \emph{coefficient functions}
and they depend on the factorization and renormalization schemes. These are
the perturbative part of the PDFs. The functions $f_i(x,\mu^2)$ contain all 
the non-perturbative physics.

Though the parton distributions are not derivable in the scope of perturbation
theory, we can see from (\ref{eq:F2glory}) that the right side cannot be 
dependent on $\mu^2$. Taking the partial derivative of both sides with respect 
to $t=\mu^2$, and ignoring the gluon part for simplicity, yields the DGLAP 
equation~\cite{Lipatov:1975qm,Gribov:1972rt,Altarelli:1977zs,Dokshitzer:1977sg}
\begin{equation}
t \frac{\partial}{\partial t} f_i(x,t) = \frac{\alpha_S(t)}{2\pi} \int_x^1
\frac{d\xi}{\xi} P\left( \frac{x}{\xi} \right) f_i(\xi,t).
\label{eq:DGLAP}
\end{equation}
This equation is analogous to the $\beta$ functions which describe the 
variation of $\alpha_S(t)$ with $t$. Including all the terms this can more 
generally be written as a $(2n_q+1)$--dimensional matrix equation in the space
of the quarks, antiquarks and gluons. This equation is a fundamental equation 
for the parton distribution functions as well as the parton shower. 

When developing an event simulation, the PDFs will define which parton is 
drawn from the hadron for the interaction. The remaining part of the hadron is
known as the beam remnant. Since the way that hadrons are formed and stay 
bound lies in the non-perturbative regime of QCD and is therefore not well 
understood, the only way to describe the remnants is through phenomenological 
models.

The beam remnant carries a portion of the beam hadron's momentum and often
will travel straight down the beam pipe, leading to undetectable physics. This
isn't always the case, however, and these remnants can lead to some detectable
physical results and as such are worth studying. There are very simple models,
such as the UA5 model~\cite{Alner:1987is} that can be used. This model is 
simply a 
parameterization of the data and does not scale to higher energies. These
simple models lack the adequate ability to describe all the effects of the 
remnants. Instead there is 
evidence to suggest that these remnant can and do interact again within one
event. These models are known as multiple interaction models\cite{JIMMY,MIA}
and are currently still an interesting point of research. The whole process of 
the beam remnants and their interactions is known as the 
\emph{Underlying Event}.

\subsection{The Parton Shower}
\label{sec:shower}
The hard process will generate a set of final-state particles for a given
set of incoming particles. As the hard process is only accurate up to a given
order, it isn't able to describe events with high parton multiplicity. As
colliders are able to achieve higher energies, these high multiplicity events 
begin to play a larger role in the events. There is a need to generate these 
higher multiplicity final states to some approximation. The \emph{Parton 
Shower} is able to generate these states in the soft and collinear regions of 
phase space to all orders. It is also in these soft and collinear regions which
the higher order matrix elements are enhanced.

In this section I explain the parton shower and how it describes the enhanced
soft and collinear regions\cite{Ellis:1996qj}. The parton shower is an 
approximate perturbative treatment of QCD at momentum transfer-squared $t$ 
greater than some infra-red cutoff $t_0$. This treatment is then easy to 
integrate with a hadronization model which begins at the cutoff, $t_0$, and 
turns the partonic final state into hadronic states.

In order to generate the higher multiplicity states we must split a parton $a$
into two partons $b+c$. We start by assuming
\begin{equation}
p_b^2 , p_c^2 \ll p_a^2 \equiv t.
\end{equation}
If we have $a$ as an outgoing parton, this corresponds to a \emph{timelike} 
shower ($t>0$). If it is an incoming parton, this is a \emph{spacelike} shower.
For the following introduction we will consider the timelike shower. It can be
shown that the spacelike shower retains the same formulation, it just requires
different kinematics.

We can define the energy fraction as
\begin{equation}
z = E_b/E_a = 1 - E_c/E_a.
\end{equation}
Now in the region of small angles, which is the region where the matrix element
is enhanced, we have
\begin{equation}
t = 2E_b E_c (1-\cos \theta) = z (1-z) E_a^2 \theta^2,
\end{equation}
and using transverse momentum conservation
\begin{equation}
\theta = \frac{\theta_b}{1-z} = \frac{\theta_c}{z}.
\end{equation}
We then find that the matrix element squared for $n+1$ partons, from a 
$g\to gg$ splitting in the small angle approximation, is
\begin{equation}
\left| {\cal M}_{n+1} \right|^2 \sim \frac{4g^2}{t} C_A F(z; \epsilon_a, 
\epsilon_b, \epsilon_c) \left| {\cal M}_n \right|^2,
\end{equation}
where $C_A = 3$ comes from $f^{ABC} f^{ABC}$ and the function $F$ is the
helicity dependent splittings. If we average this over the incoming gluon 
spins and sum it over the outgoing gluon spins we get
\begin{equation}
C_A \left< F \right> \equiv \hat{P}_{gg}(z) = C_A \left[ \frac{1-z}{z} + 
\frac{z}{1-z} + z(1-z)\right].
\end{equation}
This is the unregularized \emph{gluon splitting function} related to the 
Altarelli-Parisi kernel~\cite{PeskinSchroeder}. If we do the same thing for 
$g \to q \bar q$ and $q \to q g$ we find the unregularized splitting functions
\begin{eqnarray}
\hat{P}_{qg}(z) &=& T_R \left[ z^2 + (1-z)^2 \right], \\
\hat{P}_{qq}(z) &=& C_F \frac{1+z^2}{1-z}.
\end{eqnarray}
Here $T_R = {\rm Tr}(t^At^A)/8 = 1/2$ and $C_F = {\rm Tr}(t^At^A)/3 = 4/3$. 
Therefore we generally find
\begin{equation}
\left| {\cal M}_{n+1} \right|^2 \sim \frac{4g^2}{t} \hat{P}_{ba} \left| 
{\cal M}_n \right|^2.
\end{equation}

We then find the new differential cross section for the $n+1$ state is
\begin{equation}
d\sigma_{n+1} = d\sigma_n \frac{dt}{t} dz \frac{\as}{2\pi} \hat{P}_{ba}(z),
\end{equation}
where $\hat{P}_{ba}(z)$ is the appropriate splitting function to generate the
new state. 

The parton shower doesn't just emit one gluon in the soft or collinear region.
It is capable of multiple branchings.  We start by introducing the 
\emph{Sudakov form factor}
\begin{equation}
\Delta(t) \equiv \mathrm{exp}\left[-\int_{t_0}^t \frac{dt'}{t'} \int dz 
\frac{\alpha_s}{2\pi}\hat{P}(z) \right].
\end{equation}
Doing this we can write the DGLAP evolution equation~(\ref{eq:DGLAP}) as
\begin{equation}
t \frac{\partial}{\partial t} f(x,t) = \int_x^1 \frac{dz}{z} \frac{\as}{2\pi}
\hat{P}(z) f(x/z,t) + \frac{f(x,t)}{\Delta(t)}t \frac{\partial}{\partial t} 
\Delta(t),
\end{equation}
which is equivalent to
\begin{equation}
t \frac{\partial}{\partial t}\left( \frac{f}{\Delta}\right) = \frac{1}{\Delta}
\int_x^1 \frac{dz}{z} \frac{\as} {2\pi} \hat{P}(z) f(x/z, t).
\label{eqn:fsratio}
\end{equation}
This has the solution
\begin{equation}
f(x,t) = \Delta(t) f(x,t_0) + \int_{t_0}^t \frac{dt'}{t'} \frac{\Delta(t)}
{\Delta(t')} \int_x^1 \frac{dz}{z} \frac{\as} {2\pi} \hat{P}(z) f(x/z, t').
\end{equation}
From this equation, the Sudakov form factor is interpreted as the probability 
of evolving from $t_0$ to $t$ without branching. The infra-red singularity of 
the splitting function is still not handled, however, so we must impose an 
upper limit on $z$ such that $z<1-\epsilon(t)$. The branching for values of 
$z$ above this are interpreted as {\em unresolvable}. These are emissions of 
gluons that are so soft they are undetectable. With the cutoff, the form 
factor is interpreted as the probability of not having any {\em resolvable} 
branchings from $t_0$ to $t$.

This idea of the Sudakov form factors easily integrates into the Monte Carlo 
method. From the interpretation of the form factors we find that the 
probability of evolving from $t_1$ to $t_2$ without (resolvable) branching is 
$\Delta(t_2)/\Delta(t_1)$. For time-like branchings we can find the 
distibution of $t_2$ by solving
\begin{equation}
\label{eqn:sudfratio}
\frac{\Delta(t_1)}{\Delta(t_2)} = {\cal R}.
\end{equation}
where ${\cal R}$ is a random number in the interval $[0,1]$.

When evolving space-like partons the structure of the hadron must be 
maintained. We can see from (\ref{eqn:fsratio}) that this requires using the 
ratio $\Delta(t_i)/f(x,t_i)$. This means we must find the distribution of 
$t_2$ by solving
\begin{equation}
\label{eqn:sudratio}
\frac{f(x,t_1) \Delta(t_2)}{f(x,t_2) \Delta(t_1)} = {\cal R},
\end{equation} 
In either case if the value of $t_2$ is higher than the hard subprocess, 
$Q^2$, then we have reached the no branching condition for that parton. 

In the timelike case, if a branching does occur, we then need to calculate the 
momentum fraction, $z$. This is done by solving
\begin{equation}
\int_\epsilon^{x_2/x_1} dz \frac{\alpha_s}{2\pi} P(z) = {\cal R}' 
\int_{\epsilon}^{1-\epsilon} dz \frac{\alpha_s}{2\pi} P(z),
\end{equation}
where ${\cal R}'$ is another random number in the interval $[0,1]$. To 
construct the momentum of the products of the emission we simply need to 
generate an azimuthal angle in the interval $[0,2\pi]$. If polarization 
correlations are taken into account, then this angle won't be uniform.

The spacelike case isn't very different. In this case we are evolving backwards
and trying to maintain the structure of the beam particle which is known from
the PDF. This requires the modification to the distribution of $z$ of
\begin{equation}
\int_\epsilon^{x_2/x_1} dz \frac{\alpha_s}{2\pi} \frac{P(z)}{z} f(x_2/z,t_1) =
{\cal R}' \int_{\epsilon}^{1-\epsilon} dz \frac{\alpha_s}{2\pi} \frac{P(z)}{z}
f(x_2/z,t_1).
\end{equation}

Another important aspect of these branchings is angular 
ordering~\cite{Ellis:1996qj}. \emph{Coherent parton branching} shows that each 
successive emission must lie within a cone with half angle given by the 
previous emission. This effect is also present in QED radiation. In this case
it is easier to explain the angular ordering effect. Since the charge partner 
is created with angle, $\theta$, an emission outside of this angle would 
effectly appear to be an emission from a chargeless object. In a sense, it 
can't be determined which particle this emission came from, making it an 
incoherent emission. Therefore a coherent emission from a particle must be 
emitted within the half angle of the previous emission. A similar logic can 
be applied to QCD radiation, though it is not as straightforward due to the 
three degrees of colour charge but yields the same result. This means that each
emission is restricted to have a smaller angle then the previous emission. 

In order to have a coherent parton branching, the shower needs to evolve with a
variable related to $\theta$, rather then one related to the virtuality. This 
is given by
\begin{equation}
\zeta = \frac{p_b \cdot p_c}{E_bE_c} \simeq 1 - \cos \theta.
\end{equation}
In \HW{}~\cite{Herwig} the variable used, which ensures angular ordering, is
\begin{equation}
\tilde{t} = E^2 \zeta \geq t_0.
\end{equation}
In Pythia \cite{Pythia} an angular ordered variable is not used. Instead,
a virtuality ordered shower is used and emissions that violate angular ordering
are vetoed. Part of my research has been into a new set of evolution variables.
This new work is discussed in Chapter \ref{chap:NewVars}. 

The shower does have some more complications, however. It is possible to 
calculate matrix elements to higher order. It is desirable to use these 
matrix elements as they contain all of the terms at a particular order, whereas
the shower includes some of the terms at next-to-leading log (NLL). As these
higher order matrix elements are enhanced in the soft and collinear regions it
is desirable to use the shower in these regions and the matrix element in the 
hard regions. In order to do so properly, over- and under-counting must be 
prevented. This process is called the \emph{matrix element corrections}.

There are two types of corrections, hard and soft. The soft corrections are 
when the shower approximation is corrected so that it is closer to the 
hard matrix element. This means that a gluon from the shower is 
prevented from occuring as it has already been properly considered in the 
matrix element. The hard corrections are when the first gluon emitted is 
determined to come from the matrix element, rather than the shower. The region 
where the matrix element is used and the region where the shower is used need 
to match smoothly in order to correctly describe the physics and these 
corrections ensure the smooth matching of the two.

The matching of the parton shower and the higher order matrix elements is 
important for generating useful results and ensuring that the simulations 
describe the theory as accurately as it can. There are two different problems 
with matching higher order diagrams. The first is in ensuring that all of 
phase space is properly covered by the next-to-leading order matrix 
element~\cite{MatchingNLO}. There is also a need to ensure the shower and 
matrix element match at next-to-leading log (NLL) without double 
counting~\cite{CKKW}. The decays of heavy quarks and other heavy coloured 
particles (such as SUSY particles) can also involve the parton shower. The 
showering of these processes and the matrix element corrections to them have 
also been studied~\cite{MECorrDecays}.

\subsection{Hadronization}
After the parton shower, we are left with a set of partons that are of the same
order in virtuality as the cutoff on the parton shower, $t_0$. At this stage
the interactions between the partons become heavily influenced by 
non-perturbative effects in the low momentum-transfer, long distance 
regime. 

A hadronization model must take the partons from virtuality $t_0$ down into 
stable and unstable hadrons. We would expect that there would be more hadrons 
created when $t_0$ is larger, as they have a higher virtuality. Likewise
we would expect there to be more partons at the initial stage of hadronization
if $t_0$ is smaller.

Since hadronization models aren't that well understood, it turns out that 
carefully letting the parton shower run to lower scales produces better 
results. This implies that though the parton shower only contains perturbative 
results, these are often still more valid than hadronization models. Quarks and
gluons are not observed in collider experiments, hadrons are; therefore in 
the end we must use a hadronization model to study collider physics. 

The simplest model is to assume that each parton produces hadrons independently
of the other partons. This is known as the \emph{Independent Fragmentation Model}.
This was originally designed by Field and Feynman\cite{Feynman} to approximate
scaling of energy distributions observed in quark jets in $e^+e^-$ events at
moderate energy. This model takes a quark and pairs it with an anti-quark
in a $q\bar q$ pair drawn from the vacuum. This forms a meson and the remaining
quark fragments the same way. This continues until the leftover energy falls
below a threshold. In this model the gluon is split into $q\bar q$ pairs. The
momentum can be distributed in many ways. All the momentum can be given to one
of the pair so a gluon behaves just like a quark in this model. The momentum
could also be given by the $g\to q\bar q$ Altarelli-Parisi splitting function.
This model has several problems. As the final partons are supposed to be on 
mass-shell it can lead to momentum conservation problems. Since there are
low energy quarks remaining in this model, there is also a problem of colour 
flow.

Another model is the \emph{String Model}. This model assumes that
two colour connected partons have some colour field between them that grows
with seperation. It is usually assumed to have a uniform energy per length. 
This amounts to a linear quark confining potential. When this ``string'' 
between the two quarks contains too much energy it breaks and a $q\bar q$ pair
fill each side of the break. This continues until each string is considered
``stable''. At this point each part forms hadrons from the flavours that
are colour connected. Gluons in this model form kinks in the strings because
they carry a localized energy and momentum. It is the hadronization of these
kinked strings that generates results that match experiments better than the
independent fragmentation model. This is the model that is implemented in 
Pythia~\cite{Pythia}.

The model that is implemented in \HW{} is the the \emph{cluster hadronization 
model}. This model tries to cluster quarks together to form hadrons. The 
cluster model of hadronization relies on the colour preconfinement property of 
parton branching~\cite{Marchesini:1980tv}. This property implies that pairs of 
colour-connected neighbouring partons have a mass distribution 
that falls off rapidly at high masses. It then makes sense to cluster these 
partons into colour-singlet clusters that can decay into observable hadrons.

The gluons in a cluster model must be split in order to form clusters. This is 
done non-perturbatively as we are in the non-perturbative regime. This means 
that the gluons are just split as a two body decay. This allows the colour 
connected quarks to form colour singlet clusters. It is important to note that 
in order to do the non-perturbative splitting, the gluons have to be given an 
effective mass. The value of this mass dictates the available flavours for the 
splitting. A common value of 750 MeV is used and this allows the gluon to split
into $u$ and $d$ flavours. A new cluster model\cite{JanCluster} is also used in
the new event generator SHERPA\cite{SHERPA}.

I have developed a new model based on the old \HW{} cluster model. This new
model and the improvements it provides are discussed in Chapter 
\ref{chap:Hadronization}.
 
\subsection{Decays}
The last part to using the Monte Carlo method in collider simulations is the 
decays of the hadrons. There are thousands of decay modes of hadrons and 
unstable particles from the Particle Data Group~\cite{PDG}. Some of these modes
are quite rare, others quite common. Since there are so many modes, and most of
them are for hadrons, calculating a distribution for the decay particles is not
easy, or even always possible. 

Though exact calculations aren't always possible, there may be simple 
distributions that are better fits to reality based on certain properties, such
as parity violation. In \HWP{} a few of these simple matrix elements have 
been included, in much the same way that they were included in \HW{}. The two 
commonly used decay matrix elements are for three-body decays with the decaying
particle of mass $m_0$ decaying into particles of masses $m_1,m_2$ and $m_3$. 
For free particles that decay weakly the free massive $(V-A)^2$ matrix 
elements is used. The decay momenta, $q_i,$ are generated in the three body 
phase space with the weight ${\cal W}(m_0;m_1,m_2,m_3)$ given by
\begin{equation}
\label{eq:V-AWeight}
{\cal W}(m_0;m_1,m_2,m_3) = \frac{4 (m_0^2+m_1^2 - {\cal S})({\cal S} - 
m_2^2 + m_3^2)}{(m_0^2+m_1^2-m_2^2-m_3^2)^2},
\end{equation}
where ${\cal S}$ is $(q_2+q_3)^2$ and is given by
\begin{equation}
{\cal S} = {\cal R}(m_2+m_3)^2 + (1-{\cal R})(m_0-m_1)^2.
\end{equation}
For bound particles that decay weakly this same matrix element is used but 
it has a veto placed on it. This veto is to ensure that the decay products 
are moving away from each other fast enough to no longer be bound.

Decays aren't always confined to happening after hadronization, however. In the
SM, for example, the top quark will decay before it ever hadronizes. Also the
$\tau$ will decay before it leaves the detector. In fact in SUSY there are
many more particles that decay before they hadronize. Often these particles
are fermions and so spin correlations can play an important part in the
distributions. I present here a method which can be integrated into Monte
Carlo simulations to provide the full spin correlations to these 
particles~\cite{Richardson:2001df}.

This algorithm is difficult to explain in a completely abstract manner. Instead
an example will be presented here. Consider a $2\rightarrow n$ hard subprocess.
Here we label the incoming particles $a_1$ and $a_2$ and the outgoing particles
$b_1$ to $b_n$. The momenta of these particles are given by the matrix element
\begin{equation}
\rho^1_{\kappa_1\kappa_1'} \rho^2_{\kappa_2\kappa_2'} {\cal M}_{\kappa_1 
\kappa_2; \lambda_1\dots\lambda_n} {\cal M}^*_{\kappa_1'\kappa_2';\lambda_1'
\dots\lambda_n'}\prod_{i=1,n} D^i_{\lambda_i\lambda_i'},
\end{equation}
where $\rho$ is the spin density matrix of the incoming particles, 
$\kappa_i$ are the incoming particles helicity, ${\cal M}$ is the matrix
element of the $2\rightarrow n$ process, $\lambda_i$ is the helicity of the
$b_i$ and $D^i_{\lambda_i \lambda_i'}$ is the decay matrix for the $b_i$. 
Initially all the $D^i$ are just $\delta_{\lambda_i\lambda_i'}$ and the
spin density matrices are given as $\rho_{\kappa_1\kappa_1'} = \delta_{\kappa_1
\kappa_1'}$ for unpolarized incoming particles and
\begin{equation}
\rho_{\kappa_1\kappa_1'} = \left( \begin{array}{cc} \frac{1}{2}(1+{\cal P}_3)
& 0 \\ 0 & \frac{1}{2}(1-{\cal P}_3) \end{array} \right),
\end{equation}
for longitudinally polarized spin $1/2$ incoming particles. Here ${\cal P}_3$ 
is the component of the polarization parallel to the beam axis. 

Next a $b_j$ is chosen at random. The spin density for this particle is given 
by
\begin{equation}
\rho_{\lambda_j\lambda_j'} = \frac{1}{N_\rho} \rho^1_{\kappa_1\kappa_1'} 
\rho^2_{\kappa_2\kappa_2'} {\cal M}_{\kappa_1 \kappa_2; \lambda_1\dots
\lambda_n} {\cal M}^*_{\kappa_1'\kappa_2';\lambda_1'\dots\lambda_n'}
\prod_{i\neq j} D^i_{\lambda_i\lambda_i'},
\end{equation}
where the normalization $N_\rho$ is chosen so the trace of the spin density 
matrix is one. The decay mode of this particle is selected based on the
branching ratios. This produces particles $c_1$ to $c_m$ with helicity 
$\upsilon_i$. The momentum of these particles is given by the matrix element
\begin{equation}
\rho_{\lambda_j\lambda_j'} {\cal M}_{\lambda_j; \upsilon_1\dots\upsilon_m}
{\cal M}^*_{\lambda_j'; \upsilon_1'\dots\upsilon_m'} \prod_{i=1,m} 
D^i_{\upsilon_i,\upsilon_i'}.
\end{equation}
Another randomly selected decay product, $c_k$, is chosen from the decay of 
$b_j$. The spin density for this new decay product is
\begin{equation}
\rho_{\upsilon_k,\upsilon_k'} = \frac{1}{N_{D\rho}} \rho_{\lambda_j\lambda_j'}
{\cal M}_{\lambda_j; \upsilon_1 \dots \upsilon_m}{\cal M}^*_{\lambda_j; 
\upsilon_1 \dots \upsilon_m} \prod_{i\neq k} D^i_{\upsilon_i\upsilon_i'},
\end{equation}
where again the normalization is chosen so the trace of the spin density matrix
is one.

This process of decaying the products continues all the way up the decay chain
until a stable particle is reached. Once this occurs the decay matrix is fixed
as an identity matrix (i.e. $\delta_{\lambda_i\lambda_i'}$). Once all of the
decay products of a particle have been handled the decay matrix for the 
particle is calculated. Returning to our example, assume that the particle
$b_j$ has had all of its decay products generated. Its decay matrix would be
\begin{equation}
D_{\lambda_j\lambda_j'} = \frac{1}{N_D} {\cal M}_{\lambda_j; \upsilon_1 \dots
\upsilon_m} {\cal M}^*_{\lambda_j'; \upsilon_1 \dots \upsilon_j'} 
\prod_{i=1,m} D^i_{\upsilon_i\upsilon_i'},
\end{equation}
where this too is normalized so the trace is one.

This whole process continues for all $b_j$ until all the decay products have 
had their spin density matrices and there decay matrices calculated. Since all 
the spin information is passed up the chain via the spin density matrices and 
then passed back down the chain via the decay matrices, the whole event has the
complete spin correlations built into the decay products. For a more thorough
discussion and examples of the spin correlation effects see
\cite{Richardson:2001df}.

There are many packages available that perform selected decays using more 
advanced algorithms. These can almost always be implemented into the Monte 
Carlo simulations at this point. One example of this is EvtGen\cite{EvtGen}.
This package uses decay amplitudes, rather than probabilities, so it can
correctly generate the angular correlations in a decay chain. Many of the
decay modes in this package have been developed from experimental data and
therefore, match data much better than the simple model built into \HWP{}.

\chapter{New formalism for QCD parton showers}
\label{chap:NewVars}

\section{Introduction}
The parton shower approximation has become an important component of a wide
range of comparisons between theory and experiment in particle physics.
Calculations of observables that are asymptotically insensitive to soft 
physics are known as \emph{infrared safe} observables. These can be
performed in fixed-order perturbation theory, but the resulting
final states consist of a few isolated partons, quite different from the
multihadron final states observed experimentally. One can attempt
to identify isolated partons with hadronic jets, but then the
energy flows within and between jets are not well represented.

Currently, the only means of connecting few-parton states with
the real world is via parton showers, which generate high-multiplicity
partonic final states in an approximation
that retains enhanced collinear and soft contributions to all orders.
Such multiparton states can be interfaced to a hadronization model
which does not require large momentum transfers in order to produce
a realistic hadronic final state.  Hadronization and detector
corrections to the fixed-order predictions can then be computed,
and the results have generally been found to be in satisfactory
agreement with the data.  Infrared-sensitive quantities such as hadron
spectra and multiplicities have also been described successfully
using parton showers. This has strengthened the  belief that similar techniques
can be used to predict new physics signals and backgrounds in
future experiments.

This chapter presents a new shower evolution formalism~\cite{NewVariables},
based on an angular variable related to transverse
momentum \cite{Catani:1991rr,Catani:1991hj,Catani:2000ef,Cacciari:2001cw}.
The main aim of these new variables is to retain the direct angular ordering of
the shower while improving the Lorentz invariance of the evolution and 
simplifying the coverage of phase space, especially in the soft region.  The
old shower variables used in \HW{} used massless splitting functions which
created an artificial lower bound for the transverse momentum. This created an 
artificial ``dead cone'' in the emission from heavier quarks in which no 
emissions could lie. By allowing evolution down to zero transverse momentum 
and the use of mass-dependent splitting functions, the new shower variables 
permit a better treatment of heavy quark fragmentation which eliminates these 
the sharply-defined collinear ``dead cones''.

In the following section the new shower variables and their
associated kinematics and dynamics are defined. The appropriate argument
of the running coupling, the mass-dependent parton branching
probability, and the shower evolution cutoff are also given.  The variables are
defined slightly differently for initial- and final-state parton
branching, and depend on the colour connection of the evolving parton,
so in subsequent sections the various possible configurations of colour flow 
between initial and final jets are considered.

The formalism presented here is implemented in the new Monte Carlo event
generator \HWP\ \cite{Gieseke:2003_20} which is described in detail in Chapter
\ref{chap:Herwig}. Results for $e^+e^-$ annihilation and comparisons with LEP 
data have been presented in a separate publication~\cite{Gieseke:2003_19} and
are also given in Chapter~\ref{chap:Results}. 
The formulae in this chapter could also be used to 
construct a matching scheme for next-to-leading order (NLO) QCD calculations
and \HWP\ parton showers, similar to that developed for \HW\ showers
in \cite{Frixione:2002ik,Frixione:2003ei} and implemented in
the {\small MC@NLO} event generator~\cite{Frixione:2003vm}.

\section{New variables for parton branching} 
As mentioned in Chapter~\ref{chap:Intro} there are two types of shower
evolutions. When a parton is space-like ($t<0$) it is an initial-state
parton and is described here as part of the initial-state shower. When a
parton is time-like ($t > 0$) it is a final-state parton and is described as
part of the final-state shower. Each individual splitting of a parton is
referred to as a branching. The complete branching history of a given parton is
called its evolution and the collection of all the evolutions of all the
final- (initial-) state partons is referred to as the final- (initial-) state 
shower.

\subsection{Final-state quark branching}
\subsubsection{Kinematics}
\begin{figure}[htb]
\begin{center}
\epsfig{figure=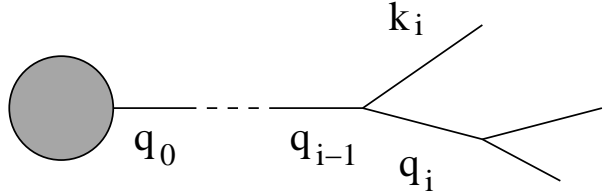,width=8cm}
\end{center}
\caption{Final-state parton branching. The blob represents the
hard subprocess.}
\label{fig_finalbr}
\end{figure}

Consider parton branching in an outgoing (heavy) quark jet. Define the quark
momentum after the $i$th gluon emission $q_{i-1}\to q_i+k_i$ (see
figure~\ref{fig_finalbr}) in the \emph{Sudakov basis} as
\begin{equation}\label{eq_qi}
q_i = \alpha_i p + \beta_i n + q_{\perp i}
\end{equation}
where $p$ is the jet's ``parent parton'' momentum ($p^2=m^2$,
the {\em on-shell} quark mass-squared),
$n$ is a lightlike ``backward'' 4-vector ($n^2=0$), and $q_{\perp i}$
is the transverse momentum ($q_{\perp i}^2=-\qt_i^2$, 
$q_{\perp i}\cdot p = q_{\perp i}\cdot n =0$).  Then
\begin{equation}\label{eq_betai}
\beta_i = \frac{\qt_i^2 + q_i^2 - \alpha_i^2 m^2}{2\alpha_i p\cdot n}\;.
\end{equation}

The momentum fraction and relative transverse momentum are now defined as
\begin{equation}\label{eq_zipti}
z_i = \frac{\alpha_i}{\alpha_{i-1}}\;,\;\;\;\;\;\;
\ptr{i} = \qtr{i} - z_i\mbox{\bf q}_{\perp i-1}\;.
\end{equation}
Then we have
\begin{equation}\label{eq_qi-1}
q_{i-1}^2 = \frac{q_i^2}{z_i}+\frac{k_i^2}{1-z_i}+
   \frac{\ptr{i}^2}{z_i(1-z_i)}\;.
\end{equation}

\subsubsection{Running coupling}
As mentioned in Chapter~\ref{chap:Intro} the QCD coupling constant, $\as$
is scale dependent. This means that we need to decide what the right scale
of a branching is. To find the optimal argument of $\as$, we consider the 
branching of a quark of virtuality $q^2$ into an on-shell quark and an 
off-shell gluon of virtuality $k^2$ \cite{Dokshitzer:1996qm}. From 
(\ref{eq_qi-1}), the propagator denominator is
\begin{equation}\label{eq_q2m2}
q^2-m^2 = \frac{1-z}{z}m^2 +\frac{k^2}{1-z}+ \frac{\pt^2}{z(1-z)}
 = \frac{1}{1-z}\left\{k^2 +\frac{1}{z}[\pt^2 + (1-z)^2 m^2]\right\}\;.
\end{equation}
The dispersion relation for the running coupling is supposed to be
\begin{equation}
\frac{\as(\mu^2)}{\mu^2} = \frac{\as(0)}{\mu^2}
+ \int_0^\infty \frac{dk^2}{k^2(k^2+\mu^2)}\rs(k^2)
\end{equation}
where $\rs(k^2)$ is the discontinuity of $\as(-k^2)$.  The first term
on the right-hand side comes from cutting through the on-shell gluon,
the second from cutting through the gluon self-energy. In our case we
have $k^2 + [\pt^2 + (1-z)^2 m^2]/z$ in place of $k^2+\mu^2$. We are interested
in soft gluon resummation ($z\to 1$) \cite{Cacciari:2001cw} and so we
ignore the factor of $1/z$ here. Thus the suggested argument of $\as$ is
$\pt^2 + (1-z)^2 m^2$.  In practice a minimum virtuality is imposed on
light quarks and gluons in the parton shower, and therefore the actual
argument is slightly more complicated (see below).

\subsubsection{Evolution variable}
The evolution variable is not simply $q^2$ since this would ignore
angular ordering. To have angular ordering, for massless parton branching,
the evolution variable should be $\pt^2/[z(1-z)]^2= q^2/z(1-z)$
\cite{Marchesini:1983bm}. For gluon emission by a massive quark we assume
this generalizes to $(q^2-m^2)/z(1-z)$. To define a resolvable emission
we also need to introduce a minimum virtuality $Q_g^2$ for gluons and
light quarks.  Therefore from (\ref{eq_q2m2}) the evolution variable is
\begin{equation}\label{eq_ftq}
\tq^2 = \frac{\pt^2}{z^2(1-z)^2} + \frac{\mu^2}{z^2}
 + \frac{Q_g^2}{z(1-z)^2}
\end{equation}
where $\mu = \max(m,Q_g)$.  For the argument of the running coupling
we use
\begin{equation}\label{eq_asarg}
z^2(1-z)^2\tq^2 = \pt^2 + (1-z)^2\mu^2 + z Q_g^2\;.
\end{equation}
The $\mu$ term allows for massive quarks to evolve down to $\pt<(1-z)m$, 
i.e.\ inside the {\em dead cone} \cite{Marchesini:1990yk,Dokshitzer:fd}.

Angular ordering of the branching $q_i\to q_{i+1}$ is defined by
\begin{equation}
\tq_{i+1} < z_i\tq_i\;.
\end{equation}
The factor of $z_i$ enters because the angle at each branching is inversely
proportional to the momentum fraction of the parent.
Similarly for branching on the gluon, $k_i\to k_{i+1}$, we require
\begin{equation}
\tilde k_{i+1} < (1-z_i)\tq_i\;.
\end{equation}

\subsubsection{Branching probability}
For the parton branching probability we use the mass-dependent
splitting functions given in ref.~\cite{Catani:2000ef}.  These are
derived in the {\em quasi-collinear limit}, in which $\pt^2$
and $m^2$ are treated as small (compared to $p\cdot n$) but
$\pt^2/m^2$ is not necessarily small.  In this limit the $q\to qg$
splitting function is
\begin{equation}
P_{qq}(z,\pt^2) = C_F\left[\frac{1+z^2}{1-z}
-\frac{2z(1-z)m^2}{\pt^2+(1-z)^2m^2}\right]\;.
\end{equation}
It is the second term of this equation which differs from the splitting
functions used in \HW{}. Note that at $\pt=0$ the factor in square brackets is
just $1-z$, i.e.\ the soft singularity at $z\to 1$ becomes a zero
in the collinear direction. The minimum virtuality $Q_g^2$ serves
only to define a resolvable emission, and therefore we omit it
when defining the branching probability in terms of the evolution
variable (\ref{eq_ftq}) as
\begin{equation}\label{eq_dPqqg}
dP(q\to qg) = \frac{\as}{2\pi}\frac{d\tq^2}{\tq^2}\,P_{qq}\,dz
= \frac{C_F}{2\pi}\as[z^2(1-z)^2\tq^2]\frac{d\tq^2}{\tq^2}\frac{dz}{1-z}
\left[1+z^2-\frac{2m^2}{z\tq^2}\right]\;.
\end{equation}

\subsection{Gluon splitting}
\label{sec:gsplit}
In the case of a final-state gluon splitting into a pair of heavy quarks of
mass $m$, the quasi-collinear splitting function derived
in \cite{Catani:2000ef} is
\begin{equation}\label{eq:gQQ2}
P_{qg}(z,\pt^2) = 
  T_R\left[1-2z(1-z)\frac{\pt^2}{\pt^2 + m^2}\right]\,. 
\end{equation}
We note that this splitting function is bounded above by its value $T_R=\half$
at the phase space boundary $\pt=\0t$, and below by $T_R/2$.
By analogy with (\ref{eq_ftq}), in this case the evolution
variable $\tq$ is related to the virtuality of the gluon or the relative
transverse momentum of the splitting by
\begin{equation}\label{eq:qtgqq}
  \tq^2 = \frac{q^2}{z(1-z)} = \frac{\pt^2+m^2}{z^2(1-z)^2}\,.
\end{equation}
In terms of the variables $\tilde q, z$, the $g\to q\qb$
branching probability then reads
\begin{equation}\label{eq:gQQ}
dP(g\to q\qb) = \frac{T_R}{2\pi}\as[z^2(1-z)^2\tq^2]\frac{d\tq^2}{\tq^2}
\left[1-2z(1-z)+\frac{2m^2}{z(1-z)\tq^2}\right]\,dz\,. 
\end{equation}

In the case of gluon splitting into gluons, the branching probability
takes the familiar form
\begin{equation}\label{eq:ggg}
dP(g\to gg) = \frac{C_A}{2\pi}\as[z^2(1-z)^2\tq^2]\frac{d\tq^2}{\tq^2}
\left[\frac{z}{1-z}+\frac{1-z}{z}+z(1-z)\right]\,dz\,. 
\end{equation}
Since we introduce a minimum virtuality $Q_g^2$ for gluons,
the relationship between the evolution variable and the relative
transverse momentum for this splitting is as in (\ref{eq:qtgqq}) but
with the heavy quark mass $m$ replaced by $Q_g$. Similarly, for gluon
splitting to light quarks we use (\ref{eq:qtgqq}) with
$\mu = \max(m,Q_g)$ in place of $m$.

\subsection{Initial-state branching}
\begin{figure}[htb]
\begin{center}
\epsfig{figure=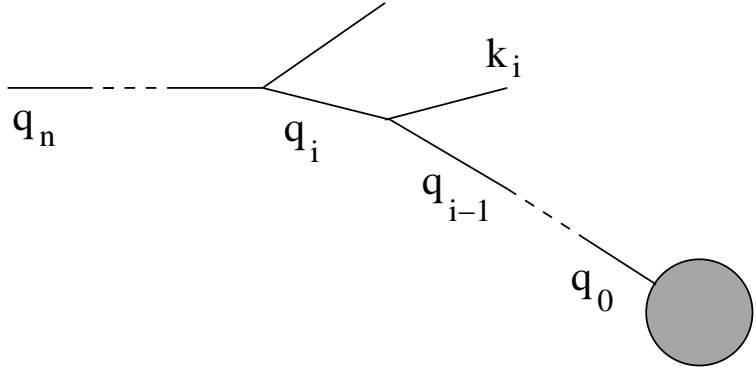,width=10cm}
\end{center}
\caption{Initial-state parton branching. The blob represents the
hard subprocess.}
\label{fig_initbr}
\end{figure}

Consider the initial-state (spacelike) branching of a partonic
constituent of an incoming hadron that undergoes some hard
collisions subprocess such as deep inelastic lepton scattering.
The momenta are defined as in (\ref{eq_qi}), with the reference
vector $p$ along the beam direction.
In this case the evolution is performed {\em backwards} from the hard
sub-process to the incoming hadron, as shown in figure~\ref{fig_initbr}.
Thus we now define in place of (\ref{eq_zipti})
\begin{equation}\label{eq_zispa}
z_i = \frac{\alpha_{i-1}}{\alpha_i}\;,\;\;\;\;\;\;
\ptr{i} = \qtr{i-1} - z_i\qtr{i}\;.
\end{equation}
Then
\begin{equation}\label{eq_qispa}
q_{i-1}^2 = z_i q_i^2 -\frac{z_i}{1-z_i}k_i^2 - \frac{\ptr{i}^2}{1-z_i}\;.
\end{equation}

We assume a massless variable-flavour-number evolution
scheme~\cite{Aivazis:1993pi,Thorne:1997ga} for constituent
parton branching, setting $m=0$ and putting all emitted
gluons at the minimum virtuality, $k_i^2=Q_g^2$.  The angular evolution
variable now relates only to the angle of the emitted gluon
and therefore we choose
\begin{equation}\label{eq_itqi}
\tq_i^2 =  \frac{\ptr{i}^2 +z_i Q_g^2}{(1-z_i)^2}\;,
\end{equation}
with ordering condition simply $\tq_{i+1} < \tq_i$.
Correspondingly, for the argument of the running coupling we now use
$(1-z)^2\tq^2$.

A different type of initial-state branching occurs in the decay of
heavy, quasi-stable coloured objects like the top quark.  Here the
momentum of the incoming heavy object is fixed and evolution is
performed forwards to the hard decay process. In this case we
cannot neglect the mass of the parton and (\ref{eq_itqi})
becomes
\begin{equation}\label{eq_dec_tqi}
\tq_i^2 =  \frac{\ptr{i}^2+z_i Q_g^2}{(1-z_i)^2}+m^2\;,
\end{equation}
while the branching probability (\ref{eq_dPqqg}) is replaced by
\begin{equation}\label{eq_dec_dP}
dP(q\to qg) = \frac{C_F}{2\pi}\as[(1-z)^2\tq^2]
\frac{d\tq^2}{\tq^2}\frac{dz}{1-z}
\left[1+z^2-\frac{2zm^2}{\tq^2}\right]\;.
\end{equation}

\subsection{Allowed regions and termination of branching}
The allowed phase space for each branching is given by requiring a real
relative transverse momentum, $\pt^2> 0$.  In final-state $q\to qg$
branching, we have from (\ref{eq_asarg})
\begin{equation}\label{eq_qqg_ps}
z^2(1-z)^2\tq^2 > (1-z)^2\mu^2 + z Q_g^2\;.
\end{equation}
This yields a rather complicated boundary in the $(\tq,z)$ plane.
However, since
\begin{equation}
(1-z)^2\mu^2 + z Q_g^2 > (1-z)^2\mu^2\,,\; z^2 Q_g^2
\end{equation}
we see that the phase space lies inside the region
\begin{equation}
  \label{eq:qtzlimits}
  \frac{m}{\tq} < z < 1-\frac{Q_g}{\tq}\;, 
\end{equation}
and approaches these limits for large values of $\tq$.  The precise
phase space can therefore be filled efficiently by generating
values of $z$  between these limits and rejecting those that violate
the inequality (\ref{eq_qqg_ps}).
The resulting threshold for $\tq$ is slightly larger than but of
the order of $m+Q_g$.

In gluon splitting, we obtain the allowed phase space range
from (\ref{eq:qtgqq}) as
\begin{equation}\label{eq:zrangegqq}
  z_- < z < z_+, \qquad 
  z_{\pm} = \frac{1}{2}\left(1\pm \sqrt{1-\frac{4\mu}{\tq}}\right) 
  \qquad\mathrm{ and }\quad
  \tq > 4\mu
\end{equation}
where $\mu=m$ for splitting into heavy quarks, or $\mu=\max(m,Q_g)$
more generally.  Therefore, analogously to (\ref{eq:qtzlimits}),
the phase space lies within the range
\begin{equation}
  \frac{\mu}{\tq} < z < 1-\frac{\mu}{\tq}\;. 
\end{equation}

Schematically, the parton shower corresponds to selecting a sequence of
$(\tq_i,z_i)$ values by solving the equations
\begin{eqnarray}
{\cal R}_1 &=& \exp\left(-\int_{\tq_i}^{\tq_{i-1}}d\tq\int_{z_-}^{z_+}dz\,
\frac{d^2P}{d\tq\,dz}\right)\nonumber \\
{\cal R}_2 &=& \int_{z_-}^{z_i}dz\,\frac{d^2P}{d\tq\,dz}\Bigg/\int_{z_-}^{z_+}
dz\,\frac{d^2P}{d\tq\,dz}
\end{eqnarray}
where ${\cal R}_{1,2}\in [0,1]$ are uniform pseudorandom numbers.
Whenever the algorithm selects a value of $\tq$ below
the threshold, branching of that parton is terminated.
The minimum virtuality $Q_g$ thus determines the scale at which soft or
collinear parton emission becomes unresolvable.  In the absence
of such a scale one eventually reaches a region where the perturbative
expression for the running coupling is divergent.

One may wish to use a parameterization of $\as$ at low scales such that
$\as(0)$ is finite.  However, a cutoff $Q_g$ is still needed
to avoid divergence of the $q\to qg$ and $g\to gg$ branching probabilities.
Alternatively one could consider parameterizing $\as$ such that
$\as(0)=0$, e.g.
\begin{equation}
\as(q^2) = \frac{q}{Q_c}\as(Q_c^2)\;\;\;\;\mbox{for}\; q<Q_c\;,
\end{equation}
where $Q_c>\Lambda$.  Then the total branching probability below
$Q_c$ is (for massless quarks)
\begin{equation}
P_c(q\to qg) = C_F\frac{\as(Q_c^2)}{2\pi}\int_0^1 2z(1+z^2)\,dz
=\frac{\as(Q_c^2)}{\pi}\;,
\end{equation}
and no explicit cutoff is required, although of course $Q_c$ is
essentially playing the same r\^ole.

After branching has terminated, the outgoing partons are put on
mass-shell (or given the virtual mass $Q_g$ if lighter) and the
relative transverse momenta of the branchings in the shower
are computed.  For final-state gluon splitting we have
\begin{equation}
|\pt| = \sqrt{z^2(1-z)^2 \tq^2 - \mu^2}, 
\end{equation}
or else, if the parent is a quark, 
\begin{equation}
|\pt| = \sqrt{ (1-z)^2 (z^2 \tq^2 - \mu^2) - z Q_g^2 }.
\end{equation}
The virtualities of the internal lines of the shower can now be
computed backwards according to (\ref{eq_qi-1}). Finally, the azimuthal
directions of the $\pt$'s can be chosen \cite{Richardson:2001df} and the
full 4-momenta reconstructed using eqs.~(\ref{eq_qi}) and (\ref{eq_betai}).

In initial-state constituent parton branching the evolution is ``guided''
by the parton distribution functions (PDFs) of the incoming parent hadron.
Since PDFs are often not tabulated below some scale $Q_s>Q_0$, one may wish
to terminate branching whenever $\tq <Q_s$ is selected. In that case
the incoming parton is assigned virtuality $q_n^2\sim -Q_s^2$ and
the spacelike virtualities of internal lines are then reconstructed
back from $q_n^2$ to $q_0^2$ using the transverse momenta deduced
from (\ref{eq_itqi}) inserted in  (\ref{eq_qispa}).

For initial-state branching in the decay of a heavy, quasi-stable
coloured object, the branching proceeds in the opposite direction but
the reconstruction of momenta is similar, using (\ref{eq_dec_tqi})
instead of (\ref{eq_itqi}).

\subsection{Treatment of colour flows}
The remaining sections of this chapter present a detailed treatment of the 
colour flows which depend on the choice of the ``backward'' vector
$n$ and on which quantities are to be held fixed during jet evolution.
Normally $n$ should be taken along the colour-connected partner of the
radiating parton, and the 4-momentum of the colour-connected system
should be preserved. The upper limits on the evolution variable $\tq$
for the  colour-connected jets should be chosen so as to cover the phase
space in the soft limit, with the best possible approximation to the
correct angular distribution. In setting these limits we neglect the
minimum virtuality $Q_g^2$, which is a good approximation at high
energies. In the remaining sections we consider separately the four cases that
the colour connection is between two final-state jets, two initial-state
(beam) jets, a beam jet and a final-state jet, or a decaying heavy parton
and a decay-product jet.

\section{Final-final colour connection}\label{sec_finfin}
Consider the process $a\to b+c$ where $a$ is a colour singlet and $b$ and
$c$ are colour-connected.  Examples are $\ee\to q\qb$ and $W\to q\qb'$.
We need to preserve the 4-momentum of $a$ and therefore we work
in its rest-frame,
\begin{equation}
p_a = Q(1,\0t,0)\;,\;\;\; p_b= \half Q(1+b-c,\0t,\lambda)
\;,\;\;\; p_c= \half Q(1-b+c,\0t,-\lambda)\;,
\end{equation}
where $p_a^2=Q^2$, $b=m_b^2/Q^2$, $c=m_c^2/Q^2$ and
\begin{equation}
\lambda = \lambda(1,b,c)\equiv\sqrt{(1+b-c)^2-4b} = \sqrt{(1-b+c)^2-4c}\;.
\end{equation}
For emission of a gluon $g$ from $b$ we write
\begin{equation}\label{eq_ffqi}
q_i = \alpha_i p_b + \beta_i n + q_{\perp i}
\end{equation}
where $\qtr{g}=\kt$, $\qtr{b}=-\kt$, $\qtr{c}=\0t$ and we choose
\begin{equation}
n = \half Q(\lambda,\0t,-\lambda)\;.
\end{equation}
Notice that, if $c$ is massive, the alignment of $n$ along $p_c$ is
exact only in a certain class of Lorentz frames.  However, if we try to
use a massive ``backward'' vector the kinematics become too complicated.

To preserve $p_a=q_b+q_c+q_g$ we require
\begin{equation}\label{eq_ffsums}
\sum\alpha_i = \sum\beta_i = \frac{2}{1+b-c+\lambda}
\end{equation}
whereas the mass-shell conditions give
\begin{eqnarray}\label{eq_ffbetas}
\beta_b &=& \frac{2}{\lambda(1+b-c+\lambda)}
\left(\frac{b+\kappa}{\alpha_b}-b\alpha_b\right) \nonumber \\
\beta_c &=& \frac{2}{\lambda(1+b-c+\lambda)}
\left(\frac{c}{\alpha_c}-b\alpha_c\right) \\
\beta_g &=& \frac{2}{\lambda(1+b-c+\lambda)}
\left(\frac{\kappa}{\alpha_g}-b\alpha_g\right) \nonumber
\end{eqnarray}
where $\kappa\equiv\kt^2/Q^2$.  Our new variables are
\begin{equation}\label{eq_fztk}
z = \frac{\alpha_b}{\alpha_b+\alpha_g}\;,\;\;\;
\tk \equiv \frac{\tq^2}{Q^2} > b\;,
\end{equation}
where from (\ref{eq_ftq}) we have
\begin{equation}\label{eq_fkt}
\kappa= (z^2\tk-b)(1-z)^2\;,
\end{equation}
and so $\sqrt{b/\tk}<z<1$. From eqs.(\ref{eq_ffsums})-(\ref{eq_fkt}) we find
\begin{eqnarray}\label{eq_ffalphas}
\alpha_b &=&\frac{z}{1+b-c+\lambda}\left(1+b-c+ z(1-z)\tk+
\sqrt{[1-b+c- z(1-z)\tk]^2-4b}\right)\;,\nonumber \\
\alpha_c &=& \frac{2}{1+b-c+\lambda}-\frac{\alpha_b}{z}\;,\\
\alpha_g &=&\frac{1-z}{z}\alpha_b\;,\nonumber
\end{eqnarray}
with the $\beta_i$'s given by (\ref{eq_ffbetas}).

\subsection{Phase space variables}
It is convenient to express the phase space in terms of the Dalitz plot
variables
\begin{equation}
x_i = \frac{2p_a\cdot q_i}{Q^2} = (1+b-c)\alpha_i + \lambda\beta_i\;.
\end{equation}
Substituting from eqs.~(\ref{eq_ffbetas}) and (\ref{eq_ffalphas}), we find
\begin{eqnarray}\label{eq_xcbg}
x_c &=& 1-b+c- z(1-z)\tk \nonumber \\
x_b &=& (2-x_c)r + (z-r)\sqrt{x_c^2-4c}\\
x_g &=& (2-x_c)(1-r) - (z-r)\sqrt{x_c^2-4c}\nonumber
\end{eqnarray}
where
\begin{equation}\label{eq_r}
r = \frac 12\left(1+\frac{b}{1+c-x_c}\right)\;.
\end{equation}
The Jacobian factor is thus simply
\begin{equation}
\frac{\partial(x_b,x_c)}{\partial(z,\tk)} = z(1-z)\sqrt{x_c^2-4c}
\end{equation}
and the quasi-collinear branching probability (\ref{eq_dPqqg})
translates to
\begin{equation}\label{eq_dPx}
dP(q\to qg) = C_F\frac{\as}{2\pi}\frac{dx_b\,dx_c}{(1-b+c-x_c)\sqrt{x_c^2-4c}}
\left[\frac{1+z^2}{1-z}-\frac{2b}{1-b+c-x_c}\right]
\end{equation}
where
\begin{equation}
z= r +\frac{x_b-(2-x_c)r}{\sqrt{x_c^2-4c}}\;,
\end{equation}
$r$ being the function of $x_c$ given in (\ref{eq_r}).

For emission from parton $c$ we write
\begin{equation}
q_i = \alpha_i p_c + \beta_i n + q_{\perp i}
\end{equation}
where now we choose
\begin{equation}
n = \half Q(\lambda,\0t,\lambda)\;.
\end{equation}
Clearly, the region covered and the branching probability
will be as for emission from parton $b$, but with $x_b$ and
$x_c$, $b$ and $c$ interchanged.

\subsection{Soft gluon region}\label{sec_finfin_sof}
For emission from parton $b$ in the soft region $1-z=\eps\to 0$ we have
\begin{equation}
x_c\sim 1-b+c-\eps\tk\;,\;\;\;
x_b\sim 1+b-c-\eps\tk'
\end{equation}
where
\begin{equation}\label{eq_soft}
\tk' = \lambda+\frac{\tk}{2b}(1-b-c-\lambda)\;.
\end{equation}
Since $\tk$ is an angular variable, we can express it in terms of
the angle $\theta_{bg}$ between the directions of the emitting
parton $b$ and the emitted gluon in the rest frame of $a$.
In the soft region we find
\begin{equation}
\tk = \frac{(1+b-c+\lambda)(1+b-c-\lambda\cos\theta_{bg})}
{2(1+\cos\theta_{bg})}
\end{equation}
Thus $\tk=b$ at $\theta_{bg}=0$ and $\tk\to\infty$ as $\theta_{bg}\to\pi$. 

For soft emission from parton $c$, the roles of $x_b$ and $x_c$,
$b$ and $c$ are interchanged. To cover the whole angular region
in the soft limit, we therefore require $\tk<\tk_b$ in jet $b$ and
 $\tk<\tk_c$ in jet $c$. We also want the slope of the boundaries to match 
as they approach the soft limit, while still covering the divergence. In this 
case this gives the condition
\begin{equation}
\frac{\tk_b}{\tk'_b} = \frac{\tk'_c}{\tk_c}
\end{equation}
and hence
\begin{equation}\label{eq_fftkbc}
(\tk_b -b)(\tk_c -c) =  \frac 14 (1-b-c+\lambda)^2\;.
\end{equation}
In particular, the most symmetric choice is
\begin{equation}\label{eq_fftks}
\tk_b = \half(1+b-c+\lambda)\;,\;\;\;
\tk_c = \half(1-b+c+\lambda)\;.
\end{equation}
The largest region that can be covered by one jet corresponds to the maximal
value of $\tk$ allowed in (\ref{eq_ffalphas}) for real $\alpha_b$,
i.e.\ for the maximal $b$ jet
\begin{equation}\label{eq_fftkmax}
\tk_b = 4(1-2\sqrt b -b+c)\;.
\end{equation}

\subsection{Example: $\ee\to q\bar q g$}
\begin{figure}[htb]
\begin{center}
\epsfig{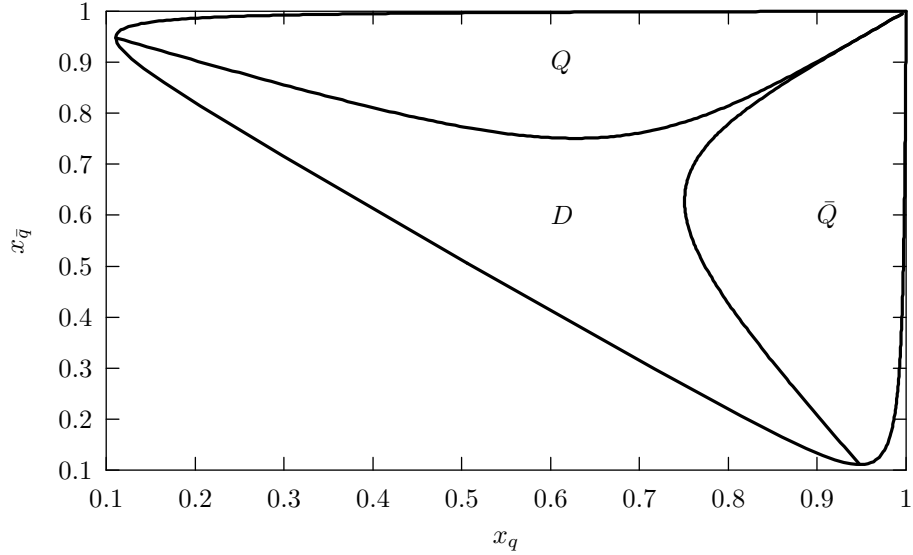}
\end{center}
\caption{Phase space for $e^+e^-\to q\qb g$ for $m_q=5$ GeV, $Q^2=m_Z^2$,
with symmetric definition of quark and antiquark jets.}
\label{fig_bfragn}
\end{figure}

Here we have $b=c=\rho$, $\lambda = \sqrt{1-4\rho}=v$, the quark
velocity in the Born process $\ee\to q\bar q$.  The phase space and
the two jet regions for the symmetrical choice (\ref{eq_fftks}) are
shown in figure~\ref{fig_bfragn}. The region D, corresponding to hard
non-collinear gluon emission, is not included in either jet and must
be filled using the ${\cal O}(\as)$ matrix element (see below).

\begin{figure}[htb]
\begin{center}
\epsfig{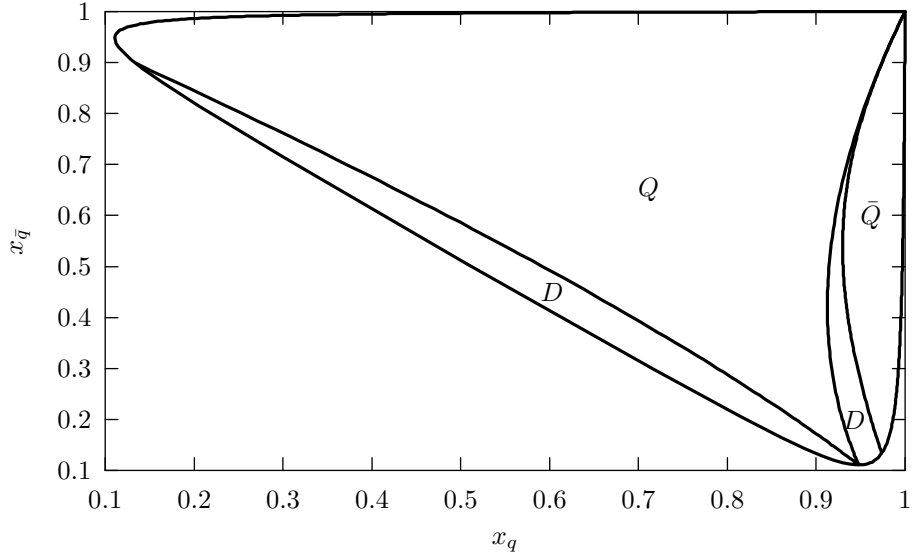}
\end{center}
\caption{Phase space for $e^+e^-\to q\qb g$ for $m_q=5$ GeV, $Q^2=m_Z^2$,
with maximal region for the quark jet.}
\label{fig_bfragn_max}
\end{figure}

For the maximal quark jet we get from (\ref{eq_fftkmax})
\begin{equation}
\tk_q = 4(1-2\sqrt\rho)\;,
\end{equation}
as shown in figure~\ref{fig_bfragn_max} together with the complementary
antiquark jet region given by (\ref{eq_fftkbc}).

\subsubsection{Exact matrix element}
The $\ee\to V\to q\qb g$ differential cross section,
where $V$ represents a vector current such as a virtual photon,
is given to first order in $\as$ by \cite{Nason:1994xx,Nason:1997pk}
\begin{equation}\label{eq_dsigdxdxb}
\frac{1}{\sigma_V}\frac{d^2\sigma_V}{dx_q\,dx_\qb} =
\frac{\as}{2\pi}\frac{C_F}{v}
\left[ \frac{(x_q+2\rho)^2+(x_\qb+2\rho)^2+\zeta_V}
{(1+2\rho)(1-x_q)(1-x_\qb)} -\frac{2\rho}{(1-x_q)^2}
-\frac{2\rho}{(1-x_\qb)^2}\right]
\end{equation}
where
\begin{equation}\label{eq_zetaV}
\zeta_V = -8\rho(1+2\rho)
\end{equation}
and
\begin{equation}\label{eq_sigmaQ}
\sigma_V = \sigma_0\left(1+2\rho\right)v
\end{equation}
is the Born cross section for heavy quark production by a vector
current, $\sigma_0$ being the massless quark Born cross section.

In the case of the axial current contribution $\ee\to A\to q\qb g$,
instead of (\ref{eq_dsigdxdxb}) we have
\begin{equation}\label{eq_dsigA}
\frac{1}{\sigma_A}\frac{d^2\sigma_A}{dx_q\,dx_\qb} =
\frac{\as}{2\pi}\frac{C_F}{v}
\left[\frac{(x_q+2\rho)^2+(x_\qb+2\rho)^2+\zeta_A}
{v^2(1-x_q)(1-x_\qb)}
-\frac{2\rho}{(1-x_q)^2}
-\frac{2\rho}{(1-x_\qb)^2}\right]\;,
\end{equation}
where
\begin{equation}\label{eq_zetaA}
\zeta_A = 2\rho[(3+x_g)^2-19+4\rho]\;,
\end{equation}
$\sigma_A$ being the Born cross section for heavy quark
production by the axial current:
\begin{equation}\label{eq_sigmaA}
\sigma_A = \sigma_0v^3\;.
\end{equation}

\subsubsection{Soft gluon distribution}
In the soft gluon region $1-z=\eps\to 0$ the branching probability
(\ref{eq_dPx}) becomes
\begin{eqnarray}\label{eq_dPsoft}
\frac{d^2P}{dx_q\,dx_\qb} &\sim& \frac{\as}{2\pi}\frac{2C_F}{v\eps^2}f_s(\tk)
\nonumber\\
f_s(\tk)&=&\frac{1}{\tk} -\frac{\rho}{\tk^2}\;.
\end{eqnarray}

\begin{figure}[htb]
\begin{center}
\epsfig{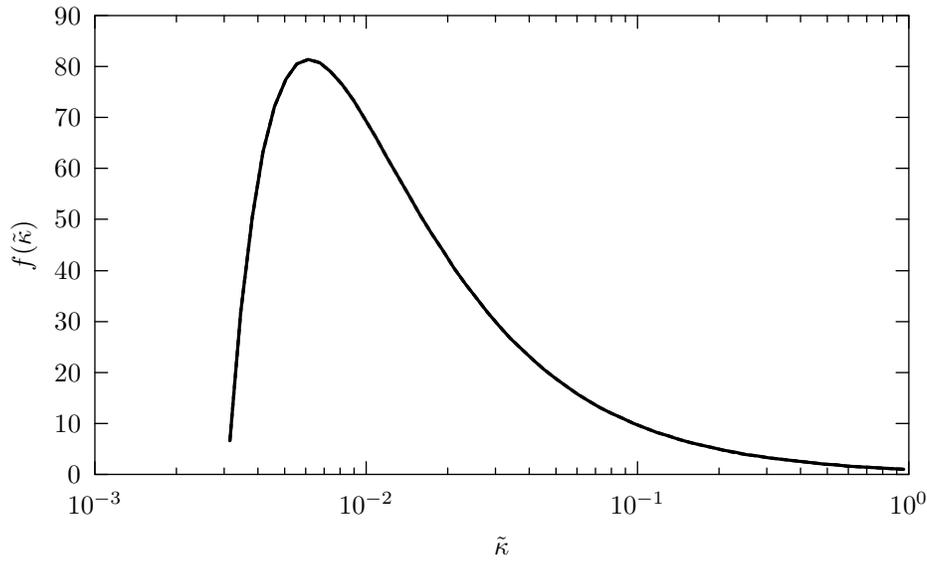}
\end{center}
\caption{The function $f(\tk)$ giving the gluon angular distribution in
the soft limit, for $m=5$ GeV, $Q^2=m_Z^2$. The exact result
({\protect\ref{eq_dPVsoft}}), solid curve, and shower approximation
({\protect\ref{eq_dPsoft}}), dashed, are not distinguishable on this scale.}
\label{fig_soft}
\end{figure}

In this limit, the exact vector and axial current matrix elements,
eqs.~(\ref{eq_dsigdxdxb}) and (\ref{eq_dsigA}) respectively, give
identical distributions:
\begin{eqnarray}\label{eq_dPVsoft}
\frac{1}{\sigma_V}\frac{d^2\sigma_V}{dx_q\,dx_\qb}&\sim&
\frac{1}{\sigma_A}\frac{d^2\sigma_A}{dx_q\,dx_\qb}
\sim \frac{\as}{2\pi}\frac{2C_F}{v\eps^2}f(\tk)
\nonumber\\
f(\tk)&=&\frac{1-2\rho}{\tk\tk'}
-\frac{\rho}{\tk^2}-\frac{\rho}{\tk'^2}
\nonumber\\
&=&f_s(\tk)\left(\frac{v}{\tk'}\right)^2\;.
\end{eqnarray}
Since from (\ref{eq_soft})
\begin{equation}
\tk' = v+\tk\left(\frac{1-v}{1+v}\right) > v\;,
\end{equation}
the parton shower approximation
(\ref{eq_dPsoft}) always overestimates the true result in the soft limit,
and so correction by the rejection method is straightforward.
For small values of $\rho$ we have
\begin{equation}\label{eq_dPVlorho}
f(\tk)= \frac{1}{\tk}-\frac{\rho}{\tk^2}
+\frac{2\rho^2}{\tk}-2\rho+{\cal O}(\rho^2)\;.
\end{equation}
Since $\tk>\rho$ we see that the error in the approximation
(\ref{eq_dPsoft}) is at most ${\cal O}(\rho)$, for any value of $\tk$
(figure~\ref{fig_soft}).

\subsubsection{Dead region contribution}
\begin{figure}[htb]
\begin{center}
\epsfig{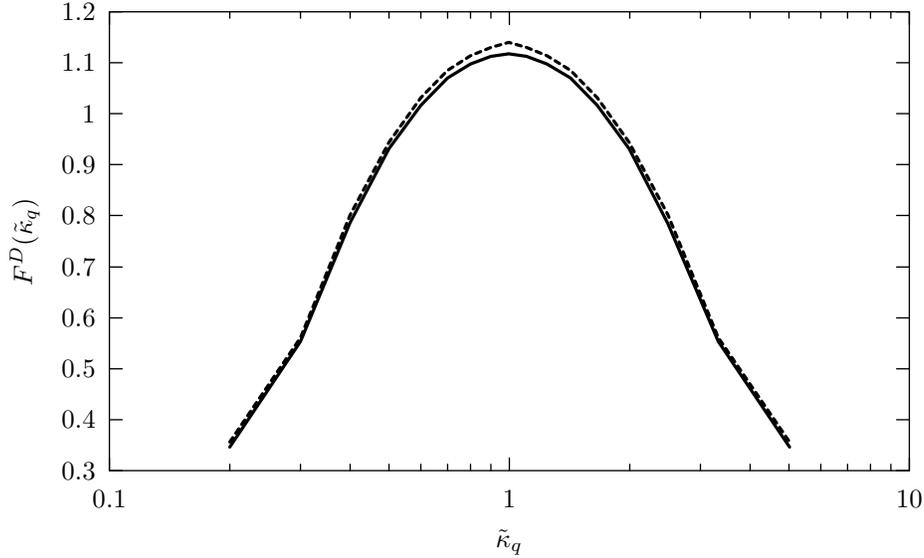}
\end{center}
\caption{The function $F^D(\tk_q)$ giving the contribution of the dead
region to the cross section, for $m=5$ GeV, $Q^2=m_Z^2$.
Solid: vector current. Dashed: axial current.}
\label{fig_fdead}
\end{figure}

The integral over the dead region may be expressed as
\begin{equation}\label{eq_dsigdead}
\frac{1}{\sigma_V}\int_D d^2\sigma_V \equiv
\frac{\as}{2\pi} C_F\, F^D_V(\tk_q)
\end{equation}
where $\tk_q$ parameterizes the boundary of the quark jet.  As shown in
figure~\ref{fig_fdead}, this is actually maximal, but still small,
at the symmetric point given by (\ref{eq_fftks}).

\begin{figure}[htb]
\begin{center}
\epsfig{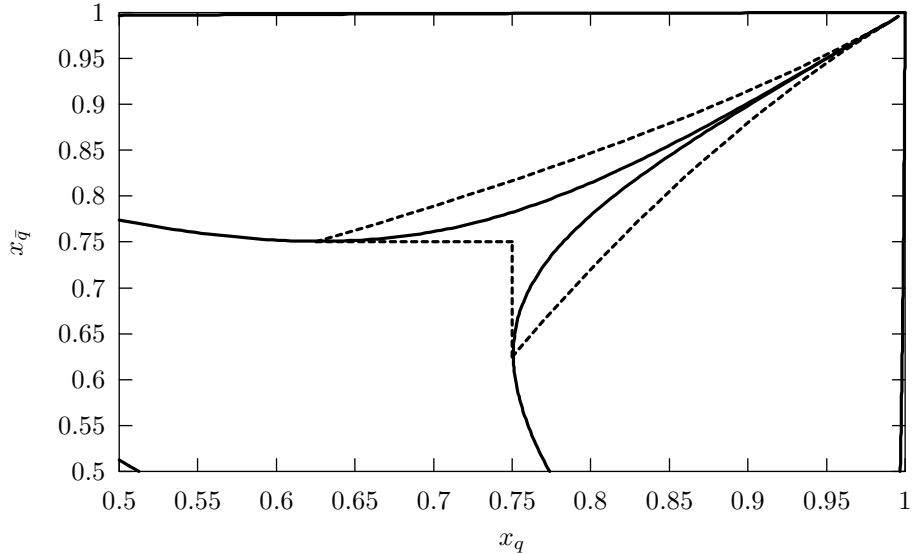}
\end{center}
\caption{The soft region, with jet boundaries (solid) and mapped region
(dashed), for $m=5$ GeV, $Q^2=m_Z^2$.}
\label{fig_soft_map}
\end{figure}

Although the integral in (\ref{eq_dsigdead}) is finite, the integrand
diverges as one approaches the soft limit $x_q=x_\qb=1$ via the narrow
``neck'' of the dead region in figure~\ref{fig_bfragn} or \ref{fig_bfragn_max}.
This could cause problems in generating $q\qb g$ configurations in the
dead region in order to apply a matrix element
correction~\cite{Seymour:1994we}.  To avoid such problems, one can
map the region $x_q,x_\qb > \frac 34$ into a region whose width vanishes
quadratically as $x_q, x_\qb\to 1$, as illustrated in 
figure~\ref{fig_soft_map}.
The mapping shown is
\begin{eqnarray}
x_q \to x'_q &=& 1-\left[\frac 14 -(1-x_q)\right] =\frac 74 -x_q\;,
\nonumber\\
x_\qb \to x'_\qb &=& 1-2(1-x'_q)\left[\frac 34 -(1-x_q)\right]
= \frac 58 +\frac 12 x_q +\frac 32 x_\qb -2x_q x_\qb
\end{eqnarray}
when $x_q > x_\qb > \frac 34$.  Within the mapped region, the integrand
then has an extra weight factor of $2(1-x'_q)$ which regularizes the soft
divergence. When $x_\qb > x_q > \frac 34$, $x_q$ and  $x_\qb$ are
interchanged in both the mapping and the weight.

\section{Initial-initial colour connection}
Here we consider the inverse process $b+c\to a$ where $a$ is a colour singlet
of invariant mass $Q$ and $b,c$ are beam jets.  The kinematics are simple
because we take beam jets to be massless: in the c.m.\ frame
\begin{equation}
p_a = Q(1,\0t,0)\;,\;\;\; p_b= \half Q(1,\0t,1)
\;,\;\;\; p_c= \half Q(1,\0t,-1)\;.
\end{equation}
For emission of a gluon $g$ from $b$ we write
\begin{equation}\label{eq_iiqi}
q_i = \alpha_i p_b + \beta_i p_c + q_{\perp i}
\end{equation}
where $\qtr{g}=\kt$, $\qtr{a}=-\kt$, $\qtr{b}=\qtr{c}=\0t$.  Notice that
in this case the recoil transverse momentum is taken by the colour singlet
$a$ so we cannot preserve its 4-momentum.  We choose to preserve its mass
and rapidity, so that
\begin{equation}
\alpha_a = \beta_a =\sqrt{1+\kappa}\;,
\end{equation}
where as before $\kappa\equiv\kt^2/Q^2$.  Now we have
\begin{eqnarray}
\beta_b  &=& \alpha_c = 0\;,\;\;\; \alpha_g\beta_g = \kappa\;,\nonumber\\
\alpha_a &=& \alpha_b - \alpha_g\;,\;\;\;
\beta_a = \beta_c - \beta_g\;,
\end{eqnarray}
and our new variables in this case are
\begin{equation}\label{eq_iztk}
z=1-\frac{\alpha_g}{\alpha_b}\;,\;\;\;
\tk \equiv \frac{\tq^2}{Q^2} = \frac{\kappa}{(1-z)^2}\;.
\end{equation}
Thus we find
\begin{eqnarray}\label{eq_iialphas}
\alpha_a &=& \beta_a =\sqrt{1+(1-z)^2\tk}\;,\nonumber\\
\alpha_b &=& \frac{1}{z}\sqrt{1+(1-z)^2\tk}\;,\nonumber\\
\beta_c &=& \frac{1+(1-z)\tk}{\sqrt{1+(1-z)^2\tk}}\;.
\end{eqnarray}

\subsection{Phase space variables}
It is convenient to express the kinematics in terms of the ``reduced''
Mandelstam invariants:
\begin{equation}
\sb=(q_b+q_c)^2/Q^2\;,\;\;\; \tb=(q_b-q_g)^2/Q^2\;,\;\;\;
\ub=(q_c-q_g)^2/Q^2\;.
\end{equation}
The phase space limits are
\begin{equation}
1 < \sb < S/Q^2\;,\;\;\; 1-\sb < \tb < 0\;,\;\;\; \ub = 1-\sb-\tb
\end{equation}
where $S$ is the beam-beam c.m.\ energy squared.
In terms of the shower variables for beam jet $b$, we have
\begin{equation}
\sb = \alpha_b\beta_c =  \frac{1}{z}[1+(1-z)\tk]\;,\;\;\;
\tb = -\alpha_b\beta_g = -(1-z)\tk\;,\;\;\; \ub = -(1-z)\sb\;.
\end{equation}
Thus curves of constant $\tk$ in the $(\sb,\tb)$ plane are given by
\begin{equation}
\tb = \frac{\tk(1-\sb)}{\tk+\sb}
\end{equation}
and the Jacobian factor for conversion of the shower variables to the
Mandelstam invariants is
\begin{equation}
\frac{\partial(\sb,\tb)}{\partial(z,\tk)} = \frac{1-z}{z}\sb\;.
\end{equation}

\begin{figure}
\begin{center}
\input{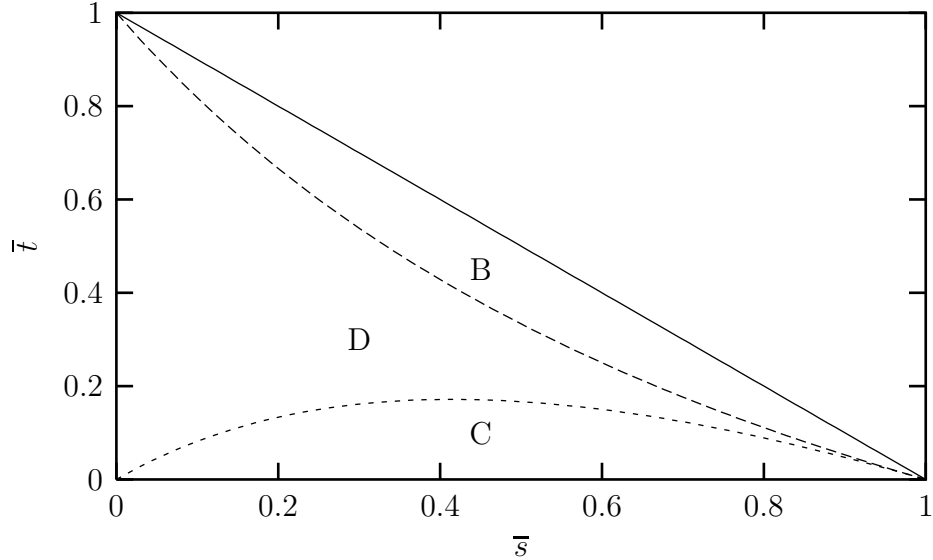}
\end{center}
\caption{Beam jets (B,C) and dead region (D) in initial-state branching.}
\label{fig_DYps}
\end{figure}

For the other beam jet $c$ we have $\tb\leftrightarrow\ub$ and thus
\begin{equation}
\tb = \frac{\sb(1-\sb)}{\tk+\sb}\;.
\end{equation}
We see that in order for the jet regions to touch without overlapping
in the soft limit $\sb\to 1$, $\tb\to 0$, we need $\tk <\tk_b$ in jet
$b$ and  $\tk <\tk_c$ in jet $c$, where $\tk_c = 1/\tk_b$.  The most
symmetrical choice is $\tk_c =\tk_b =1$, as shown in figure~\ref{fig_DYps},
but we can take $\tk_b$ or $\tk_c$ as large as we like.

\subsection{Example: Drell-Yan process}
Consider radiation from the quark in the Drell-Yan process,
$q\bar q\to g Z^0$.  In the laboratory frame we have
\begin{equation}
q_q = (Px_1,\0t,Px_1)\;,\;\;\;
q_{\qb} = (Px_2,\0t,-Px_2)
\end{equation}
where $P =\half\sqrt S$ is the beam momentum.  If we generated
the initial hard process  $q\bar q\to Z^0$ with momentum fractions
$x_q$, $x_\qb$ and we want to preserve the mass and
rapidity of the $Z^0$ we require
\begin{equation}
x_1 = x_q\alpha_b\;,\;\;\; x_2 = x_\qb\beta_c
\end{equation}
where $\alpha_b$ and $\beta_c$ are given by eqs~(\ref{eq_iialphas}).

The branching probability in the parton shower approximation is
\begin{equation}\label{eq_Pzk}
\frac{d^2 P}{dz\,d\tk}
= C_F\frac{\as}{2\pi}\,\frac{1}{\tk}\,\frac{1+z^2}{1-z}\;,
\end{equation}
which gives a differential cross section ($s=\sb Q^2$, etc.)
\begin{equation}\label{eq_dsigDYps}
\frac{1}{\sigma_0}\frac{d^2\sigma}{ds\,dt}=
\frac{D(x_1)D(x_2)}{D(x_q)D(x_\qb)}
\frac{\as}{2\pi}C_F\frac{s+u}{s^3 t u}
\left[s^2+(s+u)^2\right]
\end{equation}
where $\sigma_0$ is the Born cross section.
The functions $D(x_1)$ etc.\ are parton distribution functions in the
incoming hadrons; these factors take account of the change of
kinematics $x_q,x_\qb \to x_1,x_2$ discussed above.

The exact differential cross section for $q\bar q\to g Z^0$ to order $\as$ is
\begin{equation}\label{eq_dsigDYme}
\frac{1}{\sigma_0}\frac{d^2\sigma}{ds\,dt}=
\frac{D(x_1)D(x_2)}{D(x_q)D(x_\qb)}
\frac{\as}{2\pi}C_F\frac{Q^2}{s^3 t u}
\left[(s+t)^2+(s+u)^2\right]\;.
\end{equation}
Since $Q^2=s+t+u$ and $t\leq 0$, we see that the parton shower
approximation (\ref{eq_dsigDYps}) overestimates the exact expression,
becoming exact in the collinear or soft limit $t\to 0$.  Therefore the
gluon distribution in the jet regions can be corrected efficiently
by the rejection method, and the dead region can be filled using
the matrix element, as was done in \cite{Corcella:1999gs}.
The benefit of the new variables is that the angular distribution
of soft gluon emission requires no correction, provided the jet regions
touch without overlapping in the soft region.  As shown above, this will
be the case if the upper limits on $\tk$ satisfy $\tk_\qb = 1/\tk_q$.

\section{Initial-final colour connection}
Consider the process $a+b\to c$ where $a$ is a colour singlet and the
beam parton $b$ and outgoing parton $c$ are colour-connected.
An example is deep inelastic scattering,
where $a$ is a (charged or neutral) virtual gauge boson.
We need to preserve the 4-momentum of $a$ and therefore we work
in the Breit frame:
\begin{equation}
p_a = Q(0,\0t,-1)\;,\;\;\; p_b= \half Q(1+c,\0t,1+c)
\;,\;\;\; p_c= \half Q(1+c,\0t,-1+c)\;,
\end{equation}
where $p_a^2=-Q^2$, $p_b^2=0$, and $m_c^2=cQ^2$.  Notice that
the beam parton $b$ is always taken to be massless, but the
outgoing parton $c$ can be massive (e.g.\ in $W^+ d\to c$).

\subsection{Initial-state branching}\label{sec_ini_fin}
For emission of a gluon $g$ from the incoming parton $b$ we write
\begin{equation}
q_i = \alpha_i p_b + \beta_i n + q_{\perp i}
\end{equation}
where $\qtr{g}=\kt$, $\qtr{b}=\0t$, $\qtr{c}=-\kt$ and we choose
\begin{equation}\label{eq_ndef}
n = \half Q(1+c,\0t,-1-c)\;.
\end{equation}
To preserve $p_a=q_c+q_g-q_b$ we now require
\begin{equation}
\alpha_b-\alpha_c-\alpha_g = \beta_c+\beta_g-\beta_b= \frac{1}{1+c}
\end{equation}
whereas the mass-shell condition is
\begin{equation}
\alpha_i\beta_i\,Q^2(1+c)^2 = \qtr{i}^2 + q_i^2
\end{equation}
which gives
\begin{equation}\label{eq_iibetas}
1+c =\frac{c}{\alpha_c}+\kappa\left(\frac{1}{\alpha_c}+
\frac{1}{\alpha_g}\right)\;.
\end{equation}

The new variables for emission from the beam jet are as in
(\ref{eq_iztk}).  Substituting in (\ref{eq_iibetas}), we find
\begin{eqnarray}\label{eq_inifin_abs}
\alpha_b &=& \frac{1}{2z(1+c)}\left(1+c+(1-z)\tk +
\sqrt{[1+c+(1-z)\tk]^2-4z(1-z)\tk}\right)\;,\nonumber\\
\alpha_c &=& z\alpha_b - \frac{1}{1+c}\;,\;\;\;
\alpha_g = (1-z)\alpha_b\;,\;\;\;\beta_b=0\;,\nonumber\\
\beta_c &=& \frac{1}{1+c}\cdot
\frac{c+(1-z)^2\tk}{z(1+c)\alpha_b-1}\;,\;\;\;
\beta_g = \frac{(1-z)\tk}{(1+c)^2\alpha_b}\;.
\end{eqnarray}

\subsection{Final-state branching}\label{sec_fin_ini}
Next consider emission from the outgoing parton $c$. In this case we write
\begin{equation}
q_i = \alpha_i p_c + \beta_i p_b + q_{\perp i}
\end{equation}
To preserve $p_a=q_c+q_g-q_b$ we require
\begin{equation}
\alpha_c+\alpha_g-\alpha_b = \beta_b-\beta_c-\beta_g= 1
\end{equation}
whereas the mass-shell condition is now
\begin{equation}
\alpha_i\beta_i\,(Q^2+m_c^2) = \qtr{i}^2 + q_i^2 -\alpha_i^2 m_c^2\;.
\end{equation}
The new variables for emission from an outgoing parton are as in
eqs.~(\ref{eq_fztk},\ref{eq_fkt}) with $b$ replaced by $c$:
\begin{equation}\label{eq_fztkc}
z = \frac{\alpha_c}{\alpha_c+\alpha_g}\;,\;\;\;
\tk \equiv \frac{\tq^2}{Q^2}=\frac{1}{z^2}\left[
c+\frac{\kappa}{(1-z)^2}\right]\;.
\end{equation}
Thus in this case we find
\begin{eqnarray}\label{eq_ifalphas}
\alpha_b &=& 0\;,\;\;\; \alpha_c = z\;,\;\;\; \alpha_g = 1-z\;,\nonumber\\
\beta_b &=& \frac{1}{1+c}[1+c+z(1-z)\tk]\;,\nonumber\\
\beta_c &=& \frac{1-z}{1+c}[2c+z(1-z)\tk]\;,\nonumber\\
\beta_g &=& \frac{1-z}{1+c}[z^2\tk -2c]\;.
\end{eqnarray}

\subsection{Phase space variables}\label{sec_inifin_var}
In this process the invariant phase space variables are usually taken to be
\begin{equation}
x_p = \frac{Q^2}{2p_a\cdot q_b}\;,\;\;\;
z_p = \frac{q_c\cdot q_b}{p_a\cdot q_b}\;.
\end{equation}
In terms of the new variables for emission from the beam parton, we have
\begin{eqnarray}
x_p &=& \frac{1}{(1+c)\alpha_b} = 2z
\left(1+c+(1-z)\tk +\sqrt{[1+c+(1-z)\tk]^2-4z(1-z)\tk}\right)^{-1} \\
z_p &=& (1+c)\beta_c = \frac 12
\left(1-c-(1-z)\tk +\sqrt{[1+c+(1-z)\tk]^2-4z(1-z)\tk}\right)\;,
\end{eqnarray}
with the Jacobian
\begin{equation}\label{eq_inifin_jac_ini}
\frac{\partial(x_p,z_p)}{\partial(z,\tk)} = \frac{1}{\tk}
\left(\frac{1}{x_p}+\frac{1+c}{1-z_p}-2\right)^{-1}\;.
\end{equation}
In the soft limit $z=1-\eps$ we therefore find for the beam jet
\begin{equation}\label{eq_inifin_sof_ini}
x_p \sim \frac{1}{1+c}\left[1-\eps -\frac{\eps c\tk}{(1+c)^2}\right]
\;,\;\;\; z_p \sim 1-\frac{\eps \tk}{1+c}
\end{equation}
and
\begin{equation}\label{eq_inifin_jac_sof}
\frac{\partial(x_p,z_p)}{\partial(z,\tk)} \sim \frac{\eps}{(1+c)^2}\;.
\end{equation}

\begin{figure}
\begin{center}
\input{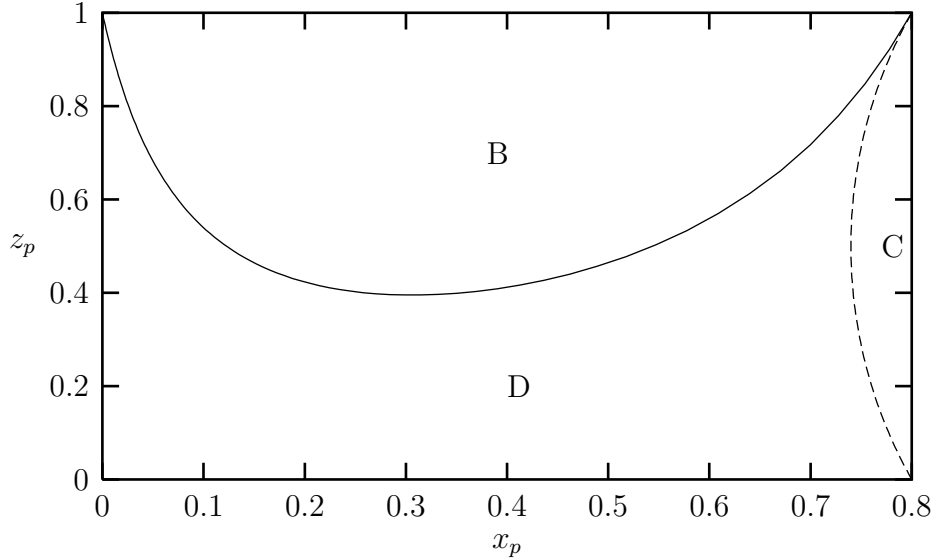}
\end{center}
\caption{Beam jet (B), outgoing jet (C) and dead region (D) in initial-final
state branching: $c=0.25$, $\tk_b=10$, $\tk_c=0.40625$.}
\label{fig_inifin_10}
\end{figure}

In terms of the variables for emission from the outgoing parton,
\begin{equation}
x_p =\frac{1}{(1+c)\beta_b} = \frac{1}{1+c+z(1-z)\tk}\;,\;\;\;
z_p = \alpha_c = z\;,
\end{equation}
so the Jacobian is simply
\begin{equation}\label{eq_inifin_jac_fin}
\frac{\partial(x_p,z_p)}{\partial(z,\tk)} = z(1-z)x_p^2\;,
\end{equation}
and in the soft limit
\begin{equation}\label{eq_inifin_sof_fin}
x_p \sim  \frac{1}{1+c}\left[1-\frac{\eps \tk}{1+c}\right]\;,\;\;\;
z_p \sim 1-\eps\;,
\end{equation}
with the Jacobian again given by (\ref{eq_inifin_jac_sof}).
For full coverage of phase space in the soft limit we require
$\tk<\tk_b$ in jet $b$ and $\tk<\tk_c$ in jet $c$, where
\begin{equation}\label{eq_inifin_kbc}
\tk_b (\tk_c -c) = (1+c)^2\;.
\end{equation}
Thus the most symmetrical choice is $\tk_b=1+c$, $\tk_c=1+2c$, as shown in
figure~\ref{fig_inifin_1}.  On the other hand, any larger or smaller
combination satisfying (\ref{eq_inifin_kbc}) is allowed, as
illustrated in figure~\ref{fig_inifin_10} for $\tk_b=10$.
\begin{figure}
\begin{center}
\input{figs/chapter2/ini-fin.tex}
\end{center}
\caption{Beam jet (B), outgoing jet (C) and dead region (D) in initial-final
state branching: $c=0.25$, $\tk_b=1.25$, $\tk_c=1.5$.}
\label{fig_inifin_1}
\end{figure}

\subsection{Example: deep inelastic scattering}
Consider deep inelastic scattering on a hadron of momentum
$P^\mu$ by exchange of a virtual photon of momentum $q^\mu$.
If the contribution to the Born cross section from scattering
on a quark of momentum fraction $x_B=Q^2/2P\cdot q$ is
represented by $\sigma_0$ (a function of $x_B$ and $Q^2$),
then the correction due to single gluon emission is given by
\begin{equation}\label{eq_dsig_dis}
\frac{1}{\sigma_0}\frac{d^2\sigma}{dx_p\,dz_p}=
\frac{C_F\as}{2\pi}\frac{D(x_B/x_p)}{D(x_B)}
\frac{1+(x_p+z_p-1)^2}{x_p(1-x_p)(1-z_p)}\;.
\end{equation}

In the soft limit $x_p,z_p\to 1$ we have, from
eqs.~(\ref{eq_inifin_sof_ini},\ref{eq_inifin_sof_fin}) with $c=0$,
\begin{equation}
(1-x_p)(1-z_p)\sim \eps^2\tk
\end{equation}
and so
\begin{equation}\label{eq_dsig_dis_sof}
\frac{1}{\sigma_0}\frac{d^2\sigma}{dx_p\,dz_p}\sim
\frac{C_F\as}{\pi}\frac{1}{\eps^2\tk}\;,
\end{equation}
whereas the parton shower approximation gives
\begin{equation}\label{eq_dsig_dis_ps}
\frac{1}{\sigma_0}\frac{d^2\sigma}{dz\,d\tk}\sim
\frac{C_F\as}{\pi}\frac{1}{\eps\tk}\;.
\end{equation}
Since the Jacobian factor (\ref{eq_inifin_jac_ini}) or
(\ref{eq_inifin_jac_fin}) in this limit is simply $\eps$, the
shower approximation is exact in the soft limit.

\subsection{Example: $q\qb \to t\bar t$}
We denote the momenta in this process by $p_a+p_b \to p_c+p_d$
and the $2\to 2$ invariants by
\begin{equation}
\sb =  2p_a\cdot p_b\;,\;\;\;
\tb = -2p_a\cdot p_c\;,\;\;\;
\ub = -2p_a\cdot p_d\;,
\end{equation}
so that $\sb+\tb+\ub =0$.
Colour flows from $q$ to $t$ and anticolour from $\qb$ to $\bar t$.
Therefore the momentum transfer $q=p_a-p_c = p_d-p_b$ is carried
by a colour singlet and we preserve this 4-momentum during
showering.

For emission from the incoming light quark or the outgoing top 
quark, we work in the Breit frame for this system, where
\begin{equation}\label{eq_qqtt_q}
q = Q(0,\0t,1)\;,\;\;\; p_a= \half Q(1+c,\0t,1+c)
\;,\;\;\; p_c= \half Q(1+c,\0t,-1+c)
\end{equation}
with $Q^2=-\tb-m_t^2$ and $c=m_t^2/Q^2$. Then the treatment
of sects.~\ref{sec_ini_fin} and \ref{sec_fin_ini} can be applied
directly, with the substitution $b\to a$ since the emitting
system is now $(a,c)$ rather than $(b,c)$.  However, the
phase space variables are no longer those of sect.~\ref{sec_inifin_var}
since they involve the momenta of the $\qb$ and $\bar t$,
which in the frame (\ref{eq_qqtt_q}) take the general form
\begin{eqnarray}
p_b &=& [\half Q\sqrt{(1+c)^2+4K},\Qt,-\half Q(1+c)]\;,\nonumber\\
p_d &=& [\half Q\sqrt{(1+c)^2+4K},\Qt,\half Q(1-c)]\;,
\end{eqnarray}
where $K=\Qt^2/Q^2$ is related to the $2\to 2$ invariants:
\begin{equation}
\sb = \half Q^2 (1+c)^2\left[1+\sqrt{1+\frac{4K}{(1+c)^2}}\right]\;,\;\;\;
\tb = -Q^2 (1+c) \;,\;\;\; \ub = -\sb-\tb\;,
\end{equation}
and so
\begin{equation}
\Qt^2 = -\sb\left[1+\frac{\sb}{(1+c)\tb}\right]\;.
\end{equation}

For emission from the incoming light quark we define as in
sect.~\ref{sec_ini_fin}
\begin{equation}
q_i = \alpha_i p_a + \beta_i n + q_{\perp i}
\end{equation}
for $i=a,c,g$, where $\qtr{a}=\0t$, $\qtr{c}=-\kt$, $\qtr{g}=\kt$,
and $n$ is as in (\ref{eq_ndef}).
Then the $\alpha_i$'s and $\beta_i$'s are given by
eqs.~(\ref{eq_inifin_abs}) with the substitution $b\to a$. The
light antiquark and the antitop are not affected and therefore
$q_b=p_b$, $q_d=p_d$. This allows the complete kinematics of
the $2\to 3$ process to be reconstructed.   The $2\to 3$
invariants can be defined as in ref.~\cite{Frixione:2003ei}:
\begin{equation}
s=2q_a\cdot q_b\;,\;\;\;
t_1 = -2q_a\cdot q_c\;,\;\;\;
t_2 = -2q_b\cdot q_d\;,\;\;\;
u_1 = -2q_a\cdot q_d\;,\;\;\;
u_2 = -2q_b\cdot q_c\;.
\end{equation}
It is convenient to express $n=p_c-cq$ so that (for $i=a,c,g$)
\begin{equation}
q_i = (\alpha_i -c\beta_i)p_a + (1+c)\beta_i p_c + q_{\perp i}\;.
\end{equation}
Then we find
\begin{equation}\label{eq_qqtt_ini}
s = \alpha_a\sb\;,\;\;\;
t_1 = \alpha_a\beta_c(1+c)\tb\;,\;\;\;
t_2 = \tb\;,\;\;\;
u_1 = \alpha_a\ub\;,\;\;\;
u_2 = \beta_c(\ub-c\tb)-\alpha_c\sb -2 \kt\cdot\Qt\;.
\end{equation} 

For emission from the outgoing top we use the results of
sect.~\ref{sec_fin_ini}, again with the substitution $b\to a$.
Thus we now have for $i=a,c,g$
\begin{equation}
q_i = \alpha_i p_c + \beta_i p_a + q_{\perp i}
\end{equation}
where the $\alpha_i$'s and $\beta_i$'s are given by
eqs.~(\ref{eq_ifalphas}) with $b\to a$, and we find that
\begin{equation}\label{eq_qqtt_fin}
s = \beta_a\sb\;,\;\;\;
t_1 = \alpha_c\beta_a\tb\;,\;\;\;
t_2 = \tb\;,\;\;\;
u_1 = \beta_a\ub\;,\;\;\;
u_2 = \alpha_c\ub-\beta_c\sb -2 \kt\cdot\Qt\;.
\end{equation}

Similar formulae to eqs.~(\ref{eq_qqtt_ini}) and (\ref{eq_qqtt_fin}),
with the replacements $a\to b$ and $c\to d$, will hold for the case of
gluon emission from the colour-connected $(\qb\bar t)$ system.
Using these relations, one can study the distribution of gluon
radiation in the parton shower approximation and compare it with the
exact $q\qb\to t\bar t g$ matrix element. Agreement will be good in
the soft and/or collinear regions but there will be regions of hard,
wide-angle gluon emission in which matrix element corrections should
be applied.  Alternatively, the above equations can be used to formulate
a modified subtraction scheme for combining fixed-order and parton-shower
results, as was done in ref.~\cite{Frixione:2003ei} for a different
parton-shower algorithm.

\section{Decay colour connection}
Consider the process $b\to ca$ where $a$ is a colour singlet and the
decaying parton $b$ and outgoing parton $c$ are colour-connected.
Examples are bottom quark decay, $b\to cW^*$, and top decay, $t\to bW$.
Here we have to preserve the  4-momentum of the decaying parton $b$
and therefore we work in its rest frame,
\begin{equation}\label{eq_dec_pbcg}
p_b = m_b(1,\0t,0)\;,\;\;\; p_c= \half m_b(1-a+c,\0t,\lambda)
\;,\;\;\; p_a= \half m_b(1+a-c,\0t,-\lambda)\;,
\end{equation}
where $a=m_a^2/m_b^2$, $c=m_c^2/m_b^2$ and now
\begin{equation}
\lambda = \lambda(1,a,c) = \sqrt{(1+a-c)^2-4a} = \sqrt{(1-a+c)^2-4c}\;.
\end{equation}

\subsection{Initial-state branching}
For emission of a gluon $g$ from the decaying parton $b$ we write
\begin{equation}\label{eq_decqi}
q_i = \alpha_i p_b + \beta_i n + q_{\perp i}
\end{equation}
where $\qtr{g}=\kt$, $\qtr{c}=-\kt$, $\qtr{b}=\0t$ and we choose
\begin{equation}
n = \half m_b(1,\0t,1)\;,
\end{equation}
i.e.\ aligned along $p_c$ in the rest frame of $b$. The mass-shell
conditions give
\begin{equation}
\beta_a = \frac{a}{\alpha_a}-\alpha_a\;,\;\;\;\;
\beta_c = \frac{c+\kappa}{\alpha_c}-\alpha_c\;,\;\;\;\;
\beta_g = \frac{\kappa}{\alpha_g}-\alpha_g\;,
\end{equation}
with $\kappa=\kt^2/m_b^2$. From momentum conservation
\begin{equation}\label{eq_mom}
\alpha_a + \alpha_c + \alpha_g = 
\frac{a}{\alpha_a}+\frac{c+\kappa}{\alpha_c}+\frac{\kappa}{\alpha_g} =1\;.
\end{equation}
Recall that in initial-state branching of a heavy object
our new evolution variable is given by (\ref{eq_dec_tqi}),
so we have
\begin{equation}
\alpha_g=1-z\;,\;\;\;\; \kappa = (\tk-1)(1-z)^2
\end{equation}
where $\tk = \tq^2/m_b^2 >1$.  Introducing for brevity the notation
\begin{equation}\label{eq_wuv}
w = 1-(1-z)(\tk-1)\;,\;\;\;
u = 1+a-c-(1-z)\tk\;,\;\;\;
v = \sqrt{u^2-4awz}\;,
\end{equation}
from (\ref{eq_mom}) we find
\begin{equation}\label{eq_alphaac}
\alpha_a = \frac{u+v}{2w}\;,\;\;\;
\alpha_c = 1-\alpha_a -\alpha_g = z -\alpha_a\;.
\end{equation}

\subsection{Final-state branching}
For radiation from the outgoing parton $c$ we write
\begin{equation}
q_i = \alpha_i p_c + \beta_i n + q_{\perp i}
\end{equation}
where $p_c$ is given by (\ref{eq_dec_pbcg}).
Since the colour-connected parton $b$ is at rest in our working frame
of reference, the choice of the light-like vector $n$ in this case is
somewhat arbitrary.  By analogy with the cases treated earlier, we
choose it to be opposite to that used for the radiation from $b$,
i.e.\ along the direction of the colour singlet $a$: 
\begin{equation}
n = \half m_b(\lambda,\0t,-\lambda)\;.
\end{equation}
The kinematics are then identical with those for final-final connection
(sect.~\ref{sec_finfin}), with the replacement $b\to c$, $c\to a$.

\subsection{Phase space variables}
As in sect.~\ref{sec_finfin}, it is convenient to use the
Dalitz plot variables, which in this case are 
\begin{equation}
x_i = \frac{2 q_i\cdot p_b}{m_b^2}\;.
\end{equation}
For emission from the decaying parton $b$ we have
$x_i = 2\alpha_i +\beta_i$ and hence, from (\ref{eq_alphaac}),
\begin{equation}\label{eq_xacg}
x_a = \frac{u+v}{2w} + \frac{u-v}{2z}\;,\;\;\;
x_c = w+z-x_a\;,\;\;\; x_g = 2-w-z = (1-z)\tk\;,
\end{equation}
with the Jacobian factor
\begin{equation}
\frac{\partial(x_a,x_g)}{\partial(z,\tk)} =
(1-z)\left[\frac{u+v}{2w^2}-\frac{u-v}{2z^2}+\frac{a(w-z)^2}{vwz}\right]\;.
\end{equation}
 
In the soft limit $z\to 1-\eps$ we find
\begin{equation}\label{eq_dec_xag}
x_a\sim 1+a-c-\eps\tk'_b\;,\;\;\;
x_g\sim \eps\tk_b
\end{equation}
where
\begin{equation}\label{eq_dec_tkp}
\tk'_b = \lambda+\frac{\tk_b}{2}(1-a+c-\lambda)\;.
\end{equation}

For emission from the outgoing parton $c$ we have, from (\ref{eq_xcbg})
with the replacement $b\to c$, $c\to a$:
\begin{eqnarray}\label{eq_decfin}
x_a &=& 1+a-c- z(1-z)\tk \nonumber \\
x_c &=& (2-x_a)r + (z-r)\sqrt{x_a^2-4a}\\
x_g &=& (2-x_a)(1-r) - (z-r)\sqrt{x_a^2-4a}\nonumber
\end{eqnarray}
where
\begin{equation}\label{eq_decr}
r = \frac 12\left(1+\frac{c}{1+a-x_a}\right)\;.
\end{equation}
In the soft limit we have from (\ref{eq_soft})
\begin{equation}
x_a\sim 1+a-c-\eps\tk_c\;,\;\;\;
x_g \sim \eps\tk'_c
\end{equation}
where
\begin{equation}
\tk'_c = \lambda+\frac{\tk_c}{2c}(1-a+c-\lambda)\;.
\end{equation}
For full coverage of the soft region we require
\begin{equation}
\frac{\tk_b}{\tk'_b} = \frac{\tk'_c}{\tk_c}
\end{equation}
which gives in this case
\begin{equation}\label{eq_tkdec}
(\tk_b -1)(\tk_c -c) =  \frac 14 (1-a+c+\lambda)^2\;.
\end{equation}
Note that, while there is no upper limit on $\tk_b$, the largest value
that can be chosen for $\tk_c$ is given by the equivalent of
(\ref{eq_fftkmax}),
\begin{equation}\label{eq_dectkmax}
\tk_c < 4(1+a-2\sqrt{c} -c)\;.
\end{equation}

\subsection{Example: top decay}
In the decay $t\to Wbg$ we have $a=m_W^2/m_t^2=0.213$ and
$c=m_b^2/m_t^2=0.026$, so for simplicity we neglect $c$.
Then for radiation from the top we have from (\ref{eq_xacg})
\begin{equation}
x_W = \frac{u+v}{2w} + \frac{u-v}{2z}\;,\;\;\;
x_g = (1-z)\tk\;,
\end{equation}
where $u,v,w$ are given by eqs.~(\ref{eq_wuv}) with $c\to 0$.
The phase space is the region
\begin{equation}
0 <  x_g < 1-a \;,\;\;\;\;
1-x_g+\frac{a}{1-x_g} < x_W < 1+a\;.
\end{equation}
Notice that for real $x_W$ we require $u^2>4awz$, i.e.
\begin{equation}
1<\tk<1 + a\left[1-\sqrt{\frac{z(1-a)}{a(1-z)}}\right]^2\;,
\end{equation}
and
\begin{equation}
1- \frac{1-a}{\tk+2\sqrt{a(\tk-1)}} < z < 1\;.
\end{equation}
Thus there is no upper limit on $\tk$, but the range of $z$ becomes
more limited as $\tk$ increases.

\begin{figure}
\begin{center}
\epsfig{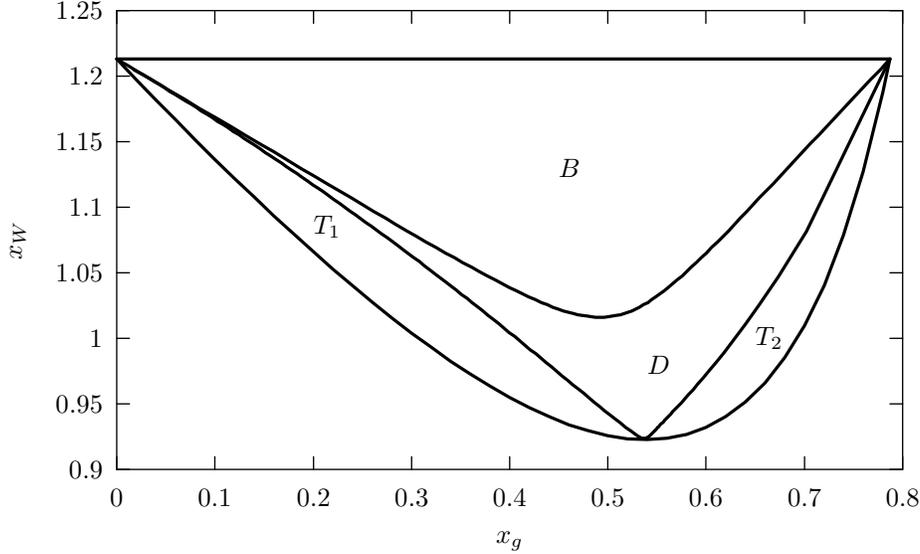}
\end{center}
\caption{Phase space for decay $t\to Wbg$, with symmetric choice of
emission regions for the $b$ ($B$) and the $t$ ($T_1$,$T_2$),
and the dead region ($D$).}
\label{fig_tdec_sym}
\end{figure}

For radiation from the $b$ we have from (\ref{eq_decfin})
\begin{eqnarray}\label{eq_topb}
x_W &=& 1+a- z(1-z)\tk \nonumber \\
x_g &=& \half(2-x_W) - (z-\half)\sqrt{x_W^2-4a}\;.
\end{eqnarray}
To cover the soft region we require $\tk<\tk_t$ for emission from
the top quark and $\tk<\tk_b$ for that from the bottom, where
(\ref{eq_tkdec}) gives
\begin{equation}\label{eq_tkb_tkt}
\tk_b = \frac{(1-a)^2}{\tk_t-1}\;.
\end{equation}
The most symmetrical choice would therefore appear to be
$\tk_b =\tk_t-1 = 1-a = 0.787$, as illustrated in figure~\ref{fig_tdec_sym}.

\begin{figure}
\begin{center}
\epsfig{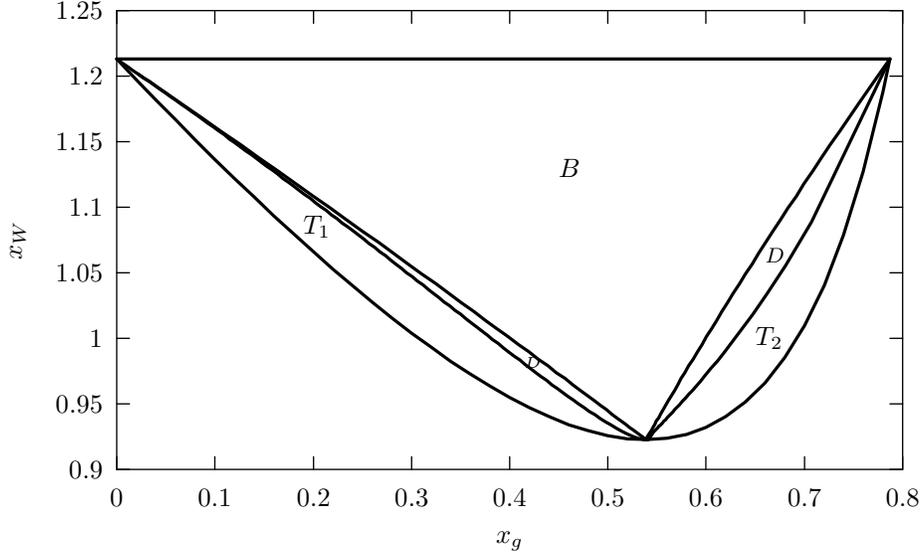}
\put(-150.0,70.0){\tiny{$D$}}
\put(-59.,110.0){\scriptsize{$D$}}
\end{center}
\caption{Phase space for decay $t\to Wbg$, with maximal region ($B$)
for emission from the $b$, together with complementary regions of
emission from the $t$ ($T_1$,$T_2$) and the dead region ($D$).}
\label{fig_tdec_brad}
\end{figure}

As mentioned above, there is no upper limit on $\tk_t$.  Thus the region
covered by gluon emission from the top quark can be as large as we like.
However, (\ref{eq_dectkmax}) tells us that the upper limit
for radiation from the $b$ is
\begin{equation}\label{eq_tkbmax}
\tk_b < 4 (1-\sqrt a)^2 = 1.16\;,
\end{equation}
and correspondingly $\tk_t > 1+\frac 14 (1+\sqrt a)^2 = 1.53$.
Figure \ref{fig_tdec_brad} shows this maximal region that can be covered
by emission from the $b$, together with the complementary regions of
emission from the $t$.

We note from figs.~\ref{fig_tdec_sym} and \ref{fig_tdec_brad} that,
for any value of $\tk_t$, the region for emission from the top quark
consists of two distinct parts that touch at the point
$x_g=1-\sqrt a$, $x_W=2\sqrt a$, where the $W$ boson is at rest:
a subregion $T_1$ which includes the soft limit $x_g\to 0$ and a hard
gluon region $T_2$.

The exact $t\to Wbg$ differential decay rate to first order in $\as$
is given in \cite{Corcella:1998rs}:
\begin{eqnarray}
\frac{1}{\Gamma_0}\frac{d^2\Gamma}{dx_W dx_g} &=& \frac{\as}{\pi}\frac{C_F}
{(1+a-x_W)x_g^2}\Biggl\{x_g-\frac{(1+a-x_W)(1-x_g)+x_g^2}{1-a}\nonumber\\
&&+x_g\frac{(x_W+x_g-1-a)^2}{2(1-a)^2}
+\frac{2a(1+a-x_W)x_g^2}{(1-a)^2(1+2a)}\Biggr\}
\end{eqnarray}
where $\Gamma_0$ is the lowest-order decay rate.  In the soft region
$x_W\to 1+a$, $x_g\to 0$ this becomes
\begin{equation}
\frac{1}{\Gamma_0}\frac{d^2\Gamma}{dx_W dx_g}\sim \frac{\as}{\pi}\frac{C_F}
{x_g}\left[\frac{1}{1+a-x_W}-\frac{1}{(1-a)x_g}\right]\;.
\end{equation}

For soft gluon emission ($1-z=\eps\to 0$) from the top quark we have from
eqs.~(\ref{eq_dec_xag},\ref{eq_dec_tkp})
\begin{equation}
x_W\sim 1+a-\eps(1-a)\;,\;\;\;
x_g\sim \eps\tk
\end{equation}
and so the exact form of the soft gluon distribution is
\begin{equation}
\frac{1}{\Gamma_0}\frac{d^2 \Gamma}{dx_g dx_W} \sim \frac{\alpha_s}{\pi}
\frac{C_F}{\eps^2} f_t(\tk)
\end{equation}
where
\begin{equation}
f_t(\tk)= \frac{\tk-1}{(1-a)\tk^2}\;.
\end{equation}
In the same region the parton shower approximation (\ref{eq_dPqqg}) gives
\begin{equation}
\frac{d^2 P}{dx_g dx_W} \sim \frac{1}{(1-a)\eps}\frac{d^2 P}{dz\,d\tk}
\sim \frac{\alpha_s}{\pi}\frac{C_F}{(1-a)\epsilon^2\tk}\left(1-
\frac{1}{\tk}\right)=\frac{\alpha_s}{\pi}\frac{C_F}{\eps^2}f(\tk)\;.
\end{equation}
Thus we see that, for emission from the top quark,
the shower approximation is exact in the soft limit.  At higher
gluon energies, inside the region $T_1$ the parton shower
overestimates the exact matrix element and can therefore be
corrected easily by the rejection method.  In the hard gluon region
$T_2$, which contributes only a small finite correction to the cross
section, the parton shower overestimates the matrix element
at lower values of $x_g$ but underestimates it at the highest values.
Therefore a combination of rejection and matrix element correction
is needed in this region.

For emission from the bottom quark in the soft limit, we use the results
of sect.~\ref{sec_finfin_sof} with the substitution $b\to 0$, $c\to a$
to obtain
\begin{equation}
x_W\sim 1+a-\eps\tk\;,\;\;\;
x_g\sim \eps\left(1-a+\frac{\tk}{1-a}\right)\;.
\end{equation}
Therefore the exact soft gluon distribution in the $b$ jet should be
\begin{equation}
\frac{1}{\Gamma_0}\frac{d^2 \Gamma}{dx_g dx_W} \sim \frac{\alpha_s}{\pi}
\frac{C_F}{\eps^2} f_b(\tk)
\end{equation}
where
\begin{equation}
f_b(\tk)= \frac{1}{(1-a)\tk}\left[1+\frac{\tk}{(1-a)^2}\right]^{-2}\;.
\end{equation}
On the other hand the parton shower approximation in this case gives simply
\begin{equation}
\frac{d^2 P}{dx_g dx_W} \sim \frac{1}{(1-a)\eps}\frac{d^2 P}{dz\,d\tk}
\sim \frac{\alpha_s}{\pi}\frac{C_F}{(1-a)\epsilon^2\tk}\;.
\end{equation}
Thus the soft gluon distribution in the $b$ jet region is overestimated
by a factor of
\begin{equation}
\left[1+\frac{\tk}{(1-a)^2}\right]^2\;,
\end{equation}
which can be corrected by the rejection method.  This factor varies from
1 to 5.2 for the symmetric choice of the $b$ jet region $\tk_b=1-a$
depicted in figure~\ref{fig_tdec_sym}. For the maximal  $b$ jet shown
in figure~\ref{fig_tdec_brad}, it rises to 8.3.  Since the shower
approximation is exact in the soft limit for emission from the top,
one can reduce the amount of soft correction required by decreasing the
$b$ jet region and increasing that for top emission,
in accordance with (\ref{eq_tkb_tkt}).  However, for large
values of $\tk_t$ the dead region moves near to the collinear
singularity at $x_W=1+a$ and a large hard matrix element correction
becomes necessary.
  
\section{Conclusions}
A new formulation of the parton-shower approximation to QCD matrix elements has
been presented. This formalism offers a number of advantages
over previous ones.  Direct angular ordering of the shower ensures
a good emulation of important QCD coherence effects, while the
connection between the shower variables and the Sudakov-like
representation of momenta (\ref{eq_qi}) simplifies the
kinematics and their relation to phase space invariants.
The use of mass-dependent splitting functions with the
new variables allows an accurate description of soft gluon
emission from heavy quarks over a wide angular region, including
the collinear direction. The separation of showering into
contributions from pairs of colour-connected hard partons
permits a general treatment of coherence effects, which should be
reliable at least to leading order in the number of colours.
Since the formulation is slightly different for initial-
and final-state showering, formulae for all colour-connected combinations 
of incoming and outgoing partons have been given.

This new shower formulation is a key element of the event generator 
\HWP\ \cite{Gieseke:2003_20}. Chapter~\ref{chap:Herwig} and 
\ref{chap:Results} describe how this new formalism is implemented and
complete results for $e^+e^-$ annihilation are presented.

\chapter{Hadronization}
\label{chap:Hadronization}

Hadronization is the process in which the perturbative partons (quarks and 
gluons) from the shower enter the non-perturbative phase and are converted 
into the observed hadrons. This is not well understood and instead is
modelled by a hadronization model. As discussed earlier there are
a few different types of hadronization models. Pythia~\cite{Pythia} uses a 
string fragmentation model whereas \HW~6.5\cite{Herwig} uses a
cluster hadronization model. Here I describe this cluster model and the 
modifications that have been made in the new \HWP~hadronization model.

\section{Cluster Formation}
\label{sec:had}
The shower terminates and leaves colour connected pairs with low virtuality. 
These partons need to combine to form hadrons. The first step of the cluster
hadronization is to form clusters out of these colour connected particles. 
These clusters are made up of quark--anti-quark pairs or 
(anti-)diquark-- (anti-)quark pairs. This section presents  each step of the 
process of cluster formation up until the decay of the clusters into hadrons.

\subsection{Gluon Splitting}
The initial step of the cluster hadronization is to split the gluons into
quark--anti-quark pairs. Since at the end of the shower the gluons are put 
on-shell, they must be given a mass, $m_g$, in order to decay into a 
quark--anti-quark pair. The default value of this mass is $0.750$ GeV. This is 
then high enough to isotropically decay into $u\bar{u}$ and $d\bar{d}$ pairs. 
Each gluon is randomly assigned a decay into either $u$ quarks or $d$ quarks 
and is decayed uniformly in $\cos \theta$ and $\phi$. After the isotropic 
decays the event is left with only colour connected (di)quarks and 
anti-(di)quarks.

\subsection{Cluster Formation}
Clusters are themselves colour-singlets. They can be made up of either two 
partons (quark--anti-quark pair) or of three partons (quarks or anti-quarks 
all of different colours). In the cluster model of \HWP{} a three parton 
cluster will occur in baryon non-conserving events, which have colour sources 
or sinks, and in events where a beam particle is a baryon. 
This can be seen because a cluster composed of three partons has baryon number 
$\pm 1$. Since drawing from the vacuum doesn't change the baryon number, the
decay of these clusters must have baryon number $\pm 1$. Unless the
cluster is derived from a beam baryon, these types of clusters must only occur
in baryon non-conserving events.

A cluster is formed simply by finding a colour-singlet pair (or triplet) of
partons. The 4-momentum of the cluster is just the sum of the momenta of the 
partons. Clusters are created for all sets of colour singlets. If a quark 
or anti-quark is created from a colour source or sink (baryon-violating)
it forms a three parton cluster with its colour neighbours. Two of these 
partons are randomly combined into a diquark for the purpose of the cluster
decay.

\begin{figure}
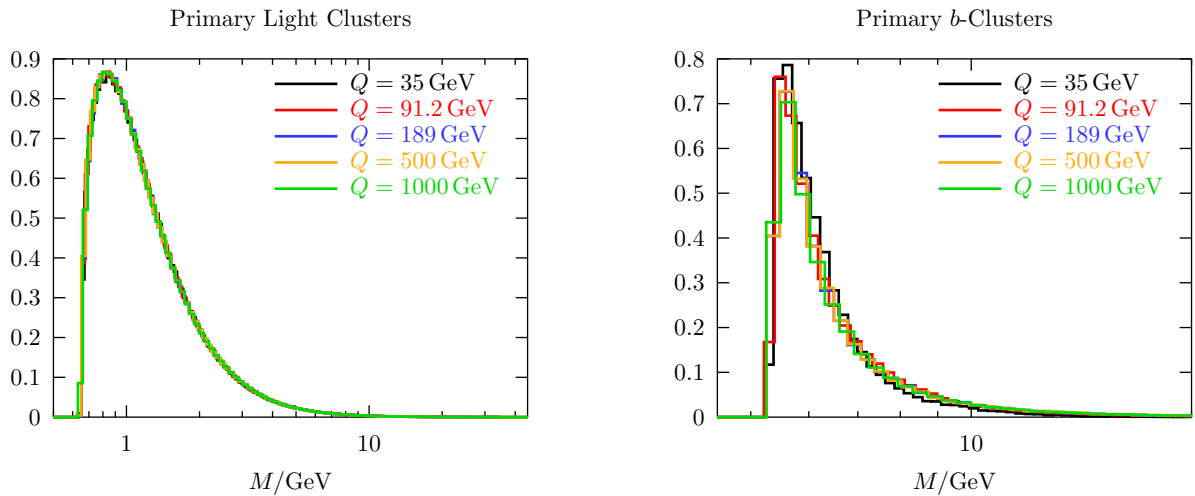

  \epsfig{file=figs/chapter3/cluana3.8,scale=0.9}\hfill
  \epsfig{file=figs/chapter3/cluana3.6,scale=0.9}
  \caption{Primary cluster mass distribution in the $\ee$ annihilation
    at various centre-of-mass energies, $Q$, for clusters containing only
    light quarks (left) and a $b$ quark (right).}
\label{plotc1}
\end{figure}

The principle of colour-preconfinement says that the invariant mass 
distribution of the clusters is independent of the centre-of-mass energy. 
This idea is needed to fully separate the perturbative regime from
the non-perturbative regime. Figure~\ref{plotc1} shows this distribution for
\HWP. The first plot in figure \ref{plotc1} shows the distribution for the light
clusters. The second plot in figure~\ref{plotc1} shows the 
distribution for the $b$ quarks only. This shows that the fall off of the
distributions is similar for each flavour.

\subsection{Cluster Fission}
\label{sec:clusterfission}
The mass of a cluster is given by $M^2=p^2$ of the cluster. In order for the
shower to combine more smoothly with the hadronization, clusters of a large 
mass are decayed into two new clusters. If the shower were to be cut
off at a larger scale then there would be fewer more energetic partons. On
the other hand if it were to be cut off at a smaller scale it would produce
many lower energy partons. Splitting the clusters with large mass into two 
clusters with smaller mass allows the hadron multiplicity to be much less 
variable with the shower cutoff. In turn this allows the tuning of the shower
to not be as dependant on the hadron multiplicity, making for more consistent
results.

A cluster is split into two clusters if the mass does not satisfy the condition
\begin{equation}
\label{eqn:thresh}
M^{\Clp} < \Clm^{\Clp} + \Sigma_c^{\Clp},
\end{equation}
where $\Clp$ and $\Clm$ are parameters and $\Sigma_c$ is the sum of 
the masses of the constituents that make up the cluster.

\begin{figure}
\centering
\includegraphics[width=5cm]{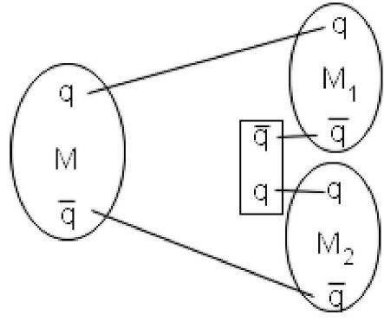}
\caption{Cluster Fission. Cluster of mass $M$ decays into two new clusters
of masses $M_1$ and $M_2$ by drawing a pair from the vacuum.}
\label{fig:fission}
\end{figure}

If a cluster is to be split a $q \bar q$ pair is chosen from the vacuum.
Only $u,d$ or $s$ flavours are chosen with probabilities given by the 
parameters $\texttt{Pwt}_i$ with $i$ being the flavour. Once a pair is chosen 
from the vacuum the cluster is decayed into two new clusters with one of the 
original partons in each cluster. A schematic of this decay is shown in 
figure~\ref{fig:fission}. In the case where the partons are not beam remnants, 
the mass distribution of the new clusters are given by
\begin{eqnarray}
M_1 &=& m_1 + \left( M - m_1 - m_3 \right) r_1^{1/P}, \\
M_2 &=& m_2 + \left( M - m_2 - m_3 \right) r_2^{1/P}
\end{eqnarray}
where $m_1$ and $m_2$ are the two components of the original cluster, $m_3$ is
the mass of the parton drawn from the vacuum, $M_1$, $M_2$ are the new cluster
masses and $r_i$ are random numbers in the interval $[0,1]$. $P$ is another 
parameter of the model. If the parton is not a $b$ quark then $P$ is 
$\texttt{PSplt1}$ otherwise it is $\texttt{PSplt2}$ (see 
table~\ref{table:HadParams}). Points are only chosen in the range confined by
\begin{equation}
\label{eq:fission_constraint}
M>M_1 + M_2;~~~M_1>m_1+m_3;~~~M_2>m_2 + m_3.
\end{equation}

\begin{figure}
\centering
\epsfig{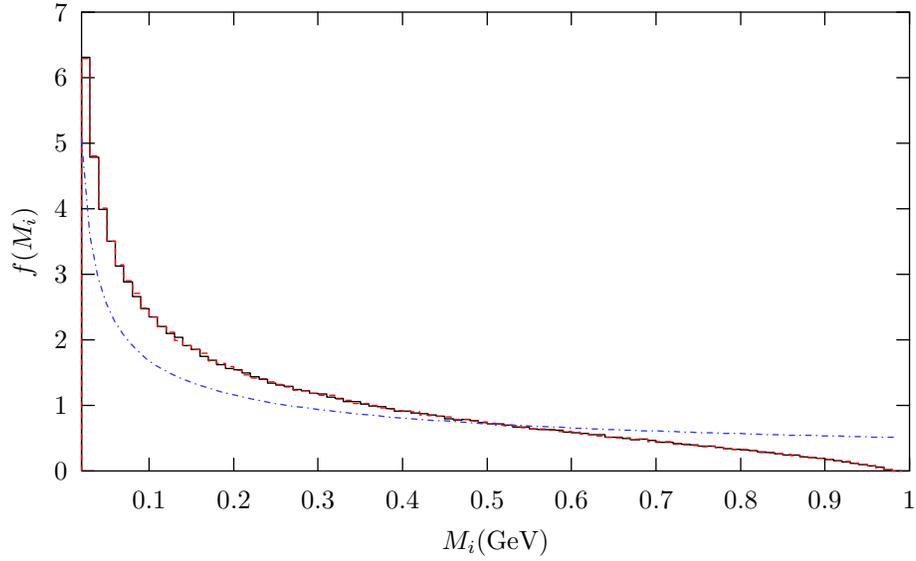}
\caption{The distribution of given in (\ref{eq:fission_dist}) 
(blue, dash-dotted) and the actual correlated distributions (black/red, solid)
when (\ref{eq:fission_constraint}) are applied. This is for a cluster mass 
of 1 GeV, $P=0.5$ and all three constituent masses $m_i = 0.01$ GeV.
\label{fig:corrdist}}
\end{figure}

The corresponding probability density is
\begin{equation}
\label{eq:fission_dist}
f(M_i) = N \frac{P (M_i-m_i)^{P-1}}{\left( M - m_i - m_3 \right)^P},
\end{equation}
where $N$ is an arbitrary normalization term.
It is important to note that because of the constraint in 
(\ref{eq:fission_constraint}) the distributions of the two clusters are
correlated and therefore do not exactly follow (\ref{eq:fission_dist}). An 
example of this is shown in fig.~\ref{fig:corrdist}

If parton $i$ is from the beam remnant and there is no underlying event model
being used, the cluster fissions with a different distribution. This 
distribution has one parameter $\Btp$, which is set to 1.0 
GeV by default. This is used to define $b = 2/\Btp$ which gives
\begin{equation}
r_{min} = e^{-b X},
\end{equation}
where $X = M - m_1 - m_2 - 2m_3$. The mass distribution is then defined
by
\begin{equation}
f(M_i) = m_i + m_3 - \frac{\log{R}}{b},
\end{equation}
where
\begin{equation}
R = \left( r_{min} + r_1 \left(1-2 r_{min}\right)\right)
    \left( r_{min} + r_2 \left(1-2 r_{min}\right)\right).
\end{equation}
$r_1$ and $r_2$ are two flatly distributed random numbers. This mass
distribution is designed to decrease rapidly to avoid splitting the cluster
many times which would produce large transverse energies. This would not
be desired as the beam remnant process is meant to be a soft process. 

\section{Cluster Decays}
The last stage of the cluster hadronization model is the cluster decays. Here
the clusters of a given flavour $(q_1, q_2)$ draw a (di)quark anti-(di)quark 
pair $(q,\bar{q})$ from the vacuum and form a pair of hadrons with the 
combination $(q_1,\bar{q})$ and $(q,q_2)$. The possible hadrons are 
selected based on spin, flavour and  phase space. The exact way of accepting
and rejecting these combinations are described in three variants: \HW
~6.5~\cite{Herwig} , Kupco's method~\cite{Kupco} and \HWP. The method of
\HWP~is described in the next section.

\subsection{\HW~6.5}
The method of cluster decays implemented in \HW~is described here. Before the
software runs, it initialized a list of data for each type of hadron. This
list contains two weights,
\begin{itemize}
\item{$w$: a spin weight for the hadron. This is $\frac{2s+1}{M_{q,q'}}$ where
$M_{q,q'}$ is $2s+1$ for the largest spin of a flavour group. There are three
of these groups which contain all the hadrons of a given flavor combination: 
(1) $u\bar u$ and $d \bar d$, (2) $s \bar s$ and (3) all the remaining 
flavours.}
\item{$s$: this is a phase space correction weight which is normally set to
unity.}
\end{itemize}

\HW~6.5 chooses a mode in several phases. First a flavour is drawn out of
the vacuum based on the probability
\begin{equation} 
P_{q} = \frac{\texttt{Pwt}(q)}{\sum_q \texttt{Pwt}(q)}.
\end{equation}
The values of $\texttt{Pwt}(q)$ are just parameters for $u,d,s,c$ and $b$. For
the diquarks, however, the weight is
\begin{equation}
P_{q_i q_j} = \frac{1}{2} \left( 1 + \delta_{ij} \right) \texttt{Pwt}_{qq}
              \texttt{Pwt}_{q_i} \texttt{Pwt}_{q_j},
\end{equation}
where $\texttt{Pwt}_{qq}$ is the diquark parameter.

The flavour combinations, $(q_1, q)$ and $(q, q_2)$, have the possibility of 
forming $N_{q_1,q}$  and $N_{q,q_2}$ different hadrons, respectively. Hadrons
$a_{q_1,q}$ and $b_{q,q_2}$ are chosen randomly from this list. These hadrons 
have spin weight $w_a$ and $w_b$ and are rejected based on these weights. 

Lastly, the resulting pair $(a,b)$ are accepted or rejected based on phase 
space. The probability is given as
\begin{equation}
P_{phase}(a,b) = \frac{p^*_{a,b} s_a s_b}{p^*_{max}}, 
\end{equation}
where $s_a$ and $s_b$ are phase space adjusting parameters and $p^*_{max}$ is 
the $p^*$ value for the lightest hadron pair of the flavours $(q_1,q),(q,q_2)$.
$p^*$ for a two-body decay is the c.m.~momentum in the decay 
$M \rightarrow m_1 m_2$ and is given by
\begin{equation}
p^* = \frac{\sqrt{M^2 - (m_1+m_2)^2}\sqrt{M^2 - (m_1-m_2)^2}}{2M},
\end{equation}
in the region of valid phase space, $M \geq m_1 + m_2$. It is set to zero 
otherwise.

In the end all of this gives (approximately) the probability of choosing 
hadrons $a_{q_1,q}$ and $b_{q,q_2}$ as
\begin{equation}
\label{hw64}
P(a_{q_1,q},b_{q,q_2}|q_1,q_2) = P_q \frac{1}{N_{q_1,q}} \frac{1}{N_{q,q_2}}
                                     \frac{w_a}{M_{q_1,q}}
                                     \frac{w_b}{M_{q,q_2}}
                                     \frac{p^*_{a,b} s_a s_b}{p^*_{max}}.
\end{equation}
Technically this isn't the probability because of the rejection scheme used in 
the algorithm but it is close and is able to illustrate the main advantages and
disadvantages of the algorithm.

The problem with this method, as described by Kupco~\cite{Kupco}, is that as 
more hadrons are added to the list then the particular flavour content of the 
new hadrons is suppressed by the growing factor $N$. In effect, the probability 
of choosing a hadron of a given flavour is proportional to the average of all 
the $p^*$'s of the flavour. In most cases a majority of hadrons are 
inaccessible due to mass contraints. This leads to a suppression of the lighter
hadrons of that flavour, even for clusters which are too light to decay into 
the new heavier states.

To look at this further we analyze the result of isospin symmetric clusters 
$u\bar u, d\bar d, u\bar d$ and $d\bar u$. For a cluster with mass just
above threshold for the production of $\pi^0 \pi^0$ and $\pi^+ \pi^-$ these 
should be produced with the ratio $\pi^0:\pi^+:\pi^- = 1:1:1$. It is easy to 
see that the ratio between the states generated by $u\bar{d}$ and 
$d\bar{u}$ are $1:1$. Instead we look here at just the part of the ratio
which differs from unity, $u\bar{u}$ and $d\bar{d}$. Using 
(\ref{hw64}) we find (after assuming $s_a = s_b = 1$ and $p^*_{\pi^0,\pi^0} 
= p^*_{\pi^+,\pi^-} = p^*_{max}$)
\begin{eqnarray}
P(\pi^0,\pi^0|u,\bar u) &=&P_u \frac{1}{N^2_{u\bar u}}
                                     \frac{w^2_{\pi^0}}{M^2_{u\bar u}},\\
P(\pi^+,\pi^-|u,\bar{u}) &=&P_d \frac{1}{N_{d\bar{u}}}
                                     \frac{1}{N_{u\bar{d}}}
                                     \frac{w_{\pi^+}}{M_{u\bar{d}}}
                                     \frac{w_{\pi^-}}{M_{d\bar{u}}},\\
P(\pi^0,\pi^0|d,\bar{d}) &=&P_d \frac{1}{N^2_{d\bar{d}}}
                                     \frac{w^2_{\pi^0}}{M^2_{d\bar{d}}},\\
P(\pi^+,\pi^-|d,\bar{d}) &=&P_u \frac{1}{N_{d\bar{u}}}
                                     \frac{1}{N_{u\bar{d}}}
                                     \frac{w_{\pi^+}}{M_{u\bar{d}}}
                                     \frac{w_{\pi^-}}{M_{d\bar{u}}}.
\end{eqnarray}
The $M$'s for $u\bar{u}$ and $d\bar{d}$ are equal and the $M$'s for 
$u\bar{d}$ and $d\bar{u}$ are equal.  We can also set the flavour 
probabilities to unity. Lastly, the spin weights of the $\pi^+$ and $\pi^-$
are equal. We now find the ratio of pions as
\begin{equation}
\pi^0:\pi^+:\pi^- = \frac{2}{N^2_{u\bar{u}}}\frac{w^2_{\pi^0}}{M_{uu}} :
                   \frac{1}{N^2_{u\bar{d}}}\frac{w^2_{\pi^+}}{M_{du}} :
                   \frac{1}{N^2_{d\bar{u}}}\frac{w^2_{\pi^-}}{M_{ud}}.
\end{equation}

\begin{figure}
\centering
\includegraphics{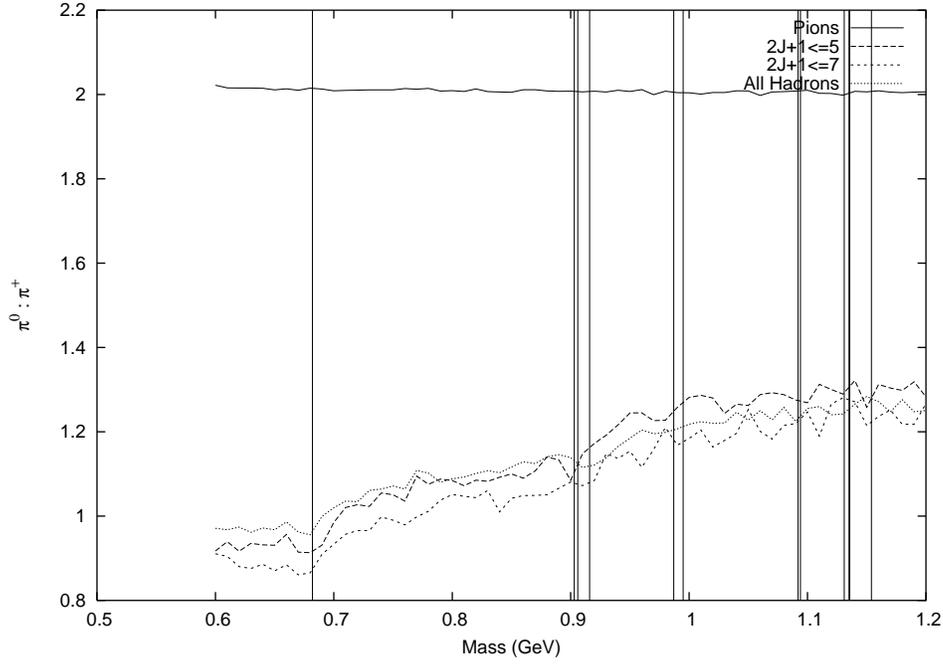}
\caption{The ratio of $\pi^0$ to $\pi^+$ as the cluster mass is increased and
with different number of hadrons in the $u\bar{u}$ list. The vertical bars
indicate the threshold for a new set of hadrons to be produced.}
\label{plota}
\end{figure}

Obviously, these ratios are dominated by the number of hadrons of a particular
flavour content. It just happens that in \HW~6.5 there was the right number
of $u\bar{u},d\bar{d}$ and $u\bar{d},d\bar{u}$ hadrons in
the list to give approximately the correct ratios. Figure \ref{plota} shows
the ratio for $\pi^0:\pi^+$ as the cluster mass increases for $u\bar{u}$
and $d\bar{d}$ clusters. It can be seen that the ratios are dictated by 
the number of $u\bar{u},d\bar{d}$ hadrons that are in the list.

\subsection{Kupco Method}
\label{sec:kupco}
Kupco~\cite{Kupco} was the first person to point out the problem of suppression
of hadrons when new modes of the same flavours were added. He realized that 
the problem was that the probabilities were proportional to an average of
the $p^*$'s of a particular flavour content. 

In order to remedy the problem a new set of probabilities for choosing a 
decay mode was used. Instead of splitting the probability into independant
parts, as was done in the original version, one weight was created for each
hadronic mode.
\begin{equation}
\label{eqn:kupcowt}
W(a_{q_1,q},b_{q,q_2}|q_1,q_2) = P_q w_a w_b s_a s_b p^*_{a,b}.
\end{equation} 
The probability is then
\begin{equation}
P(a_{q_1,q},b_{q,q_2}|q_1,q_2) = \frac{W(a_{q_1,q},b_{q_2,q}|q_1,q_2)}
                      {\sum_{c,d,q'} W(c_{q_1,q'},d_{q_2,q'}|q_1,q_2)}.
\end{equation}
The addition of new hadrons to the list now increases the probability of 
choosing that particular flavour. Because a majority of the hadrons are quite 
heavy the $p^*$ for any mode with the new hadron is zero in many cases. 
Therefore, this new hadron has no effect on the choice of mode for lighter 
clusters. 

Again, lets look at the example of decaying $u\bar{u}$ and $d 
\bar{d}$ clusters with a cluster mass just above threshold for a 
$\pi^+ \pi^-$ decay. 
\begin{eqnarray}
W(\pi^0,\pi^0|u\bar{u}) &=& P_u w_{\pi^0}^2 s_{\pi^0}^2 
                                     p^*_{\pi^0,\pi^0},\\
W(\pi^0,\pi^0|d\bar{d}) &=& P_d w_{\pi^0}^2 s_{\pi^0}^2 
                                     p^*_{\pi^0,\pi^0},\\
W(\pi^+,\pi^-|u\bar{u}) &=& P_d w_{\pi^+} w_{\pi^-} s_{\pi^+} s_{\pi^-}
                                     p^*_{\pi^+,\pi^-},\\
W(\pi^+,\pi^-|d\bar{d}) &=& P_u w_{\pi^+} w_{\pi^-} s_{\pi^+} s_{\pi^-} 
                                     p^*_{\pi^+,\pi^-}.
\end{eqnarray}
With the same assumptions as before ($s_i=1,p^*_{\pi^0,\pi^0}=p^*_{\pi+,\pi^-},
P_u = P_d$) we get
\begin{eqnarray}
P(\pi^0,\pi^0|u\bar{u})\  = &P(\pi^0,\pi^0|d\bar{d})& =\  
             \frac{w_{\pi^0}^2}{w_{\pi^0}^2 + w_{\pi^+} w_{\pi^-}},\\
P(\pi^+,\pi^-|u\bar{u})\  = &P(\pi^+,\pi^-|d\bar{d})& =\ 
             \frac{w_{\pi^+} w_{\pi^-}}{w_{\pi^0}^2 + w_{\pi^+} w_{\pi^-}}.
\end{eqnarray}
Finally this gives us the ratio
\begin{equation}
\pi^0 : \pi^+ : \pi^- = w_{\pi^0}^2 : \frac{1}{2} w_{\pi^+} w_{\pi^-}
                                    : \frac{1}{2} w_{\pi^+} w_{\pi^-},
\end{equation}
and as the $w_i$'s are static {\it a priori} weights, we can choose them to
be $w_{\pi^0} = \frac{1}{\sqrt{2}} \sqrt{w_{\pi^+} w_{\pi^-}}$ which will give
the correct ratio of $1:1:1$.

Now if we allow the cluster mass to increase so that the $\pi^0 \eta$ mode
is accessible, we get the same weights for the $\pi^0 \pi^0$ and the $\pi^+
\pi^-$ modes and the new $\pi^0 \eta$ mode has a weight of
\begin{equation}
W(\pi^0,\eta|u\bar{u}) = P_u w_{\pi^0} w_{\eta} s_{\pi^0} s_{\eta}
                                  p^*_{\pi^0,\eta},
\end{equation}
and similarly for the $d\bar{d}$ cluster. The sum of the weights, without 
some common factors and using the same assumptions, is
\begin{equation}
\Sigma = w_{\pi^0}^2 + 2 w_{\pi^0} w_{\eta} 
         \frac{p^*_{\pi^0,\eta}}{p^*_{\pi^0,\pi^0}} + w_{\pi^+} w_{\pi^-}.
\end{equation}
This is the same for both the $u\bar{u}$ and $d\bar{d}$ clusters.
Therefore the probabilities are now
\begin{eqnarray}
P(\pi^0,\pi^0) &=& \frac{w_{\pi^0}^2}{\Sigma},\\
P(\pi^0,\eta) &=& \frac{2 w_{\pi^0} w_{\eta}}{\Sigma} 
                  \frac{p^*_{\pi^0,\eta}}{p^*_{\pi^0,\pi^0}},\\
P(\pi^+,\pi^-) &=& \frac{w_{\pi^+} w_{\pi^-}}{\Sigma}.
\end{eqnarray}
To find the ratio we need to make sure we count the particles correctly.
Doing so we find
\begin{equation}
\pi^0 : \pi^+ : \pi^- = 2 w_{\pi^0}^2 + 2 w_{\pi^0} w_{\eta} 
                      \frac{p^*_{\pi^0,\eta}}{p^*_{\pi^0,\pi^0}} 
                      : w_{\pi^+} w_{\pi^-} : w_{\pi^+} w_{\pi^-}.
\end{equation}
After forcing the right ratio near the $\pi^+ \pi^-$ threshold by setting the
$w_i$'s, we now don't have the right ratio after the $\pi^0 \eta$ threshold. 

\begin{figure}[ht]
\includegraphics{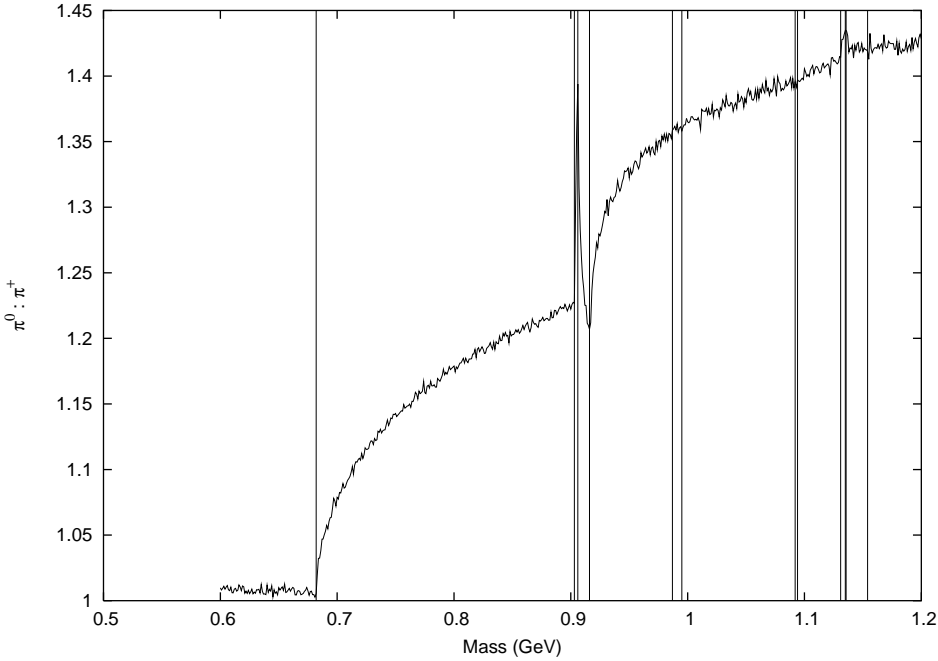}
\caption{The ratio of $\pi^0$ to $\pi^+$ as the cluster mass is increased. When
new hadrons become accessible the ratios differ from one, sometimes 
drastically.}
\label{plotb}
\end{figure}

Figure \ref{plotb} shows the ratio of $\pi^0 : \pi^+$ as the cluster mass is
increased. After a new production threshold is reached the ratio changes. The 
change can become quite dramatic and the average ratio between thresholds 
differs from unity.

Though the ratios vary as the cluster mass changes, this seems overall to be
an improvement from the original cluster decays. We no longer have the problem
that the addition of new hadrons to the lists causes decreased probability. 
We now have increased the probability of choosing a flavour content when new 
hadrons of that content are added. In doing so, however, we have created a new 
problem. In baryon-conserving processes, we only have clusters with a 
quark--antiquark pair, rather then some diquark--quark combination. This means 
two baryons must be created at a time by drawing a diquark--antidiquark pair 
out of the vacuum. Doing so requires a cluster with a high mass, which in turn 
means a new meson added to the list is likely to be available. Consider a 
$u\bar{u}$ cluster with high mass. If we have 5 hadrons with 
$u\bar{d}$ flavour and 2 baryons with $uud$ flavour we will have only 4 
possible combinations of the $uud$ flavour baryons but 25 modes of the 
$u\bar{d}$ flavour. If all the modes are accessible this makes the baryon
probabilities lower, as they are all normalized by the sum of all accessible 
modes. If we then add a 6th hadron to the $u\bar{d}$ list, we now have 36 
combinations. If, again, these are all accessible that decreases the 
probability of creating a baryon even more. This problem is addressed and 
solved in the new method implemented in \HWP~and described in the next 
section.

\section{\HWP{}}
\label{sec:Herwig++}
The nonperturbative splitting, cluster formation and cluster fission
described in section \ref{sec:had} has been implemented in \HWP.
The method for decaying clusters is similar to the Kupco method described in 
section \ref{sec:kupco} but has been changed to account for the lack of baryon
production. Results of the hadronization process in \HWP~are given in this 
section. Also the changes to the cluster decay model are presented here.

The concept of colour preconfinement says that the mass distribution of the 
colour singlet systems, after the parton shower, are independent of the 
centre-of-mass energy of the hard process. Figure \ref{plotc1} shows the 
cluster mass distribution generated from \HWP~using the default parameter set.
It can be seen that this distribution is indeed independent of $Q$, the 
c.m.~energy of the hard process.

\begin{figure}
\centering
\epsfig{file=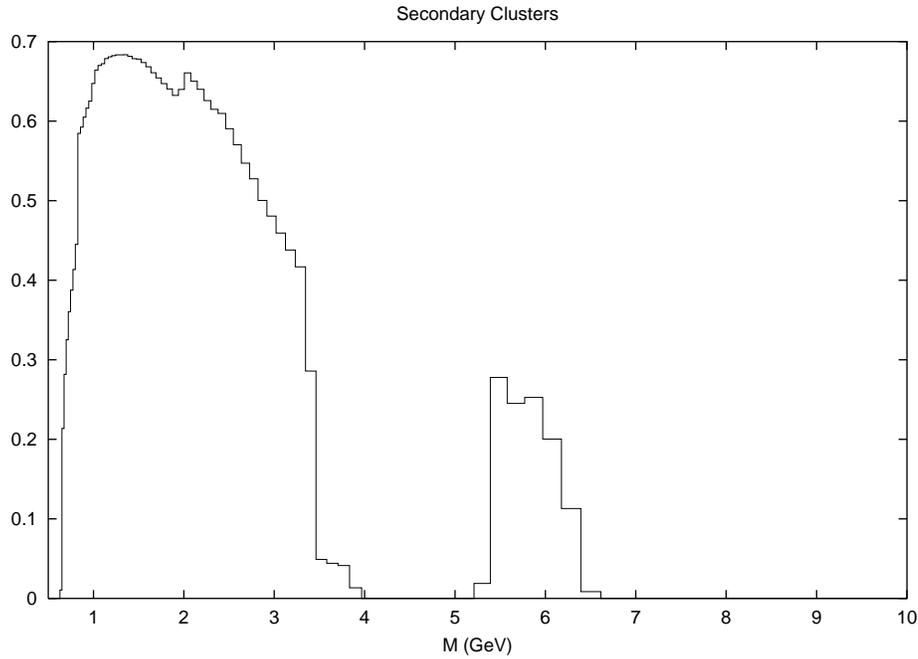}
\caption{The cluster mass distribution after cluster fission. This is using
the default parameter set. It can be seen that the heavy clusters no longer
exist and are instead folded into the lighter part of the distribution.}
\label{plotc2}
\end{figure}

The cluster fission stage described in section~\ref{sec:clusterfission}
truncates this distribution by breaking apart 
clusters that are above the threshold in (\ref{eqn:thresh}). 
Figure \ref{plotc2} shows the new distribution after this process. The two
bumps around 2 GeV and 5.5 GeV are from the charm and bottom clusters. This 
figure is for the default parameter set ($Cl_{pow} = 2.0, Cl_{max} = 3.2$ 
GeV). As can be seen, the heavier part of the distribution is folded into the 
lighter part.

\subsection{\HWP~Cluster Decay Algorithm}
The algorithm used in \HWP~for cluster decays is very similar to the method
proposed by Kupco. The weights of a particular decay mode is given by 
(\ref{eqn:kupcowt}). As discussed previously, the problem with Kupco's method
is that there are not enough baryons produced. This is because of the quantity
of possible decay modes in the meson sector compared to the quantity of modes
in the baryon sector. To fix this problem a way to separate the two sectors
was needed.

The new algorithm separates the meson sector from the baryon sector. The 
weights from (\ref{eqn:kupcowt}) are still used, but the sum of weights used
for normalizing is only summed over modes in either the meson or baryon sector.
If a cluster mass is high enough to decay into the lightest baryon pair then 
the parameter $\texttt{Pwt}_{qq}$ is used to decide whether to use the baryon 
sector or meson sector. There is a $\left(1+\texttt{Pwt}_{qq}\right)^{-1}$  
probability of using the meson sector and a 
$\frac{\texttt{Pwt}_{qq}}{1+\texttt{Pwt}_{qq}}$ probability of using the 
baryon sector. This change not only allows for more baryons to be created but 
also gives more direct control of how many baryons are produced through the 
diquark parameter $\texttt{Pwt}_{qq}$. 

\begin{table}
\centering
\begin{tabular}{l|l|l}
\hline
Parameter & \HWP & Old Model \\
\hline \hline
$\delta$ & 2.3 GeV & 2.3 GeV \\
$m_g$ & 0.750 GeV & 0.750 GeV\\
$\alpha_s (1 {\rm GeV})$ & 0.630882 GeV & 0.630882 GeV \\
\hline
$\Clm$ & 3.2 GeV & 3.0 GeV \\
$\Clp$ & 2.0 GeV & 2.0 GeV \\
\texttt{PSplt1} & 1 & 1 \\
\texttt{PSplt2} & 0.33 & 0.33 \\
\texttt{B1Lim} & 0.0 & 0.0 \\
\texttt{ClDir1} & 1 & 1 \\
\texttt{ClDir2} & 1 & 1 \\
\texttt{ClSmr1} & 0.40 & 0.0 \\
\texttt{ClSmr2} & 0.0 & 0.0 \\
$\texttt{Pwt}_d$ & 1.0 & 1.0 \\
$\texttt{Pwt}_u$ & 1.0 & 1.0 \\
$\texttt{Pwt}_s$ & 0.85 & 1.0 \\
$\texttt{Pwt}_c$ & 1.0 & 1.0 \\
$\texttt{Pwt}_b$ & 1.0 & 1.0 \\
$\texttt{Pwt}_{qq}$ & 0.55 & 0.65 \\
Singlet Weight & 1.0  & 1.0\\
Decuplet Weight & 0.7 & 1.0\\
\hline \hline
\end{tabular}
\caption{The parameters for \HWP. The first group are
shower parameters, the second are all of the hadronization parameters. The 
meaning of all the parameters is given in chapter~\ref{chap:Herwig}.}
\label{tab:Had_defaults}
\end{table}

\subsection{Results}
In this section the results of the different hadronization schemes of \HWP~
for $e^+e^-$ events at 91.2 GeV are presented. The values of the 
parameters in \HWP~used for this study are given in table 
\ref{tab:Had_defaults}. This section
presents only a few results related to the new hadronization method. Full 
results of \HWP~including event shapes and jet physics are given in 
chapter~\ref{chap:Results}.

Table~\ref{tab:mult} shows the results of the new cluster hadronization 
algorithm in comparison to the old algorithm. The column labeled `Old Model' 
is the result of using the old algorithm with the new parton shower in \HWP{}. 
The column \HWP{} is the result of using the new cluster hadronization model 
with the new parton shower variables. The last column, labeled Fortran, is the 
data generated using the Fortran \HW{} program, version 6.5. The data in this
table are combined and updated from a variety of sources, see 
ref.~\cite{Knowles:1997dk}. 

Neither of the \HWP{} implementations in the table have been tuned but the
results in the Fortran column are the result of tuning. 10000 events were
used for the `Old Model' data and the \HWP~results are for 100000 events. As 
the models are different a new set of parameters are needed. The 
parameters for this are also shown in table~\ref{tab:Had_defaults}. The 
results given in the last column, Fortran, are taken from~\cite{Webber:1999ui}.

\begin{table}
\centering
\begin{tabular}{llllll} \hline
Particle & Experiment & Measured & Old Model & \HWP{} & Fortran
\\\hline
\hline
All Charged & M,A,D,L,O & 20.924 $\pm$ 0.117  & $20.22^*$   & $20.814$     & $20.532^*$\\
\hline                                                                    
$\gamma$  & A,O & 21.27 $ \pm$ 0.6 & $23.032$               & $22.6745$    & $20.74$ \\ 
$\pi^0$   & A,D,L,O & 9.59 $ \pm$ 0.33 & $10.273$           & $10.0777$    & $9.88$ \\ 
$\rho(770)^0$ & A,D & 1.295 $ \pm$ 0.125 & $1.235$          & $1.31596$    & $1.07$ \\ 
$\pi^\pm$ & A,O & 17.04 $ \pm$ 0.25 & $16.297$              & $16.9461$    & $16.74$ \\ 
$\rho(770)^\pm$ & O & 2.4 $ \pm$ 0.43 & $1.994$             & $2.14345$    & $2.06$ \\ 
$\eta$ & A,L,O & 0.956 $ \pm$ 0.049 & $0.886$               & $0.892915$   & $0.669^*$\\ 
$\omega(782)$ & A,L,O & 1.083 $ \pm$ 0.088& $0.859$         & $0.91576$    & $1.044$ \\ 
$\eta'(958)$ & A,L,O & 0.152 $ \pm$ 0.03 & $0.13$           & $0.135935$   & $0.106$ \\ 
\hline                                                                     
$K^0$ & S,A,D,L,O & 2.027 $ \pm$ 0.025 & $2.121^*$          & $2.06214$    & $2.026$ \\ 
$K^*(892)^0$ & A,D,O & 0.761 $ \pm$ 0.032 & $0.667$         & $0.681661$   & $0.583^*$ \\ 
$K^*(1430)^0$ & D,O & 0.106 $ \pm$ 0.06 & $0.065$           & $0.078918$   & $0.072$ \\ 
$K^\pm$ & A,D,O & 2.319 $ \pm$ 0.079 & $2.335$              & $2.28551$    & $2.250$ \\ 
$K^*(892)^\pm$ & A,D,O & 0.731 $ \pm$ 0.058 & $0.637$       & $0.657485$   & $0.578$ \\ 
$\phi(1020)$ & A,D,O & 0.097 $ \pm$ 0.007 & $0.107$         & $0.113829$   & $0.134^*$ \\ 
\hline                                                                     
$p$ & A,D,O & 0.991 $ \pm$ 0.054 & $0.981$                  & $0.946807$   & $1.027$ \\ 
$\Delta^{++}$ & D,O & 0.088 $ \pm$ 0.034 & $0.185$          & $0.092217$   & $0.209^*$ \\ 
$\Sigma^-$ & O & 0.083 $ \pm$ 0.011 & $0.063$               & $0.071454$   & $0.071$ \\ 
$\Lambda$ & A,D,L,O & 0.373 $ \pm$ 0.008 & $0.325^*$        & $0.384086$   & $0.347^*$\\ 
$\Sigma^0$ & A,D,O & 0.074 $ \pm$ 0.009 & $0.078$           & $0.091162$   & $0.063$ \\ 
$\Sigma^+$ & O & 0.099 $ \pm$ 0.015 & $0.067$               & $0.077027$   & $0.088$ \\ 
$\Sigma(1385)^\pm$ & A,D,O & 0.0471 $ \pm$ 0.0046 & $0.057$ & $0.031159^*$ & $0.061^*$\\
$\Xi^-$ & A,D,O & 0.0262 $ \pm$ 0.001 & $0.024$             & $0.028565$   & $0.029$ \\ 
$\Xi(1530)^0$ & A,D,O & 0.0058 $ \pm$ 0.001 & $0.014^*$     & $0.007782$   & $0.009^*$\\ 
$\Omega^-$ & A,D,O & 0.00125 $ \pm$ 0.00024 & $0.001$       & $0.001439$   & $0.0009$ \\ 
\hline                                                                     
$f_2(1270)$ & D,L,O & 0.168 $ \pm$ 0.021 & $0.113$          & $0.150273$   & $0.173$ \\ 
$f_2'(1525)$ & D & 0.02 $ \pm$ 0.008 & $0.003$              & $0.011739$   & $0.012$ \\ 
$D^\pm$ & A,D,O & 0.184 $ \pm$ 0.018 & $0.322^*$            & $0.318519^*$ & $0.283^*$ \\ 
$D^*(2010)^\pm$ & A,D,O & 0.182 $ \pm$ 0.009 & $0.168$      & $0.180251$   & $0.151^*$\\ 
$D^0$ & A,D,O & 0.473 $ \pm$ 0.026 & $0.625^*$              & $0.570354^*$ & $0.501$ \\ 
$D_s^\pm$ & A,O & 0.129 $ \pm$ 0.013 & $0.218^*$            & $0.194775^*$ & $0.127$ \\ 
$D_s^{*\pm}$ & O & 0.096 $ \pm$ 0.046 & $0.082$             & $0.066209$   & $0.043$ \\ 
$J/\Psi$ & A,D,L,O & 0.00544 $ \pm$ 0.00029 & $0.006$       & $0.003605^*$ & $0.002^*$\\ 
$\Lambda_c^+$ & D,O & 0.077 $ \pm$ 0.016 & $0.006^*$        & $0.022621^*$ & $0.001^*$\\ 
$\Psi'(3685)$ & D,L,O & 0.00229 $ \pm$ 0.00041 & $0.001^*$  & $0.001775$   & $0.0008^*$\\ 
\hline
\hline
\end{tabular}
\caption{Multiplicities per event at 91.2 GeV. We show results 
  from \HWP{} with the implementation of the old cluster hadronization
  model (Old Model) and the new model (\HWP{}), and from \HW\ 6.5
  shower and hadronization (Fortran).  Experiments
  are ALEPH(A), DELPHI(D), L3(L), OPAL(O), MK2(M) and SLD(S). The $*$
  indicates a prediction that differs from the measured value by
  more than three standard deviations.
  \label{tab:mult}} 
\end{table}

As we can see from the multiplicity results \HWP~is able to obtain 
multiplicities that are as good as, if not better than, the old \HW~6.5 
results. The $\chi^2$ of the data sets are given in table \ref{tab:chi}. 

\begin{table}
\scriptsize
\centering
\begin{tabular}{llll}
\hline
 & Old Model & \HWP~& \HW~6.5 \\ \hline \hline
All Data from Table \ref{tab:mult} & $\chi^2 = 543.84/35 = 15.54$
                                   & $\chi^2 = 277.16/35 = 7.92$ 
                                   & $\chi^2 = 490.52/35 = 14.01$ \\                             
\hline
\hline
\end{tabular}
\caption{$\chi^2$ results for the different cluster decay methods.}
\label{tab:chi}
\end{table}

We also want to make sure that the momentum distributions of the hadrons match
those from the data. $x_p$ is defined as
\begin{equation}
x_p = \frac{2 \left| \vec{p} \right|}{E_{max}},
\end{equation}
where $E_{max}$ is the centre-of-mass energy. Presented here are the results 
for $x_p$ for all charged particles and $\xi^{uds}_p = \ln \frac{1}{x^{uds}_p}$ 
for all charged particles in uds events. Also the momentum distributions of 
$\pi^\pm$, $K^\pm$ and $p^\pm$ are shown. All results are compared to data from
the OPAL collaboration~\cite{Ackerstaff:1998hz, Akers:1994ez}. Again we can
see that the results from \HWP{} are in good agreement with the data. More
detailed results are given in chapter~\ref{chap:Results}.

\begin{figure}
\label{plotd1}
\centering
\includegraphics{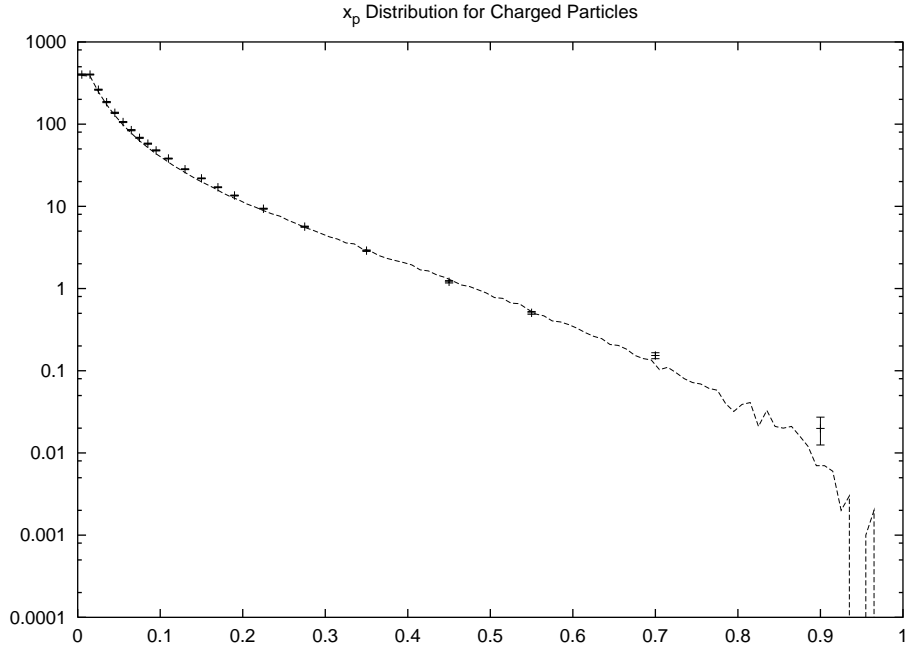}
\caption{Plot of $x_p$ for all flavours. Data from OPAL collaboration.}
\end{figure}

\begin{figure}
\label{plotd2}
\centering
\includegraphics{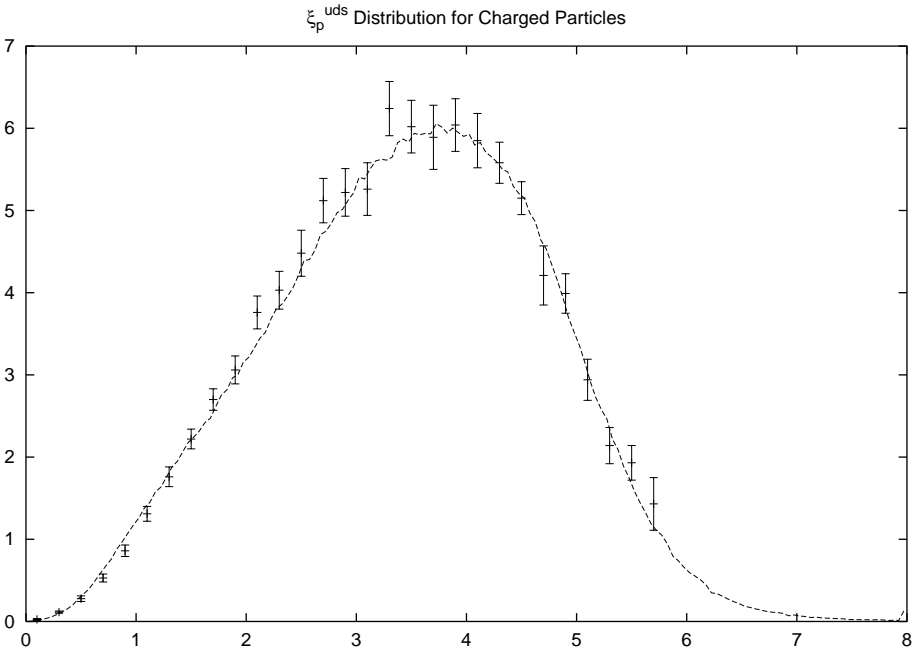}
\caption{Plot of $\ln 1/x_p$ for $uds$ flavours. Data from OPAL collaboration.}
\end{figure}

\begin{figure}
\label{plotd3}
\centering
\includegraphics{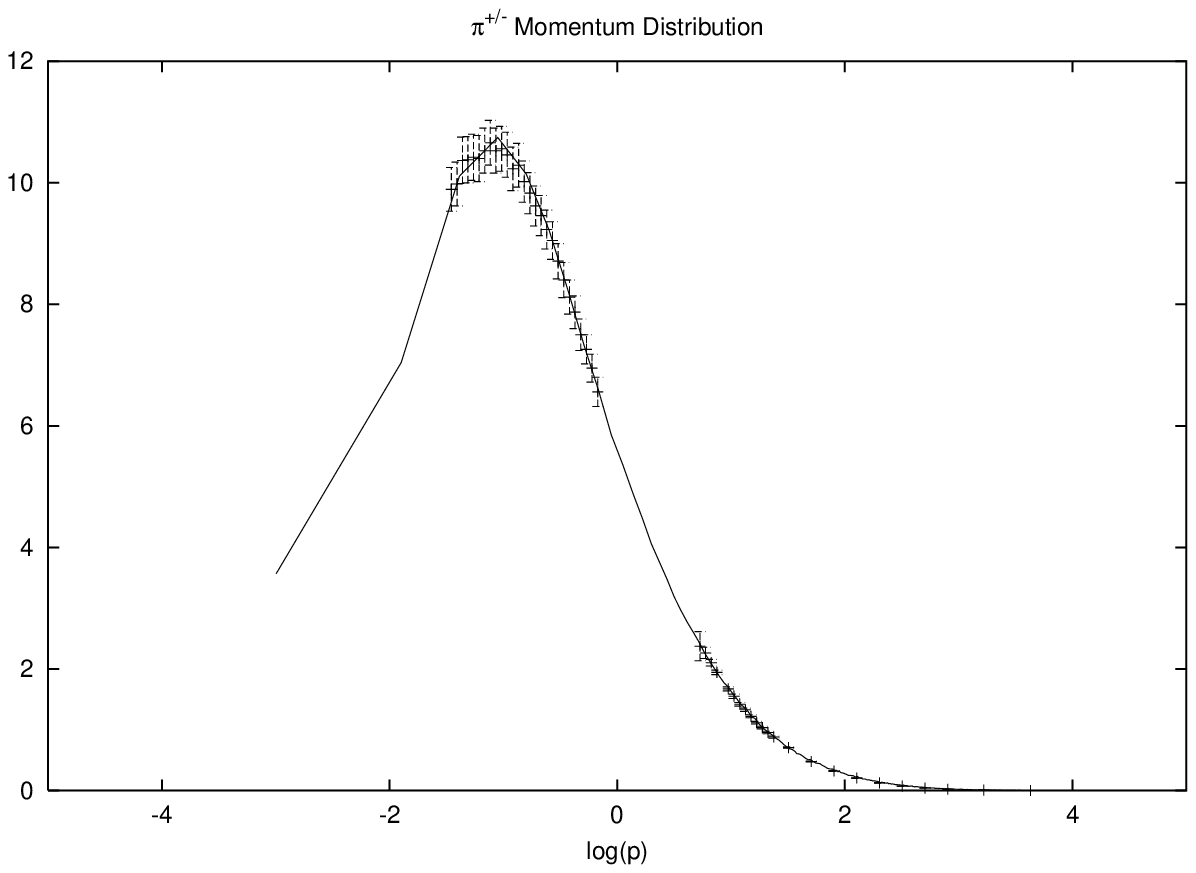}
\caption{Plot of $\log p$ for $\pi^\pm$. Data from OPAL collaboration.}
\end{figure}

\begin{figure}
\label{plotd4}
\centering
\includegraphics{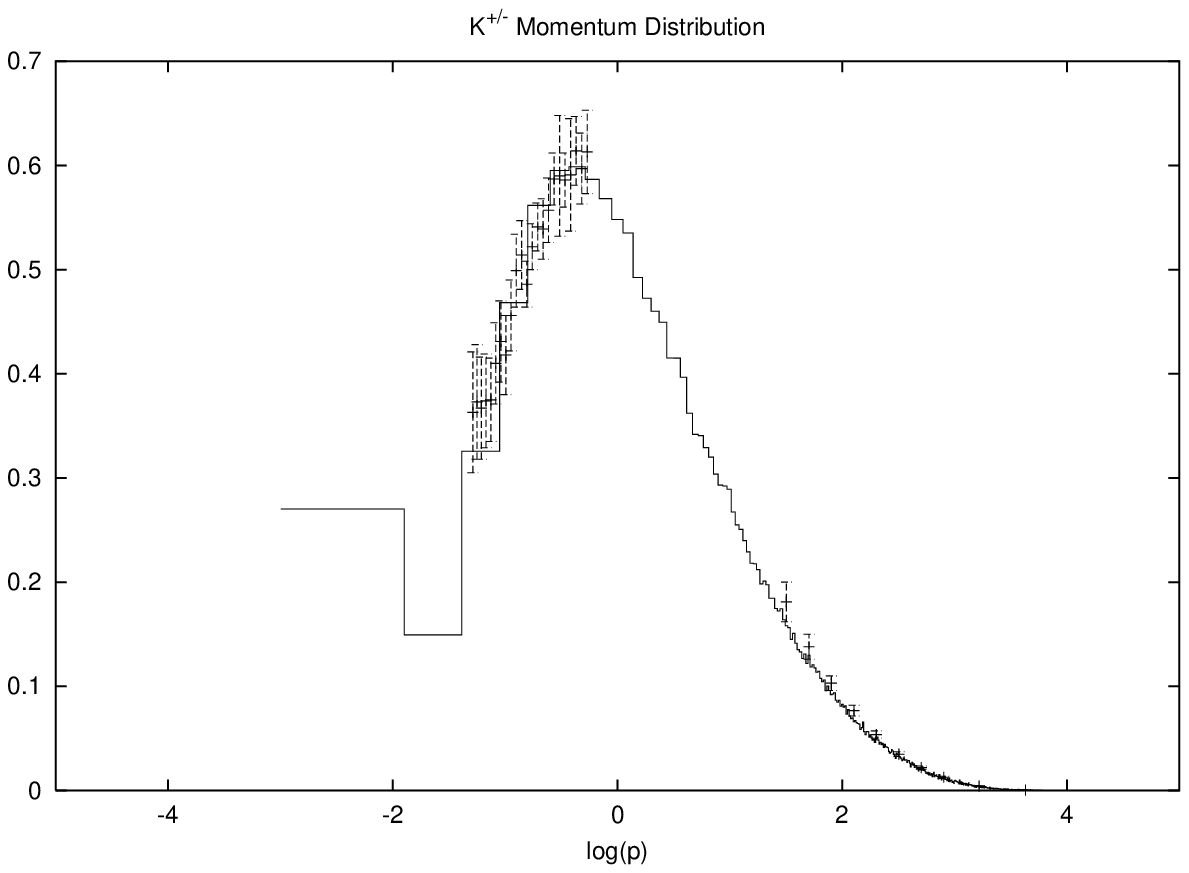}
\caption{Plot of $\log p$ for $K^\pm$. Data from OPAL collaboration.}
\end{figure}

\begin{figure}
\label{plotd5}
\centering
\includegraphics{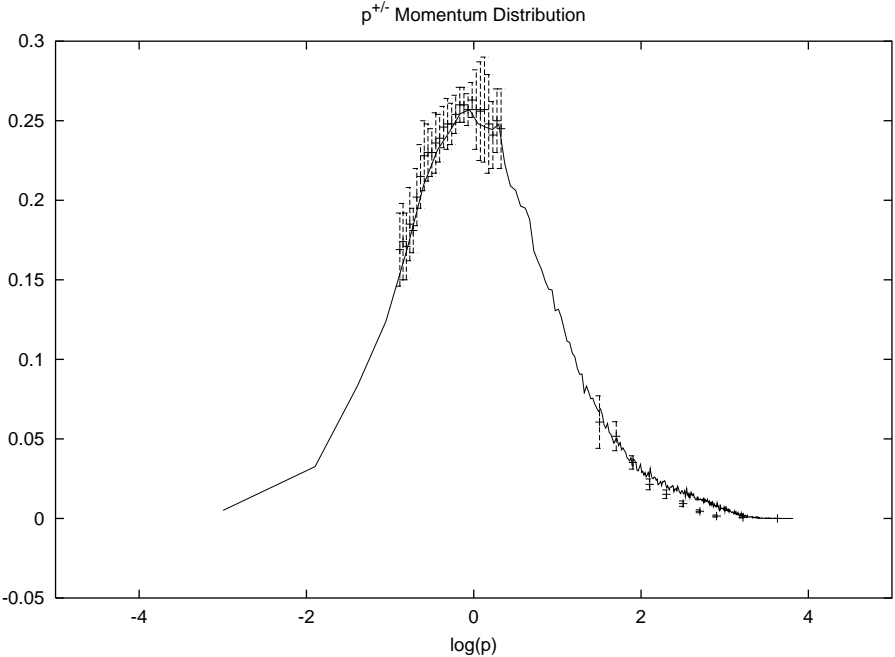}
\caption{Plot of $\log p$ for $p^\pm$. Data from OPAL collaboration.}
\end{figure}

\chapter{Herwig++}
\label{chap:Herwig}
The new generation of high energy colliders such as the Large Hadron
Collider (LHC) or a future linear collider (NLC) require new tools for
the simulation of signals and backgrounds.  The widely used event generators
\HW{} \cite{Herwig64} and \pythia{} \cite{Pythia} underwent tremendous
development during the LEP era and have reached the limit of
reasonable maintenance.  Therefore these programs 
(\Pythia7{}) \cite{Pythia:proof} as well as new projects, like \SHERPA{}
\cite{SHERPA}, are being completely (re-)developed in the object-oriented
programming language C++.

Chapters \ref{chap:NewVars} and \ref{chap:Hadronization} introduced two new
theoretical improvements to the original \HW~\cite{Herwig} Monte Carlo event 
generator. In this chapter I present the implementation of these two 
improvements in the new \HWP~\cite{Gieseke:2003_20} event generator as well as 
the implementation of the various other parts of the event generator. The 
\HWP{} event generator is built on top of \ThePEG~\cite{ThePEG}. \ThePEG{} is 
an administrative library which defines tools and data structures which are 
commonly used by Monte Carlo event generators. \ThePEG{} and \HWP{} also use
\CLHEP{}~\cite{CLHEP}. This is a package that provides general HEP functions.

This chapter will contain some discussions about object-oriented programming.
As this thesis is not intended to be a discussion of object-oriented 
programming and design, detailed discussion of these matters is given only as
a reference~\cite{C++}. For the purposes of this chapter a few keywords and 
very simple conceptual definitions of these terms are given here. Again, this 
is not intended as a complete description, just a guide for understanding the 
text of this chapter.

\emph{Classes} are the main components of an object-oriented program. 
A very simplistic view of a class is as a function or algorithm (with 
accompanying data). In a sense a
class just serves to implement some functionality or apply some algorithm. A 
class is much more complex and diverse than this, but for the current purposes
this should suffice. 

A class can be a particular type of another class. For 
example, a class could define how a particle behaves. It could contain all the 
data as well as some special functions, such as boosts, that are useful when
working with a particle. During the shower more information is needed for a 
particle, such as the 
Sudakov basis quantities $\alpha$ and $\beta$ in (\ref{eq_qi}). Rather 
then completely redefining a ShowerParticle to behave almost identically to a 
regular Particle, except for the new data, we can instead \emph{inherit} the 
old Particle class into the new ShowerParticle class. This means that all of 
the original data and functionality of the Particle class is also present in 
the ShowerParticle class, but we can now add the new data and functions for 
the ShowerParticle. This is useful because in the code we can just pass all 
the Particles and ShowerParticles around together as one set of Particles. 
Then if we want to perform special functions on the ShowerParticles we can 
identify which of the Particles is really a ShowerParticle and apply our 
operations. 

A function in a class can be defined as a \emph{virtual} function. 
Classes that inherit a class with a virtual function can redefine that 
function. These functions can actually be defined to be non-existent. If this 
is done the class is known as an \emph{abstract} class. This allows for the 
definition of an \emph{interface} without defining the implementation. For 
example, there are many things that
all matrix elements must have; all matrix elements must have a set
of incoming and outgoing particles. But the actual values of the matrix 
elements, for a given set of incoming and outgoing particles with momenta, are 
different. This is a perfect candidate for abstraction. A class \texttt{ME} 
can be defined that provides the definition of a function \texttt{double 
me2(incoming,outgoing)} but does not define it. Then any class that inherits 
\texttt{ME} can implement the function. If we now think about an algorithm 
that uses a matrix element, 
such as computing a cross section, we can define this algorithm independent of 
what matrix element we want to use. It just needs to know that the matrix
element class will return a double when given a set of incoming and outgoing
particles (or momenta).

This brief discussion of a few terms from object-oriented programming should
be adequate to comprehend the rest of this thesis. The rest of this chapter 
describes the implementation of the various parts of the Monte Carlo event
generator: hard subprocess, PDFs, parton shower, underlying event, 
hadronization, and decays. 

\section{Hard Subprocess}
As discussed in the first chapter, there is only a small set of matrix 
elements currently implemented into \HWP. For $e^+e^-$ annihilation there is
just one. It is the $e^+e^- \rightarrow \gamma^*/Z^0 \rightarrow q \bar q$ 
matrix element. Though this is not itself a QCD matrix element it inherits the
\texttt{ME2to2QCD} class from \ThePEG. This is because the \texttt{ME2to2QCD}
class has a function which returns the number of accessible flavours for a 
given scale, which is needed for the matrix element at a given scale. 

There is a much larger set of matrix elements for $pp$ collisions. All of these
inherit the \texttt{ME2to2QCD} class. Provided with \ThePEG~is the set of 
matrix elements: \texttt{MEqq2qq}, \texttt{MEQG2QG}, \texttt{MEGG2GG},
\texttt{MEQQ2QQ}, \texttt{MEGG2QQ}, \texttt{MEQQ2qq}, \texttt{MEQQ2GG} and
\texttt{MEQq2Qq}. The lower and upper case $q$ are used to decipher processes
that have different flavours. Also implemented with \HWP~is a Drell-Yan
matrix element $q \bar q \rightarrow \gamma^*/ Z^0 \rightarrow \ell^+ \ell^-$.
This type of process is particularly useful for studying the initial-state 
shower.

\begin{figure}[htb]
\centering
\epsfig{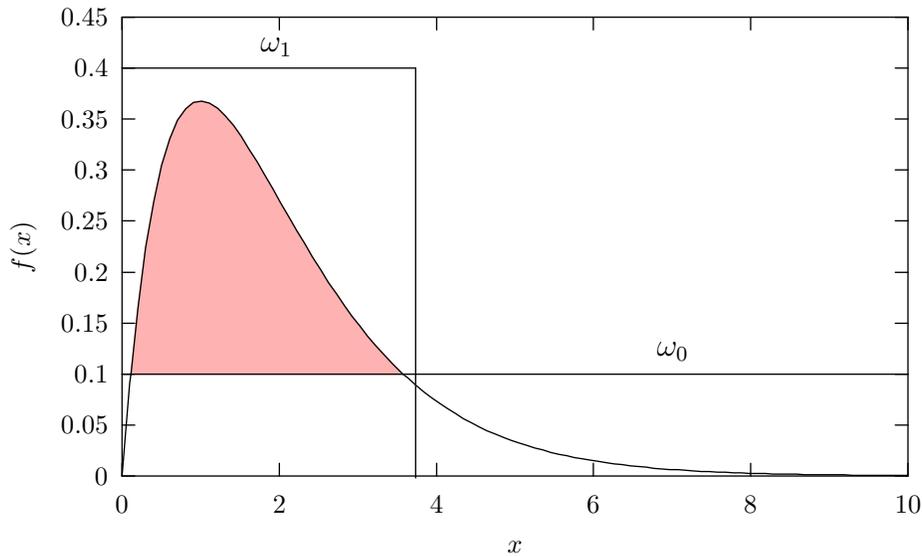}
\put(-100.,75.){$\omega_0$}
\put(-250.,190.){$\omega_1$}
\caption{This figure shows a function being integrated which has been divided
into two regions, $\omega_0$ and $\omega_1$.}
\label{fig:ACDC}
\end{figure}

Implementation of the hard subprocess requires both a matrix element and a 
phase space sampler. The method of sampling phase space is important for the
efficiency of the Monte Carlo. As discussed before, matrix elements\footnote{By
`matrix element' we actually mean `square modulus of the matrix element'} can 
have several peaks and valleys. Using a uniform sampling for this is extremely 
inefficient and advanced samplers can improve the efficiency drastically. The 
default sampler used in \HWP~is known as the ACDC sampler. 
This is a component of \ThePEG. ACDC is an acronym of Auto-Compensating
Divide-and-Conquer Phase Space Generator~\cite{LeifPrivate}. This algorithm 
uses a
divide-and-conquer scheme to divide the phase space into uniform sections which
have different maxima. Figure~\ref{fig:ACDC} shows an example of a function
which has been divided into two sections, each of which is sampled uniformly. 

The way the algorithm works is to generate points uniformly in phase-space. 
When a point is generated in which the matrix element is larger than the
value being sampled under, $\omega_0$, the space is divided into two regions. A
region is found in which the function is larger than $\omega_0$ and this
region is then sampled under $\omega_1$. The auto-compensating part of the the 
algorithm compensates for the fact that the peak has been undersampled and 
oversamples the new region until it is consistent. The shaded region of 
fig.~\ref{fig:ACDC} shows this new region which is auto-compensated.

\section{PDF}
Chapter \ref{chap:Intro} introduced the concept of the parton distribution 
functions. As mentioned earlier there are different ways to parameterize the
non-perturbative component of the PDF. Two such implementations are the 
Gl\"uck-Reya-Vogt (GRV)~\cite{GRV} PDFs and the Martin-Roberts-Stirling-Thorne 
(MRST)~\cite{MRST} PDFs. Both of these have been implemented in \HWP.

The implementation of the PDFs requires the implementation of four virtual 
functions inherited from the \texttt{PDFBase} class. These functions are:
\begin{itemize}
\item{\texttt{bool canHandleParticle(tcPDPtr particle) const}; this function is
used to specify what hadrons this PDF can work with.} 
\item{\texttt{cPDVector partons(tcPDPtr particle) const}; this function is
used to specify what partons can be extracted from the given hadron.} 
\item{\texttt{ApproxMap approx(tcPDPtr particle, const PDFCuts \&) const}; this
function is used to specify, approximately, the upper limits of the parton 
densities, of each parton, for the given hadron. The \texttt{PDFCuts} class is
used to give as input the kinematical region that the PDF will be used in.}
\item{\texttt{double xfl(tcPDPtr particle, tcPDPtr parton, Energy2 partonScale,
double l, Energy2 particleScale = 0.0*GeV2) const}; this function is the actual
implementation of the PDF. This function returns the momentum fraction as
$l = \log\left(1/x\right)$.}
\end{itemize} 
Optionally, the PDF can also be specified with the functions:
\begin{itemize}
\item{\texttt{double xfx(tcPDPtr particle, tcPDPtr parton, Energy2 partonScale,
\newline double x, double eps = 0.0, Energy2 particleScale = 0.0*GeV2) const};
\newline  this
function is used to specify the PDF and return the momentum fraction as $x$. 
By default, this just returns $\exp\left({-\texttt{xfl}(\dots)}\right)$.}
\item{\texttt{double xfvl(tcPDPtr particle, tcPDPtr parton, Energy2 
partonScale, \newline double l, Energy2 particleScale = 0.0*GeV2) const}; 
\newline this function
returns just the valence part of the PDF with momentum fraction $l=\log\left(
1/x\right)$. By default this just returns 0.} 
\item{\texttt{double xfvx(tcPDPtr particle, tcPDPtr parton, Energy2 
partonScale, \newline double x, double eps = 0.0, Energy2 particleScale = 0.0*GeV2) 
const}; \newline this function is the same as \texttt{xfvl} except it returns the 
momentum fraction as $x$ instead of $l$. It also returns 0 by default.}
\end{itemize}

The GRV PDFs have been implemented as part of \ThePEG. The GRV94L and
GRV94M PDFs have been implemented. These are two implementations of the GRV
method where optimal fits of the distributions of the partons have been made
to different data.

The MRST PDFs have also been implemented as part of \HWP. This implementation 
is a wrapper to a previous C++ implementation~\cite{Jeppe}. The original 
implementation was able to read in the data from a file. The files are 
standardized and when new data is analyzed and integrated into the 
distribution, new data files are created. The new versions are easily 
integrated into the event generator by simply reading in these data files.  

In order for the event generator to function correctly something must be done
with the remnant of the incoming hadron left behind when a parton interacts. 
These are handled by \texttt{RemnantHandler}s. 
\ThePEG~has a default remnant handler, \texttt{BaryonRemnant}, but this is 
designed to be integrated with the \Pythia7~string fragmentation model. This 
remnant handler works in two ways. If a valence parton is chosen by the PDF 
then the only remnant is the colour connected diquark remaining in the beam 
particle. If a sea parton is used in the hard subprocess then a colour 
connected parton is produced along with a colourless hadron. The implementation
of \texttt{BaryonRemnant} in \ThePEG~requires a $p_\perp$-generator and a 
$z$-generator. Though the $z$-generator can be designed to work with other 
hadronization models, it is really a feature of the string fragmentation model
and isn't something that is needed in \HWP.

Instead, since \HWP~uses a cluster hadronization model, we use a different 
remnant handler. This remnant handler, also called \texttt{BaryonRemnant}, is 
really nothing
more than a place-holder in \HWP; the initial-state shower in \HWP~is 
designed to evolve back to a valence parton. This class simply generates a 
diquark, with either spin 1 or spin 0, with a flavour depending on the parton 
drawn by the
PDF. This remnant is often wrong because the initial-state shower evolves 
backwards to a valence parton but the generation of this valence parton is 
not known at the time the remnant is created. Correctly handling the remnant 
is instead implemented as part of the backwards evolution and the original
remnant created by \texttt{BaryonRemnant} is replaced\footnote{
\texttt{BaryonRemnant} is only implemented as it is a requirement of \ThePEG.}

\section{Parton Shower}
In chapter~\ref{chap:Intro} I have developed the theoretical background of the 
parton shower and how it takes the matrix elements calculated perturbatively
and evolves down to a stage where hadronization models are valid. Chapter
\ref{chap:NewVars} presented the development of a set of evolution variables
used in the \HWP~shower. In this section I present the implementation specific
features of the \HWP~shower. This includes how the shower is initialized, how 
it terminates and how the momenta of the particles are set. 

\subsection{Hard Matrix Element Corrections}
Given an $n$--jet process it is sometimes possible to calculate the matrix 
element for $n+1$ jets and this matrix element can be used in the Monte Carlo. 
There are already soft and collinear
jets of order $n+1$ from the shower of the $n$-jet matrix element and these
jets must be matched with the $n+1$-jet matrix element to avoid double 
counting. This matching is known as \emph{hard matrix element corrections}.

Hard matrix element corrections have only been implemented for the 
process $e^+ e^- \rightarrow \gamma^* / Z^0 \rightarrow q \bar q$. To make a 
hard correction a pair of three-body phase space variables, $x, \bar x$ are
generated according to the original two-jet matrix element. Emissions in the
dead region of fig.~\ref{fig_bfragn} are only accepted according to the 
three-jet matrix element. In the
case of $e^+ e^- \rightarrow q \bar q$ only 3\% of the events are
corrected by the hard matrix element. 

If a hard emission is added the $q\bar q$-final state is replaced by a 
$q \bar q g$-final state and the orientation of either the quark or the 
anti-quark is kept with weights $x^2$ and $\bar x^2$, respectively. This 
results in properly oriented three-jet events, apart from finite mass effects
\cite{Kleiss:1986re}. This procedure takes into account the most important 
subleading higher-order corrections that are not given by the parton shower.

The \texttt{ShowerVariables} class has a variable, \texttt{MECorrMode}, which 
sets whether the hard matrix elements are on or off. 

\subsection{Initial Conditions}
As we saw in chapter~\ref{chap:Intro} there is an angular ordering principle 
that restrict the showering to occur only in the cone of half angle between two
colour connected partons. Since the shower is restricted to a cone in
relation to one of its colour partners, the first step of the shower is to
determine which colour partners the soft and collinear showers will occur 
between. There is a flag, \texttt{Approach}, used in the \texttt{PartnerFinder}
class that sets the way the shower partner is chosen. If this flag is zero, 
then the partner is set completely randomly amongst the partons that
are colour-connected to it and the partners of all partons are set
independently 
of each other. This means if we have particle $a$, which chooses its shower 
partner to be particle $b$, particle $b$ does not have to choose $a$ as its 
shower partner. Instead, if it had colour partners $a$ and $c$ it would 
randomly choose between these. On the other hand, if \texttt{Approach} is set 
to 1 then it randomly selects a shower partner and sets both particles to be 
shower partners with each other.

A partner is chosen for each gauge `charge' of the parton. For example,
a quark has a QED charge and a QCD charge. A colour partner and an electric
charge partner are both set. This allows the QED showers to compete directly
with QCD ones.

As shown in chapter~\ref{chap:NewVars}, the evolution of a particle is carried
out in the Sudakov basis,
\begin{equation}
q = \alpha p + \beta n + q_\perp,
\end{equation}
where $p$ is the momentum of the particle which is evolving and $n$ is a
lightlike ($n^2=0$) vector with 3-momentum in the `backwards' direction, which
is conventionally set to that of the colour partner of the particle in their
c.m.~frame. $q_\perp$ is
in the transverse direction and satisfies $q_\perp \cdot p = q_\perp \cdot n =
0$.

Once the partner is chosen the initial value of the evolution variable,
$\tilde{q}$, is set. The value of this variable depends on whether 
both partners are initial-state, final-state or a combination. If one parton
is initial-state and one is final-state the values are
\begin{eqnarray}
\tilde{q}_i &=& \sqrt{\left( p_i + p_f \right)^2+m_f^2}, \\
\tilde{q}_f &=& \sqrt{\left( p_i + p_f \right)^2 + 2 m_f^2},
\end{eqnarray}
where $p_i$ is the momentum of the initial-state parton and $p_f$ is the
momentum of the final-state parton. This 
corresponds to (\ref{eq_inifin_kbc}) for the most symmetrical choice, where
$\tilde{q}^2 = \tilde{\kappa} Q^2$. For final-final shower partners the 
initial conditions are
\begin{eqnarray}
\tilde{q}_1 &=& \sqrt{\left(p_1 + p_2\right)\cdot\left( p_1 + n_2 \right)}, \\
\tilde{q}_2 &=& \sqrt{\left(p_1 + p_2\right)\cdot\left( p_2 + n_1 \right)},
\end{eqnarray}
where $p_1$ and $p_2$ are the momenta of the incoming partons. 
$n_1$ has three momentum equal to $\bm{p}_1$ in the c.m.~frame and $n_1^2 = 
0$; similarly for $n_2$. This corresponds to (\ref{eq_fftks}) for the 
symmetric choice. Lastly, for the choice of two initial-state particles the 
initial conditions are
\begin{equation}
\tilde{q}_1 = \tilde{q}_2 = \sqrt{\left( p_1 + p_2 \right)^2}.
\end{equation}
This corresponds to the symmetric case of $\tilde{\kappa}_1=\tilde{\kappa}_2
=1$, shown in fig.~\ref{fig_DYps}.

\subsection{Initial-State Shower}
The initial-state shower evolution begins with the two incoming partons
that have been chosen from the PDF. These partons are considered as on-shell
partons in the hard matrix element calculation and the initial $\tilde{q}$
are set as described in the previous section.

Once the initial value is set for the evolution variable, each parton then 
evolves independently of each other. The evolution of one
parton proceeds using the \emph{veto} algorithm. For each possible 
type of branching, $a\rightarrow bc$, a new $\tilde{q}$ and a $z$ are generated
based on ratios of the Sudakov form factor
\begin{equation}
\label{eqn:sudi}
R^i_{ba}\left(\tilde{q}_i, \tilde{q}_{i+1}\right) = 
\frac{f_b(x/z,\tilde{q})\Delta_{ba} \left( \tilde{q}_C, \tilde{q}_i \right) }
{f_b(x,\tilde{q})\Delta_{ba} \left( \tilde{q}_C, \tilde{q}_{i+1} \right)},
\end{equation}
with $f_b$ the PDF and $\Delta_{ba}$ the Sudakov form factor
\begin{equation}
\Delta_{ba} \left( \tilde{q}_C, \tilde{q}\right) = \exp \left\{ - \int_
{\tilde{q}_C}^{\tilde{q}} \frac{d\tilde{q}^2}{\tilde{q}^2} \int dz
\frac{\as\left(z,\tilde{q}\right)}{2\pi} P_{ba}\left(z,\tilde{q}\right) 
\Theta\left(\bm{p}_\perp > 0\right) \right\}.
\label{eqn:vetoalg}
\end{equation}
$\tilde{q}_C$ is the lower cutoff and by default is set to be the 
non-perturbative gluon mass, $m_g=750$~MeV. The running coupling, $\as(z,
\tilde{q})$, depends on the evolution scale and the momentum fraction. The 
argument is $(1-z)^2\tilde{q}^2$ for reasons given in 
chapter~\ref{chap:NewVars}. The $P_{ba}$ are the quasi-collinear splitting 
functions for the massive partons~\cite{HeavySplitting}. For QCD branchings 
these are
\begin{eqnarray}
P_{qq}(z,\tilde{q}) &=& C_F \left[ \frac{1+z^2}{1-z} - \frac{2 m_a^2}{z(1-z)
\tilde{q}^2} \right], \\
P_{qg}(z,\tilde{q}) &=& T_R \left[ 1 - 2z(1-z) + \frac{2 m_a^2}{z(1-z)
\tilde{q}^2} \right], \\
P_{gq}(z,\tilde{q}) &=& C_F \left[ \frac{1 + (1-z)^2}{z} - \frac{2 m_a^2}{
z(1-z)\tilde{q}^2} \right], \\
P_{gg}(z,\tilde{q}) &=& C_A \left[ \frac{z}{1-z} + \frac{1-z}{z} + z(1-z)
\right],
\end{eqnarray}
and $m_a = 0$ for initial-state radiation.
For the QED case, we change $\as$ to the fine structure constant $\alpha_{\rm 
em}$ and use the branching for $q\rightarrow q \gamma$. Ignoring the parton
mass this is
\begin{equation}
P^\gamma_{qq}(z,\tilde{q}) = e_a^2 \frac{1+z^2}{1-z},
\end{equation}
where $e_a^2$ is the electric charge of the parton, in units of elementary 
charge.

The $\Theta(\bm{p}_\perp > 0)$ function in (\ref{eqn:vetoalg}) is used to 
ensure that it is possible
to reconstruct transverse momentum, $\bm{p}_\perp$, from the evolution 
variables. $\tilde{q}$ determines the relative transverse momentum. For quark 
branching this is
\begin{equation}
\left| \bm{p}_{\perp i} \right| = \sqrt{(1-z_i)^2 \tilde{q}_i^2 - z_i Q_g^2}.
\end{equation}
For the gluon branching, in the initial state shower, this is
\begin{equation}
\left| \bm{p}_{\perp i} \right| = \sqrt{z^2(1-z)^2 \tilde{q}^2 - Q_g^2}.
\end{equation}

Eqn. (\ref{eqn:sudi}) gives the probability of no branching above the scale
$\tilde{q}_{i+1}$. $1-R^i_{ba}(\tilde{q}_i, \tilde{q}_{i+1})$ is therefore the
probability for the next branching to happen above $\tilde{q}_{i+1}$. The
derivative with respect to $\tilde{q}_{i+1}$ is then the probability density
for the next branching to happen at the scale $\tilde{q}_{i+1}$.

Since (\ref{eqn:sudi}) is not directly solvable, the veto algorithm is used. 
In this algorithm each part of the distribution is sampled independently by a
function which is always greater than the desired one, and a veto is placed on
the emission if the ratio of the actual function to the approximated function
is larger then some random number. This gives several veto points
\begin{equation}
w_1 = \Theta(\bm{p}_\perp > 0),~~~w_2 = \frac{P_{ba}(z,\tilde{q})}{g_{ba}(z)},
~~~w_3 = \frac{\alpha(z,\tilde{q})}{\alpha_{\rm max}},~~~w_4 = \frac{f_b(x/z,
\tilde{q})}{f_b(x,\tilde{q})},
\label{eqn:vetop}
\end{equation}
where
\begin{eqnarray}
g_{qq}(z) = \frac{2C_F}{1-z},\\g_{qg}(z)=T_R,\\g_{gq}(z)=\frac{2C_F}{z},\\
g_{gg}(z) = C_A\left[ \frac{1}{1-z} + \frac{1}{z} \right],\\
g^\gamma_{qq}(z) = \frac{2 e_a^2}{1-z}.
\end{eqnarray}

The trick to the veto algorithm, to ensure that things are correctly sampled,
is that first a $\tilde{q}$ is generated according to 
\begin{equation}
\label{eqn:qtil}
\tilde{q}^2 = \tilde{q}_s^2 \exp \left( {\frac{2 \pi}{\alpha_{\rm max} \left( 
I(z_+) - I(z_-)\right)}} \ln {\cal R} \right),
\end{equation}
with $\tilde{q}_s$ the current scale, $z_+$ and $z_-$ the upper and lower 
bounds on $z$, respectively and $I(z_0)$ the value of $g_{ba}$ integrated over
$z$ and from zero to $z_0$. A $z$ is then generated according to the 
approximate functions, $g_{ba}$ between $z_+$ and $z_-$. Each veto is then
tested. If a veto fails a new $\tilde{q}$ is generated. This time, however, 
instead of $\tilde{q}_s$ in (\ref{eqn:qtil}) it is equal to the $\tilde{q}$ 
that was vetoed. If this $\tilde{q}$ is smaller than $\tilde{q}_C$ then there
is no more branchings. The largest value of $\tilde{q}$ generated from each
of the branchings decides which of the branchings to use. 

Once a splitting has been chosen a new initial state parton and a new
final state parton are created. The final state parton is taken as an 
on-shell parton. This will be put off-shell during the final-state evolution.
The momentum fraction, $z$, from the Sudakov form factor and the new scale 
$\tilde{q}$ are passed to the new initial state parton so it can split. This 
process is repeated until no new scale is chosen below $\tilde{q}_C$ for any 
of the branchings. From the interpretation of the Sudakov form factor, this 
means that there is no more branching and the evolution of this parton has 
terminated.

As discussed before, beam remnants aren't handled properly by the remnant 
handler. Instead they are handled here. When the initial-state evolution of
a parton terminates a set of forced branchings are imposed. There are
three types of termination points: a valence quark, a gluon, or a sea quark.
If the evolution terminates on a valence quark, then a diquark\footnote{If
a beam particle was a meson this would just be a (anti-)quark} of the 
appropriate flavours is produced as the beam remnant. If instead the 
evolution terminates on a gluon, a forced splitting of the gluon from a quark
of a valence flavour is imposed. The momentum fraction $z$ is distributed 
according to the splitting function. The $\tilde{q}$ is just distributed
by $dP = d\tilde{q}/\tilde{q}$ between $\tilde{q}_C$ and $\tilde{q}_s$. The 
remnant is again just a (di)quark of the flavour(s) that remain in the beam 
particle. The most complex case is when the evolution terminates on a sea 
quark. In this case two forced splittings are imposed. The first is to force a 
splitting into a gluon. Again the momentum fraction is distributed according to
the appropriate splitting function. The $\tilde{q}$ is also distributed as 
$dP = d\tilde{q}/\tilde{q}$ between $\tilde{q}_C$ and $\tilde{q}_s$. The new 
gluon is then forced to split into a valence quark in the same manner described
above.

This process is repeated for both incoming partons. Once they have evolved and
their remnants are correctly set the momenta of all the partons needs to be
set. The initial condition is that the beam particles are coming in with known
momenta. The Sudakov variable $\alpha$ of the first parton is then set to 
unity, the 
$\beta$ is set to zero and its $x$ is known from the backwards evolution. Each 
child of the parton then has its $\alpha$ set to $z\alpha_i$, for the initial 
state particle and $(1-z)\alpha_i$ for the final state partner produced during 
the backward evolution. The $\beta$ for the on-shell final state partons is
\begin{equation}
\beta'_{i+1} = \frac{m^2 + \bm{p}'^{2}_{\perp i+1}}{2 \alpha_i p \cdot n}.
\end{equation}
The $p$ and $n$ vectors are the Sudakov basis vectors for the shower and 
\begin{equation}
\bm{p}'_{\perp i+1} = z \bm{p}_{\perp i} - \bm{p}_\phi \sqrt{z^2 \tilde{q}^2 - 
(1-z)Q_g^2}.
\end{equation}
Here $\bm{p}_\phi$ is a 2 vector given by $(\cos \phi, \sin \phi)$. By default 
$\phi$ is distributed uniformly in the region $\left[ 0, 2\pi \right]$ but 
improvements, such as the spin correlations described in 
chapter~\ref{chap:Intro}, can be implemented.  The values for
the initial-state parton $i+1$ are 
\begin{eqnarray}
\beta_{i+1} &=& \beta_i - \beta'_{i+1},\\
\bm{p}_{\perp i+1} &=& \bm{p}_{\perp i} - \bm{p}'_{\perp i+1}.
\end{eqnarray}
 
All of the $\alpha$'s and $\beta's$ are set until the parton involved in the 
hard subprocess is reached. This fully defines the momentum of the
partons. Unfortunately, this does not guarantee momentum conservation at the
hard subprocess. Instead
the momentum of the partons must be ``shuffled'' in order to impose momentum
conservation. Rescaling the momentum of the partons in the hard subprocess,
and correspondingly boosting all the partons involved in the evolution of each 
initial-state partons, allows for momentum conservation. This rescaling can 
affect other properties, however, and we want to constrain the value of the 
rescaling so that certain properties are retained. 

In proton-proton collisions we want to conserve the rapidity, $Y$, and the 
c.m. energy squared, $M^2$, of the hard process, while rescaling each 
incoming parton independently.  Each parton has its momentum shifted by $k_1$ 
and $k_2$, given in the Sudakov base by
\begin{eqnarray}
q_1 &=& k_1 \alpha_1 p_1 + \frac{\beta_1}{k_1} n_2 + p_{1\perp}, \nonumber \\
q_2 &=& \frac{\alpha_2}{k_2} p_2 + k_2 \beta_2 n_1 + p_{2\perp}.
\end{eqnarray}
These shifts, $k_1$ and $k_2$, are applied so that the virtuality of the 
partons is conserved. Conserving rapidity and the c.m. energy squared requires
\begin{eqnarray}
\frac{\beta_1}{k_1} + k_2 \beta_2 &=& k_1 \alpha_1 + \frac{\alpha_2}{k_2},\\
1+\frac{(p_{1\perp} + p_{2\perp})^2}{M^2} &=& \left(k_1 \alpha_1 +
 \frac{\alpha_2}{k_2}\right) \left( k_2 \beta_2 + \frac{\beta_1}{k_1} \right).
\end{eqnarray}
Both of these give a quadratic solution for each $k$ value. This leads to 
four different combinations of values which must be considered. Two 
combinations of these four have negative $k$'s. This corresponds to flipping
the event, e.g. the parton which started from the $-z$ direction ends up on
the $+z$ direction and vice versa. In some cases the reconstruction of the
shower fails. In these cases the shower is \emph{vetoed} and started again 
from the on-shell incoming partons. 

\subsection{Final-State Shower}
The final-state shower is similar to the initial state shower. The partons
start as on-shell partons but are taken off shell by the evolution. The initial
scale of a parton is set according to its shower partner and the partons 
evolve down to the scale $\tilde{q}_C$ in the same manner as the initial-state 
shower, except the probability of emission is modified. Instead of $R^i_{ba}$ 
we have
\begin{equation}
R^f_{ba} = \frac{\Delta_{ba}(\tilde{q}_C,\tilde{q}_i)}{\Delta_{ba}(\tilde{q}_C,
\tilde{q}_{i+1})}.
\end{equation}
This means that the extra veto, $w_4$ from (\ref{eqn:vetop}), is not applied 
to final-state 
branchings. Though the generation of the branchings is similar, the kinematics 
of the final-state shower is different from that of the initial-state shower. 
From the results in chapter~\ref{chap:NewVars} we find the relative transverse 
momentum as
\begin{eqnarray}
\left| \bm{p}_{\perp i} \right| &=& \sqrt{(1-z_i)^2 z_i^2 (\tilde{q}_i^2 - 
\mu^2) - z_i Q_g^2},\\
\left| \bm{p}_{\perp i} \right| &=& \sqrt{z_i^2(1-z_i)^2 \tilde{q}_i^2 -\mu^2},
\end{eqnarray}
for quark branching and gluon branching respectively and 
$\mu = {\rm max}(m_a, Q_g)$. 

All of the partons created during the evolution also shower starting at the 
$\tilde{q}$ that they were produced at. Every parton showers until the 
condition of no more branching is met. Once this is reached
the parton is put on its mass shell so that momentum reconstruction can be 
done. 

The partons produced in the hard subprocess have their $\alpha$ set to unity 
and their $\beta$ set to 0. Each of children then have their Sudakov variable
$\alpha$ set to 
$z \alpha_i$ or $(1-z) \alpha_i$, set by the convention of the splitting 
function. The $\beta$ for the parton corresponding to the $z$ momentum fraction
is
\begin{equation}
\beta_{i+1} = \frac{q_{i+1}^2 + \bm{q}_{_\perp i+1}^2 - \alpha_{i+1}^2 p^2}{2 
\alpha_{i+1} p \cdot n},
\end{equation}
where $p$ and $n$ are the Sudakov basis vectors of the shower. The $\beta$
corresponding to the $(1-z)$ momentum fraction is just
\begin{equation}
\beta'_{i+1} = \beta_i - \beta_{i+1}.
\end{equation}

Since the final partons are on their mass shells, the initial partons aren't.
This means instead of having $p_j^2 = m_j^2$ they have acquired some 
virtuality, $p_j^2 = q_j^2$. The original momenta, $p_j = \left( \sqrt{
\bm{p}_j^2 + m_j^2},\bm{p}_j\right)$ in the centre-of-mass frame, define some
properties of the hard subprocess. In the case of $e^+ e^-$ annihilation we
want to preserve the centre-of-mass energy
\begin{equation}
\sqrt{s} = \sum_{j=1}^n \sqrt{m_j^2 + \bm{p}_j^2}
\end{equation}
while keeping the sum of momenta equal to zero. This requires momentum
reshuffling. To conserve $\sqrt{s}$ we rescale the momentum of each jet by
a common factor, $k$, that is determined from the equation
\begin{equation}
\sqrt{s} = \sum_{j=1}^n \sqrt{q_j^2 + k \bm{p}_j^2}.
\label{eq:solvek}
\end{equation}
This effectively creates a Lorentz transformation which is applied to all 
partons in the final state. For $n=2$ outgoing particles from the hard
process (\ref{eq:solvek}) can be solved for $k$ explicitly. For $n > 2$ this
is done numerically.

\subsection{Soft Matrix Element Corrections}
We saw earlier that a hard correction is added in to populate the `dead region'
of the $(n+1)$--body matrix element. The parton shower can also be improved in 
the shower regions of phase space by restricting relatively hard gluons that 
are produced in the shower. These gluons are no longer in the domain of 
of validity of the quasi-collinear approximation. 

Each of the $\bm{p}_\perp$ values that are generated during the evolution of 
one jet
is tracked. If a new $\bm{p}_\perp$ is larger than any previously generated
during the evolution a soft matrix element correction is applied to 
it~\cite{Mike}. This correction assumes all other emissions are infinitely 
soft and we can treat the emission
as part of a three-body phase space (e.g. $q \bar q g$). This allows us to
compute the three-body variables, $(x,\bar x)$, from the parton shower 
variables $(\tilde{q},z)$ and the respective Jacobian. The ratio of the
hard matrix element and the shower is then compared to a random number and
the emission is vetoed if the ratio is smaller. This requires that the 
shower approximation is larger than the matrix element everywhere in phase 
space. Otherwise
the ratio must have some factor applied to ensure that the ratio is always less
than 1. This has been studied for several cases in chapter~\ref{chap:NewVars}
and the relevant ratios have been derived.

\subsection{Parameterization of $Q_g$}
The cutoff $Q_g$ is introduced to regularize the soft gluon singularities in
the splitting functions. The relative transverse momentum, $\bm{p}_\perp$ is
related to the Sudakov variables of the parton branching by
\begin{equation}
\bm{p}_\perp = \bm{q}_{\perp i+1} - z \bm{q}_{\perp i}.
\end{equation}
$z$ is required to correspond to a real value of $\bm{p}_\perp$. For a gluon
splitting this is explicitly
\begin{equation}
z_- < z < z_+,~~~~~z_\pm = \frac{1}{2} \left( 1 \pm \sqrt{1-\frac{4 \mu}{
\tilde{q}}} \right),
\end{equation}
with $\tilde{q} > 4 \mu$. For quark splittings $z$ is the solution of a cubic
but is always in the range
\begin{equation}
\frac{\mu}{\tilde{q}} < z < 1 - \frac{Q_g}{\tilde{q}}.
\end{equation}
This allows $z$ to be generated within these regions and simply rejected if it
lies outside phase space. 

\begin{figure}
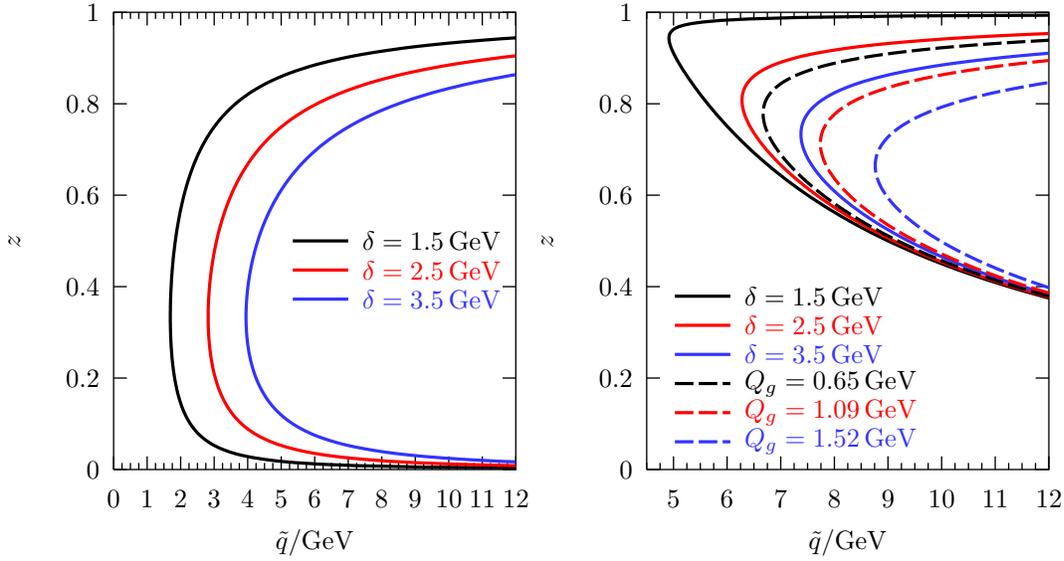

  \epsfig{file=figs/chapter4/thr.10}
  \epsfig{file=figs/chapter4/thr.11}
  \caption{Available phase space of light (left) and $b$--quarks 
    (right) for $q\to qg$ splitting for various values of $Q_g$ and
    depending on the parameterization in terms of $\delta$,  
    eq.~\eqref{eqn:Qgp}.  The dashed lines on the right correspond
    to a $Q_g$ which is not mass dependent, obtained by setting $m=0$
    in (\ref{eqn:Qgp}).}
  \label{fig:Qgparam}
\end{figure}

$Q_g$ is parameterized according to
\begin{equation}
\label{eqn:Qgp}
Q_g = \frac{\delta - 0.3 m}{2.3},
\end{equation}
where $\delta$ is the parameter \texttt{cutoffKinScale} in the class
\texttt{ShowerVariables} and $m$ is the mass of the parton splitting. This 
cutoff is used to give the gluons a minimum virtuality which ensures
that they are able be put on a mass shell with non-zero mass. As we can see 
from figure~\ref{fig:Qgparam}, the form of this function is also to ensure 
that $b$ quarks don't have an extra artificial cutoff that over-restricts the 
phase space of splitting $b$ quarks.

\section{Hadronization}
Chapter~\ref{chap:Hadronization} described in detail the hadronization scheme
used in \HWP. This section will discuss the features of the code and the
implementation of the specific features. Table~\ref{table:HadParams} is a
table with all the relevant parameters of the hadronization.

The main driver of the code is the \texttt{ClusterHadronization} class. This 
class organizes and directs which classes are used next. This class also 
handles the unusual case of a hadronization veto. This can occur during the
light cluster reshuffling, which will be described below. In order to 
direct the algorithm this class has references to \texttt{PartonSplitter},
\texttt{ClusterFinder}, \texttt{ColourReconnector}, \texttt{ClusterFissioner},
and the two decayer classes \texttt{LightClusterDecayer} and 
\texttt{ClusterDecayer} classes. The class
\texttt{ColourReconnector} by default doesn't do anything. This class
is designed to allow for a different colour configuration in the forming of 
the clusters, for example this is done in \SHERPA~\cite{SHERPA}.

The algorithm begins by taking the gluons, which are on a mass-shell, and
non-perturbatively splitting these into $q\bar q$ pairs. The possible flavours
depend on what the mass of the gluon is. This is done for all final-state
gluons produced in the final-state shower (including the final-state evolution
of the gluons produced during the initial-state shower). The splitting is
done by the class \texttt{PartonSplitter}. This class has a reference
to the \texttt{GlobalParameters} object that defines all the global parameters
used throughout \HWP.

\begin{table}
\begin{tabular}{cclc}
\hline
Parameter & Class & Description & Default Value \\ 
\hline \hline
$\Clm$ & \texttt{ClusterFissioner} & The maximum cluster mass & 3.2 \\
     &                             & before a cluster fissions. & \\
$\Clp$ & \texttt{ClusterFissioner} & The cluster mass exponent that & 2.0 \\
     &                             & controls cluster fissions. & \\
\texttt{PSplt1} & \texttt{ClusterFissioner} & This is the mass splitting & 1.0
\\        &                           & parameter for $udsc$ flavours.&\\
\texttt{PSplt2} & \texttt{ClusterFissioner} & This is the mass splitting & 1.0
\\              &                           & parameter for $b$ quarks. &\\
\hline
\texttt{B1Lim} & \texttt{LightClusterDecayer} & Parameter to set the limit of
& 0.0 \\       &                              & light $b$ clusters before &\\
               &                              & forcing decay to one hadron.&
\\  \hline
\texttt{ClDir1} & \texttt{ClusterDecayer} & Flag to turn on or off the & 1 \\
                &                         &the smearing for non-$b$ quarks.&\\
\texttt{ClDir2} & \texttt{ClusterDecayer} & Flag to turn on or off the & 1 \\
                &                         & the smearing for $b$ quarks.&\\
\texttt{ClSmr1} & \texttt{ClusterDecayer} &Gaussian smearing of non-$b$ quarks
& 0.0 \\ 
\texttt{ClSmr2} & \texttt{ClusterDecayer} &Gaussian smearing of $b$ quarks 
& 0.0 \\ \hline
$\texttt{Pwt}_d$ & \texttt{HadronSelector} & Weight of $d$ quarks & 1.0 \\
$\texttt{Pwt}_u$ & \texttt{HadronSelector} & Weight of $u$ quarks& 1.0 \\
$\texttt{Pwt}_s$ & \texttt{HadronSelector} & Weight of $s$ quarks& 1.0 \\
$\texttt{Pwt}_c$ & \texttt{HadronSelector} & Weight of $c$ quarks& 1.0 \\
$\texttt{Pwt}_b$ & \texttt{HadronSelector} & Weight of $b$ quarks& 1.0 \\
$\texttt{Pwt}_{qq}$ & \texttt{HadronSelector} & Weight of diquarks& 1.0 \\
\texttt{SngWt} & \texttt{HadronSelector} & Weight of baryon singlets & 1.0 \\
\texttt{DecWt} & \texttt{HadronSelector} & Weight of baron decuplets & 1.0 \\
\texttt{DKMode} & \texttt{HadronSelector} & Which hadron decay method to use
& 2 \\ \hline \hline
\end{tabular}
\caption{This is a table of all of the relevant hadronization parameters.
Most of the parameters are discussed in chapter~\ref{chap:Hadronization} and
the rest are discussed in this section.}
\label{table:HadParams}
\end{table}

Once the gluons have all been split into $q\bar q$ pairs clusters are 
formed out of the colour connected partons. This is done by the 
\texttt{ClusterFinder} class. This class simply searches all coloured partons
and creates a colour singlet state out of the colour connected partons. There
is also a feature to work with partons that are created by colour sources or
sinks which stem from baryon violating processes. This feature has not been
tested though and this needs to be done once baryon violating processes are 
included into \HWP. After the clusters are formed, if a 
\texttt{ColourReconnector} class was defined to change things, it would be
called to do so at this stage.

After the \texttt{ClusterFinder} class has created all of the colour singlet
clusters, they are passed to the \texttt{ClusterFissioner} class. As was
described in chapter~\ref{chap:Hadronization}, the heavy clusters are decayed
into lighter clusters. This occurs in three ways. If the mass drawn for one
of the new clusters in not heavy enough to form the lightest possible pair
of hadrons, given the set of flavours, the decay is instead forced to $C
\rightarrow H + C'$ where $H$ is a hadron and $C'$ is a new cluster.
$H$ is the lightest hadron of the flavour of the cluster whose mass is too 
light. The mass generated for the light cluster is changed to equal the mass of 
the lightest hadron.  If 
the mass of the cluster $C'$ produced in the decay would then violate the 
phase space bounds it is set to have the largest mass available. The next 
special case is when both clusters are too light to form a hadron pair. This 
decay,
$C\rightarrow H_1 + H_2$, follows the same procedure as the previous one. Each
hadron is the lightest hadron of the appropriate flavours and the masses are
set to the mass of the respective hadron. If this violates the phase space
bounds, then the cluster isn't decayed via this mechanism. In fact, if this
still violates phase space it will be handled by the 
\texttt{LightClusterDecayer} later. The last possible way for a heavy cluster 
to be decayed is directly into two new clusters, $C \rightarrow C'_1 + C'_2$. 
In any case, if any of the new clusters is still too heavy, it also decays 
via this algorithm.

After the heavy clusters have been handled, the clusters which are too light
are treated. This is done by the \texttt{LightClusterDecayer}. This takes
clusters which are too light to decay into a pair of hadrons and turns them
directly into a single hadron. It may be possible that a cluster has formed 
which is even too light to decay directly into the lightest hadron. In this 
case the momentum of this cluster is `reshuffled' with a cluster which is 
nearby in $x$-space. The algorithm only `borrows' momentum from a given cluster
once during the process. If it is impossible to find enough momentum to borrow
then the event is vetoed and begun again from the \texttt{ClusterFissioner} 
stage. In the rare case that after vetoing several times there is still not a 
possible configuration to decay all the clusters then the whole event is 
vetoed.

The last step of the hadronization is turning the clusters into hadrons. This
is done by the \texttt{ClusterDecayer} class. This has a reference to a 
\texttt{HadronSelector} class. The \texttt{HadronSelector} class has 
been implemented with a flag, \texttt{DKMode}, which is used to select
which of the methods discussed in chapter~\ref{chap:Hadronization} to use.
By default this flag is set to 2, which is the new method. If it was changed
to 0 the implementation of the \HW{} method would be used. A value of 1 is for
the Kupco method. Once a pair of hadrons has been selected by one of the 
methods then the cluster is decayed. The decay products are generated in the
direction of the constituent quark. This direction can have a Gaussian 
smearing applied to it to. If the flag \texttt{ClDir} is set and 
\texttt{ClSmr} is larger then 0.0001, then cluster decay is in the direction
\begin{equation}
\cos \theta = \cos \theta_q \cos \theta_{\rm smr} - \sin \theta_q 
\sin \theta_{\rm smr},
\end{equation}
where $\theta_q$ is the $\theta$ of the original constituent quark and
\begin{equation}
\cos \theta_{\rm smr} = 1 + \texttt{ClSmr} \log {\cal R},
\end{equation}
with ${\cal R}$ a random number. The azimuthal angle is distributed 
uniformly in $[-\pi,\pi]$.

\section{Decays}
In chapter~\ref{chap:Intro} the general scheme of the decays was presented.
Since the algorithms implemented in \HWP~are very basic, in this section
I will explain the way to setup the decay modes in \HWP~and how one would
implement a new set of decay modes.

In \HWP~the default decays are defined in the file
\texttt{Hw64Decays.in}. In this file all of the decay modes are included.
An entry of a decay mode takes the form
$$\mbox{\texttt{decaymode} $a$-$>$$b,c,d,\dots;$ $BR$ 0/1 Class}$$
for the decay $a\rightarrow b+c+d+\dots$, $BR$ is the branching ratio and
the 0/1 option specifies whether to turn this mode off or on, respectively.
The last argument specifies which class to use to perform the decay. By 
default \HWP~has only the \texttt{HwDecayer} class, the \texttt{HeavyDecayer}
class and the \texttt{QuarkoniumDecayer} class. All of these classes use a
parameter \texttt{MECode} to specify the matrix element code to use.

The class \texttt{HwDecayer} is the most common class used in the default
\HWP~decays. This just implements the decays as described in 
chapter~\ref{chap:Intro} based on the \texttt{MECode} parameter. A value of
100 uses the free massive $(V-A)(V-A)$ matrix element, for example $\tau^- 
\rightarrow \bar \nu_e e^- \nu_\tau$. 101 is the code to use the bound massive 
matrix element. This would be used for a decay such as $K^- \rightarrow 
\bar \nu_e e^- \pi^0$. Any other value of the parameter uses an isotropic 
$n$-body decay.

The class \texttt{HeavyDecayer} takes heavy mesons and decays the heavy
parton weakly. The $W$ produced in the decay is also decayed all in
one step. This produces 4 partons, two from the decay of the $W$, one 
from the production of the $W$ and the remaining spectator parton. For
example a chain could look like, $B^0 \rightarrow d \bar c W^+ \rightarrow
d \bar c \bar d u$. The decay products have the colour connections set
properly and the administrative class \texttt{HwDecayHandler} directs the 
program to reshower and rehadronize these partons.

The last decayer, the \texttt{QuarkoniumDecayer}, is designed to work
with heavy mesons and baryons that decay into gluons and quarks. Possible
modes are to 3 gluon states, 2 gluons and a photon, just 2 gluons or a $q
\bar q$ state. This class handles a \texttt{MECode} of 130 by using 
a positronium matrix element. Decays from this class also lead to showering 
and hadronization of the partonic decay products. An example of this type of 
decay is $\eta_c \rightarrow g g$.

Implementation of a new decay mode requires the definition of only two 
functions. A new decay matrix element must inherit the \texttt{Decayer} class
from \ThePEG. The two functions that must be implemented are
\begin{itemize}
\item{\texttt{bool accept(const DecayMode \&) const}; this function is used
to indicate if a particular decay mode can be handled. For example if you
implemented a decay matrix element for $\tau^\pm$ decays, if the incoming
particle was not a $\tau^\pm$, this function would simply return false.}
\item{\texttt{ParticleVector decay(const DecayMode \&, const Particle \&) 
const}; this function returns the decay products with their momentum set. The 
input is the decay mode selected based on the branching ratios, and the 
instance of the particle which is decaying.}
\end{itemize}

Improving the decay modes of \HWP~is currently underway. This is a lengthy 
process and as long as experimentalists study the decays of particles, 
improving the decay mode matrix elements, as well as the branching ratios, 
will always be possible. But as will be shown in the next chapter, even with
the simple decay matrix elements, fits to particle spectra are good.

\chapter{Results}
\label{chap:Results}

\section{Introduction}
This chapter presents results from the \HWP\ . The results given here are
for $e^+e^-$ annihilation events, as this is the first step in the 
redevelopment of \HW{}.  In order to have full control of the basic physics 
steps that are simulated, it was thought to be very important to put the new 
generator on a firm basis with respect to LEP and SLC results before upgrading 
it to be able to deal with initial-state showers and the other requirements for
the simulation of lepton-hadron and hadron-hadron collisions.  Therefore,  
thorough tests have been performed on the predictions of the generator against 
a wide range of observables that have been measured at LEP and SLC. We have 
also explored the sensitivity to the most important parameters and cutoffs.  
These results are not the result of a high-precision tuning: the main aim here 
is rather to show the results of the program and that it is able to give 
results as acceptable as those generated by its predecessor \HW{} for a 
reasonable choice of parameters.

\subsection{Main features of the code}
As was discussed in detail in the last three chapters, the main stages of the 
simulation of events is the same as in \HW{} \cite{Herwig64}.
However, in comparison to its predecessor, \HWP{} features a new parton
shower and an improved cluster hadronization model.  At present,
hadronic decays are implemented in the same fashion as they were in
\HW{}. 

As discussed in chapter~\ref{chap:Herwig}, the program is based on the 
Toolkit for High Energy Physics Event Generation (\ThePEG{}) \cite{ThePEG} and 
the Class Library for High Energy Physics (\CLHEP{}) \cite{CLHEP}.  They are 
utilized in order to take advantage of the extended general functionality they 
can provide. The usage of \ThePEG{} unifies the event generation framework with
that of \Pythia7{}.  This will provide benefits for the user, as the
user interface, event storage etc.\ will appear to be the same.  The
implementations of the physics models, however, are completely
different and independent from each other.

The simulation of $\ee$ annihilation events starts with an initial hard process
$\ee\to\gamma^*/Z^0\to\qq+\gamma\gamma$.  The final state photons simulate QED
radiation from the initial state, so that a radiative return can be
properly simulated.  For these results we are only interested
in the details of the QCD parton shower in the final state.  The
final-state parton shower starts with a quark and antiquark that carry
momenta $p_q$ and $p_{\bar q}$, respectively, and have an invariant
mass squared of $Q^2=(p_q+p_{\bar q})^2$.  For the $\ee$ results, the only 
detail we are concerned with in relation to initial-state radiation is that the
centre-of-mass frame of the $\qq$--pair is slightly boosted with
respect to the collider laboratory frame and that $Q$ may be different
from the $\ee$ centre-of-mass energy.  We have made sure that the
applied cuts on the energy of the annihilating $\ee$ subsystem are the
same as those used in the experimental analyses.

Currently, proton-proton collisions are being studied. These results are 
simulated by starting with the Drell-Yan hard process 
$q\bar q\to\gamma^*/Z^0\to \ell^+ \ell^-$, with $\ell^\pm$ a charged 
lepton. These results concentrate on the effects of the initial-state radiation
from the incoming quarks and gluons. 

\subsubsection{Parton shower} 
As has been shown in the preceding chapters, the partonic evolution from the 
large scale of the hard collision process down to hadronic scales via the 
coherent emission of partons, mainly gluons, is simulated on the basis of the 
Sudakov form factor.  Starting from the hard process scale $Q_0$, subsequent 
emissions at scales $Q_i$ and momentum fractions $z_i$ are randomly generated 
as a Markov chain on the basis of the soft and collinear approximation to 
partonic matrix elements.  Chapter \ref{chap:NewVars} has shown that for \hpp{}
we have chosen a new framework of variables, generically called 
$(\tilde q, z)$.  Here, $\tilde q$ is a scale that appears
naturally in the quasi-collinear approximation of massive partonic matrix
elements and generalizes the evolution variable of \HW{} to the
evolution of massive quarks.  $z$ is a relative momentum
fraction; the evolution is carried out in terms of the Sudakov
decomposition of momenta in the frame where the respective colour
partners are back-to-back.  As in \HW, the use of the new variables
allows for an inherent angular ordering of the parton cascade, which
simulates coherence effects in soft gluon emission.  The details of
this underlying formalism have been described in chapter~\ref{chap:NewVars}.

The most important parameter of the parton shower that we will be
concerned with in this chapter is the cutoff parameter $Q_g$, which
regularizes the soft gluon singularity in the splitting functions and
determines the termination of the parton shower. This is set by $\delta$ in
(\ref{eqn:Qgp}).  Less important but
relevant in extreme cases is the treatment of the strong coupling
constant at low scales.  We have parametrized $\alpha_S(Q)$ below a
small scale $Q_{\rm min} > \Lambda_{QCD}$ in different ways.  We keep
$Q_{\rm min}$ generally to be of the order of 1\,GeV, where we expect
non-perturbative effects to become relevant.  Below that scale
$\alpha_S (Q)$ can optionally be
\begin{itemize}
\item set to zero, $\alpha_S(Q<Q_{\rm min}) = 0$, 
\item frozen, $\alpha_S(Q<Q_{\rm min}) = \alpha_S(Q_{\rm min})$, 
\item linearly interpolated in $Q$, between 0 and $\alpha_S(Q_{\rm min})$,
\item quadratically interpolated in $Q$, between 0 and 
  $\alpha_S(Q_{\rm min})$\,.
\end{itemize}
We put the final partons of the shower evolution on their constituent
mass shells, since the non-perturbative cluster hadronization will
take over at this scale, so we usually have kinematical constraints
that keep $Q$ above $Q_{\rm min}$, in which case the treatment below
$Q_{\rm min}$ is irrelevant.  Typically, $\alpha_S(Q_{\rm min}) \sim
1$ here.

\subsubsection{Hadronization and decay} 
As discussed in chapter~\ref{chap:Hadronization}, the partonic final state is 
turned into a hadronic final state within the framework of the cluster 
hadronization model of \HW\ \cite{Webber:1983if}.  All three methods of cluster
decays have been implemented in \HWP{}, but the new cluster hadronization 
model is used for the results given in this chapter.  The
emerging hadrons are possibly unstable and eventually decay.  The
decay matrix elements and modes correspond to those in \HW{}.

\begin{table}
\centering
\begin{tabular}{lll}
\hline
Parameter & \HWP{} & Initial\\
\hline \hline
$\alpha_s(M_Z)$  & 0.118 & 0.114 \\
$\delta/$GeV     & 2.3 & ---\\
$m_g/$GeV        & 0.750 &---\\
$Q_{\rm min}$/GeV in $\alpha_s(Q_{\rm min})$\quad{} 
& 0.631\qquad{}
&---\\
\hline
$\Clm/$GeV                               & 3.2 & 3.35\\
$\Clp$                               & 2.0 &---\\
PSplt1 & 1&---\\
PSplt2 & 0.33 &---\\
B1Lim  & 0.0&---\\
ClDir1 & 1&---\\
ClDir2 & 1&---\\
ClSmr1 & 0.40 &---\\
ClSmr2 & 0.0&---\\
${\rm Pwt}_d$ & 1.0&---\\
${\rm Pwt}_u$ & 1.0&---\\
${\rm Pwt}_s$ & 0.85 & 1.0\\
${\rm Pwt}_c$ & 1.0&---\\
${\rm Pwt}_b$ & 1.0&---\\
${\rm Pwt}_{qq}$ & 0.55 & 1.0\\
Singlet Weight & 1.0&---\\
Decuplet Weight & 0.7 & 1.0\\
\hline \hline
\end{tabular}
\caption{The parameters for \hpp{} used in this study. The first group are
shower parameters, the second are all of the hadronization
parameters. In the third column we show initial values of our study,
taken from \HW{}. }
\label{tab:Herwig_defaults}
\end{table}

\section{$\ee$ Annihilation}
\label{sec:ee}
This section presents the results for $\ee$ annihilation events. The 
properties of different measurements are discussed and the comparison of 
\HWP{} to data
is presented. Histograms for all the distributions have been booked in the same
bins as the experimental data.  For a given bin $i$ we then
compare the data $D_i$ value with the \HWP{} Monte Carlo result
$M_i$.  Given the data errors $\delta D_i$ (statistical plus systematic, added
in quadrature)
and Monte Carlo errors $\delta M_i$ (statistical only), we can calculate a
$\chi^2$ for each observable. We keep
the statistical error of the Monte Carlo generally smaller than the
experimental error.  In distributions where the normalization is not
fixed, such as momentum spectra, we allow the normalization of the
Monte Carlo to be free to minimize $\chi^2$.  The normalization is
then tested separately against the average multiplicity.
In all other cases we normalize histograms to unity.  

As we do not want to put too much emphasis on a single observable or a
particular region in phase space where the data are very precise,
in computing $\chi^2$ we set the relative
experimental error in each bin to $\max(\delta D_i/D_i,5\%)$.
This takes into account the fact that the Monte Carlo is only
an approximation to QCD and agreement with the data within 5\% would
be entirely satisfactory. The general trend for the preferred range of a single
parameter was however never altered by this procedure.

After normalization the ratio 
\begin{equation}
  \label{eq:binratio}
  R_i = \frac{M_i-D_i}{D_i} \pm \left( \frac{\delta M_i}{D_i} 
    \oplus \frac{M_i \delta D_i}{D_i^2}\right)
\end{equation}
is computed for each bin in order to see precisely where the model
fails.  This ratio as well as the relative experimental error
and the relative contribution of each bin to the $\chi^2$ of an
observable is plotted below each histogram.

\subsection{Strategy}
\label{sec:multcomment}
We have taken $\chi^2$ values for hadron multiplicities into account
in the same way as we weighted the event shapes.  In general the
multiplicities of individual particle species are sensitive to a
completely different set of parameters.  The general strategy was to
get a good value for the total number of charged particles with a
reasonable set of values for the parton shower cutoff parameter
$\delta$ and the maximum cluster mass parameter $\Clm$. Once this was fixed, 
the hadronization parameters that determine the multiplicities of individual 
particle species were determined. Following this we compared this `preferred' 
set of parameters with the `default' set from \HW{}.  The resulting parameter 
set is shown in Table \ref{tab:Herwig_defaults}. 

A wide range of observables have been tested in order to study the aspects of
the model. Event shape variables and multiplicities are considered in order
to test the dynamical aspects of the parton shower and hadronization models, 
which are closely linked at their interface by the parton shower cutoff
parameter, $\delta$. Ideally, the models should combine smoothly at scales
where $Q_g\sim$ 1 GeV. Many of the figures shown in this chapter contain three 
sets of plots per figure. In order from top to bottom these are
\begin{itemize}
\item{the actual distribution. The \HWP{} result is plotted as a histogram
  together with the experimental data points;}
\item{the ratio $R_i$ (\ref{eq:binratio}) together with an error band showing
  the relative statistical and systematic errors;}
\item{the relative contribution of each data point to the total $\chi^2$ of
  each plot.}
\end{itemize}

\subsection{Hadron multiplicities}
The charged particle multiplicity distribution and the overall multiplicities
of a wide range of hadron species have been taken to test the overall flow of
quantum numbers through the different stages of the simulation. This also
allows a thorough test of the new hadronization model, developed in
chapter~\ref{chap:Hadronization}, against the measured observables.

Table~\ref{tab:mult} showed the results of the new cluster hadronization 
algorithm in comparison to the old algorithm. Even before systematic tuning, 
we can see that the overall results are in better agreement with the data than
those of \HW{}, with fewer results that differ from the data by more than 
three standard deviations (indicated by a star in the table). We can also see
from table~\ref{tab:chi} that the difference between the models is quantified 
by their $\chi^2$.

\begin{figure}
\centering
  \epsfig{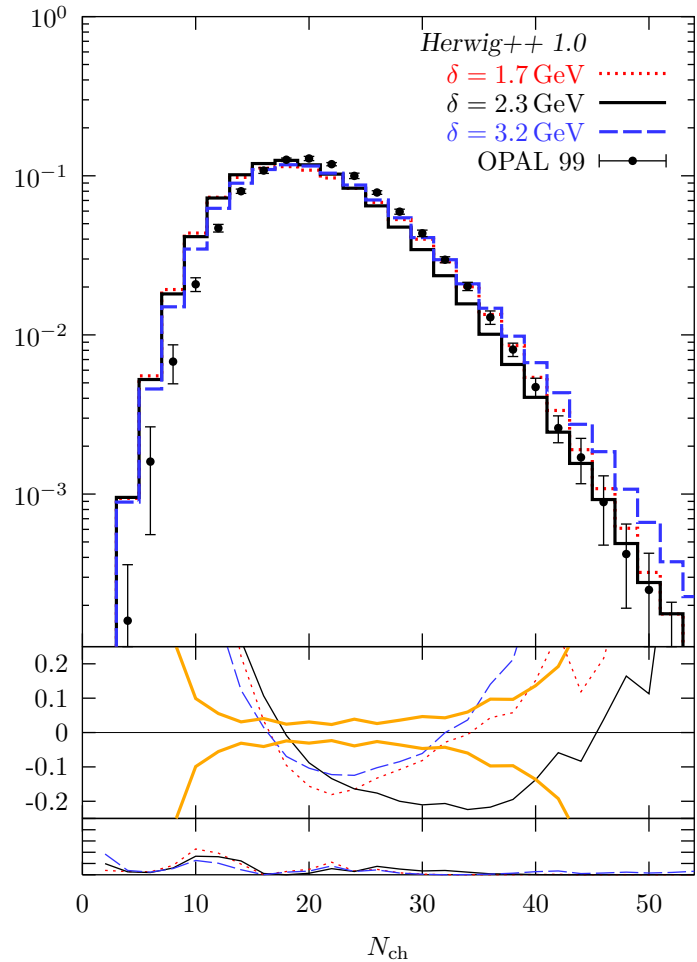}
  \caption{The distribution of the charged particle multiplicity. 
    \label{fig:nchdist}}
\end{figure}

A very well measured property, and therefore important to get accurate, is the
distribution of charged particle multiplicity. Figure~\ref{fig:nchdist} shows 
the results of \HWP{} compared to OPAL data~\cite{Acton:1991aa} and is found to
be in fairly good agreement. There is an excess of lower multiplicity events, 
however. It is also shown that varying $\delta$ doesn't greatly alter
this distribution, another confirmation that the interface between the 
parton shower and the hadronization is consistent.

\subsection{Jet multiplicity} 
This measurement is the multiplicity of (mini-)jets in $\ee$-collisions for 
different values of the jet resolution $\ycut$. We use the Durham-- or 
$k_\perp$--clustering scheme \cite{Durham} throughout this chapter for jet 
observables. To be specific, for a given final state the jet measure 
\begin{equation}
\label{eq:yij}
y_{ij} = \frac{2\min(E_i^2, E_j^2)}{Q^2}(1-\cos\theta_{ij})
\end{equation}
is calculated for every particle pair $(i, j)$. The particles with
minimal `distance' are clustered such that the momentum of the
clustered pseudo-particle is the sum of the four-momenta of the
constituents. The jet multiplicity is then the number of
pseudo-particles remaining when all $y_{ij}>\ycut$. This inclusive observable 
has been predicted and measured at LEP energies and will test the dynamics
of the parton shower as well as the interface between parton shower
and hadronization. We use the \texttt{KtJet}-package \cite{ktjet} that
implements the above jet-finding algorithm in C++ and have written a
simple wrapper around it in order to use it with the particle record of \HWP{}.

\begin{figure}
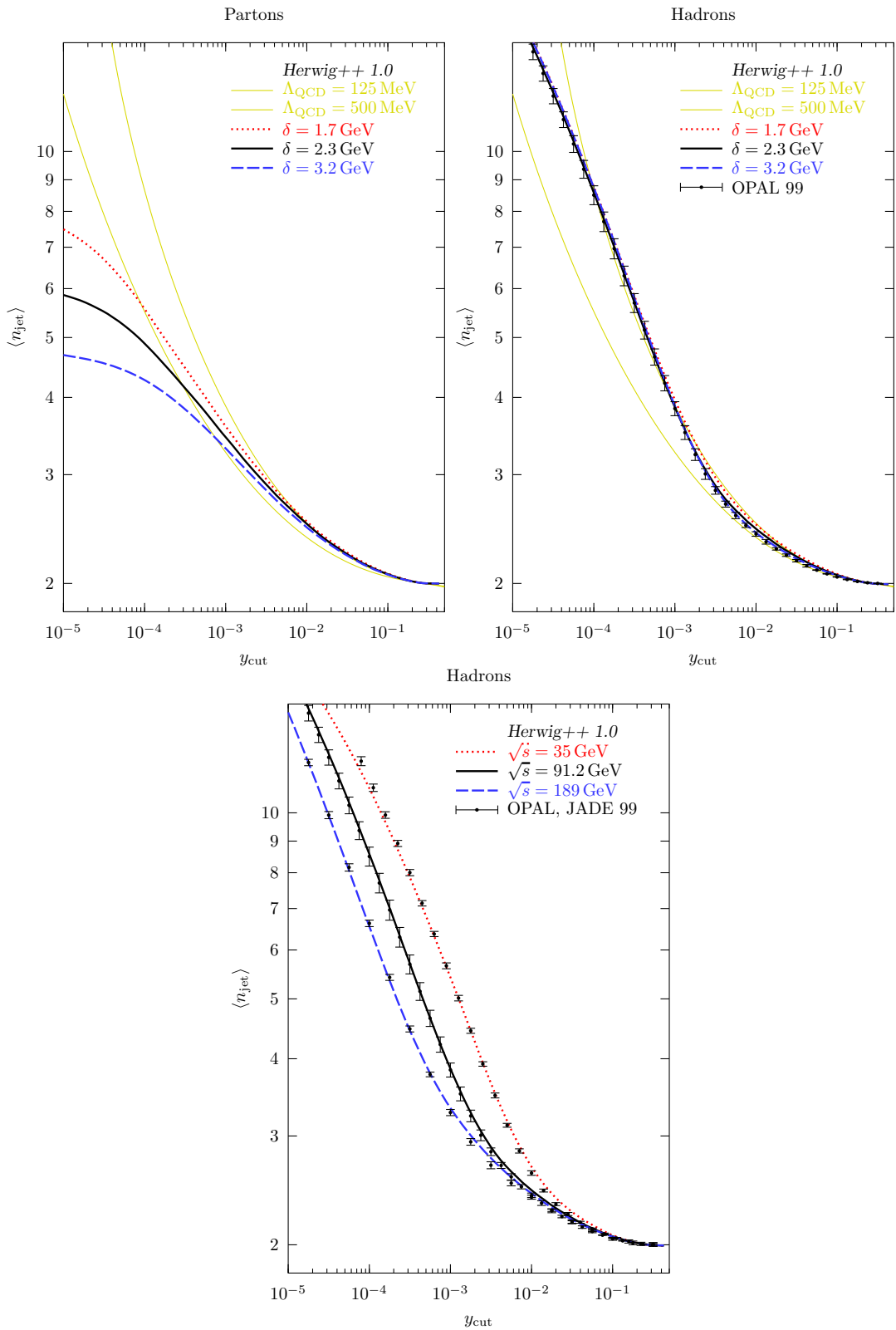

\centering
  \epsfig{file=figs/chapter5/njets2.4, scale=0.8}
  \epsfig{file=figs/chapter5/njets2.5, scale=0.8}\\
  \epsfig{file=figs/chapter5/njets2.7, scale=0.8}
  \caption{Jet multiplicities for different values of the cutoff parameter
    $\delta$ and different centre-of-mass energies.}
  \label{fig:njet}
\end{figure}

Figure~\ref{fig:njet} shows the average number of jets $\left< n_{\rm jets}
\right>$ at the $Z^0$--pole, as a function of the Durham jet resolution, 
$y_{\rm cut}$, for various values of the cutoff parameter $\delta$. At the 
parton level (top  left) the jet multiplicity varies substantially toward 
smaller values of $y_{\rm cut}$, saturating at the number of partons that are 
present in a single event. The order of magnitude of the visible saturation 
scales is characterized for each flavour by different cutoff values $Q_g$ as 
$y_{\rm sat} = Q_g^2/Q^2$. For example, at $Q=91.2$ GeV and $\delta=2.3$ GeV,
the saturation scale for light quarks is of the order $10^{-4}$ while for
$b$-quarks it is of the order $10^{-5}$.

During hadronization, low parton multiplicities lead
to large mass clusters which, as described before, tend to decay into low mass 
clusters below the cutoff mass, $\Clm$. This has been fixed to its default 
value for the results given in this chapter. Figure~\ref{fig:njet} (top right) 
shows that the hadronization compensates for lower partonic multiplicities, 
giving a result which is insensitive to $\delta$ at the hadron level. This 
means that we have a smooth interface between the perturbative and 
non-perturbative dynamics of the lower end of the parton shower and the cluster
hadronization model. On the hadron level we describe LEP data from 
OPAL~\cite{OPALnjet} well.

In order to test the sensitivity of \HWP{} against the variation of the 
centre-of-mass energy we calculate the jet multiplicities at PETRA and LEP II
energies as well as LEP I energies (\ref{fig:njet}, bottom). The comparison to
JADE~\cite{JADEnjet} and OPAL~\cite{OPALnjet} data shows a good agreement. For 
the generation of all the Monte Carlo data we applied the same cutoffs on the
energy of the partonic subsystem as was done in the experiments.

The other curves in figure~\ref{fig:njet} show the prediction for the jet
multiplicity \cite{NjetPrediction} from the resummation of leading logarithms. 
Note that the parameter $\Lambda_{\rm QCD}$ in the resummed calculation is not 
$\Lambda_{\overline{\rm MS}}$. For a value of $\Lambda_{\rm QCD} = 500$~MeV
we can see that there is good agreement with the data and the \HWP{} results
throughout the perturbative region, $y_{\rm cut} > 10^{-4}$.

\subsection{Jet fractions and $Y_n$}
These measures give a closer look `into' the jets. This is done by
considering the rates of jets at a given value of $y_{\rm cut}$ in the
Durham scheme. The jet fraction is given by 
\begin{equation}
R_n = \frac{N_{n-{\rm jet}}}{N_{\rm evts}}, 
\end{equation}
for $n=2$ up to $n=6$ jets. Presented here are also the distributions of $Y_n$,
the $y_{\rm cut}$-values at which an $n+1$--jet event is merged into an
$n$-jet event in the Durham clustering scheme. The results are presented here 
for $n=2$ up to $n=6$, without $n=5$. These distributions will not only probe 
the dynamics of the parton shower but also the hadronization model; at the 
lowest values of $y_{\rm cut}\sim (\tilde{q}_C/Q)^2$ the dynamics is dominated
by the hadronization.

\begin{figure}
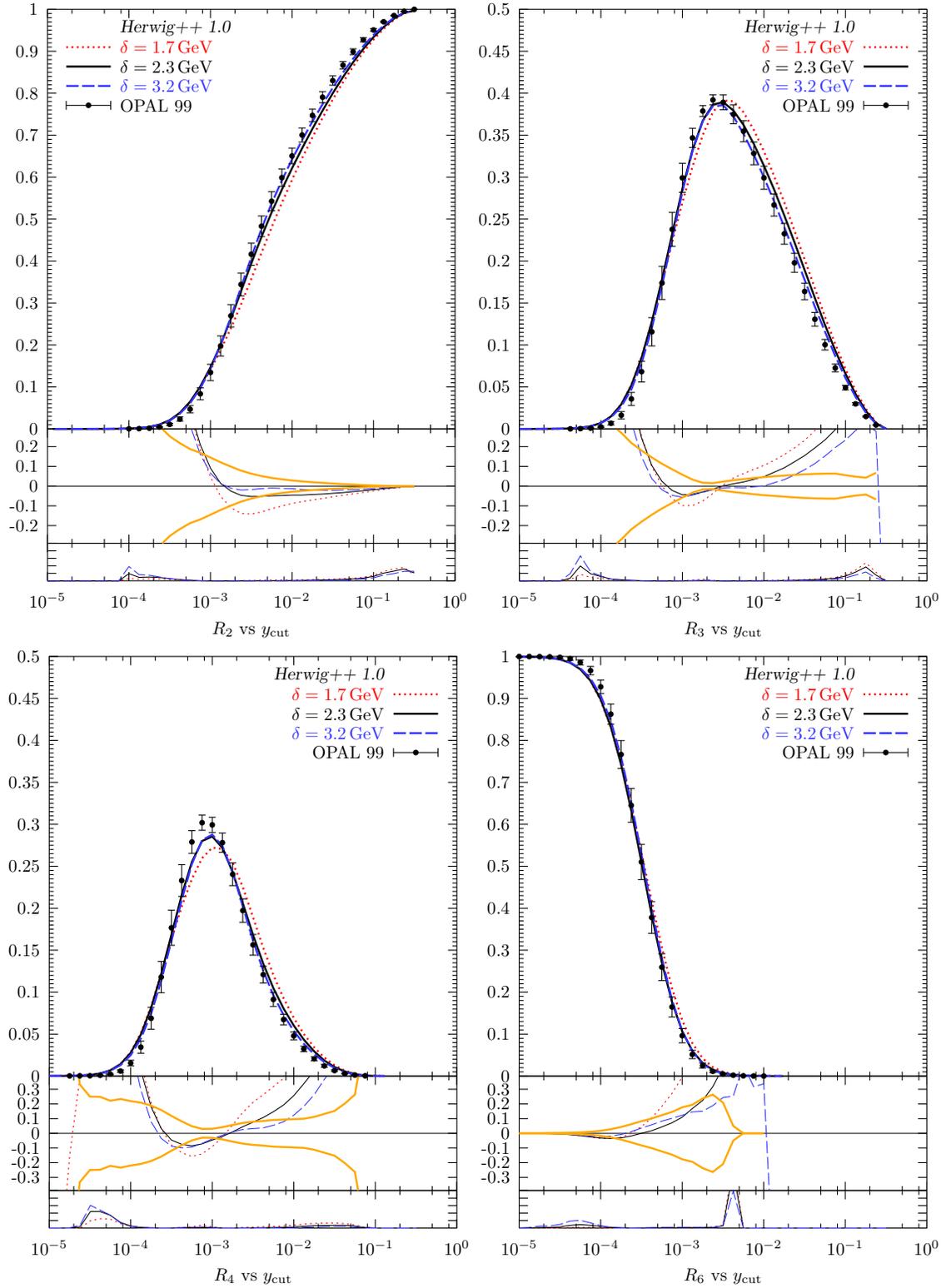

  \epsfig{file=figs/chapter5/es.61,scale=0.8}
  \epsfig{file=figs/chapter5/es.63,scale=0.8}\\[5pt]
  \epsfig{file=figs/chapter5/es.65,scale=0.8}
  \epsfig{file=figs/chapter5/es.69,scale=0.8}
  \caption{Jet rates in the Durham algorithm for different values of
  the cutoff $\delta$.\label{fig:ratesqg}} 
\end{figure}

Figure~\ref{fig:ratesqg} compares the results from \HWP{} with LEP data from
\cite{OPALnjet} and shows good agreement. On the hadron level these predictions
are not very sensitive to the cutoff parameter $\delta$. The results of the
$R_5$ are not shown here but show similarly good agreement. The $R_6$ plot 
contains all jets $n=6$ and greater.

\begin{figure}
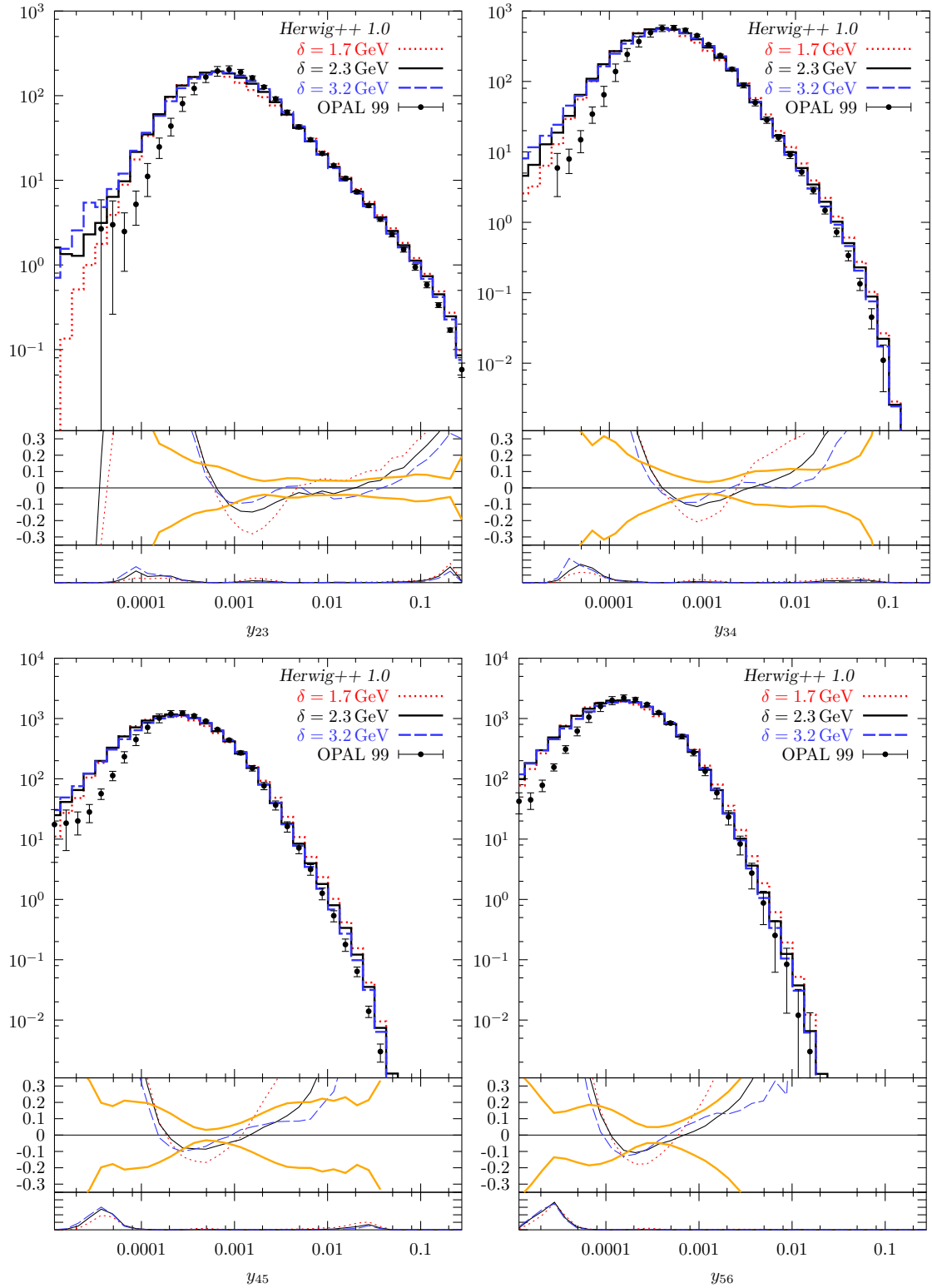

  \epsfig{file=figs/chapter5/es.51,scale=0.8}
  \epsfig{file=figs/chapter5/es.53,scale=0.8}\\[5pt]
  \epsfig{file=figs/chapter5/es.55,scale=0.8}
  \epsfig{file=figs/chapter5/es.57,scale=0.8}
  \caption{Durham $Y_n$ distributions for different values of the 
    cutoff $\delta$.\label{fig:Ynqg}}
\end{figure}

The Durham $Y_n$ distributions are given in figure~\ref{fig:Ynqg}. These are 
histograms of the $y_{\rm cut}$ values at which the $n+1$--jet event in the 
Durham jet clustering scheme is merged into an $n$--jet event. This resolved 
more of the internal structure of the jets than the $n$--jet rates alone.
Overall, the agreement between the model and the data is good. There is a
tendency to exceed the data at low $Y_n$. This is a problem that was also
present in \HW{}.

\subsection{Event shapes}
Event shape distributions have been measured to very high accuracy at LEP
and aim at resolving the properties of the parton shower quite thoroughly. All
of the event shapes given here, except $S, P$ and $A$, defined below, 
are `infrared safe'. 
This means that they can be computed in perturbation theory.
In order to test the dynamics of the parton shower in \hpp{} in more
detail we consider a set of commonly used event shape variables.  Not
only the collinear region of the parton shower is probed in greater
detail but also the regions of phase space which are vetoed as matrix
element corrections.  We compare all results to DELPHI data
\cite{DELPHIshapes}.

The thrust is a well studied property. The definition of the thrust is given
by
\begin{equation}
T = \max_{\bm n} \frac{\sum_\alpha \left| {\bm p}_\alpha \cdot
{\bm n} \right|}{\sum_\alpha \left| {\bm p}_\alpha \right| }.
\label{eqn:thrustaxis}
\end{equation}
Finding the vector ${\bm n}$ is a computational intensive task. This can
instead be reduced using physical arguments to a simpler procedure. We
start by separating the sum into two different parts: those where 
${\bm p}_\alpha \cdot {\bm n} > 0$ and those where 
${\bm p}_\alpha \cdot {\bm n} < 0$. If we first take out the normalization 
factor
\begin{equation}
N = {\sum_\alpha \left| {\bm p}_\alpha \right| },
\end{equation}
we get
\begin{equation}
T = \frac{1}{N} \max_{\bm n} \left(
\sum_{{\bm p}_\alpha \cdot {\bm n} > 0} 
    \left| {\bm p}_\alpha \cdot {\bm n} \right| + 
\sum_{{\bm p}_\alpha \cdot {\bm n} < 0} 
    \left| {\bm p}_\alpha \cdot {\bm n} \right| \right).
\end{equation}
The magnitudes of the dot products can be removed and this gives
\begin{equation}
T = \frac{1}{N} \max_{\bm n} \left(
\sum_{{\bm p}_\alpha \cdot {\bm n} > 0} {\bm p}_\alpha \cdot {\bm n} - 
\sum_{{\bm p}_\alpha \cdot {\bm n} < 0} {\bm p}_\alpha \cdot {\bm n} \right).
\end{equation}
Since the ${\bm n}$ is independent of the sum it can be taken outside the 
summation yielding
\begin{equation}
T = \frac{1}{N} \max_{\bm n} \left( {\bm P}_+({\bm n}) - {\bm P}_-({\bm n})
\right) \cdot {\bm n},
\end{equation}
where ${\bm P}_+({\bm n})$ is the sum of all momenta in the same hemisphere
as ${\bm n}$ and ${\bm P}_-({\bm n})$ is the sum of all momenta in the other
hemisphere. Momentum conservation says (in the c.m. frame) that 
${\bm P}_+ = -{\bm P}_-$ so the
thrust is given by
\begin{equation}
T = \frac{2}{N} \max_{\bm n} {\bm P}_+({\bm n}) \cdot {\bm n}.
\label{eqn:Tdecomp}
\end{equation}

\begin{figure}
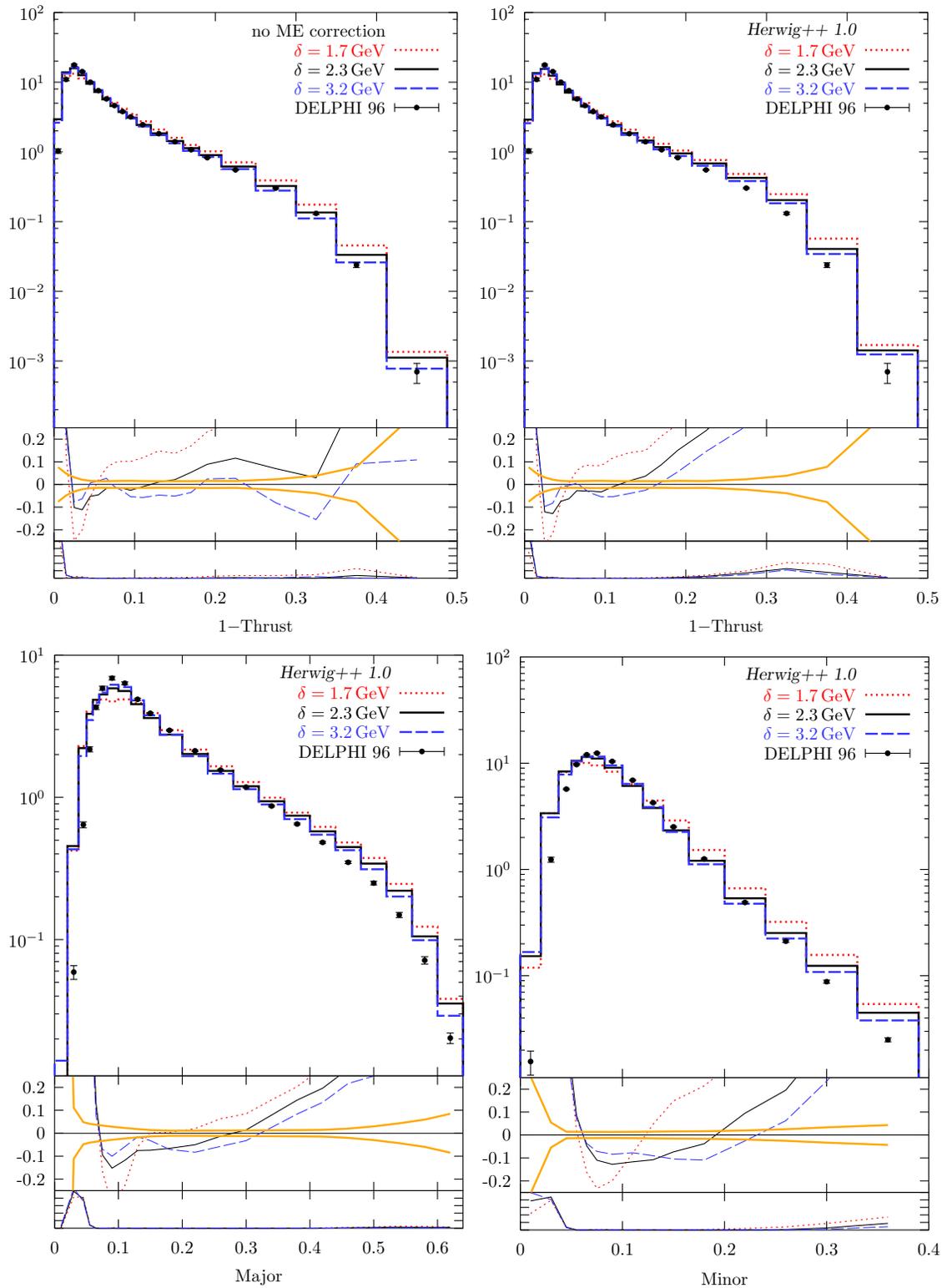

  \epsfig{file=figs/chapter5/es.0,scale=0.79}
  \epsfig{file=figs/chapter5/es.1,scale=0.79}\\[5pt]
  \epsfig{file=figs/chapter5/es.3,scale=0.8}
  \epsfig{file=figs/chapter5/es.5,scale=0.8}
  \caption{Thrust without (top left) and with (top right) matrix
    element corrections switched on, thrust major and thrust minor
    (bottom).
    \label{fig:tocd}} 
\end{figure}

From this it is obvious that the maximum value for a given ${\bm n}$ is when
the vector lies parallel to ${\bm P}_+({\bm n})$. Previously, to find the 
vector ${\bm n}$ that maximizes (\ref{eqn:thrustaxis}) we would have to 
consider all possible combinations of momenta, which has complexity of order 
$N!$. After deriving (\ref{eqn:Tdecomp}) this has been reduced to considering 
all sets of two vectors which define a plane. This plane divides the
space into two hemispheres from which the thrust can be computed by simply
summing all the momenta in one of the hemispheres and taking the magnitude. 
Since the two vectors chosen to define the plane do not unambiguously lie in
either hemisphere, each possibility must be considered. This has reduced the
problem to simply iterating over all sets of two momenta and recording which
set produces the maximum. This is now only order $N^2$.

There are two other measures of the data related to the 
thrust. One is called the thrust major and the other the thrust minor. These
are defined as
\begin{eqnarray}
M &=& \max_{{\bm n}\perp{\bm n}_T} \frac{\sum_\alpha \left| {\bm p}_\alpha 
\cdot {\bm n} \right|}{\sum_\alpha \left| {\bm p}_\alpha \right| },\\
m &=& \frac{\sum_\alpha \left| {\bm p}_\alpha \cdot {\bm n}_m \right|}
{\sum_\alpha \left| {\bm p}_\alpha \right| },
\end{eqnarray}
where $M$ is the thrust major, $m$ is the thrust minor and ${\bm n}_m = 
{\bm n}_T \times {\bm n}_M$.

The thrust is a measure used to describe how `pencil-like' the event is. High
thrust means that the event is more 2-jet like, where as lower thrust means
that the event is much more planar or spherical, thus it has more than 2 
jets. The thrust major is used to describe the major component of the momentum 
in a plane perpendicular to the thrust axis. High values of the thrust major
are usually indicative of planar events. The thrust minor, therefore, is 
used to describe the remaining degree of freedom of the momenta. High values 
of the thrust minor correspond to spherical events. These are events that 
have at least 4 jets.

In fig.~\ref{fig:tocd} we show the distribution of thrust and
thrust-major and thrust-minor.  These variables are all obtained from
the equations given above.  The thrust distribution is shown with and
without matrix element corrections switched on. The prediction without
matrix element corrections is very much better than that of \HW,
owing to the improved shower algorithm.  It is 
interesting that the matrix element corrections seem to generate
almost too much transverse structure, leading to event shapes that are 
less two-jet-like.  On the other hand, there is also a slight excess
of events close to the two-jet limit.

\begin{figure}
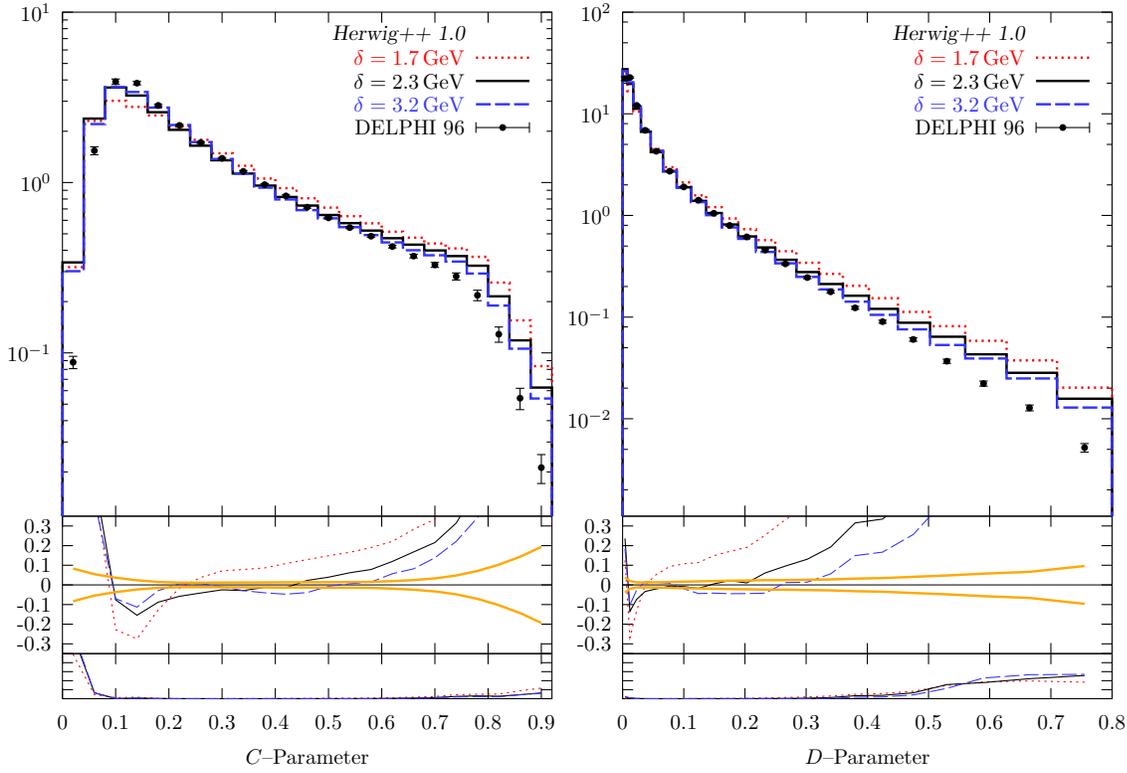

  \epsfig{file=figs/chapter5/es.15,scale=0.8}  
  \epsfig{file=figs/chapter5/es.17,scale=0.8}  
  \caption{$C$ parameter and $D$ parameter distribution. \label{fig:cd}} 
\end{figure}

Though the thrust describe 2-jet like events quite well, multi-jet
events are less understood in terms of these measurements. 
The $C$ and $D$ parameters are used to describe three- and four-jet-like
events. These are given by combinations of the eigenvalues of the linear 
momentum tensor 
\begin{equation}
L_{ij} = \frac{\sum_\alpha \left( {\bm p}_\alpha \right)_i 
\left( {\bm p}_\alpha \right)_j / \left| {\bm p}_\alpha \right|}
{\sum_\alpha \left| {\bm p}_\alpha \right|}.
\end{equation}
The definition of $C$ and $D$ is then
\begin{eqnarray}
C &=& 3\left( \lambda_1 \lambda_2 + \lambda_2 \lambda_3 + \lambda_3 \lambda_1 
\right), \\
D &=& 27 \lambda_1 \lambda_2 \lambda_3,
\end{eqnarray}
where $\lambda_i$ are the eigenvalues of $L_{ij}$ and 
\begin{equation}
\lambda_1 + \lambda_2 + \lambda_3 = 1.
\end{equation}
Both of these parameters have a coefficient defined so that the range of the
parameter is $[0,1]$.

It is remarkable how well distributions like $C$ and $D$ parameter
(fig.~\ref{fig:cd}) which are sensitive to three- and four-jet-like
events are described by our model even though we are limited to three
jet matrix elements plus showers.  Here again we have in fact a small
excess at high values.

\begin{figure}
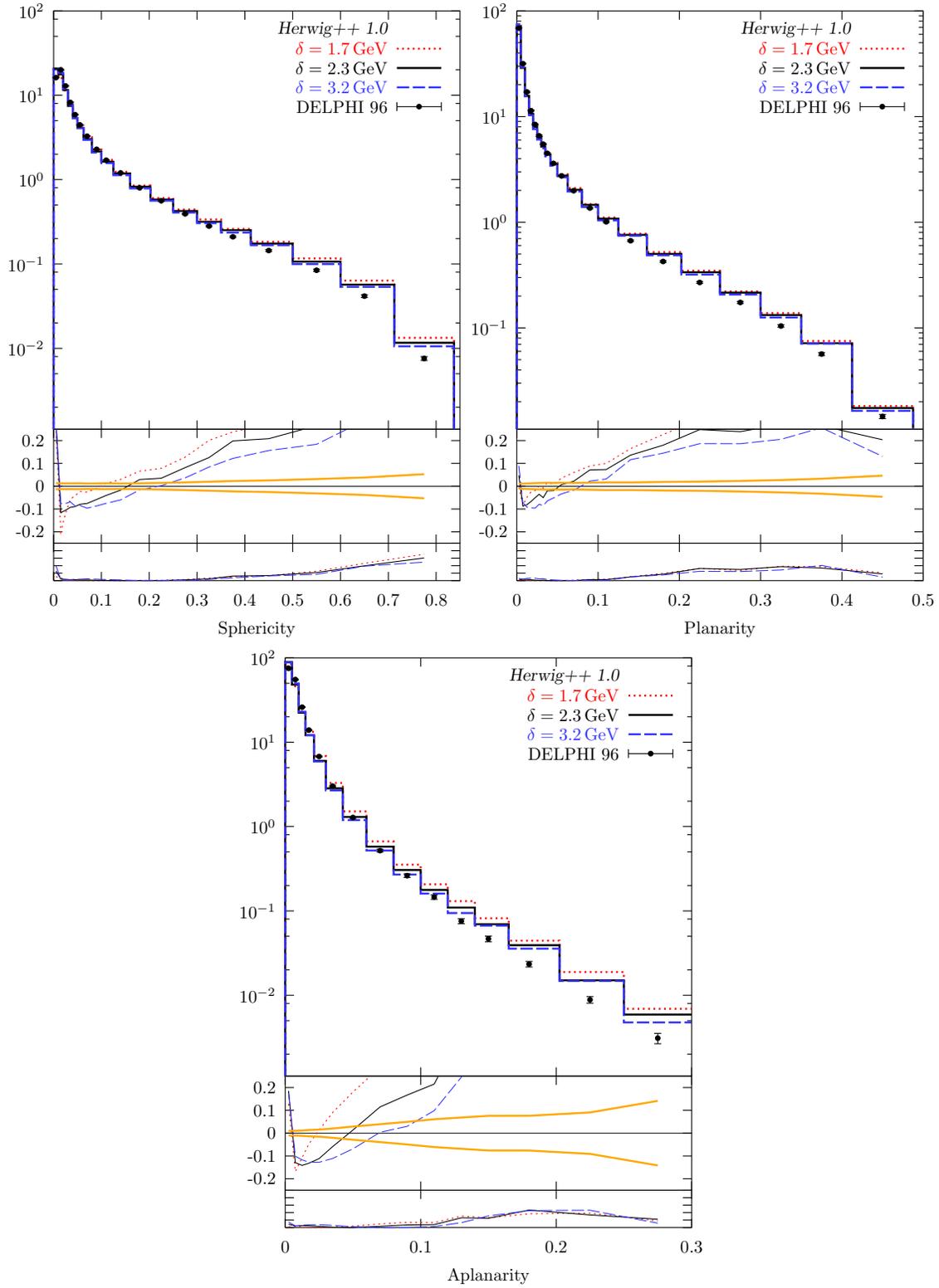

\centering
  \epsfig{file=figs/chapter5/es.9,scale=0.8}  
  \epsfig{file=figs/chapter5/es.11,scale=0.8}\\[5pt]
  \epsfig{file=figs/chapter5/es.13,scale=0.8}
  \caption{Sphericity, planarity, and aplanarity parameter distribution.
  \label{fig:sp}}
\end{figure}

We show also in fig.~\ref{fig:sp} the distributions which are obtained from a
quadratic momentum tensor
\begin{equation}
Q_{ij} = \frac{\sum_\alpha \left( {\bm p}_\alpha \right)_i 
\left( {\bm p}_\alpha \right)_j} {\sum_\alpha {\bm p}_\alpha^2}.
\end{equation}
The three measures that arise from this are sphericity ($S$), planarity ($P$)
and aplanarity ($A$). These are given by 
\begin{eqnarray}
S &=& \frac{3}{2} \left( \lambda_2 + \lambda_3 \right),\\
P &=& \lambda_2 - \lambda_3,\\
A &=& \frac{3}{2} \lambda_3,
\end{eqnarray}
where $\lambda_i$ is the $i$th eigenvalue and they obey the relations
\begin{equation}
\lambda_1 > \lambda_2 > \lambda_3,~~~~\lambda_1 + \lambda_2 + \lambda_3 = 1.
\end{equation}
These distributions put more emphasis on high momenta.
As the names imply these distributions indicate the events that are spherical,
planar or aplanar. The sphericity axis is just ${\bm n}_s = {\bm v}_1$
where ${\bm v}_i$ is the eigenvector of the $i$th eigenvalue. This is made
obvious by seeing that sphericity is really defined as $\frac{3}{2} 
(1-\lambda_1)$. Therefore, events which have high sphericity have momentum 
which tend to be along the sphericity axis, ${\bm n}_s$. As was the case for 
the thrust-related distributions, we tend to have slightly wide events. 

\begin{figure}
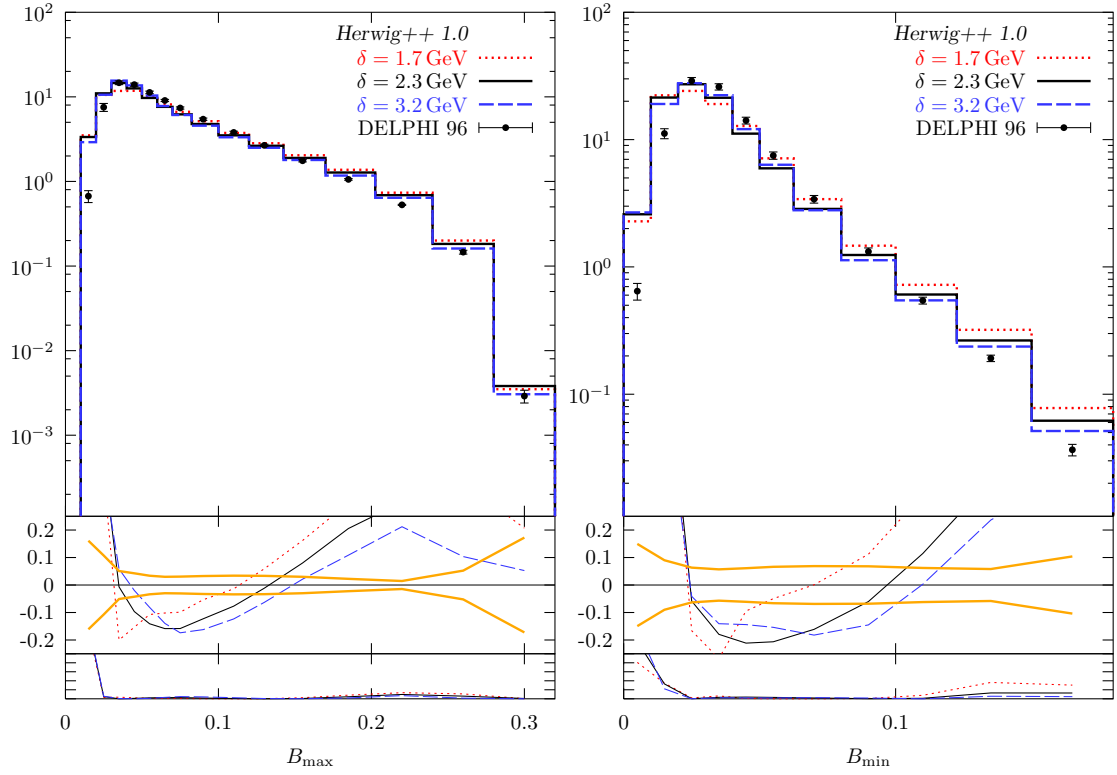

  \epsfig{file=figs/chapter5/es.25,scale=0.8}
  \epsfig{file=figs/chapter5/es.27,scale=0.8}
  \caption{The wide and narrow jet broadening measures
  $B_{\rm max}$ and $B_{\rm min}$. 
  \label{fig:bmaxbmin}} 
\end{figure}

\begin{figure}
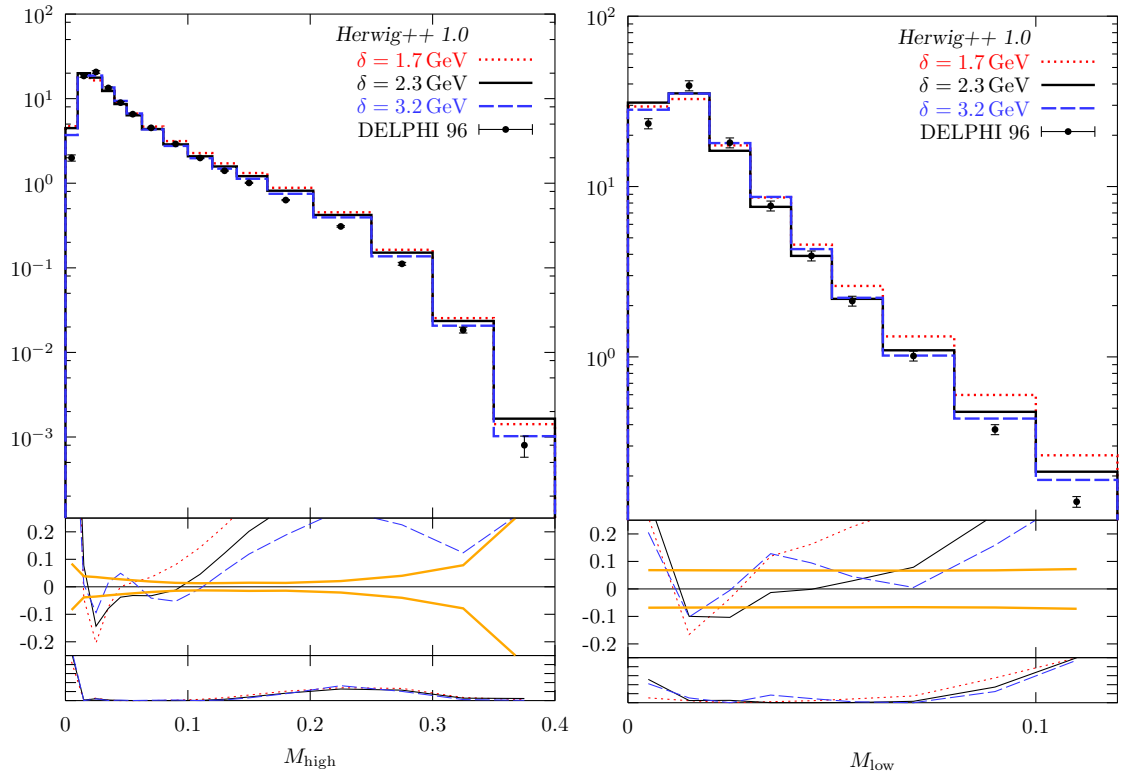

  \epsfig{file=figs/chapter5/es.19,scale=0.8}
  \epsfig{file=figs/chapter5/es.21,scale=0.8}
  \caption{The high and low hemisphere masses. 
  \label{fig:mhighmlow}} 
\end{figure}

In addition we consider the jet broadening measures $B_{\rm max}$ and 
$B_{\rm min}$ and the hemisphere jet masses (fig.~\ref{fig:bmaxbmin} and
fig.~\ref{fig:mhighmlow}). 
The jet broadening measures are defined as~\cite{Catani:1992jc,Burby:2001uz}
\begin{equation}
B_{\rm max} = \max_{i=1,2} \frac{\sum_{{\bm p}_k \in H_i} \left| {\bm p}_k
\times {\bm n}_T \right|}{2 \sum_k \left| {\bm p}_k \right|},
\end{equation}
where ${\bm n}_T$ is the thrust axis and $H_i$ indicates one of the two 
hemispheres defined by the plane normal to ${\bm n}_T$. If ${\bm p}_k \cdot
{\bm n}_T > 0$ then ${\bm p}_k$ is in hemisphere, $H_1$, otherwise it is
in hemisphere $H_2$. $B_{\rm min}$ is then
\begin{equation}
B_{\rm min} = \min_{i=1,2} \frac{\sum_{{\bm p}_k \in H_i} \left| {\bm p}_k
\times {\bm n}_T \right|}{2 \sum_k \left| {\bm p}_k \right|}.
\end{equation}
$B_{\rm max}, B_{\rm min}$ measure the scalar momentum transverse to the 
thrust axis for the wider and narrower jet hemispheres respectively. We can 
see from figure \ref{fig:bmaxbmin} that there
is good agreement between the model and the data. We have also looked at two 
other jet broadening measures, $B_{\rm diff} = B_{\rm max} - B_{\rm min}$ and 
$B_{\rm sum} = B_{\rm max} + B_{\rm min}$. These are not shown here as they 
contain the same information as $B_{\rm max}$ and $B_{\rm min}$. 

The last jet measure we show here is that of the hemisphere masses. These
also use the same definitions of the two hemispheres with respect to the 
thrust axis as the jet broadening measures, but these measure the total 
momentum squared within a hemisphere. The high hemisphere mass is
\begin{equation}
M_{\rm high} = \frac{1}{s} \max_{i=1,2} \left( \sum_{{\bm p}_k \in H_i} p_k
\right)^2,
\end{equation}
and the lower hemisphere mass is
\begin{equation}
M_{\rm low} = \frac{1}{s} \min_{i=1,2} \left( \sum_{{\bm p}_k \in H_i} p_k
\right)^2,
\end{equation}
where $s$ is the c.m. energy squared.
In both cases we can see from figure~\ref{fig:mhighmlow} that the agreement 
between model and data is good.

\subsection{Four jet angles}
\label{sec:fourj}
\begin{figure}
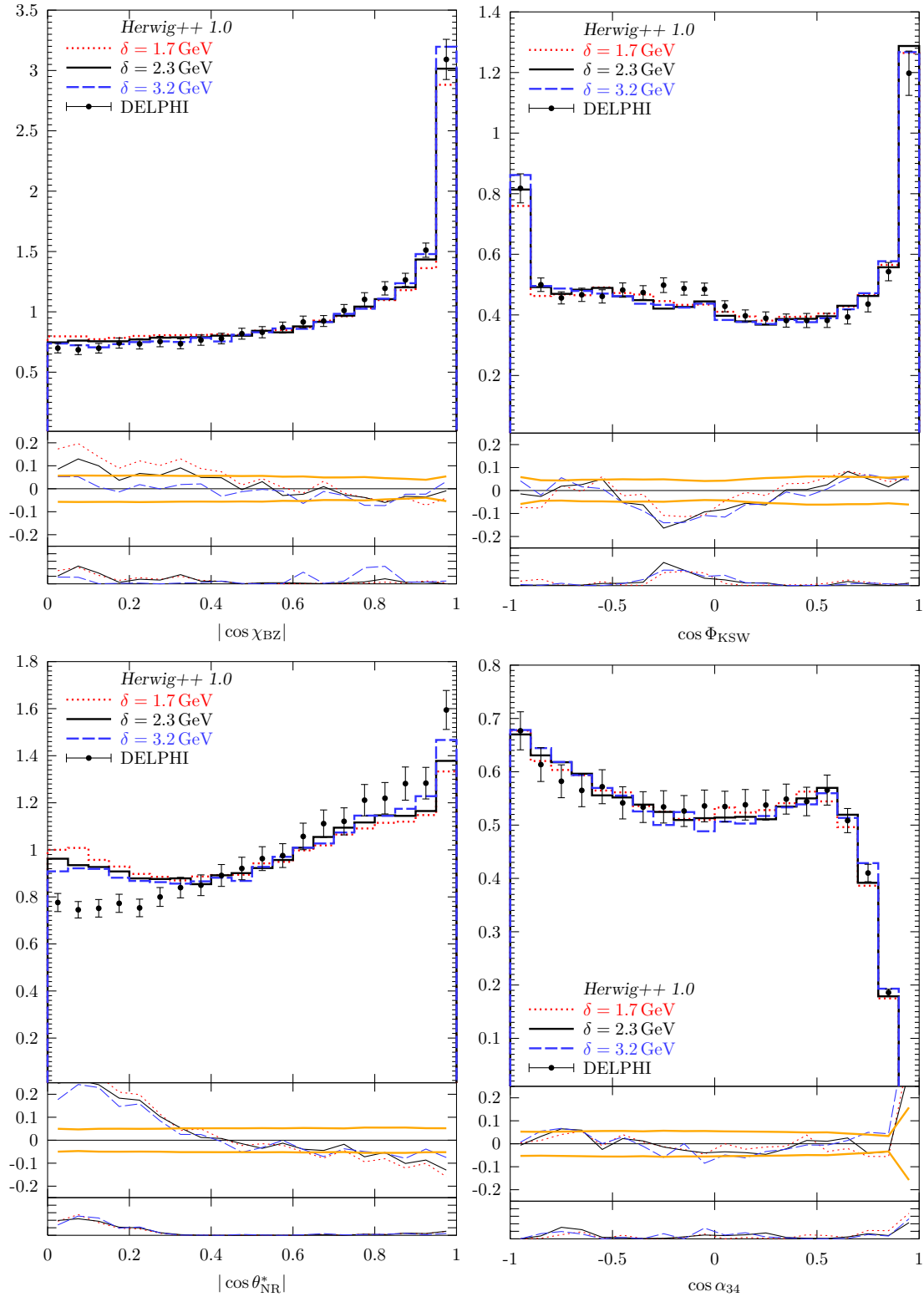

  \epsfig{file=figs/chapter5/es.71,scale=0.8}
  \epsfig{file=figs/chapter5/es.73,scale=0.8}\\[5pt]
  \epsfig{file=figs/chapter5/es.75,scale=0.8}
  \epsfig{file=figs/chapter5/es.77,scale=0.8}
  \caption{Four jet angle distributions.  The points are
  from preliminary DELPHI data.\label{fig:fourj}}
\end{figure}

We show the four-jet angles in fig.~\ref{fig:fourj}.  They
are considered only for events where we have a four-jet event at
$\ycut=0.008$.  Each of the different angles measures a property of
the four-jet configuration. 

If we denote the cosine of the angle between two vectors, ${\bm a}$ and
${\bm b}$ as $C({\bm a},{\bm b})$, we can define the four different jet angles
as
\begin{eqnarray}
\cos \chi_{\rm BZ} &=& C\left( ({\bm p}_1\times{\bm p}_2),
                               ({\bm p}_3\times{\bm p}_4) \right), \\
\cos \Phi_{\rm KSW} &=& \cos \left[ \frac{\alpha_1 + \alpha_2}{2} \right], \\
\cos \Theta_{\rm NR} &=& C \left( ({\bm p}_1 - {\bm p_2}), 
                                  ({\bm p}_3 - {\bm p}_4) \right), \\
\cos \alpha_{34} &=& C \left( {\bm p}_3, {\bm p}_4 \right),
\end{eqnarray}
where 
\begin{eqnarray}
\cos \alpha_1 &=& C\left( ({\bm p}_1\times{\bm p}_4),
                          ({\bm p}_2\times{\bm p}_3) \right), \\
\cos \alpha_2 &=& C\left( ({\bm p}_1\times{\bm p}_3),
                          ({\bm p}_2\times{\bm p}_4) \right).
\end{eqnarray}
${\bm p}_1$ is the three-momentum of the hardest jet and ${\bm p}_4$ 
is the three-momentum of the softest jet.

Despite the fact that we do not have any matching to
higher order matrix elements, as was proposed in \cite{CKKW} and
implemented in \cite{Apacic}, the agreement between model and
data~\cite{DELPHIfourj} is remarkably good.  We expected the
implementation of hard and soft matrix element corrections in \HWP{}
to improve the description of these observables but we did
not find very significant differences with or without the application
of matrix element corrections.

\subsection{Single particle distributions}
\label{sec:singlep}

\begin{figure}
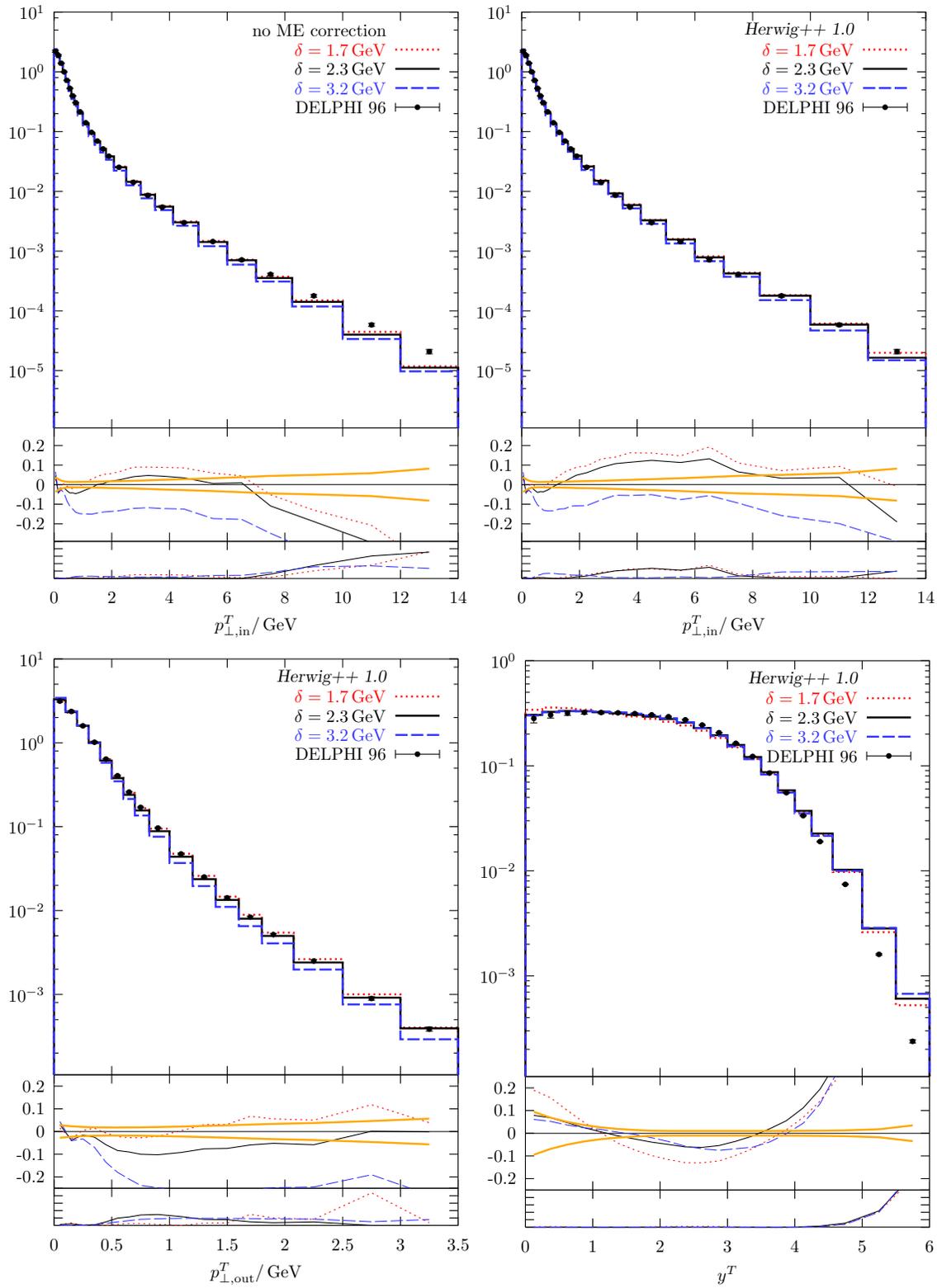

  \epsfig{file=figs/chapter5/es.32,scale=0.8}
  \epsfig{file=figs/chapter5/es.33,scale=0.8}\\[5pt]
  \epsfig{file=figs/chapter5/es.35,scale=0.8}
  \epsfig{file=figs/chapter5/es.37,scale=0.8}
  \caption{Momentum distributions of charged particles with respect to
    the thrust axis, $p_{\perp, {\rm in}}^T$ (with and without matrix element
    corrections), $p_{\perp, {\rm out}}^T$ and $y^T$.
    \label{fig:ptyT}}
\end{figure}

Using the thrust axis defined in (\ref{eqn:thrustaxis}), one can define the 
transverse momentum of a particle with respect to this axis. We can then define 
a plane given by the thrust axis and the thrust major axis, ${\bm n}_M$ as the 
event plane. Therefore, the plane given by the thrust and the thrust minor,
${\bm n}_m$, is considered out of the event plane. In fig.~\ref{fig:ptyT} we 
have single charged particle distributions. The distributions are of the 
transverse momentum within the event plane, $p_{\perp, {\rm in}}^T$, and the 
transverse momentum out of the event plane, $p_{\perp, {\rm out}}^T$. The 
momenta in the event plane are shown with and without matrix element 
corrections.  In contrast to the thrust distribution we find that the matrix 
element corrections actually improve the distribution.  Furthermore, 
$p_{\perp, {\rm out}}^T$ and the rapidity along the thrust axis, $y_T$, are 
rather well described.  A similar technique can be used to define the 
transverse momentum with respect to the sphericity axis ${\bm n}_s$. We do not 
show these but they have similar features to the distributions with respect to 
the thrust axis. 

\begin{figure}
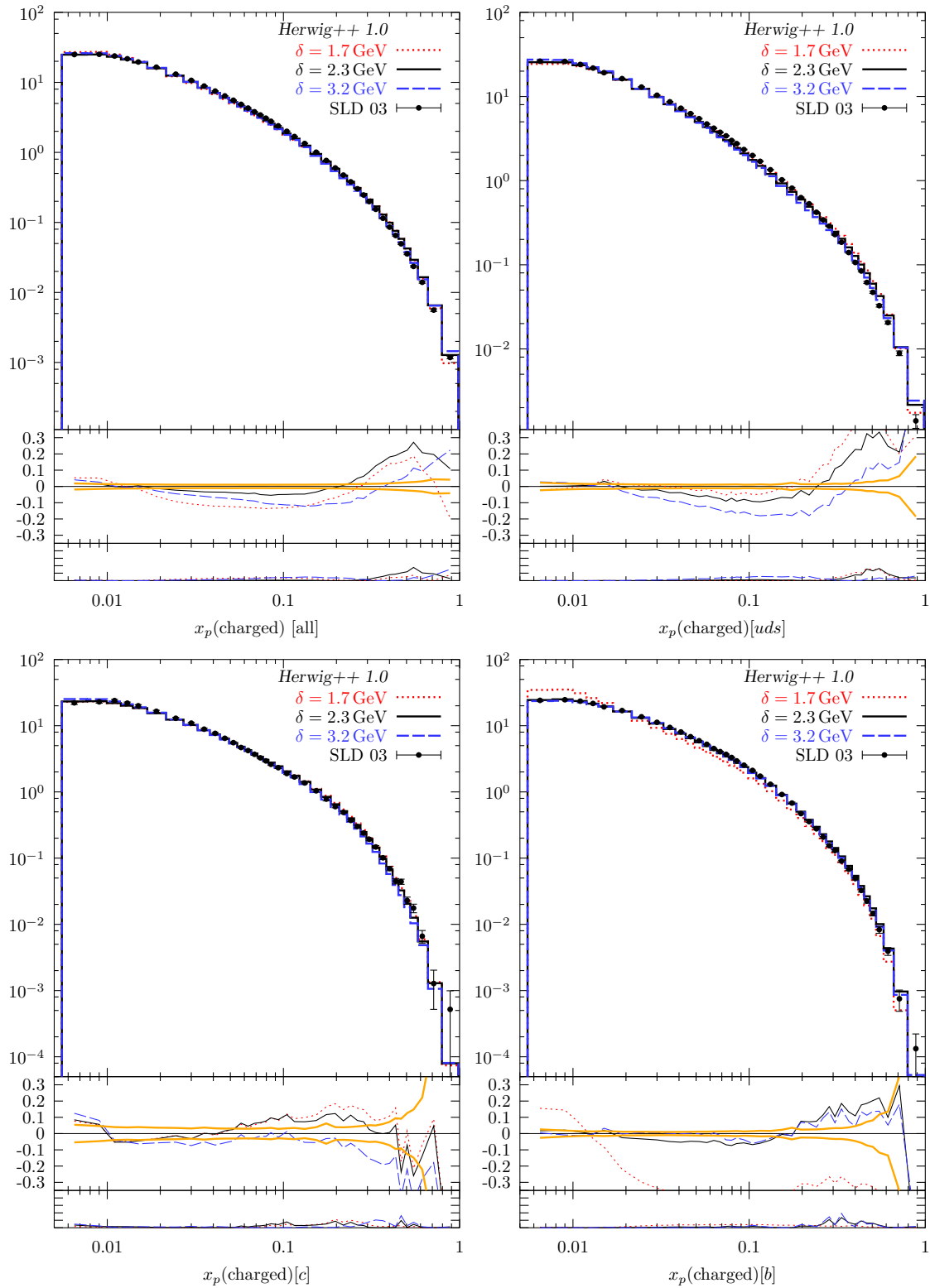

  \epsfig{file=figs/chapter5/es.81,scale=0.8}
  \epsfig{file=figs/chapter5/es.83,scale=0.8}\\[5pt]
  \epsfig{file=figs/chapter5/es.85,scale=0.8}
  \epsfig{file=figs/chapter5/es.87,scale=0.8}
  \caption{The scaled momentum distribution $x_p$ of charged
    particles for all events as well as for $uds$,  $c$ and $b$
    events separately. \label{fig:xpch}} 
\end{figure}

As we saw in chapter~\ref{chap:Hadronization}, we can consider the distribution
of scaled momentum $x_p=2|\bm p|/Q$ of charged particles. The results in 
chapter~\ref{chap:Hadronization} were given to show that the hadronization 
technique could describe these distributions quite well. In 
fig.~\ref{fig:xpch} these distributions are given to show their dependence on
the shower cutoff, $\delta$.  In addition to the full distribution we also 
consider the results from light ($uds$), $c$ and $b$ events.
\footnote{The flavour of the quark-antiquark produced in the initial hard 
process}  In all cases we compare with data from SLD \cite{SLD03}.
The charged particle distribution is well described in all
four cases, in fact somewhat better for heavy primary quarks.

\subsection{Identified hadron spectra}
\label{sec:identhad}

\begin{figure}
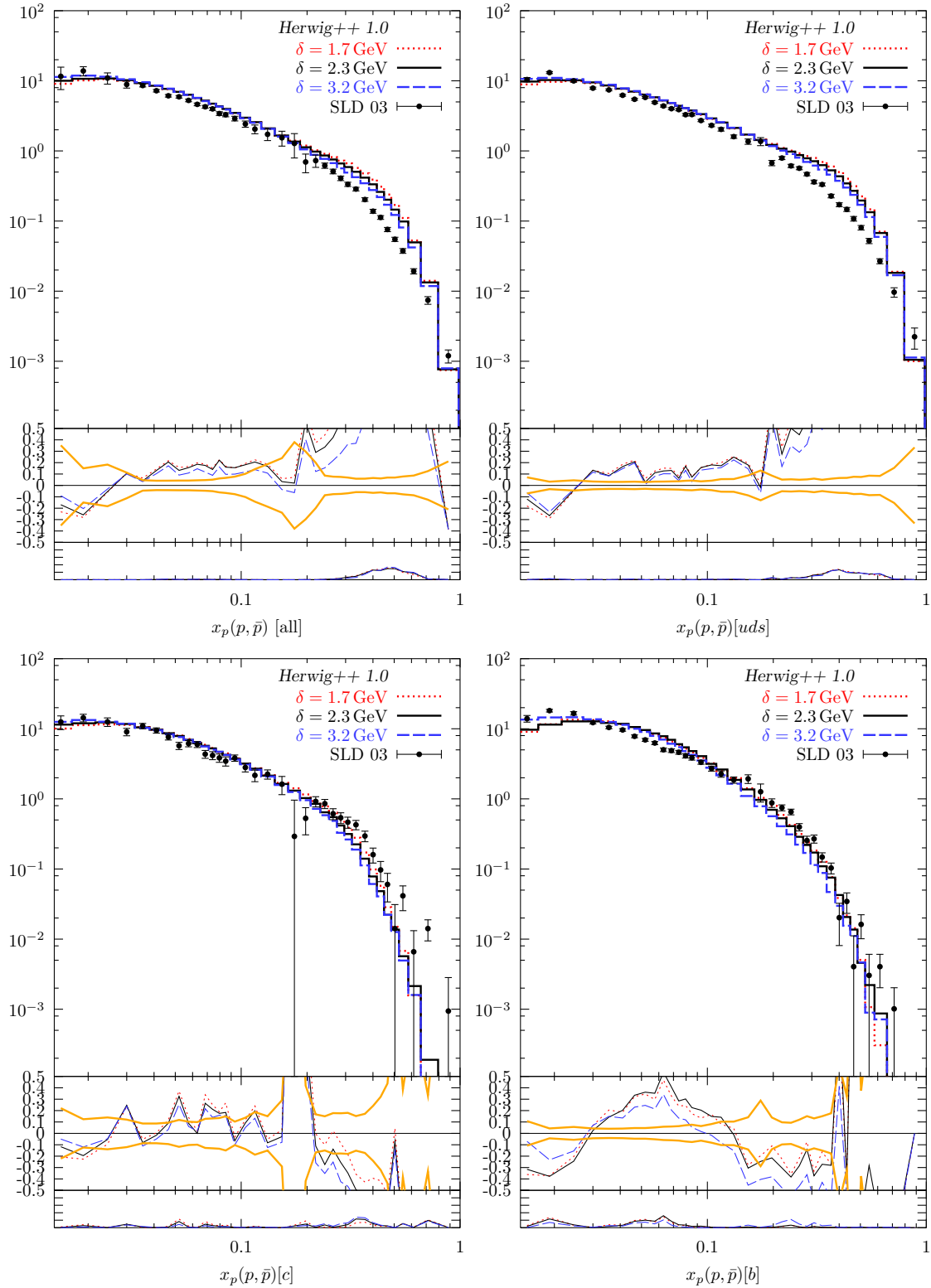

  \epsfig{file=figs/chapter5/es.105,scale=0.8}
  \epsfig{file=figs/chapter5/es.107,scale=0.8}\\[5pt]
  \epsfig{file=figs/chapter5/es.109,scale=0.8}
  \epsfig{file=figs/chapter5/es.111,scale=0.8}
  \caption{The scaled momentum distribution $x_p$ of protons, shown
    separately for all events as well as for $uds$,  $c$ and $b$
    events. \label{fig:xpp}} 
\end{figure}

As in the case of all charged particles we can compare identified
particle spectra from events of different flavour to SLD data \cite{SLD03}.
Data for $\pi^\pm$ (not shown, being almost equivalent to all charged
particles), $K^\pm$ and ($p, \bar p$) is available.  In
fig.~\ref{fig:xpp} we see the data for ($p, \bar p$) spectra from
events of different flavour.  For large values of $x_p$ we clearly
overshoot the data in light flavoured events, producing the `bump'.  This is 
somewhat compensated by the heavy quark events which in turn seem to prefer
lower values of $x_p$.  The origin of the `bump' is not well understood. We 
believe that this feature is related to the hadronization, being similar to but
smaller than that seen in \HW, but this has not been shown conclusively.

\begin{figure}
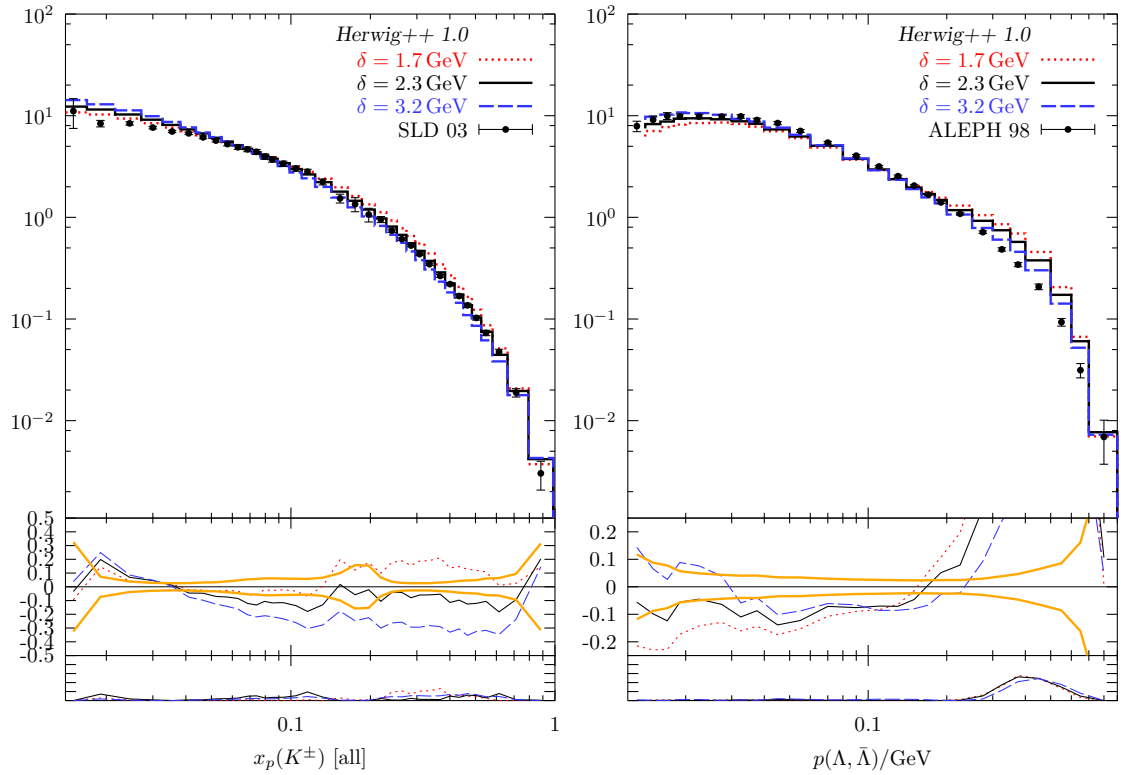

  \epsfig{file=figs/chapter5/es.97,scale=0.8}
  \epsfig{file=figs/chapter5/es.125,scale=0.8}
  \caption{Distribution of scaled Kaon momentum and $\Lambda, 
    \bar \Lambda$ momentum. 
  \label{fig:flavmom}}
\end{figure}

Fig.~\ref{fig:flavmom} shows distributions for $K^\pm$ and $\Lambda,
\bar \Lambda$.  Both are rather better described than the proton
spectra but the distribution of $\Lambda, \bar \Lambda$ tends to have
a similar, though smaller, `bump' in comparison to data from
ALEPH~\cite{ALEPHLambda}.

\subsection{B fragmentation function} 
\begin{figure}
  \centering
  \epsfig{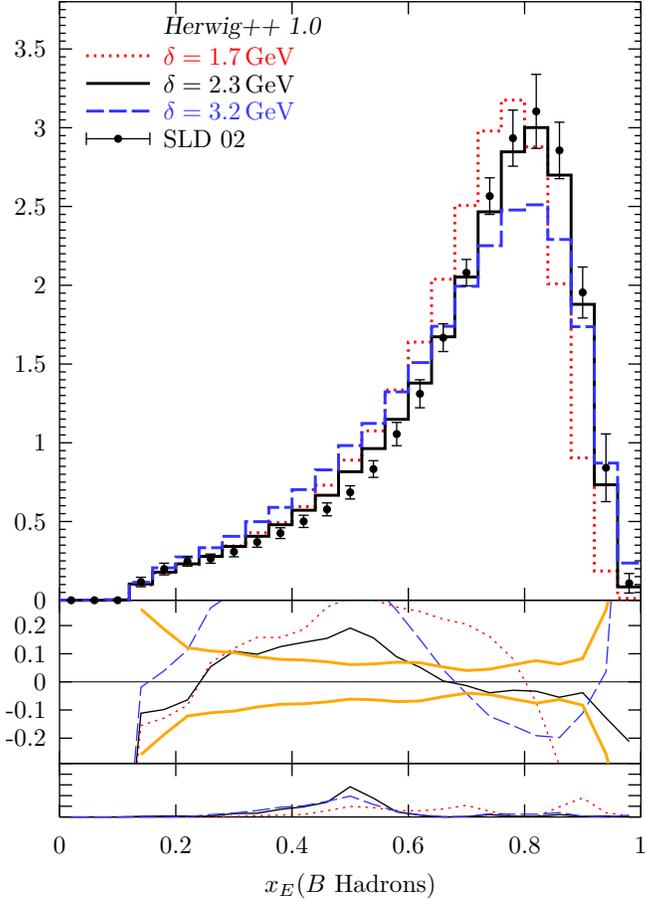}
  \caption{The $B$--hadron fragmentation function. For different
    values of the cutoff $\delta$.\label{fig:bfraghad}} 
\end{figure}

The hadron fragmentation function is defined as the distribution of
\begin{equation}
x_E = \frac{2 E_h}{E_{\rm cm}},
\end{equation}
where $E_h$ is the energy of the hadron.
Using this we can look at the fragmentation function of different flavours
or species of hadrons. In fig.~\ref{fig:bfraghad} we consider the $B$ hadron 
fragmentation function in comparison to data from SLD \cite{SLDbfrag}.  This is
the fragmentation function for all weakly decaying $B$ hadrons. We have also 
considered data from ALEPH \cite{ALEPHbfrag} for comparison, though these are 
not shown here. 
We can describe the data quite well without any additional tuning of the
hadronization model to this data.  The parton shower formulation in
terms of the new variables \cite{NewVariables} and taking quark masses
in the splitting functions into account clearly improves the
description of heavy quark events.

From figure~\ref{fig:bfraghad} we tend to bias the fit towards the $\delta$ of
2.3 GeV, as the improved treatment of the $b$ quarks was the main motivation 
for deriving the new variables and using the massive splitting functions.
\HW{} couldn't describe the data as well even with the extra flags added to the
hadronization model to parameterize $B$ hadrons differently.

\subsection{Overall results of $\ee$ annihilation}
In Tab.~\ref{tab:chi2} we show a list of $\chi^2$ values for all
observables that were studied during our analysis, including those
not shown in the plots.  The most sensitive parameters were the
cutoff value $\delta$ and the use of (hard plus soft) matrix element
corrections.  The table shows three values of $\delta$:  our preferred
value of $\delta=2.3\,$GeV as well as the lowest and highest values
that we considered.  

The results should be interpreted with care.  The overall trend
suggests that we should prefer a large cutoff scale.  However, we
have just averaged over all possible observables.  Taking a closer
look, we may want to weight different observables in a different way.

\begin{table}
\scriptsize
\centering
\begin{tabular}{lcrrrrrr}
\hline
\hline
&&\multicolumn{3}{c}{ME corrections off} 
&\multicolumn{3}{c}{ME corrections on}\\
Observable &Ref.& $\delta=1.7\,$GeV& $\delta=2.3\,$GeV &$\delta=3.2\,$GeV
& $\delta=1.7\,$GeV& $\delta=2.3\,$GeV &$\delta=3.2\,$GeV\\
\hline
\hline
$1-T$                      &\D    & 44.65& 33.06& 22.29& 70.34& 44.14& 25.99\\
$M$                        &\D    &246.25&265.81&198.37&257.99&242.19&174.50\\
$m$                        &\D    &150.74&155.11&137.43&167.05&150.10&120.59\\
$O$                        &\D    &  7.41&  5.60&  5.14& 21.48& 19.27& 12.78\\
\hline
$S$                        &\D    &  4.42&  3.48&  4.07& 22.99& 13.09&  7.89\\
$P$                        &\D    &  4.48&  5.53&  6.54& 10.27&  7.34&  5.35\\
$A$                        &\D    & 19.52& 10.75&  7.17& 41.86& 18.26&  9.62\\
\hline
$C $                       &\D    & 66.86& 59.08& 39.56& 79.85& 64.93& 42.32\\
$D $                       &\D    & 84.23& 28.40& 12.36&145.14& 51.44& 23.96\\
\hline
$M_{\rm high}             $&\D    & 25.78& 18.82& 12.38& 37.80& 25.39& 11.34\\
$M_{\rm low}              $&\D    & 15.25&  5.36&  2.50& 30.29&  9.93&  4.78\\
$M_{\rm diff}             $&\D    &  7.28&  5.18&  7.25& 17.62& 12.26&  4.42\\
\hline
$B_{\rm max}              $&\D    & 54.48& 50.47& 38.91& 59.06& 49.27& 33.32\\
$B_{\rm min}              $&\D    & 53.25& 55.83& 53.18& 63.65& 56.69& 49.91\\
$B_{\rm sum}              $&\D    &102.29& 97.12& 74.60&116.81& 98.56& 69.94\\
$B_{\rm diff}             $&\D    &  8.28&  5.56&  4.70& 17.91& 13.77&  5.94\\
\hline
$p_{\perp, {\rm in}}^T    $&\D    &  2.48&  3.14& 11.52&  3.22&  1.78&  4.17\\
$p_{\perp, {\rm out}}^T   $&\D    &  0.25&  3.25& 21.65&  0.71&  1.69& 16.11\\
$y^T                      $&\D    & 34.52& 59.78& 66.05& 32.88& 49.36& 55.54\\
\hline
$p_{\perp, {\rm in}}^S    $&\D    &  2.53&  3.18& 11.76&  2.21&  1.40&  4.28\\
$p_{\perp, {\rm out}}^S   $&\D    &  0.37&  3.78& 22.64&  1.01&  2.02& 16.67\\
$y^S                      $&\D    &  9.04& 17.42& 24.85&  7.53& 13.80& 21.35\\
\hline
$D_2^D                    $&\D    &  9.37&  3.65&  3.76& 25.14& 11.81&  5.17\\
$D_3^D                    $&\D    & 25.85&  6.32&  2.14& 46.56& 15.68&  5.39\\
$D_4^D                    $&\D    & 43.90& 10.56&  2.69& 77.40& 22.61&  7.16\\
\hline
$y_{23}                   $&\O    &  8.75&  6.19&  5.36& 12.17&  8.41&  6.27\\
$y_{34}                   $&\O    & 10.20&  9.49&  9.07& 11.42&  9.31&  8.64\\
$y_{45}                   $&\O    & 15.53& 14.33& 11.78& 17.07& 14.47& 11.66\\
$y_{56}                   $&\O    & 16.02& 17.62& 15.13& 15.28& 16.56& 13.82\\
\hline
$\langle N_{\rm jets}\rangle$&\O & 12.84&  3.38&  0.62& 27.84& 13.11&  5.81\\
$R_2                      $&\O    &  9.75&  6.64&  6.18& 19.71& 13.40&  9.55\\
$R_3                      $&\O    & 10.46&  8.51&  9.36& 23.45& 15.76& 12.02\\
$R_4                      $&\O    & 13.47& 11.06& 10.36& 15.45& 12.69& 10.29\\
$R_5                      $&\O    & 25.53& 25.18& 23.43& 27.88& 26.38& 22.40\\
$R_6                      $&\O    & 10.37&  1.80&  0.67& 18.98&  4.31&  1.41\\
\hline
$\cos(\chi_{\rm BZ})      $&\DF   &  2.90&  1.04&  0.48&  2.60&  1.05&  0.45\\
$\cos(\Phi_{\rm KSW})     $&\DF   &  2.30&  1.99&  2.56&  1.30&  1.63&  1.71\\
$\cos(\theta_{\rm NR}^*)  $&\DF   &  7.68&  4.82&  2.72&  8.57&  5.52&  3.74\\
$\cos(\alpha_{34})        $&\DF   &  1.41&  1.47&  1.71&  0.51&  0.46&  0.74\\
\hline
$N_{ch}                   $&\OC   & 21.86& 25.68& 12.90& 19.88& 22.55& 13.07\\
\hline
$x_p({\rm ch}) [\rm all]  $&\Sx    &  5.32&  5.65&  3.49&  4.77&  4.10&  3.02\\
$x_p({\rm ch}) [uds]      $&\Sx    & 15.72&  8.49&  6.13& 12.70&  6.69&  5.82\\
$x_p({\rm ch}) [c]        $&\Sx    &  3.95&  2.33&  2.17&  2.95&  1.72&  2.72\\
$x_p({\rm ch}) [b]        $&\Sx    & 35.05&  3.23&  1.79& 35.80&  2.46&  1.15\\
\hline
$x_p(\pi^\pm) [\rm all]   $&\Sx    &  8.29&  9.31&  6.18&  7.17&  7.52&  5.52\\
$x_p(\pi^\pm) [uds]       $&\Sx    & 28.30& 15.99& 10.47& 23.71& 13.19&  9.45\\
$x_p(\pi^\pm) [c]         $&\Sx    &  4.65&  3.04&  1.38&  3.67&  2.24&  1.60\\
$x_p(\pi^\pm) [b]         $&\Sx    & 49.13&  3.13&  1.56& 49.37&  3.69&  2.05\\
\hline
$x_p(K^\pm) [\rm all]     $&\Sx    &  4.99&  2.01& 15.38&  3.67&  2.84& 17.41\\
$x_p(K^\pm) [uds]         $&\Sx    &  6.46& 16.98& 36.45&  6.79& 19.23& 38.67\\
$x_p(K^\pm) [c]           $&\Sx    & 21.01&  2.22&  3.35& 17.71&  1.74&  4.24\\
$x_p(K^\pm) [b]           $&\Sx    &  8.56&  7.07&  4.34&  7.63&  5.74&  4.97\\
\hline
$x_p(p, \bar p) [\rm all]$ &\Sx    &143.34& 96.18& 42.90&135.35& 80.31& 34.27\\
$x_p(p, \bar p) [uds]$     &\Sx    &145.35&100.90& 52.78&135.61& 85.09& 43.31\\
$x_p(p, \bar p) [c]$       &\Sx    &  2.26&  2.38&  2.86&  2.34&  2.51&  2.87\\
$x_p(p, \bar p) [b]$       &\Sx    & 11.26& 13.54&  8.12& 10.98& 13.31&  8.34\\
\hline
$p(\Lambda, \bar \Lambda)$ &\A    & 58.02& 28.32&  9.47& 53.88& 24.92&  7.69\\
\hline
$x_E (B)$                  &\SB   &  8.93&  0.95&  8.16&  9.41&  1.35&  9.93\\
$x_E (B)$                  &\AB   & 15.40&  1.74&  7.35& 15.77&  2.02&  8.23\\
\hline
\hline
$\langle\chi^2\rangle/{\rm dof}$&& 32.75& 25.54& 20.93& 39.22& 26.72& 19.84\\
\hline
\hline
\end{tabular}
\caption{$\chi^2$ values for all observables we studied and a relevant subset of
  parameters. \label{tab:chi2}}
\end{table}

In more detail, the general trend is the following: event shapes,
jet rates and differential jet rates prefer a low cutoff. The single
particle distributions along the thrust and sphericity axes prefer a
small cutoff value. The $y_{nm}$ distributions prefer either a high or
a low cutoff value.  The spectra of identified particles tend to
prefer the high cutoff value with some exceptions for light quark
events.  The $B$ fragmentation function clearly prefers the
intermediate value.  

In addition, as indicated in sect.~\ref{sec:multcomment}, we found
that the measured yields of identified particles clearly prefer the
value $\delta=2.3\,$GeV.

\section{Conclusions}
\label{sec:conclusions}
We have achieved a complete event generator for $\ee$ annihilation into
hadrons.  The main physics features, in comparison to the previous
versions of \HW, are an
improved parton shower, capable of properly describing the
perturbative splitting of heavy quarks, and an improved cluster
hadronization model.

We have tested our model against a wide range of data from $\ee$
colliders and are able to give a good general description of the data.

For many observables the description of the data has been improved
with respect to \HW{}.  The new parton shower has a number of
remarkable features. The need for matrix element corrections has
decreased.  The main reason for this is the use of improved
splitting functions, which give a far better approximation of the
matrix elements in the region of collinear gluon emissions.  We can
describe observables involving light or heavy quark splitting with a
unique set of parameters. The new hadronization model also improves
the description of identified particle spectra and multiplicities.

The detailed analysis of
our results leaves us with a \emph{recommendation}: the set of
parameters that is shown in table~\ref{tab:Herwig_defaults}.  This set
of parameters is understood as a weighted compromise in order give a
good overall description of the data we have considered so far.
We did not aim at a complete tuning of the model, but rather wanted
to study its ability to  describe the broad features
of the data, which turned out to be very successful.

Work is currently under way testing the parton shower on initial
state radiation. A model for the soft underlying event in
hadron--hadron collisions is also under development. The aim is to 
have the code tested and debugged so a complete event generator for
the simulation of Tevatron and LHC events is available.

\chapter{Effective Potential Analysis: \EFF{}}
\label{chap:Effective}
\section{Supersymmetry}
\label{sec:susy}
Though the SM presently describes physical phenomena extremely well there is
reason to believe that it will eventually be insufficient to describe physics
at higher scales. The scale tested to date is of the order of 100 GeV. 
The Planck scale, $M_P = (8\pi G_{\rm Newton})^{-1/2} = 2.4 \times 10^{18}$ 
GeV, is the scale at which gravitational effects will be of the same order as 
the other interactions. This scale is 16 orders of magnitude higher than the 
currently observed phenomena. On an intuitive basis this is suggestive that 
there must be new phenomena occurring between these scales. This is commonly 
referred to as the hierarchy problem.

\begin{figure}
\centering
\input{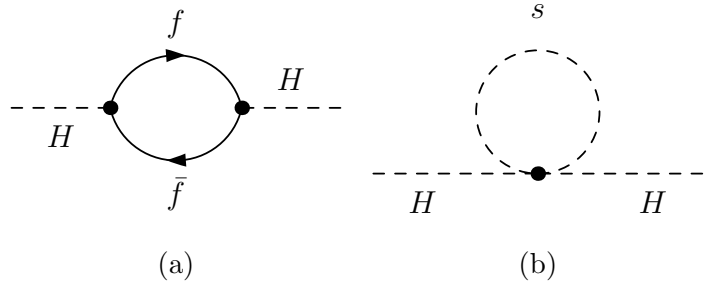}
\caption{These diagrams are the one loop corrections to the Higgs mass for
fermions and scalars.\label{fig:HiggsCorrection}}
\end{figure}

There is also another problem, known as the fine tuning problem. This is
due to the Higgs boson. The mass of the Higgs boson is restricted to be of the 
order $100~{\rm GeV}$. Figure \ref{fig:HiggsCorrection} shows the diagrams 
that correct this mass at one loop. From fig. \ref{fig:HiggsCorrection}a we get
the correction due to Dirac fermions with mass $m_f$ \cite{MartinPrimer}
\begin{equation}
\label{eq:fermCorrection}
\Delta m_H^2 = \frac{\left| \lambda_f \right|^2}{16 \pi^2} \left[ -2 
\Lambda_{UV}^2 + 6m_f^2 \ln \left( \Lambda_{UV}/m_f \right) + \dots \right],
\end{equation}
where $\Lambda_{\rm UV}$ is an ultraviolet momentum cutoff at the scale of new
physics and $\lambda_f$ is the coupling derived from the term $-\lambda_f 
\overline{f} \Phi f$. This problem directly affects only the correction to the
Higgs scalar boson $({\rm mass})^2$ because the fermions and gauge bosons do
not have the quadratic sensitivity to $\Lambda_{\rm UV}$. If 
$\Lambda_{\rm UV}$ is of the order of $M_P$ then this correction is about 30 
orders of magnitude larger than the expected mass-squared of the Higgs boson. 
This will indirectly affect the SM particles and their known masses as 
corrections to these masses have some dependence on $m_H$. Therefore, this 
cannot be the case.

The Higgs mass also receives a contribution from 
fig.~\ref{fig:HiggsCorrection}b
from the scalar particles. For a real scalar field, this contribution takes 
the form \cite{MartinPrimer}
\begin{equation}
\label{eq:scalarCorrection}
\Delta m_H^2 = \frac{\lambda_S}{16 \pi^2} \left[ \Lambda_{UV}^2 - 2m_S^2 
\ln \left( \Lambda_{UV}/m_S \right) + \dots \right].
\end{equation}
Due to the opposite signs in (\ref{eq:scalarCorrection}) versus 
(\ref{eq:fermCorrection}) there
could be a particular combination of particles and masses that exactly cancel
the divergences. Though plausible, it seems extremely unlikely that all of the 
parameters that are currently in the SM plus any particles occurring with a 
higher mass are tuned exactly to cancel this.

Instead a systematic cancellation of the divergences can occur. We see that if
we have two complex scalar fields for each fermion field which couple to the 
Higgs boson exactly as $\lambda_S = \left| \lambda_f \right|^2$ then they 
exactly cancel the quadratic terms. This symmetry between fermions and bosons 
is known as supersymmetry \cite{MartinPrimer, WeinbergIII, Haber}.

Of course this symmetry must be broken because we do not observe bosonic states
of the same mass as a fermion partner. The scale at which it is broken is
$M_{\rm SUSY}$ and the SUSY particles will have masses of the order of 
$M_{\rm SUSY}$. In SUSY models \cite{MartinPrimer} this value is expected to be
at most 1 TeV in order to allow for a Higgs VEV which will give the correct
values of $m_Z$ and $m_W$.

At a scale of 1 TeV, the experiments at LHC should be able to discover SUSY 
particles. Research is still ongoing into new ways to use the general purpose 
experiments ATLAS\cite{ATLAS} and CMS\cite{CMS} to search for SUSY 
particles (for example~\cite{SUSYSign,Barr:2003rg}).

\subsection{Superpotential}
\label{sec:superpotential}
The fermion and boson that are supersymmetric partners are combined into a 
multiplet called a supermultiplet. There are two kinds of supermultiplets, a 
chiral and a gauge supermultiplet. A chiral supermultiplet is a supermultiplet 
formed by the matter fields. The gauge fields and their superpartners form 
gauge supermultiplets. These are the objects that enter into the supersymmetric
Lagrangian. This in turn can be written in terms of the \emph{superpotential}, 
which will be described in this section.

It can be shown that the most general set of renormalizable SUSY interactions 
for a field $\psi_i$ and its SUSY partner can be written as 
\begin{equation}
\label{eq:lint}
{\cal L}_{\rm int} = -\frac{1}{2}W^{ij}\psi_i \psi_j + W^i F_i + {\rm c.c.},
\end{equation}
where the $W$'s are explained below, the $F$ is an auxiliary term and the 
c.c. indicates complex conjugate. The auxiliary terms are used as a 
bookkeeping device and are eliminated by the equations of motion. Doing so 
shows that $F_i = -W^*_i$.

We can now define the superpotential as
\begin{equation}
W = \frac{1}{2}M^{ij} \phi_i \phi_j + \frac{1}{6} y^{ijk} \phi_i \phi_j \phi_k,
\end{equation}
where $M^{ij}$ is a mass matrix for the fermions and $y^{ijk}$ is a Yukawa 
coupling of scalar, $\phi_k$, and fermions $\psi_i,\psi_j$. This is related to 
the $W$'s in eq. (\ref{eq:lint}) by
\begin{eqnarray}
W^{ij} &=& \frac{\delta^2 W}{\delta\phi_i \delta \phi_j} \\
W_i    &=& \frac{\delta W}{\delta \phi_i}.
\end{eqnarray}

By replacing the $F$ terms and adding the kinetic term, we find the complete
Lagrangian for a chiral supermultiplet is
\begin{equation}
{\cal L}_{\rm chiral} = {\cal L}_{\rm kin} - \frac{1}{2}\left( W^{ij} \psi_i 
\psi_j + W^{*ij} \psi^\dagger_i \psi^\dagger_j \right) - V(\phi,\phi^*),
\end{equation}
where $V(\phi,\phi^*) = W^iW^*_i$ is the scalar potential and 
${\cal L}_{\rm kin}$ is the standard kinetic terms for a fermion and a scalar.

The gauge field interactions also have to be written in terms of their
superpartners. For a gauge boson, $A^a_\mu$, and its superpartner fermion, 
$\lambda^a$, this is
\begin{equation}
{\cal L}_{\rm gauge} = -\frac{1}{4} F^a_{\mu\nu}F^{a\mu\nu} - i 
\lambda^{\dagger a}\overline{\sigma}^\mu D_\mu \lambda^a + \frac{1}{2}D^aD^a,
\label{eqn:Lgauge}
\end{equation}
where $F^a_{\mu\nu}$ is the same term as in the SM and $D^a$ is another 
auxiliary field. $\sigma^\mu$ and $\bar \sigma^\mu$ are given by 
\begin{equation}
\sigma^\mu = ({\bm 1}, \sigma^i),~~~\bar \sigma^\mu = ({\bm 1}, -\sigma^i),
\end{equation}
and are related to the standard $\gamma$-matrices by
\begin{equation}
\gamma^\mu = \left( \begin{array}{cc} 0 & \sigma^\mu \\ \bar \sigma^\mu & 0
\end{array} \right).
\end{equation}

It can be shown that there are two other terms that can be added and the 
Lagrangian will remain invariant under supersymmetric transformations. When 
added to the Lagrangian they give
\begin{equation}
{\cal L} = {\cal L}_{\rm chiral} + {\cal L}_{\rm gauge} - \sqrt{2} g \left[
\left( \phi^* T^a \psi \right) \lambda^a + \lambda^{\dagger a} \left( 
\psi^\dagger T^a \phi \right) \right] + g \left( \phi^* T^a \phi \right) D^a,
\label{eqn:lagr1}
\end{equation}
where $T^a$ is the generator of the group $\lambda^a$ is a mediator of. 
It is from this equation that we can find the equations of motion to remove
$D^a$. We find that $D^a = -g \left( \phi^* T^a \phi \right)$ and the 
$\frac{1}{2}D^a D^a$ from (\ref{eqn:Lgauge}) combines with the last term in 
(\ref{eqn:lagr1}). Since
both $D^a$ and $F_i$ can be expressed purely in terms of scalar fields they
can be used to write the complete scalar potential as
\begin{equation}
V(\phi^*,\phi) = F^*_i F^i + \frac{1}{2} \sum_a D^aD^a = -W^*_i W^i + 
\frac{1}{2}  \sum_a g_a^2 \left( \phi^* T^a \phi \right)^2.
\end{equation}

\subsection{Minimal Supersymmetric Standard Model}
The Minimal Supersymmetric Standard Model (MSSM) is the model that contains
all of the possible $N=1$ SUSY interactions and special cases of this model are
generally analyzed phenomenologically. This model has one superpartner for 
every SM particle and an extra $SU(2)_L$ Higgs doublet. The extra Higgs
doublet is due to the requirement that the superpotential be an analytic 
function. This prevents the addition of a Yukawa term like $\bar Q 
\phi^\dagger u_R$ and instead a second Higgs doublet is needed. Table 
\ref{table:MSSMparticles} shows the particle content
in terms of the interaction eigenstates. We also introduce the convention that
all superpartners of fermions are names with the letter \emph{s} preceding
the name and all superpartners of bosons have \emph{ino} appended
to the name. For example the partner of an electron is a selectron and the 
partner of a B is a Bino. Also, by convention the symbol for a
superpartner field is to put a tilde on top of the field symbol, e.g. $B^\mu
\rightarrow \tilde{B}$. Bosons with a superscript $\mu$ are vector bosons,
whereas those without are scalar bosons. 

\begin{table}
\begin{center}
\begin{tabular}{c|c|c|c|c}
Fermion & Boson       & ${\rm SU(3)}_c$ & ${\rm SU(2)}_L$ & ${\rm U(1)}_Y$ \\
\hline
$Q_L$ & $\tilde{Q}_L$ & $\bm{3}$        & $\bm{2}$        & $\frac{1}{6}$ \\
$u_R$ & $\tilde{u}_R$ & $\bm{3}$        & $\bm{1}$        & $\frac{2}{3}$ \\
$d_R$ & $\tilde{d}_R$ & $\bm{3}$        & $\bm{1}$        & -$\frac{1}{3}$\\
$\ell$& $\tilde{\ell}$& $\bm{1}$        & $\bm{2}$        & -$\frac{1}{2}$\\
$e_R$ & $\tilde{e}_R$ & $\bm{1}$        & $\bm{1}$        & -1 \\ 
$\tilde{H}_1$ & $H_1$ & $\bm{1}$        & $\bm{2}$        & $\frac{1}{2}$ \\
$\tilde{H}_2$ & $H_2$ & $\bm{1}$        & $\bm{2}$        & $-\frac{1}{2}$ \\
\hline
$\tilde{g}$ & $g^\mu$ & $\bm{8}$        & $\bm{1}$        &  0 \\
$\tilde{W}$ & $W^\mu$ & $\bm{1}$        & $\bm{3}$        &  0 \\
$\tilde{B}$ & $B^\mu$ & $\bm{1}$        & $\bm{1}$        &  0 \\
\hline
\end{tabular}
\end{center}
\caption{The interaction states of MSSM and the respective gauge charges. 
    	There are also three generations of (s)quarks and (s)leptons.}
\label{table:MSSMparticles}
\end{table}

Just like in the SM, several of the interaction states mix to form the mass
states. Table \ref{table:MSSMmixing} shows which interaction states mix to 
form which mass states; the mass states are denoted by the word for the
physical particles that are expected to be observed. The mixture of the 
(s)quarks given in the table is to signify that the left and right handed
(s)quarks mix, not the up- and down-type (s)quarks. The Higgs sector isn't 
quite as straightforward. Each Higgs doublet 
contains a charged part and a neutral part. The real parts of the two neutral 
Higgs doublets mix to form the CP even Higgs states, often denoted by $h$ and 
$H$. Here the lower case is the lighter
Higgs, the one that is expected in the SM. The imaginary parts of the neutral
Higgs doublets mix to form the CP odd Higgs, $A$. The remaining charged parts 
of both doublets then mix to form two charged Higgs bosons. Also, often the
convention $H_u$ and $H_d$ is used\cite{MartinPrimer}. By the definitions given
here this is equivalent to $H_1 \equiv H_u$ and $H_2 \equiv H_d$. The mass 
states of the four neutralinos are denoted by $\chi^0_{1,2,3,4}$ in order
of mass. The four mass state charginos are denoted by $\chi^\pm_{1,2}$.

\begin{table}
\begin{center}
\begin{tabular}{cccc}
Mixing Fermions & Mass States & Mixing Bosons & Mass States \\
\hline
$Q_L, u_R, d_R$ & quarks & $\tilde{Q}_L, \tilde{u}_R, \tilde{d}_R$& squarks \\
$\ell, e_R$ & electrons, neutrinos & $\tilde{\ell}, \tilde{e}_R$ 
       & selectrons,sneutrino \\
$\tilde{B}, \tilde{W_3}, \tilde{H}_1, \tilde{H}_2$ & neutralinos, 
$\tilde \chi^0_i$ 
       & $B^\mu, W_3^\mu$ & $Z^0$, photon\\
$\tilde{W}_1, \tilde{W}_2, \tilde{H}_1, \tilde{H}_2$ & charginos,
$\tilde \chi^\pm_i$
       & $W_1^\mu, W_2^\mu$ & $W^\pm$\\
&& $H_1,H_2$ & CP-even/odd Higgs, Charged Higgs \\
$\tilde{g}$ & gluino & $g^\mu$ & gluon \\
\hline
\end{tabular}
\end{center}
\caption{The interaction states that mix and yield the mass states. The mass
states are given by name.}
\label{table:MSSMmixing}
\end{table}

\subsubsection{Soft Breaking}
As none of the superpartners of the SM particles have been observed this
implies that supersymmetry must be a broken symmetry. If we refer back to the
fine tuning argument we see that in order for the quadratically divergent parts
to still cancel, the dimensionless couplings must cancel (i.e. $\lambda_S 
= \left| \lambda_f \right|^2$). This leads us to only consider ``soft'' 
breaking of supersymmetry. This means that we can write the Lagrangian as
\begin{equation}
{\cal L} = {\cal L}_{\rm SUSY} + {\cal L}_{\rm soft}
\end{equation}
where ${\cal L}_{\rm SUSY}$ are all the terms preserving supersymmetry and
${\cal L}_{\rm soft}$ are all the terms that softly break supersymmetry. To 
ensure renormalizability and to maintain the natural cancellation of quadratic
divergences, ${\cal L}_{\rm soft}$ must have only mass terms and 
couplings with \emph{positive} mass dimension.

It is the existence of the soft term that allows all of the superpartners
of the SM particles to be heavier than the top quark. In the absence of
electroweak symmetry breaking, all of the SM fields would be massless. This
isn't true for the superpartners. The scalars can have a 
mass term in the Lagrangian of the form $m^2\left|\phi\right|^2$. The gauginos
and Higgsinos also do not require electroweak breaking in order to acquire a
mass due to the fact that they are fermions in a real representation of their 
gauge group. 

If the largest mass term occurring in ${\cal L}_{\rm soft}$ is $m_{\rm soft}$.
We can see that the corrections to the Higgs mass-squared, due to the SUSY 
particles, must vanish as $m_{\rm soft} \rightarrow 0$. This means that these
corrections cannot be quadratic in $\Lambda_{UV}$ and the corrections must be 
of the form
\begin{equation}
\Delta m_H^2 = m_{\rm soft}^2 \left[ \frac{\lambda}{16\pi^2} \ln \left(
\Lambda_{UV} / m_{\rm soft} \right) + \dots \right].
\end{equation}
If we take $\lambda$ of the order 1 and $\Lambda_{UV}$ of order $M_P$, we find
that the lightest SUSY particles should be about 1 TeV, as stated previously. 
This is what leads to the optimism that SUSY will be discovered at the LHC.

The actual soft breaking terms added to MSSM are
\begin{equation}
{\cal L}_{\rm soft} = -\frac{1}{2} \left( M_\lambda \lambda^{\dagger a} 
\lambda^a + {\rm c.c.} \right) - \left( m^2 \right)^i_j \phi^{*j} \phi_i 
- \left( \frac{1}{2} b^{ij} \phi_i \phi_j + \frac{1}{6} a^{ijk} \phi_i \phi_j
\phi_k + {\rm c.c.} \right).
\label{eqn:lsoft}
\end{equation}
There is another term that could be added without losing the renormalizability
of the theory. This term is $-\frac{1}{2} c_i^{jk}\phi^{*i} \phi_j \phi_k +
{\rm c.c.}$  but this is often not included as it can lead to
quadratic divergences in some models. It is also evident that the terms in 
(\ref{eqn:lsoft}) do break supersymmetry because they are not expressions of
the supermultiplets. This is because mass terms for the fermions can be 
reabsorbed into a redefinition of the superpotential and the 
$\left( m^2 \right)^i_j$ and $c^{jk}_i$ terms. 

To complete the definition of the MSSM we must also specify the 
superpotential. This is given as
\begin{equation}
W_{\rm MSSM} = Y^u_{ij} \overline{u} H_1 Q - Y^d_{ij} \overline{d} H_2 Q
- Y^e_{ij} \overline{e} H_2 \ell + \mu H_1 H_2.
\end{equation}
Here the fields $Q, \ell, H_1, H_2, \overline{u}, \overline{d}, \overline{e}$
are the chiral supermultiplets, not just the SM fields. We can see the family
Yukawa matrices and the Higgs fields which are used to give the masses. We
also see the supersymmetric equivalent to the $\mu$ term in the SM.

This section has been a simple introduction to SUSY models, SUSY
Lagrangians and MSSM. For a more detailed discussion of the development of all
these topics see \cite{MartinPrimer,WeinbergIII,Haber}. Next the concept
of the effective potential and its uses is given and finally the software
\EFF{} is discussed which uses the effective potential to study $N=1$ SUSY
models.

\section{Effective Potential}
The effective potential was originally introduced by Euler and Heisenberg 
\cite{EulerHeisenberg} and further expanded by Schwinger~\cite{Schwinger}. This
was later applied to studies of spontaneous symmetry breaking by Goldstone, 
Salam and Weinberg~\cite{spontsymm}. The development of the effective 
potential given in this section, as well as a complete summary of the
effective potential and its uses is given in~\cite{Quiros}.

The effective potential is capable of providing quite a lot of information 
about a theory, while working with a simpler expression.
Here is presented the development of the effective potential and in subsequent
sections the various things that can be derived from an effective potential are
given. These are all capabilities of the software, \EFF, which is discussed in 
the last section.

We start by giving the theoretical derivation of the effective potential. With
this derivation it can be shown that the the one-loop contribution to the
effective potential can be derived in a model independent way. Unfortunately,
the higher order corrections cannot be developed in a model independent way.
Because of the model dependence, the two-loop effective potential is not 
implemented in \EFF{}, though future extensions of the code could provide this
functionality.

\subsection{Generating Functionals}
As an example we start with a theory described by the scalar field $\phi$ with
a Lagrangian density ${\cal L}\left\{ \phi(x) \right\}$. The action is then
\begin{equation}
S\left[ \phi \right] = \int d^4 x {\cal L}\left\{ \phi(x) \right\}.
\end{equation}

The generating functional is the vacuum-to-vacuum expectation value
$\left< 0_{out} \right| \left. 0_{in} \right>_j$ and is given by the 
path-integral representation,
\begin{equation}
Z\left[ j \right] = \left< 0_{out} \right| \left. 0_{in} \right>_j \equiv 
\int {\cal D} \phi \exp \left\{ i \left( S[\phi] + \phi j \right) \right\},
\label{eqn:functional}
\end{equation}
where 
\begin{equation}
\phi j = \int d^4 x \phi(x) j(x).
\end{equation}

Using (\ref{eqn:functional}) we define
\begin{equation}
Z[j] \equiv \exp\left\{i W\left[ j \right]\right\},
\label{eqn:confunctional}
\end{equation}
where $W \left[ j \right]$ is known as the connected generating functional. 
The effective action $\Gamma\left[ \bar \phi \right]$ is just the Legendre 
transformation of (\ref{eqn:confunctional}) and is given by
\begin{equation}
\Gamma \left[ \bar \phi \right] = W[j] - \int d^4 x 
\frac{\delta W\left[ j \right]}{\delta j(x)} j(x),
\end{equation}
where
\begin{equation}
\bar \phi (x) = \frac{\delta W\left[ j \right]}{\delta j(x)}.
\end{equation}

$\bar \phi(x)$ is the weighted average of the fluctuations of the field.
In a translationally invariant theory, which are the ones \EFF{} is designed to
deal with, $\bar \phi (x)$ is a constant
\begin{equation}
\bar \phi (x) = \phi_c.
\end{equation}

The effective potential can then be defined as
\begin{equation}
\Gamma \left[ \phi_c \right] = - \int d^4 x V_{\rm eff} \left( \phi_c \right),
\end{equation}
which can be written as an expansion as
\begin{equation}
V_{\rm eff} \left( \phi_c \right) = -\sum_{n=0}^{\infty} \frac{1}{n!} \phi_c^n
\Gamma^{(n)}\left( p_i = 0 \right),
\label{eqn:effpot}
\end{equation}
where $\Gamma^{(n)}$ are the one-particle irreducible (1PI) Green functions.
Minimizing the effective potential over the constant fields, $\phi_c$, 
gives the vacuum state of the theory~\cite{PeskinSchroeder}.

\subsection{One-Loop Potential}
The tree-level effective potential is identical to the classical effective 
potential. This is simply
\begin{equation}
V_{\rm tree} =  - \left. {\cal L} \right|_{\phi_i(x) \rightarrow \phi_{ic}}.
\end{equation}
The one-loop contribution can be written in closed form for any theory 
containing fields of spin $0,\frac{1}{2},$ or $1$. 

We show here the one-loop correction for a model with one self-interacting
scalar field described by the Lagrangian
\begin{equation}
{\cal L} = \frac{1}{2} \partial^\mu \phi \partial_\mu \phi - V_0(\phi),
\end{equation}
where $V_0$ is the tree-level potential given by
\begin{equation}
V_0 = \frac{1}{2} m^2 \phi^2 + \frac{\lambda}{4!} \phi^4.
\end{equation}

As we expressed in (\ref{eqn:effpot}), the one-loop correction to the 
tree-level effective potential is given by the sum of all 1PI diagrams with a 
single loop and zero external momenta. The $n$-th diagram has $n$ propagators, 
$n$ vertices and $2n$ external legs. The $n$ propagators contribute a factor of
$i^n (p^2-m^2 + i\epsilon)^{-n}$, as we saw in chapter~\ref{chap:Intro}. Each 
pair of the external lines contributes a factor of $\phi_c^{2n}$ and each 
vertex a factor of $-i\lambda/2$. There is also a global symmetry factor of 
$1/2n$. 

This gives
\begin{eqnarray}
V_1(\phi_c) &=& i \sum_{n=1}^{\infty} \int \frac{d^4p}{(2\pi)^4}\frac{1}{2n}
\left[ \frac{\lambda \phi_c^2/2}{p^2 - m^2 + i\epsilon} \right]^n \nonumber \\
&=& -\frac{i}{2} \int \frac{d^4p}{(2\pi)^4}\log \left[ 1 - \frac{\lambda 
\phi_c^2/2}{p^2 - m^2 + i\epsilon} \right].
\label{eqn:V1norm}
\end{eqnarray}
In~(\ref{eqn:confunctional}) we did not include a normalization factor of 
$\exp \left\{ i W[0]\right\}$. This is needed to correctly 
generate~(\ref{eqn:V1norm}). Without
it a shift is introduced to the effective potential. After a Wick rotation and
using the unnormalized generating functional we find
\begin{equation}
V_1(\phi_c) = \frac{1}{2} \int \frac{d^4p}{(2\pi)^4} \log \left[ p^2 +
  m^2(\phi_c)\right],
\end{equation}
where the momentum is now in Euclidean space and the mass is the shifted mass
given by
\begin{equation}
m^2(\phi_c) = \frac{d^2 V_0(\phi_c)}{d\phi_c^2}.
\end{equation}
This is finally
\begin{equation}
V_1(\phi_c) = \frac{1}{64 \pi^2} m^4(\phi_c) \left( \ln \frac{m^2(\phi_c)}
{\mu^2} - \frac{3}{2} \right),
\end{equation}
in the $\overline{\rm DR}$ renormalization scheme \cite{DRscheme} where $\mu$ 
is the renormalization scale. This result can be trivially generalized to the 
case of $N_s$ complex scalar fields each described by the Lagrangian,
\begin{equation}
{\cal L} = \partial^\mu \phi^a \partial_\mu \phi_a^\dagger - V_0(\phi^a,
\phi_a^\dagger).
\end{equation}
The one-loop contribution in the $\overline{\rm DR}$ renormalization scheme is
\begin{equation}
V_1(\phi^a,\phi^\dagger_a) = \frac{1}{64 \pi^2} {\rm Tr} \left[ M_s^4(\phi^a, 
\phi_a^\dagger) \left( \ln \frac{M_s^2(\phi^a, \phi_a^\dagger)}{\mu^2} - 
\frac{3}{2} \right) \right],
\label{eqn:scont}
\end{equation}
where $M_s$ is the mass matrix of the fields given by
\begin{equation}
\left( M_s^2 \right)^a_b = \frac{\partial^2 V}{\partial \phi_a^\dagger \partial
\phi^b}.
\end{equation}

Eq. (\ref{eqn:scont}) can be generalized to fermions obeying the Dirac equation
and gauge bosons \cite{Quiros,Wein} as
\begin{equation}
V_1 = \sum_s \frac{1}{64 \pi^2} (2 s + 1)(-1)^{2s} {\rm Tr_s} \left[ M^4(
\phi^a, \phi_a^\dagger) \left( \ln \frac{M^2(\phi^a, \phi_a^\dagger)}{\mu^2} - 
\frac{3}{2} \right) \right],
\end{equation}
where the sum is over the spins, $0,\frac{1}{2},$ and $1$, and ${\rm Tr}_s$ is 
the trace is over the mass matrix of spin, $s$. Defining the ``Supertrace''
as
\begin{equation}
{\rm STr} f(X) = \sum_i (2s_i + 1)(-1)^{2 s_i} f(X_i),
\end{equation}
which is the spin-weighted trace, we have the one-loop contribution in the 
$\overline{\rm DR}$ renormalization scheme as
\begin{equation}
V_1 = \frac{1}{64 \pi^2}{\rm STr} \left[ M^4 \left( \ln \frac{M^2}{\mu^2} - 
\frac{3}{2} \right) \right].
\label{eqn:eff1loop}
\end{equation}

\section{Mass Matrices}
\label{sec:masscorr}
The fields defined as representations of the groups in a field theory are 
interaction eigenstates of the theory. These
are states in which the interactions are directly governed by the respective
couplings. The particles that are observed are the mass eigenstates of the
theory. These states are a mixture of interaction eigenstates, and thus the
interactions of these states are a mixture of the couplings in the theory.
The interaction states that mix to form the mass states, as well as the mixing
angles, can be derived from the mass matrices. In should be noted that the
mass matrices are not derivable from the effective potential. Rather, the
value of the VEVs chosen from the minimization of the effective potential
determine the mass spectrum of the particles. Derivation of the masses 
requires the full potential.

For complex scalars the mass matrix is defined as
\begin{equation}
\left( M_s^2 \right)^a_b = \left. \frac{\partial^2 V}{\partial \phi_a^\dagger 
\partial \phi^b} \right|_{\phi_i(x) \rightarrow \phi_{ic}},
\label{eqn:mms}
\end{equation}
where $\phi_{ic}$ is the expectation value of the the scalar field $\phi_i(x)$.
Similarly for vectors the mass matrix is
\begin{equation}
\left( M_V^2 \right)^a_b = \left. \frac{\partial^2 V}{\partial {\cal A}_a 
\partial {\cal A}^b} \right|_{\phi_i(x) \rightarrow \phi_{ic}}.
\label{eqn:mmv}
\end{equation}
The mass matrix for fermions is not defined as a mass squared matrix. Instead
for complex fermions it is given by
\begin{equation}
\left( M_f \right)^a_b = \left. \frac{\partial^2 V}{\partial \psi_a^\dagger 
\partial \psi^b} \right|_{\phi_i(x) \rightarrow \phi_{ic}}.
\label{eqn:mmf}
\end{equation}

Diagonalizing these mass matrices gives
\begin{equation}
M_{s,V}^2 = U^T D_{s,V}^2 V,~~~~~M_f = U^T D_f V,
\end{equation}
where $U$ and $V$ are unitary matrices which define the mixing angles of the
interaction states. $D$ is the diagonal matrix whose elements are the masses
of the mass eigenstates. 

Eqs. (\ref{eqn:mms}-\ref{eqn:mmf}) don't indicate whether the potential is 
to be taken at tree-level, one-loop or some other order. That
is because the masses also have higher order corrections. If the tree-level
potential is used, it yields the tree-level mass matrices. Likewise, if the
one-loop potential is used it yields the one-loop mass matrix. For the exact
mass term the exact potential would be needed.

As we saw earlier, the one-loop effective potential can be written in terms of 
the mass matrices only. The matrices in (\ref{eqn:eff1loop}) are the tree
level matrices. From (\ref{eqn:mms}-\ref{eqn:mmf}) we can see that the one-loop
masses are derived from the full one-loop potential. This would require the
calculation of the full one-loop potential, which is much more complicated.
Instead the one-loop corrections to the mass matrices can be computed in 
another way. If we refer back to chapter~\ref{chap:Intro}, the masses of a 
field are related to the two-point Green's function of that field. In fact, 
it can be shown that the one-loop correction to the mass is just the one-loop 
correction to the two-point Green's function \cite{PeskinSchroeder}. 

Before we present the one-loop corrections to the mass matrices, the idea of
a ghost field must be explained. We saw that in chapter~\ref{chap:Intro} that
the propagator for a gauge boson is given as $\frac{-i g_{\mu \nu}}{p^2
+ i \eps}$. This is not quite true for non-Abelian groups. Instead using the
Faddeev-Popov method to quantize the field and the gauge condition
\begin{equation}
G(A) = \partial^\mu A^a_\mu(x) - \omega^a(x) = 0,
\end{equation}
where $\omega^a$ is a Gaussian weight, we find the propagator is
\begin{equation}
\left< A_\mu^a(x) A_\nu^a(x) \right> = \frac{-i}{p^2 + i \eps} \left(
g_{\mu \nu} - (1-\xi) \frac{p_\mu p_\nu}{p^2}\right).
\end{equation}
The Faddeev-Popov quantization inserts a determinant of the form $\det\left(
\frac{\delta G(A^\alpha)}{\delta \alpha}\right)$ where $\alpha$ is now used to
represent the infinitesimal form of the gauge transformation. This determinant
is zero for an Abelian group but non-zero for a non-Abelian group. Instead
Faddeev and Popov showed it can be expressed as
\begin{equation}
\det \left(\frac{\delta G(A^\alpha)}{\delta \alpha}\right) = \det \left(
\frac{1}{g} \partial^\mu D_\mu \right) = \int {\cal D}c\, {\cal D}\bar c\,
\exp\left[ i \int d^4 x\, \bar c \left(-\partial^\mu D_\mu\right)c\,\right],
\end{equation}
where $c$ and $\bar c$ are now anti-commuting fields which are scalars under
Lorentz transformations. These new fields are known as ghost fields and
introduce new Feynman rules that must be considered to ensure that the value 
of an $S$-matrix calculation is gauge-invariant.

For a spontaneously broken symmetry, we impose the gauge condition
\begin{equation}
G(A) = \frac{1}{\sqrt{\xi}} ( \partial_\mu A^\mu - \xi g \upsilon \phi)=0,
\end{equation}
where $\phi$ is the Goldstone boson associated with the breaking and 
$\upsilon$ is the VEV of the field which leads to the breaking. Using this
form of the gauge fixing condition gives the massive gauge boson the 
propagator
\begin{equation}
\left< A_\mu A_\nu \right> = \frac{-i}{p^2 - m^2}\left[ g_{\mu \nu} -
\frac{p_\mu p_\nu}{p^2 - \xi m^2}(1-\xi) \right],
\end{equation}
which is the same value for the gauge boson propagator shown before for the
massless case. This also gives the Goldstone boson, $\phi$, a mass
of $\xi m_A^2$ where $m_A$ is the mass of the gauge boson. The ghost also
acquires the same mass, $\xi m_A^2$, and this must be taken into account when
computing Feynman rules of a process. 

The one-loop corrections for a field can be given in a gauge independent form 
by explicitly keeping $\xi$. That has been done for the one-loop corrections
given here. In \EFF{}, the ghost contributions are hard to automate. Instead 
we work in the Landau gauge, $\xi \to 0$ where the masses of the Goldstone 
bosons remain zero and the ghost terms do not contribute to the calculations.

The one-loop corrections can be written in terms of the Passarino-Veltman 
\cite{Passarino:1979jh,Denner:1993kt,Pierce:1997zz} functions. These are 
defined by
\begin{eqnarray}
A_0(m^2; \mu^2) &=& 16 \pi^2 \mu^{4-d} \int \frac{i d^d q}{(2\pi)^d} \frac{1}
{q^2 - m^2 + i\epsilon}, \nonumber \\
B_0(p^2, m_1^2, m_2^2; \mu^2) &=&  16 \pi^2 \mu^{4-d}\int \frac{i d^d q}
{(2\pi)^d} \frac{1}{\left[ q^2 - m_1^2 + i\epsilon \right] \left[ (q-p)^2 - 
m_2^2 + i\epsilon \right]}, \nonumber \\
p_\mu B_1(p^2, m_1^2, m_2^2; \mu^2) &=& 16 \pi^2 \mu^{4-d}\int \frac{i d^d q}
{(2\pi)^d} \frac{q_\mu}{\left[ q^2 - m_1^2 + i\epsilon \right] \left[ (q-p)^2 -
m_2^2 + i\epsilon \right]}, \nonumber \\
p_\mu p_\nu B_{11}(p^2, m_1^2, m_2^2; \mu^2) &+& g_{\mu \nu} B_{00}(p^2,m_1^2,
m_2^2; \mu^2) \\
&=& 16 \pi^2 \mu^{4-d}\int \frac{i d^d q}
{(2\pi)^d} \frac{q_\mu q_\nu}{\left[ q^2 - m_1^2 + i\epsilon \right] \left[ 
(q-p)^2 - m_2^2 + i\epsilon \right]}, \nonumber 
\end{eqnarray}
where $\mu$\footnote{Not to be confused with $\mu$ for the Lorentz index} is 
the renormalization scale and the integrals are regularized in $d = 4 - 2 \eps$
dimensions. The first two of these can be integrated and expressed (expanding 
to ${\cal O}(\eps)$) as
\begin{eqnarray}
A_0(m^2; \mu^2) &=& m^2 \left( \frac{1}{\hat \eps} - 1 + \ln \frac{m^2}{\mu^2} 
\right) + {\cal O}(\eps), \\
B_0(p^2, m_1^2, m_2^2; \mu^2) &=& \frac{1}{\hat \eps} - \ln \frac{p^2}{\mu^2} -
f_B(x_+) - f_B(x_-)i + {\cal O}(\eps),
\end{eqnarray}
where $1/\hat \eps = 1/\eps - \gamma_E + \ln 4\pi$,
\begin{equation}
x_\pm = \frac{s \pm \sqrt{s^2 - 4p^2(m_1^2 - i\epsilon)}}{2p^2},~~~f_B(x) =
\ln(1-x) - x \ln\left(1-\frac{1}{x}\right) - 1,
\end{equation}
and $s = p^2 - m_2^2 + m_1^2$. The other Passarino-Veltman functions 
can be decomposed into the two scalar functions, $A_0$ and 
$B_0$~\cite{Denner:1993kt}. There
are more functions for higher loop corrections that are not given here. These
can be decomposed into their respective higher-loop scalar functions (e.g.
$C_0$, $D_0$, \dots). 

The corrections are summed over all internal fields and the couplings are given
from the Lagrangian. There are two ways that the one-loop corrections can
be applied. First one could find the diagonal form of the tree-level matrix. 
This defines the mass eigenstates and therefore the couplings of these states.
The corrections can then be directly applied to the tree-level masses
given the couplings defined in this manner. This is not exactly the one-loop
correction, however. This will give a close approximation to the one-loop
masses, but will only give the tree-level mixings. If instead the one-loop
corrections are applied to the undiagonalized matrix, with corrections to the
off-diagonal elements, then the diagonalization of this corrected matrix
will give the correct one-loop masses and mixings. Both techniques have been 
implemented in \EFF{}, but the latter is much more time intensive and if 
only the masses are desired, applying the corrections to the mass eigenstates 
is a good approximation.

\subsection{Scalar One-Loop Corrections}
\begin{figure}
\centering
\input{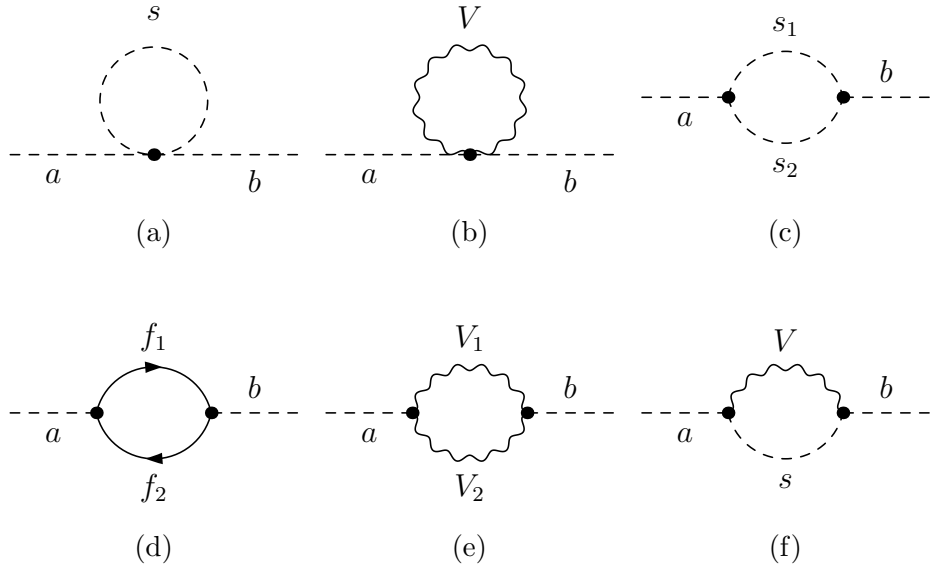}
  \caption{The different diagrams that contribute to the one-loop correction 
    of the scalar mass matrix element $(M_s^2)_{ab}$.\label{fig:scalcorr}}
\end{figure}

We first give the one-loop corrections to the scalar masses. Figure
\ref{fig:scalcorr} shows the corrections to the scalar mass matrix term 
$(M_s^2)_b^a$. These diagrams are summed over all possible intermediate states
given in the theory. The results given here have been calculated with the 
help of the computer software {\small FeynCalc}~\cite{Mertig:1991an} and
{\small FormCalc}~\cite{Hahn:1998yk}.

The form for the contribution of the scalar four-point vertex, given by the 
first diagram in figure~\ref{fig:scalcorr} is
\begin{equation}
\left(\Sigma^{(assb)}\right)_b^a(p^2, m^2; \mu^2) = \frac{i C^{(assb)}}{32 
\pi^2} A_0(m^2; \mu^2),
\end{equation}
where $C^{(assb)}$ is the four-point coupling of the fields $a$ and $b$ with
an arbitrary scalar field $s$. The second diagram in figure~\ref{fig:scalcorr}
is the contribution of the four-point vertex given with a vector loop. This 
correction is
\begin{equation}
\left(\Sigma^{(aVVb)}\right)_b^a(p^2, m^2; \mu^2) = \frac{i C^{(aVVb)}}{32 
\pi^2}\left( 2m^2 - 3 A_0(m^2) + \xi A_0(\xi m^2) \right),
\end{equation}
The diagram given by fig. \ref{fig:scalcorr}c is of the scalar three-point 
vertex. This is given by
\begin{equation}
\left(\Sigma^{(as_1s_2b)}\right)_b^a(p^2, m_1^2, m_2^2; \mu^2) = 
\frac{-C^{(as_1s_2)} C^{(s_1s_2b)}}{32 \pi^2} B_0(p^2, m_1^2, m_2^2)
\end{equation}
where $C^{(as_1s_2)}$ is the coupling of field $a$ to the two fields in the
loop and $C^{(s_1s_2b)}$ is the coupling of field $b$. 

There is no fermion four-point diagram as this type of term would have a mass
dimension greater than four in the Lagrangian. Instead there is only a 
three-point correction. This term (fig. \ref{fig:veccorr}d) is given by the 
expression
\begin{eqnarray}
\left(\Sigma^{(af_1f_2b)}\right)_b^a(p^2, m_1^2, m_2^2; \mu^2) &=&
\frac{C^{(af_1f_2)} C^{(f_1f_2b)}}{16 \pi^2} \left\{  A_0(m_1^2; \mu^2) + 
A_0(m_2^2; \mu^2) + \right. \nonumber \\
&& \left. \,((m_1+m_2)^2 - p^2) B_0(p^2,m_1^2,m_2^2; \mu^2) \right\}.
\end{eqnarray}
The vector three-point term, given by the fifth diagram in 
figure~\ref{fig:scalcorr}, is 
\begin{eqnarray}
\left(\Sigma^{(aV_1V_2b)}\right)_b^a(p^2, m_1^2, m_2^2; \mu^2) &=&
\frac{C^{(aV_1V_2)} C^{(V_1V_2b)}}{32 \pi^2} \left\{ 2 + (\frac{p^2}{m_2^2}-3)
B_0^0  - \frac{p^2}{m_2^2} B_0^2 + \right. \nonumber \\
&& \, \xi \frac{A_0^2}{m_2^2} -  \xi \frac{A_0^4}{m_2^2} + \left( \xi^2 
\frac{m_1^2}{m_2^2}  - \xi \right) B_0^1 - \xi^2 \frac{m_1^2}{m_2^2} B_0^{12} +
\nonumber \\
&& \frac{p^2}{m_1^2 m_2^2} \left( B_{00}^2 - B_{00}^0 + B_{00}^1 - 
B_{00}^{12} \right) + \nonumber \\
&& \frac{p^4}{m_1^2 m_2^2} \left( B_{11}^2 - B_{11}^0 + B_{11}^1 - 
B_{11}^{12} \right) + \nonumber \\
&& \left. \, 2 \xi \frac{p^2}{m_2^2} (B_1^1 - B_1^{12}) \right\},
\end{eqnarray}
where
\begin{eqnarray}
A_0^1 & \equiv & A_0(m_1^2; \mu^2),~~A_0^2 \equiv A_0(m_2^2; \mu^2),~~
A_0^3 \equiv A_0(\xi m_1^2; \mu^2),~~A_0^4 \equiv A_0(\xi m_2^2; \mu^2), 
\nonumber \\
B_x^0 & \equiv & B_x(p^2,m_1^2,m_2^2; \mu^2),~~B_x^1 \equiv B_x(p^2, m_2^2,
\xi m_1^2; \mu^2), B_x^2 \equiv  B_x(p^2,m_1^2,\xi m_2^2; \mu^2),
\nonumber \\
B_x^{12} & \equiv & B_x(p^2, \xi m_1^2,\xi m_2^2; \mu^2).
\label{eq:abbrv}
\end{eqnarray}
Remarkably, in the 't Hooft-Feynman gauge ($\xi \rightarrow 1$) this reduces to
\begin{equation}
\left(\Sigma^{(aV_1V_2b)}\right)_b^a(p^2, m_1^2, m_2^2; \mu^2) =
\frac{C^{(aV_1V_2)} C^{(V_1V_2b)}}{16 \pi^2} \left( 1 - 2 B_0(p^2, m_1^2, 
m_2^2; \mu^2) \right).
\end{equation}
The last one-loop corrections to the scalar mass matrix, $\left( M_s^2 
\right)_b^a$, is given by the last diagram in figure~\ref{fig:scalcorr}. 
This is the correction given by the vector-scalar three-point vertex, given 
by
\begin{eqnarray}
\left(\Sigma^{(aVsb)}\right)_b^a(p^2, m_V^2, m_s^2; \mu^2) &=&
\frac{-C^{(aVs)} C^{(Vsb)}}{16 m_V^2 \pi^2} \left\{ (p^2 - m_s^2) A_0^1
- (p^2 - \xi m_V^2 - m_s^2) A_0^3 + \right. \nonumber \\ 
&& \, \left( m_V^2 (p^2 + m_s^2) - (p^2 - m_s^2)^2 \right) B_0^0 - \nonumber \\
&& \left. \, (p^2 - m_s^2)^2 B_0^{1} - 2 m_V^2 p^2 B_1^0 \right\},
\end{eqnarray}
where the vector corresponds to loop field 1 in the previous abbreviations
and the scalar field is loop field 2. 

\subsection{Vector One-Loop Corrections}
\begin{figure}
\centering
\input{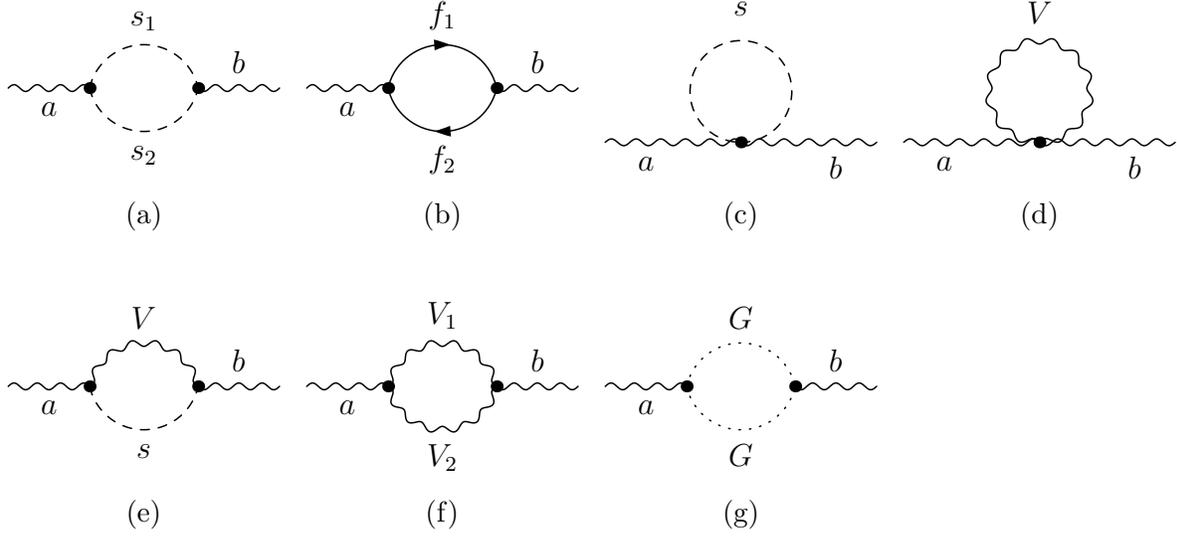}
  \caption{The different diagrams that contribute to the one-loop correction 
    of the vector mass matrix element $(M_V^2)_{ab}$.\label{fig:veccorr}}
\end{figure}

The one-loop corrections to the vector fields can also be generated using the
same two programs {\small FeynCalc} and {\small FormCalc}. 
Figure~\ref{fig:veccorr} shows the diagrams which give corrections to the 
vector mass.

The first diagram in figure~\ref{fig:veccorr} leads to a correction of the
form
\begin{equation}
\left(\Pi^{(as_1s_2b)}_{\mu \nu} \right)_b^a(p^2, m_1^2, m_2^2; \mu^2) =
\frac{C^{(as_1s_2)} C^{(s_1s_2b)}}{8 \pi^2} B_{00}(p^2, m_1^2,
m_2^2; \mu^2) \left( g_{\mu \nu} - \frac{p_\mu p_\nu}{p^2}\right).
\end{equation}
The contribution from the fermion loop is gauge independent and is given by
\begin{eqnarray}
\left(\Pi^{(af_1f_2b)}_{\mu \nu} \right)_b^a(p^2, m_1^2, m_2^2; \mu^2) &=&
-\frac{C^{(af_1f_2)} C^{(f_1f_2b)}}{16 \pi^2} \left\{ A_0(m_1^2; 
\mu^2) + A_0(m_2^2; \mu^2) + \right. \nonumber \\
&& \, \left( (m_1-m_2)^2 - p^2 \right) B_0(p^2, m_1^2, m_2^2; \mu^2) -
\nonumber \\
&& \left. \, 4 B_{00}(p^2, m_1^2, m_2^2; \mu^2) \right\}
\left( g_{\mu \nu} - \frac{p_\mu p_\nu}{p^2}\right),
\end{eqnarray}
and fig. \ref{fig:veccorr}c shows the scalar four-point correction. This is
\begin{equation}
\left(\Pi^{(asb)}_{\mu \nu} \right)_b^a(p^2, m_1^2, m_2^2; \mu^2) =
\frac{i C^{(aSSb)}}{32 \pi^2} A_0(m_S^2; \mu^2) 
\left( g_{\mu \nu} - \frac{p_\mu p_\nu}{p^2}\right).
\end{equation}

The next two diagrams are of the four-point vector loop and the
three-point vector-scalar loop. These are given by
\begin{equation}
\left(\Pi^{(aVVb)}_{\mu \nu} \right)_b^a(p^2, m^2; \mu^2) =
\frac{i C^{(aVVb)}}{32 \pi^2} \left( 2m^2 - 5 A_0(m^2; \mu^2)
- \xi A_0(\xi m^2; \mu^2) \right)
\left( g_{\mu \nu} - \frac{p_\mu p_\nu}{p^2}\right),
\end{equation}
and
\begin{equation}
\left(\Pi^{(aVsb)}_{\mu \nu} \right)_b^a(p^2, m_V^2, m_s^2; \mu^2) =
\frac{i C^{(aVs)} C^{(Vsb)}}{32 \pi^2 m_V^2} \left( m_V^2 B_0^2
- B_{00}^2 + B_{00}^{12} \right)
\left( g_{\mu \nu} - \frac{p_\mu p_\nu}{p^2}\right),
\end{equation}
where we have used the same abbreviations (\ref{eq:abbrv}) and the vector
corresponds to field 1 in the notation and the scalar is field 2. The remaining
two diagrams are for the vector three-point correction and the ghost terms.
The vector three-point correction is quite complicated and is given by
\begin{eqnarray}
\left(\Pi^{(aV_1V_2b)}_{\mu \nu} \right)_b^a(p^2, m_1^2, m_2^2; \mu^2) &=&
\frac{C^{(aV_1V_2)} C^{(V_1V_2b)}}{32 \pi^2} \left\{ \frac{2}{3}p^2
- 2 (m_1^2+m_2^2) + \right. \nonumber \\
&& \, (\rho_2 - \kappa_1 + \xi) A_0^2 - (\rho_2 - \kappa_1 - \xi) A_0^4 
+ \nonumber \\
&& \, (m_1^2(1-\kappa_1) + p^2(2 \kappa_1 + 5 -  \rho_2) B_0^0 + \nonumber \\
&& \, (m_1^2(\kappa_1 + \xi^2) + p^2(\rho_2 -  2 \kappa_1))B_0^2 + \nonumber \\
&& \, (\kappa_1 + 11 - 2\rho_2 - 5 \rho_1 + \rho_1 \rho_2) B_{00}^0 + 
\nonumber \\
&& \, (2 \rho_2 - \kappa_1 - \rho_1 \rho_2) B_{00}^2 + 
(5 \rho_1 - \rho_1 \rho_2 - \xi) B_{00}^1 + \rho_1 \rho_2 B_{00}^{12} + 
\nonumber \\
&& \, - 2 p^2 B_1^0 + 4 \xi p^2 B_1^1  + \nonumber \\
&& \left. \, 4 \rho_1 p^2 \left( B_{11}^1 - B_{11}^0 \right) \right. \bigg\}
\left( g_{\mu \nu} - \frac{p_\mu p_\nu}{p^2}\right),
\label{eqn:V3pt}
\end{eqnarray}
where the abbreviations (\ref{eq:abbrv}) are used and $\kappa_1 = \frac{m_1^2}
{m_2^2}$, $\rho_1 = \frac{p^2}{m_1^2}$ and $\rho_2 = \frac{p^2}{m_2^2}$. In the
't~Hooft-Feynman gauge this reduces to
\begin{eqnarray}
\left(\Pi^{(aV_1V_2b)}_{\mu \nu} \right)_b^a(p^2, m_1^2, m_2^2; \mu^2) &=&
\frac{C^{(aV_1V_2)} C^{(V_1V_2b)}}{32 \pi^2} \left\{ \frac{2}{3}p^2
- 2 (m_1^2+m_2^2) + \right. \nonumber \\
&& \, 2 \left( m_1^2 + 3 p^2 \right) B_0(p^2, m_1^2, m_2^2; \mu^2) - 2 \kappa_1
A_0(m_2^2; \mu^2) + \nonumber \\
&& \, (10 + \rho_1) B_{00}(p^2, m_1^2, m_2^2; \mu^2) + \nonumber \\ 
&& \left. \, 2 p^2 B_1\left(p^2, m_1^2, m_2^2; \mu^2 \right) \frac{}{}\right\}
\left( g_{\mu \nu} - \frac{p_\mu p_\nu}{p^2}\right).
\end{eqnarray}
It must also be noted that the gauge bosons from an unbroken group are 
massless. These bosons only couple to themselves and the correction is
proportional to $p^2$. The correction is to be evaluated at $p^2=m^2$, which
for bosons from an unbroken group is zero.

The remaining one-loop correction to the vector mass matrix is
from the ghost terms. This is the last diagram in figure~\ref{fig:veccorr} and
is given by
\begin{equation}
\left(\Pi^{(aGGb)}_{\mu \nu} \right)_b^a(p^2, m_G^2; \mu^2) =
\frac{C^{(aGG)} C^{(GGb)}}{32 \pi^2} \xi 
B_{00}(p^2, \xi m_G^2, \xi m_G^2; \mu^2) 
\left( g_{\mu \nu} - \frac{p_\mu p_\nu}{p^2}\right),
\end{equation}
from which it is easy to see that the ghost contribution is zero in the
Landau gauge.

\subsection{Fermion One-Loop Corrections}

\begin{figure}
\centering
\input{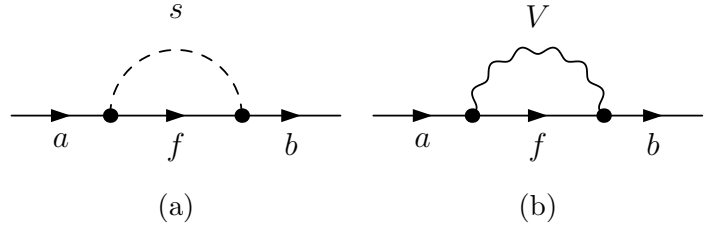}
  \caption{The different diagrams that contribute to the one-loop correction 
    of the fermion mass matrix element $(M_f)^a_b$.\label{fig:fermcorr}}
\end{figure}

There are only two mass corrections to the fermion masses. The diagrams of
these are given in fig. \ref{fig:fermcorr}. The first one is the correction
due to a scalar-fermion three-point coupling. This is
\begin{equation}
\left( \Sigma^{(asfb)} \right)_b^a(p^2, m_s^2, m_f^2; \mu^2) =
\frac{-C^{(asf)} C^{(sfb)}}{16 \pi^2} \Fs{p} B_1(p^2,m_s^2,m_f^2;\mu^2).
\end{equation}
The mass correction is given by $\delta_m = \left. \Sigma(\Fs{p}) 
\right|_{~\Fs{~p}=m}$. This can be approximated by $\delta_m \approx 
\left. \Sigma(\Fs{p}) \right|_{~\Fs{~p}=m_0}$.

Lastly, the correction given by fig. \ref{fig:fermcorr}b is from a vector
boson three-point coupling. This is
\begin{equation}
\left( \Sigma^{(aVfb)} \right)_b^a(p^2, m_V^2, m_f^2; \mu^2) =
\frac{C^{(aVf)} C^{(Vfb)}}{8 \pi^2}\Fs{p} \left[ B_1^0
 - (1-\xi) ( B_{00}^1 + p^2 B_{11}^1) \right],
\end{equation}
where the abbreviations given earlier are used and the index 1 refers to the 
vector and the index 2 is the fermion. 

\section{Renormalization Group Equations}
The basic concept of renormalization stems from the observation that the
divergences in the one-loop graphs amount to shifts in the parameters of 
the action. For example, as we saw in the last section they change the
mass of a field. Renormalization is the procedure of cancelling the divergences
from these shifts by introducing counterterms into the Lagrangian. These
counter terms are defined such that they exactly cancel the divergent
quantity in the one-loop correction but leave the finite parts 
untouched~\cite{Collins}.

We can first consider the $\phi^4$ theory. The Lagrangian is given by
\begin{equation}
{\cal L} = (\partial \phi_0)^2/2 - m_0^2 \phi_0^2 / 2 - g_0 \phi_0^4/4!
\end{equation}
We can then rescale the field by $\phi_0 = Z^{1/2} \phi$, so that in terms of
the new `renormalized field' $\phi$ we have
\begin{equation}
{\cal L} = Z (\partial \phi)^2/2 - Z m_0^2 \phi^2 / 2 - g_0 Z^2 \phi^4/4!
\end{equation}
We can then write this in terms of the counterterms
\begin{eqnarray}
{\cal L} &=& (\partial \phi)^2/2 - m^2 \phi^2 / 2 - g \phi^4/4! \nonumber \\
&& + \delta_Z (\partial \phi)^2/2 - \delta m^2 \phi^2/2 - \delta g \phi^4 /4!
\label{eqn:counter}
\end{eqnarray}
We can see that $\delta_Z = Z-1$, $\delta m^2 = Z m_0^2 -  m^2$ and 
$\delta g = Z^2 g_0 - g$. These counterterms are adjusted to cancel the
divergences of each term.

To renormalize a theory renormalization conditions must be imposed. The general
conditions that are imposed are to set the contributions from the 1PI diagrams
to the two-point Green's functions and the derivatives of these diagrams to 
zero at a spacelike momentum, $p^2 = -\mu^2$. $\mu$ is then 
known as the \emph{renormalization scale}. The conditions are imposed at a
spacelike momentum in order to avoid threshold singularities. 
Similarly, the 1PI contributions to the three- and four-point Green's functions
are defined such that the coupling at that scale is exactly the physical 
coupling (e.g. $g$ from (\ref{eqn:counter})).

The renormalization scale is completely arbitrary. The theory can just as
easily be defined at another renormalization scale. This implies that the
theory and the redefined couplings should depend on the scale in such a way 
that physical calculations are independent of the scale. For example if
we take an $n$-point Green's function we require
\begin{equation}
\delta G^{(n)} = \frac{\partial G^{(n)}}{\partial \mu} \delta \mu + 
\frac{\partial G^{(n)}}{\partial \lambda} \delta \lambda = n \delta \eta
G^{(n)}.
\end{equation}
This is for either a massless field or if the mass of the field is considered
as on of the vertices that needs to be renormalized.
$\delta \lambda$ is the counterterm corresponding to the coupling $\lambda$
and $\delta \eta$ is the rescaling of the external field in the $n$-point
Green's function. Defining
\begin{equation}
\beta_\lambda \equiv \mu \frac{\partial \lambda}{\partial \mu},~~~~\gamma_\eta
\equiv - \mu \frac{\partial \eta}{\partial \mu},
\end{equation}
we can now give the Callan-Symanzik 
equation~\cite{Callan:1970yg,Symanzik:1970rt}. This is
\begin{equation}
\left[ \mu \frac{\partial}{\partial \mu} + \beta_\lambda \frac{\partial}
{\partial \lambda} + n \gamma_\eta \right] G^{(n)} = 0.
\label{eqn:CalSym}
\end{equation}

The $\beta$ and $\gamma$ functions are the same between different Green's 
functions. Knowing this we can find the $\gamma$ function from the
2-point Green's function of the field. The 2-point Green's function can be
expressed in terms of the loop diagrams and the counterterms. Doing so $\gamma$
can be shown to be
\begin{equation}
\gamma = \frac{1}{2} \mu \frac{\partial}{\partial \mu} \delta_Z.
\end{equation}
The counterterm can be written in the form (at lowest order)
\begin{equation}
\delta_Z = A \log \frac{\Lambda^2}{\mu^2} + {\rm finite},
\end{equation}
in order to cancel the divergent logarithm in $G^{(2)}$. This means that we 
can find $\gamma$ simply by determining $A$, the coefficient of the divergent
piece of the 1PI diagrams. Since $\gamma$ is related to the field 
renormalization it is known as the \emph{anomalous dimension} of the field.

Knowing the anomalous dimensions, the $\beta$ functions can be derived by using
the appropriate Green's function ($G^{(n)}$ for an $n$-point coupling). The
$n$-point Green's function is given by the sum of the tree-level diagram,
the 1PI loop diagrams of the vertex, the vertex counterterm and the external 
leg corrections. This can be expressed for the $n$-point coupling $g$, at 
one-loop order, as
\begin{equation}
G^{(n)} = \left( \prod_i \frac{i}{p_i^2} \right) \left[ -i g - i B \log 
\frac{\Lambda^2}{-p^2} - i \delta_g + (-i g) \sum_i \left(A_i \log 
\frac{\Lambda^2}{-p_i^2} - \delta_{Z_i} \right) \right],
\end{equation} 
plus the finite terms. Applying the Callan-Symanzik equation we find
\begin{equation}
\beta_g = \mu \frac{\partial}{\partial \mu} \left(- \delta_g + \frac{1}{2} g \sum_i
\delta_{Z_i} \right).
\end{equation}
We can see that the second term on the right is just the sum of the anomalous
dimensions of all the external particles in the Green's function. At one-loop,
$\delta_g$ is given by an expression of the form
\begin{equation}
\delta_g = -B \log \frac{\Lambda^2}{\mu^2} + {\rm finite}.
\end{equation}
This means that we just have to compute the divergent pieces of the 1PI 
corrections to the coupling and $\beta_g$ is given by
\begin{equation}
\beta_g = -2 B + g \sum_i \gamma_i.
\label{eqn:fullbeta}
\end{equation}

Most theories have many couplings and fields associated with them. This 
requires the procedure that produced (\ref{eqn:CalSym}) to be applied to a
set of couplings, $\lambda_I$, and a set of fields $\phi_i$. 
\begin{equation}
\left[ \mu \frac{\partial}{\partial \mu} + \beta_I \frac{\partial}{\partial
\lambda_I} + \gamma_i \phi_{i} \frac{\partial}{\partial \phi_{i}} \right] 
{\cal O}(\lambda_I, m_i; \mu) = 0,
\label{eqn:CallanSymanzik}
\end{equation}
We saw earlier that using the two-, three- and four-point Green's functions,
eq. (\ref{eqn:CallanSymanzik}) can be applied to determine the form of all 
$\beta_I$ and all $\gamma_i$, known as the \emph{renormalization group 
equations}~\cite{Cabibbo:1979ay,Ellis:1991wk,Ford:1993mv}. This technique 
requires the calculation of 
several 1PI diagrams and is quite computationally intensive. Instead there are 
some couplings which can be computed either completely for an arbitrary theory,
or by applying the Callan-Symanzik equation to a different set of observables.
 
The running of the gauge group couplings can be computed from many
different vertices. They all require the boson self-energy which is given by 
the diagrams in fig. \ref{fig:veccorr}. The divergences of these diagrams 
define the anomalous dimension of the gauge boson field. These sum to give 
\begin{equation}
\gamma_b = \frac{g^2}{(4\pi)^2} \left[ \left( \frac{13}{6}-\frac{\xi}{2}\right)
C_2(G) - \frac{4}{6} \sum_i C(r_i) - \frac{1}{6} \sum_j C(r_j) \right],
\end{equation}
where $i$ is the sum over fermions, $j$ is the sum over scalars, 
$C_2(G) = f^{acd}f^{acd}$ and $C(r) = \Tr [ t_r^a t_r^a ]$. 
One can then use the fermion-fermion-vector coupling, as in 
\cite{PeskinSchroeder}, or one could just as easily use the 
gauge boson three- or four-point couplings (in non-Abelian theories). The
final result is gauge independent and given in general by the expression
\begin{equation}
\beta_g = - \frac{g^3}{(4\pi)^2} \left[ \frac{11}{3} C_2(G) - \frac{4}{3}
\sum_i C(r_i) - \frac{1}{3} \sum_j C(r_j) \right],
\end{equation}
where again the index $i$ indicates the fermions and $j$ the scalars. For the
case of Weyl spinors, the sum of the fermions has an additional factor of a
half (e.g. $\frac{2}{3}$ rather than $\frac{4}{3}$), and similarly for real
scalars.

\begin{figure}
\centering
\epsfig{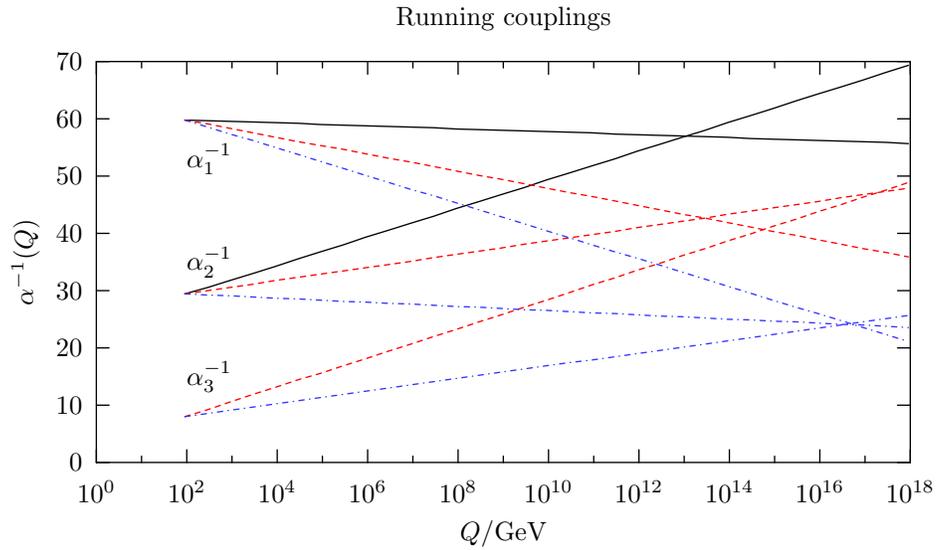}
  \caption{The scale dependence of the different gauge couplings. The black 
    (solid) lines are for the EW model with one fermion generation. The red
    (dashed) lines are for the SM with all three fermion generations. The blue 
    (dot-dashed) lines show the the couplings for the set of particles in the 
    MSSM.\label{fig:SMrunning}}
\end{figure}

The SM has Yukawa couplings, gauge couplings and couplings of the Higgs
boson ($\lambda$ and $\mu$). These are all dependent on the scale
\footnote{Here $Q$ is used instead of $\mu$ to avoid confusion with the Higgs
coupling, $\mu$.}, $Q$, at which they are evaluated. A common argument for 
beyond standard model (BSM) physics is that given the particles in the SM,
the couplings, $\alpha$, $\alpha_{\rm weak}$ and $\as$, will never meet at
a single ultimate scale, known as the unification scale. Though there is
no proof that indicates these should unify, it is believed to be a desired
and expected result of a theory. Figure~\ref{fig:SMrunning} shows the running
of the different $\alpha$'s for the particles in the EW theory with one 
generation (black, solid), the SM with three generations (red, dashed) and the 
MSSM with three generations (blue, dot-dashed). We can see from this figure 
that the couplings from the MSSM all meet at a single scale. The derivation of 
the $\beta$ functions and the running of the couplings in this figure were all 
done by entering the models into \EFF{}. In unification theories, the $U(1)$
coupling is multiplied by $\sqrt{5/3}$ and this is also done here.

In \EFF{} the observable that is used to define some of the couplings is
the effective potential~\cite{Quiros,Forshaw:2003kh}. The one-loop effective 
potential $V_0+V_1$, where $V_1$ is given in (\ref{eqn:eff1loop}), can be used
to define the 
RGEs for all the couplings that appear in the potential. This is the set 
of couplings for the fields that may acquire a non-zero VEV. If we apply the 
Callan-Symanzik equation to the one-loop effective potential, where now
the derivatives with respect to the fields are replaced by derivatives
with respect to the VEVs, this yields
\begin{equation}
\left[ \beta_I \frac{\partial}{\partial \lambda_I} + \gamma_i \phi_{ic}
\frac{\partial}{\partial \phi_{ic}} \right] V_0 = 
\frac{1}{32 \pi^2}{\rm STr}~M^4 .
\end{equation}
Using this one can compare the powers of the VEVs ($\phi_{ic}$) on the left 
hand side of the equation to those of the right and derive the relations of the
$\beta$ functions, given the anomalous dimensions, $\gamma_i$.

\begin{figure}
\input{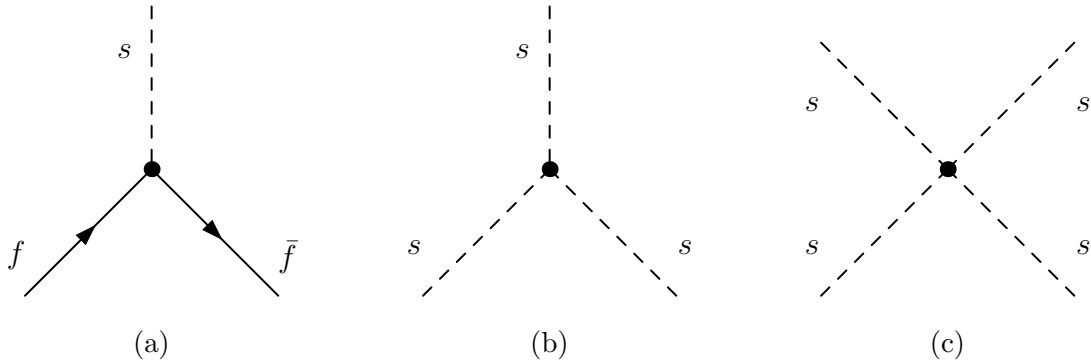}
\centering
  \caption{The three- and four-point diagrams that must have their 1PI
	diagrams computed to define the complete set of $\beta$ functions for
	a theory.\label{fig:tafpc}}
\end{figure}

To compute the remaining $\beta$ functions, we must use (\ref{eqn:fullbeta}). 
This requires us to compute the 1PI diagrams for the three- and four-point
vertices given in fig.~\ref{fig:tafpc}. Since we have already derived the 
$\beta$ functions of the gauge group couplings we do not need to consider the 
vertices with a gauge boson as an external leg.

\section{\EFF{}}

\EFF{} is a computer program that is able to generate the effective potential,
the mass spectrum and the RGEs of a theory. It is originally based on the
work in \cite{James} and uses the {\small GiNaC}\cite{GiNaC} computer 
algebra system to do algebraic manipulations. This package provides for 
evaluations and simplifications of complex algebraic structures within the 
C++ environment. As this is a C++ library, it also allows for expansion. New
functions and objects can be created that can then have their own set of
functions that are used during simplifications and evaluations. It is the 
ability to integrate the computer algebra with the standard C++ functionality 
and to be able to define new objects that interact with the computer algebra
package that make {\small GiNaC} an ideal library for building \EFF{} on.

My contribution to \EFF{} covers a broad spectrum. I have restructured and
written the code to work with the algebra package, {\small GiNaC}. The original
one-loop mass corrections in the code were only for scalar particles. These
were taken from the literature and did not all use the same gauge and as a 
complete correction, were wrong. As has been
seen earlier, these have been rederived in an arbitrary gauge for all different
spins and implemented in \EFF{} in the Landau gauge. Also, I have done  
the development and implementation of the RGEs in \EFF{}.

\EFF{} has currently been designed to work with $N=1$ SUSY. \EFF{} itself
is a package that can be used to define a SUSY model. Once this model is
defined a user can interact with it. It is these different interactions that
are discussed in this section.

\subsection{Model Definition}

A model in \EFF{} is defined by a set of gauge groups, fields and additional
interactions. These are then used to define a Lagrangian and an effective
potential. The Lagrangian can be used to define the couplings of the theory
and the mass matrices. As we saw in the last section, the effective potential
as well as the Green's functions can define the RGEs of the theory. 

A gauge group is just that. In \EFF{} this is a class that defines the 
structure constants and the generators. This also defines the dimension of the
indices in the fundamental and adjoint representations. It is also possible to
have several gauge groups of the same group structure defined. For example if 
a user wants to test a theory with two $SU(2)$ groups this can be done. Each 
gauge group has an extra flag to indicate the `line' that it is part of. If 
someone wants to use two gauges with the same group structure, they just have 
to define a new line for the new group. When indices are being contracted, 
these lines are compared and only when they are identical will a contraction 
occur. \EFF{} currently has the $SU(3)$, $SU(2)$ and $U(1)$ groups implemented,
but it is easy to expand and add new groups. An example of this is given in the
appendix.

Once the groups have been defined, the gauge fields can be added. These are
the fields which mediate the interactions. It is possible to create gauge 
bosons and their superpartners, gauge fermions. When the fields are created
the relevant terms can be added to the Lagrangian. This is given in section
\ref{sec:susy} but is repeated here for convenience
\begin{equation}
{\cal L}_{\rm gauge} = -\frac{1}{4}F^{a \mu \nu}F^a_{\mu \nu} - i 
\lambda^{\dagger a}\bar \sigma^\mu D_\mu \lambda^a,
\label{eqn:shortGauge}
\end{equation}
where $\lambda^a$ is the gauge fermion and $F^a_{\mu \nu}$ is the standard
tensor formed by a gauge boson. In $N=1$ SUSY there are no other terms which
depend only on the gauge mediators. It is possible to extend \EFF{} to provide 
for $N \neq 1$ SUSY. If this were done then a new set of terms will also need 
to be added to the Lagrangian.

After the gauge groups and the gauge fields are defined the matter fields
can be given. These are fields that can have charges under more than one 
group. These fields also define the other gauge invariant terms. For a given
fermion, $\psi$, and superpartner scalar, $\phi$, we have the terms
\begin{equation}
{\cal L}_{\rm matter} = \psi^\dagger \bar \sigma^\mu D_\mu \psi + 
\left(D^\mu \phi\right)^\dagger D_\mu \phi - \sqrt{2} g_i \left[ \left(
\phi^* T^a \phi \right) \lambda^a + {\rm c.c.} \right] + \frac{1}{2} g_i^2
\left( \phi^* T^a \phi \right)^2,
\label{eqn:shortMatter}
\end{equation}
where $\lambda$ is again the gauge mediating fermion, $D_\mu$ is the covariant
derivative for the groups the fields are charged under, and $g_i$ is the 
coupling of the $i$-th gauge group. This expression is summed over all the 
gauge groups of the matter fields. It is also at this point that the VEVs of 
the scalar fields are defined. A set of parameters can be defined and 
combinations of these parameters can be given as the VEVs. For example, in MSSM
one often works with the VEVs in terms of $\upsilon$ and $\tan \beta$, not the 
actual VEVs, $\upsilon_1$ and $\upsilon_2$. This can also be done in \EFF{} so
the set of parameters that one works with is consistent with the set of 
parameters given in the literature. An example of this is given in the 
appendix. Unfortunately, some of the routines in \EFF{} require the actual
VEVs to be entered. These include the RGEs and the minimization of the 
potential. If neither of these functions is needed, however, then the set of
parameters used in a paper can also be used in \EFF{}.

Once the fields of the model are given, the superpotential can be defined. 
In \EFF{} this is not given as a function of superfields, as the definition
of the superpotential is, but the user must specify this in terms of the 
relevant fields (scalars, fermions, etc.). This is a feature of the code that
could be improved to provide the appropriate superfield formalism. Once the
superpotential is defined then the following terms are added to the Lagrangian
\begin{equation}
{\cal L}_{\rm W} = \left( W^{ij} \psi_i \psi_j + {\rm c.c.} \right) -W^*_i W^i,
\end{equation}
where we saw before that
\begin{equation}
W^{ij} = \frac{\partial^2 W}{\partial \phi_i \partial \phi_j},~~~
W^i = \frac{\partial W}{\partial \phi_i}.
\end{equation}

After the definition of the superpotential the last remaining ingredient of a
model is the symmetry breaking terms. These are usually the soft breaking terms
of a SUSY model but \EFF{} can be used to work with non-SUSY models as well.
In this type of model one would define any other gauge invariant terms
that are desired for the model but have not been added up to this point (e.g. 
the Yukawa terms).

\subsection{Mass Matrices}
Once the VEVs have been defined the code can generate the tree-level mass 
matrices. \EFF{} is able to do this and produce the matrices in analytic form. 
It is important to note that in \EFF{} the real and imaginary parts of the 
fields are treated independently. The separation of these parts has been 
implemented to provide a mechanism for studying CP violating models. This 
separation causes several complications when comparing to the literature 
though. 

Often in the literature the states are given as the mass 
eigenstates. These are a combination of real and imaginary parts of the 
the interaction states. It is these real and imaginary combinations that must 
be compared to the results of \EFF{} for a true comparison. Often this is a 
tedious task. Instead it is usually more convincing, and easier, to check the
numerical results of \EFF{} against the literature.

Another complication that arises when implementing formulae in \EFF{} is 
that Weyl spinors are used, not Dirac ones which are often used in literature.
This affects the degrees of freedom of the fields in \EFF{}. Usually one
considers a fermionic field to have a spin factor of $-2$ from the spin 
coefficient $(2s+1)(-1)^{2s}$. Instead since \EFF{} uses Weyl spinors, this 
has a factor of $\frac{1}{2}$ attached to it, giving $-1$. The difference
between Dirac spinors and Weyl spinors must also be taken into account when 
implementing the one-loop corrections given in section \ref{sec:masscorr}. 

As mentioned earlier, \EFF{} uses the Landau gauge ($\xi = 0$) to avoid the
need for ghost contributions. The one-loop corrections to the mass matrices
have been implemented in \EFF{} in this gauge. These corrections are given
in numerical form only. There are two ways in which the corrections can be
applied. The first is that the tree-level matrices are diagonalized and the
mixing matrices are set. The corrections are then applied directly to the
mass eigenstates. This is only an approximation to the one-loop corrections 
and provides only the tree-level mixing matrices. The other way in which 
the one-loop corrections can be applied is to add the one-loop corrections of 
the interaction eigenstates to the tree-level mass matrix. This new matrix is
then diagonalized and this process defines the one-loop mixing matrices and
the one-loop masses. \EFF{} produces the tree-level matrices and the 
corresponding diagonal form in order to perform all its other calculations. 
The full one-loop correction requires performing an additional diagonalization 
for each set of parameters one would want to compute the masses at. Since the
diagonalization of the matrices is one of the most CPU intensive calculations
performed by \EFF{}, doing the full one-loop corrections is very time 
consuming.

\subsection{Effective Potential}
At this point the model has been completely defined and the tree-level masses
have been computed. It is now possible to find the effective potential and 
minimize it over a set of parameters. The analytic tree-level effective 
potential is generated by \EFF{} and can be returned. \EFF{} also provides a 
function \texttt{ex extremizePotential(vector<numeric*>)} that goes through the
parameters given in the input vector and minimizes the potential over these
parameters using the Powell~\cite{NumericalRecipes} minimization routine. The
resulting minimum potential is returned and the values of the input parameters
are set to the values that give a minimum. For example the tree-level effective
potential of the electroweak model is given in figure~\ref{fig:higgs_pot}. 
Passing the VEV parameter $\upsilon$ into this routine returns 
$-1.69 \times 10^8$ GeV$^4$ for the potential and sets the value of 
$\upsilon$ to 246.0 GeV. 

The one-loop potential can also be minimized in \EFF{}. This uses the last
calculation of the mass matrix (whether it is tree-level or one-loop) to
compute the one-loop correction given in (\ref{eqn:eff1loop}). This can then be
minimized in the same manner as the tree-level potential. 

\begin{figure}
\centering
\epsfig{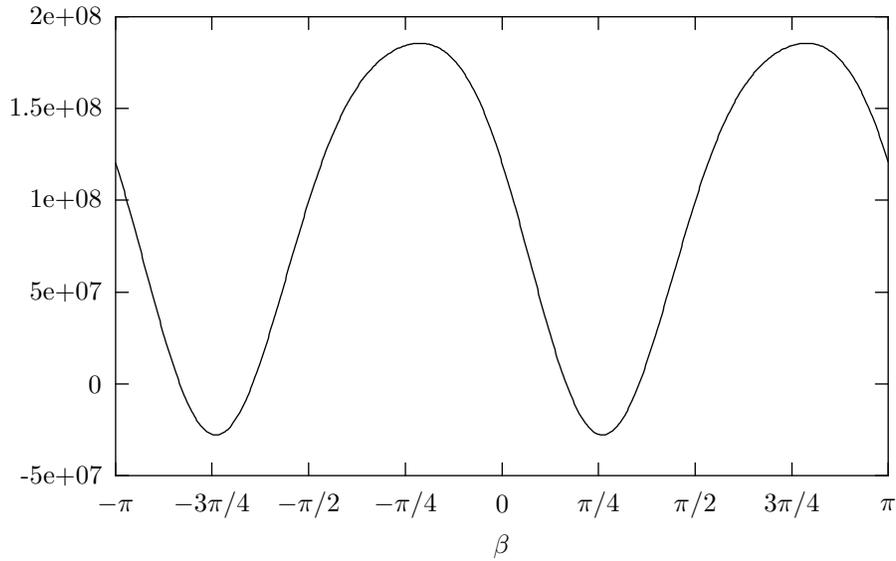}
\caption{The effective potential of the MSSM for the parameter set 
	$\{\mu=30.0~\GeV, m_{H_1}^2 = 100.0~\GeV^2, m_{H_2}^2 = -609.0~\GeV^2,
	b = 7000.0~\GeV^2 \}$ plotted in terms of $\beta$.
	This potential can be minimized in \EFF{} over the parameters,
	$\upsilon_1$ and $\upsilon_2$.\label{fig:MSSMpot}}
\end{figure}

Figure~\ref{fig:MSSMpot} shows a plot of the MSSM potential from \EFF{} with 
the given parameter set $\{\mu=30.0~\GeV, m_{H_1}^2 = 100.0~\GeV^2, 
m_{H_2}^2 = -609.0~\GeV^2, b = 7000.0~\GeV^2\}$ and the two gauge group 
couplings are set from the 
SM. In this plot we set $\upsilon_1^2 + \upsilon_2^2 = \upsilon^2 = 
(246.0~{\rm GeV})^2$ and $\tan \beta = \frac{\upsilon_2}{\upsilon_1}$. The 
tree-level effective potential of this model is derived in \EFF{} and is 
given by
\begin{equation}
V_{\rm eff} = \frac{1}{32} 
\left(g'^2 + g_W^2\right) \left(\upsilon_1^4 + \upsilon_2^4\right)
-\frac{\upsilon_1 \upsilon_2 b}{2}+\frac{\upsilon_1^2 m^2_{H^1}}{2}  
+\frac{\upsilon_2^2 m^2_{H^2}}{2} +\frac{\upsilon_1^2 \mu^2}{2} 
+\frac{\upsilon_2^2 \mu^2}{2}.
\end{equation}
From this equation we can see with the right set of parameters, the $b$ term
will push the potential below zero and require a particular combination of the
VEVs to minimize.

\subsection{Renormalization Group Equations}
The last feature of a model that is generated automatically by \EFF{} is the
RGEs. Each parameter of the model can have its default value set at its own
scale. The RGEs then evolve these parameters so that the values are all used
consistently at any desired scale.

As was shown earlier the $\beta$ functions for each parameter are generated 
using the one-loop effective potential and the two-, three- and four-point 
Green's functions and the anomalous dimension of the external fields. 

The solution to the boundary value problem can be given as follows. Start with
a scale $x=M_x$. The boundary conditions at this scale can be given by
\begin{equation}
f(p(x)) = 0.
\label{eq:p(x)}
\end{equation}
where $p$ is the set of parameters which are defined at scale $x$. 
The boundary condition at a new scale, $z$, is
\begin{equation}
g(q(z)) = 0,
\label{eq:g(z)}
\end{equation}
and again, $q$ is the set of parameters defined at scale $z$.
The full set of parameters of the model are given by $\left\{ p(t), q(t) 
\right\}$ where $p(t)$ is constrained at $t=x$ and $q(t)$ is constrained at
$t=z$. The parameters at the two scales are related by two functions, $R^p$
and $R^q$ by
\begin{eqnarray}
q(z) = R^q(q(x),p(x)) \nonumber \\
p(z) = R^p(q(x),p(x)),
\label{eq:p(z)q(z)}
\end{eqnarray}
where these functions are determined by the RGEs. Using these equations we can 
numerically solve for the different parameters.

There are three numerical methods that have been considered in 
\EFF{}~\cite{James}. The
first is known as the ``shooting'' method~\cite{NumericalRecipes}. Using this 
method one computes $g(q(z))$ numerically by use of the function $R^q$. This 
amounts to numerically finding the solution to
\begin{equation}
g(R^q(q(x),p(x))) = 0,
\end{equation}
where $p(x)$ is the boundary values of the parameters at scale $x$. This just 
requires solving this equation over the set of parameters $q(x)$.

The ``shooting'' method cannot be used in SUSY, however, as SUSY requires the
input to be given at three scales: $M_Z, M_S$ and $M_x$. Another reason that
this method cannot be used for SUSY theories is that the boundary 
conditions of a SUSY theory cannot be conveniently expressed as is done in 
(\ref{eq:g(z)}).

An alternative method~\cite{NumericalRecipes} is the ``drift'' method. This 
method is a modified version of the ``shooting'' method. Here a guess 
$q_x=q(x)$ is taken and $p(x)$ is determined from (\ref{eq:p(x)}). $q(z)$ and 
$p(z)$ are then given 
from (\ref{eq:p(z)q(z)}). The boundary conditions at scale $z$ are imposed to 
give $q(z)$. These are then run back up to the scale $x$ by the inverse of 
(\ref{eq:p(z)q(z)}). This gives a new set of values, $q'(x)$ and $p'(x)$. The
conditions at scale $x$ are then imposed again. This process defines a 
recurrence relation $q'(x) = Q(q_x)$ of which the desired values of $p$ and $q$
will be a fixed point. The problem with this method is that the physical
values might not be the only fixed point and there is no guarantee that this
fixed point is stable.

A third method, which has been implemented in \EFF{}, handles the case where 
the fixed point is not stable. In this case we can instead numerically solve 
the equation
\begin{equation}
Q(q_x) - q_x = 0,
\end{equation}
using the Newton-Raphson~\cite{NumericalRecipes} method. This method can be
rather slow but if one makes reasonable guesses for some of the less
important parameters the parameter space can be reduced and convergence 
is quicker. 

\begin{figure}
\centering
\epsfig{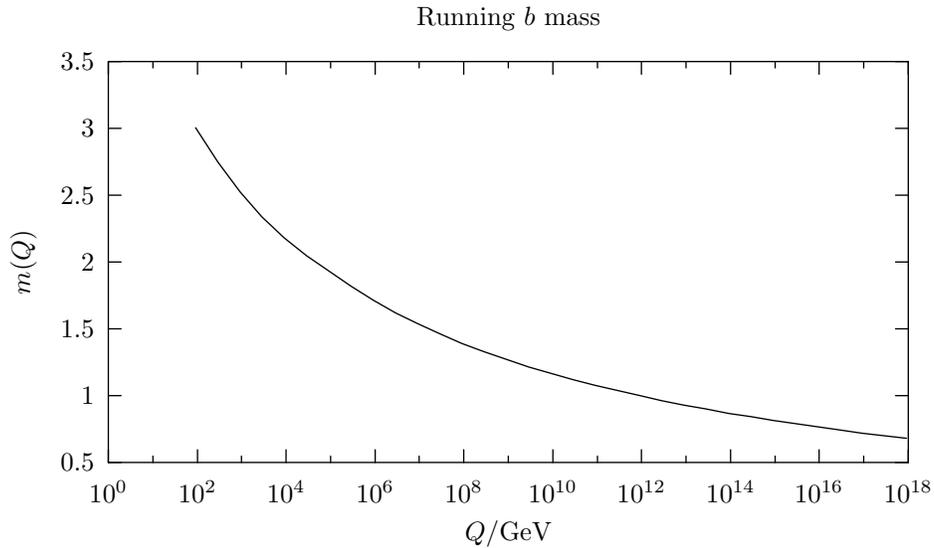}
  \caption{The running of the $b$ mass. This is given by the $\beta$ function
	of the Yukawa coupling.
    \label{fig:brunning}}
\end{figure}

We have already seen the running of the gauge group couplings in 
fig.~\ref{fig:SMrunning} produced by \EFF{}. These were generated by setting
the fields and groups of the model. \EFF{} then computes the contribution
to the $\beta$ function from each field of the model. The coupling is
evolved according to the $\beta$ function generated.

Figure \ref{fig:brunning} shows the evolution of the $b$ mass given by the
$\beta$ function for the Yukawa coupling in the SM. This has been evolved 
using $\beta$ functions produced by \EFF{} and the
Runge-Kutta~\cite{NumericalRecipes} method of differential equation evolution. 
The initial condition for this figure is $m_b(m_Z) = 3.0$ GeV. Again, the
$\beta$ function is generated by \EFF{}. Each of the anomalous dimensions
of the Yukawa coupling are added together. This is the sum from 
(\ref{eqn:fullbeta}). The $B$ factors have been calculated from the 1PI 
contributions of the vertices given in fig.~\ref{fig:tafpc}. 

\subsection{Future Extensions of \EFF{}}
Before we discuss the extensions of \EFF{} we first summarize what it can 
already do and how a user could use \EFF{} without any extensions.

\EFF{} is able to generate the full Lagrangian of a SUSY model based on the
groups, fields, superpotential and SUSY breaking terms. This Lagrangian is
used to generate the effective potential, mass matrices and RGEs of the 
model. At this stage a user can input their parameters over a range of scales
and evolve them to the same scale. The mass spectrum at tree-level or one-loop
can be evaluated given the parameter set. The user could then automate the
study of the mass spectrum over a range of parameters. 

The code of \EFF{} allows for several extensions. The most useful one in
producing physical results is allowing additional diagrams to be coded. The
library provides a class \texttt{Diagram} which provides several useful
functions. This class allows a user to pass in the external fields of a diagram
and a set for each of the internal propagators. The class provides functions
which find the couplings of a set of fields given as either interaction states,
mass states or combinations of the two. This is done by computing the coupling
of mass states based on their mixing matrices. The couplings can then be
used in the computation of a diagram by inheriting the \texttt{Diagram} class 
and implementing the virtual function \texttt{ex function()} function. The
diagram is then evaluated by iterating over all the internal propagator fields
and summing the result of each diagram by a call to the \texttt{ex evaluate()}
function. Implementing ones own diagrams would allow \EFF{} to be used to
produce results of physical calculations (like cross-sections) over a range of
parameters.

The implementation of a users own diagrams isn't the only possible extension.
The classes \texttt{GaugeField} and \texttt{MatterField} implement the set of
expressions given in (\ref{eqn:shortGauge}, \ref{eqn:shortMatter}). If someone
wanted to consider a model which needed other terms (such as $N \ne 1$ SUSY)
they would need to inherit these classes and override the function 
\texttt{ex interaction()}. They could then add new terms that depend on the
fields. Unfortunately, this isn't all that would be needed for $N \ne 1$ SUSY.
There would be a need to inherit the \texttt{Model} class and change several
functions. But this is still a feasible task.

One may also consider extra dimensional models. These too could be implemented
in \EFF{}. One would have to implement a mechanism for producing a four 
dimensional effective Lagrangian from the input (groups, fields, 
superpotential, plus any extra interactions) but this too would be a feasible
task. It may be that an implementation such as this could only be done for
a specific type of compactification scheme. But would still be useful for
studying both SUSY and non-SUSY models derived from an effective four 
dimensional Lagrangian.

Another extremely useful extension that would be desirable is implementing
a method of deriving the Feynman rules from the Lagrangian. All of the elements
are in place to do such a task but would require quite a lot of work. These
rules could be output in a format suitable for reading into an matrix element
generator, like FormCalc, or one could even implement a matrix element 
generator as another extension of \EFF{}. This would be a very complicated
project, however.

\EFF{} in its current form can mainly be used to study the mass spectrum of a
model once the VEVs that minimize the effective potential are determined. This
mass spectrum depends on the input parameter set which can be given at several
scales and the RGEs of the model can be used to match them at the scale the
mass spectrum is desired at. Future extensions of the software would allow this
mass spectrum to be used in calculations of cross-sections and other
properties. More substantial extensions could allow the software to be used on
$N \ne 1$ SUSY and extra dimensional models. Finally the development of
Feynman rules for a model could be implemented with an eye towards use in 
matrix element generators. 

\chapter*{Conclusion}

This thesis has presented two new software packages, \HWP{} and \EFF{}. 
\HWP{} is a Monte Carlo event generator. In its current state, this generator
is able to generate $\ee$-annihilation events. Chapter~\ref{chap:NewVars} has 
developed a new set of variables for the parton shower and 
chapter~\ref{chap:Hadronization} has presented an improved hadronization 
model. A full discussion of the implementation specific features of \HWP{} have
been given in chapter~\ref{chap:Herwig}. Results of the new models and the 
package as a whole have been shown in chapter~\ref{chap:Results}. These results
give us great confidence that the new shower and hadronization are a better
description of the physics.

Currently, we are working on \HWP{} to generate initial-state showers. This
will enable us to describe hadron-hadron and lepton-hadron events. The aim is
to have this fully functional for use with Tevatron and LHC studies. Once the
initial-state shower is fully functional a new underlying event model can be
implemented as well as improving the hadronic decays. Work is already underway
to include spin correlations into the shower and decays. New matrix elements
for various processes can be implemented and SUSY particles and processes can
be included. As the history of \HW{} has shown, there will continue to be a lot
of potential for future research in \HWP{}.

\EFF{} is a model building program. This program has been designed to work with
$N = 1$ SUSY models but future extensions could move it beyond just these
models. This program automates the process of determining the mass spectrum
of a model for a given parameter set. The mass spectrum generated with 
\EFF{} can be read into \HWP{} to define the masses of the supersymmetric
particles in \HWP{}. This would provide an automated way of studying
SUSY at colliders over a range of parameters. Currently, only special purpose 
programs,
like SOFTSUSY~\cite{Allanach:2001kg}, are able to do something similar. 
Future extensions of \EFF{} may provide more functionality and make it an
extremely powerful and useful tool for studying models.

\appendix
\chapter{\HWP{}}

This appendix contains a few relevant issues pertaining to the use and future
development of \HWP{}. I first give an example analysis program to count the
number of $\pi^0$ in an event. Following that is a brief discussion of how to 
change the parameters in \HWP{}. Finally, I give the skeleton structure of
how one would implement one's own matrix element.

\section{Counting Pions}
This section provides a description of the functions needed to run a 
simple \HWP{} program. The program given here will simply count the number of
neutral pions in the final state.

\HWP{} can be run inside an exception handling clause. In C++ this is the
statement \\
\indent \indent \texttt{try \{ ...your code here ...\}} \\
\indent \indent \texttt{catch(...) \{ 
\emph{// do something with the exception} \}}\\
\HWP{} and \ThePEG{} will throw exceptions to explain why the program has
terminated. If these aren't caught the message associated with the end of
the program won't be known and the reason for the termination will remain
a mystery. Therefore, the main program to run \HWP{} will be enclosed in 
such a statement and the message from the exception will be printed. The
exact form we use is\\
\indent \indent \texttt{try \{ ...generate events/analyze ...\}} \\
\indent \indent \texttt{catch(std::exception \& e) \{}\\ 
\indent \indent \indent \texttt{cerr << e.what() << endl;}\\ 
\indent \indent \indent \texttt{return 1;}\\
\indent \indent \texttt{\} catch(...) \{}\\
\indent \indent \indent \texttt{cerr << "Unknown exception$\backslash$n";}\\
\indent \indent \indent \texttt{return 2;}\\
\indent \indent \texttt{\}}

We now move on to discussing what is put in the ``generate events/analyze''
section. A helper class \texttt{HerwigRun} has been developed for \HWP{}. This
class takes the command line arguments (\texttt{int argc, char **argv}) and
sets up \HWP{} to either initialize, read or run. These different stages
are discussed in the next section. The command to create a \texttt{HerwigRun}
class is\\
\indent \indent \texttt{Herwig::HerwigRun hw(argc,argv);}\\
Once this class is created generating an event is straightforward. We
first must check the status of the program, given by the functions
\texttt{isRunMode()} and \texttt{preparedToRun()}. We can then iterate over
the number of events to generate given by the function \texttt{getN()}. To
generate an event we simply call \texttt{generateEvent()}. The code to do all
of this is \\
\indent \indent \texttt{if (hw.isRunMode() \&\& hw.preparedToRun()) \{}\\
\indent \indent \indent \texttt{for(int i = 0; i<hw.getN(); i++) \{}\\
\indent \indent \indent \indent \texttt{hw.generateEvent();}\\

At this point we have now generated an event. This is where the analysis code
must be implemented. The \texttt{HerwigRun} class also offers a function to
retrieve the particles from the last event generated. 
\texttt{ThePEG::tPVector getFinalState(int i)} is this function. 
If \texttt{i} is not given (e.g. \texttt{getFinalState()}) 
then the final state particles of the event are returned. If the \texttt{i}
argument is given this function returns the particles in the final state given
by the step \texttt{i}. 

Returning to our example of counting the number of $\pi^0$ particles in the
final state we pass the vector of final state particles to a counting function
which we define, \texttt{int countPions(ThePEG::tPVector particles)}. This
function is fairly straightforward and is given by\\
\indent \indent \texttt{int countPions(ThePEG::tPVector particles) \{}\\
\indent \indent \indent \texttt{ThePEG::tPVector::iterator it;}\\
\indent \indent \indent \texttt{int count = 0;}\\
\indent \indent \indent \texttt{for(it=particles.begin(); it!=particles.end();
it++) \{}\\
\indent \indent \indent \indent \texttt{if((*it)->id() == ThePEG::ParticleID::pi0)
count++;}\\
\indent \indent \indent \texttt{\}}\\
\indent \indent \indent \texttt{return count;}\\
\indent \indent \texttt{\}}\\

This function has relied on the class \texttt{ThePEG::Particle} which defines
several functions of the particle. Here we have used the \texttt{long id()} 
function to determine the PDG code of the particle. There are many more 
functions that would be useful for analysis. The user is encouraged to read
the documentation of the \texttt{Particle} class for more details.

The \texttt{HerwigRun} class also defines functions which return some of
the elements from \texttt{ThePEG}. These are things like the \texttt{Step}
or the \texttt{CollisionHandler} from which the user is able to obtain all
the information of the event. 

\section{Repository}
The repository is a feature of \ThePEG{} that allows the parameters of the
program to be changed without having to recompile the code. These parameter
files can then be exchanged within collaborations to ensure that researchers
are working with the same parameters.

The repository can be thought of as the input files to \HWP{}. 
These are written
in a human readable form, though the repository has a simple syntax of its
own. The input files can then be saved in machine readable form for faster 
access on subsequent runs. In \HWP{} there is a default input file,
\texttt{HerwigDefaults.in}. This sets up all the relevant objects and instructs
the program to allocate memory for the various parts of the event generator.
This initial setup can be saved in a machine readable file, 
\texttt{HerwigDefaults.rpo}. The \texttt{HerwigRun} class reads in the command
line argument \texttt{init}. This instructs the class to create the 
\texttt{.rpo} from the \texttt{.in} file. \texttt{HerwigDefaults.in} also 
reads \texttt{Shower.in}, \texttt{Hadronization.in} and \texttt{Decays.in}
as well as the particles in \texttt{ThePEGParticles.in} and 
\texttt{HerwigParticles.in} and the decay modes from \texttt{HwDecays.in}.

The \texttt{HerwigDefaults.in} currently defines two types of generators. One
for LHC-like events and another for LEP-like events. These two generators can
then have their parameters changed (such as c.m.~energy or various cuts)
by editing the files \texttt{LHC.in} or \texttt{LEP.in}, respectively. These
modifications are read in by the \texttt{read} command line argument. This
reads in the appropriate \texttt{.in} file and produces a \texttt{.run} file.
The \texttt{.run} is checked to ensure that all the relevant objects and 
links have been set so that the event generator can run.

The last stage is to actually run some events. Using the command line argument
\texttt{run} reads in the appropriate \texttt{.run} file and generates events.
There is a parameter \texttt{NumberOfEvents} in each generator object that
sets the maximum number of events to run. The \texttt{HerwigRun} class provides
for a new number of events to be given at runtime. This number must be less 
than or equal to the parameter given to the \texttt{NumberOfEvents} field.
To instruct the program to use a new number simply use the command line
argument \texttt{-N \#} when starting the program.

To summarize there are three ways in which a program that uses a
\texttt{HerwigRun} class can by used. These are given as\\
\indent \indent \texttt{Program\_name init}\\
\indent \indent \texttt{Program\_name read Generator\_file.in}\\
\indent \indent \texttt{Program\_name run Generator\_file.run -N \#}\\
There are more commands that can be given in the command line but these
mostly change some of the lower level instructions. The full set of commands 
is\\
\indent \indent \texttt{Program\_name init|read|run  [-N num-events] [-seed random-generator-seed] [-d debug-level] [-dHw herwig-debug-level] [-l load-path] [-L first-load-path] [-r repo-file] [-i initialization file] [run-file]}

To change the value of a parameter you find the line in the input file that
this is governed by. For example if we wanted to turn the initial-state 
radiation off we would look in the \texttt{Shower.in} file. This file has the
line\\
\indent \indent \texttt{set theSplittingGenerator:OnOffISRMode 1}\\
If we simply edit this file and change the 1 into a 0 this will turn the
initial-state radiation off. Most of the commands are straightforward like
this. There are other commands in the repository than \texttt{set}. A relevant
subset is \texttt{create,set,mkdir,cd,library}. The \texttt{create} command
is used to define a new object given by a class. The \texttt{set} command is 
used to change one of the parameters of that object. The repository is 
structured like a file system. You can create the objects you want in 
directories in order to keep things ordered. The \texttt{mkdir} command creates
a new directory and the \texttt{cd} command changes to the given directory.
The last command is the \texttt{library} command. This is used to dynamically
load a library. This must be done to define objects that are in a library that
hasn't been loaded yet. For example, if we want to define an 
\texttt{AmegicInterface} object we would have to load the 
\texttt{libHwAmegic.so} library first.

\section{Matrix Element Development}
In this last section I show how a user would implement their own matrix element.
The class \texttt{MEBase} defines the lowest level of matrix element 
abstraction. To implement a matrix element one must at least inherit this
class and define the functions\\
\indent \indent \texttt{unsigned int orderInAlphaS() const}\\
\indent \indent \texttt{unsigned int orderInAlphaEW() const}\\
\indent \indent \texttt{double me2() const}\\
\indent \indent \texttt{Energy2 scale() const}\\
\indent \indent \texttt{generateKinematics(const double r)}\\
\indent \indent \texttt{CrossSection dSigHatDR() const}\\
\indent \indent \texttt{void getDiagrams() const}\\
\indent \indent \texttt{Selector<const ColourLines*> colourGeometries(tcDiagPtr diag) const}\\
This list looks quite daunting. \ThePEG{} has already generated these functions 
for special types of matrix elements. For example the class \texttt{ME2to2Base}
as able to define the \texttt{generateKinematics()}, 
\texttt{dSigHatDR()} as well
as the \texttt{scale()} for all $2 \to 2$ processes. A further subclass
\texttt{ME2to2QCD} defines \texttt{orderInAlphaS()} and 
\texttt{orderInAlphaEW()} functions as well as providing routines for general
functions like determining the number of active flavours at a given scale.

If we take an example to inherit the \texttt{ME2to2QCD} class this means we
only need to define the functions\\
\indent \indent \texttt{double me2() const}\\
\indent \indent \texttt{void getDiagrams() const}\\
\indent \indent \texttt{Selector<const ColourLines*> colourGeometries(tcDiagPtr diag) const}

The function \texttt{double me2() const} is self explanatory. This is just the
value of the matrix element for the set of currently generated points. To
access the set of points there is a function \texttt{meMomenta()} which
returns a vector of \texttt{Lorentz5Momentum} objects. This vector contains
the momentum generated for all the incoming and outgoing particles.

The \texttt{getDiagrams() const} class is used to indicate to the matrix
element what are the diagrams used to produce the matrix element (squared).
There is a function \texttt{add(DiagPtr)} which adds a diagram to the list.
There is a helper method defined to define a diagram. For example\\
\indent \indent \texttt{new\_ptr((Tree2toNDiagram(2), q, qb, 1, gamma, 3, l, 
3, lb, -1))}\\
define a $2 \to 2$ process where the \texttt{q} and the \texttt{qb} are the
incoming particles. The 1 indicates that the \texttt{gamma} particle is
the child of the first incoming particle. The \texttt{3,l,3,lb} series
indicates that both the \texttt{l} and \texttt{lb} particles are the 
children of the \texttt{gamma} particle. The last argument of -1 indicates
that this is the end of the diagram definition. This is a flag that must be
negative but can have different values. The different values are used to 
determine the different types of diagrams. If one wanted to weight the 
different diagrams they would need to overload the \texttt{Selector<
DiagramIndex> diagrams(const DiagramVector \&) const} function. In this
function one can iterate over the diagrams and add them into the 
\texttt{Selector} object with a weight. When a diagram is selected the weights
are taken into account.

The last function that must be implemented is the 
\texttt{Selector<const ColourLines *> colourGeometries(tcDiagPtr diag) const}
function. This function allows you to set the different colour connections. 
These can also be given a weight which the particular colour connection 
implemented is chosen from.

Having implemented all of these functions the matrix element is defined and 
can be used in \HWP{}. The typical place to use the matrix element is in the
hard subprocess. It is possible to use a matrix element in other processes, 
such as in the decays, but that requires more development and working with the
code on a lower level.

\chapter{\EFF{}}

This appendix is devoted to coding issues of \EFF{}. I give here an example
of coding the MSSM model in \EFF{} and how one can use the model to 
get the mass of the neutralinos. I also show how to evolve the RGEs to provide
the scale dependence of the parameters. At the end I give a skeleton outline
of how one would implement one's own groups and diagrams.

\section{MSSM}
The main driver of \EFF{} is the class \texttt{Model}. This class is what
directs all the calculations and provides all the components. This 
class that must be inherited to implement ones own model. The methods
that need to be defined are\\
\indent \indent \texttt{void createGaugeGroups()}\\
\indent \indent \texttt{void createGaugeFields()}\\
\indent \indent \texttt{void createMatterFields()}\\
\indent \indent \texttt{void addOtherTerms()}\\
\indent \indent \texttt{ex superPotential()}

The \texttt{createGaugeGroups()} function specifies what groups are used
in the model. For example in MSSM and the SM we have only the $SU(3)_c$,
$SU(2)_L$ and $U(1)_Y$ groups. These can be implemented via the function
\begin{verbatim}
void MSSM::createGaugeGroups() {
  addGaugeGroup(new U1Group("U1", "{g'}", this, U1Y));
  addGaugeGroup(new SU2Group("SU2", "{g_W}", this, SU2w));
  addGaugeGroup(new SU3Group("SU3", "{g_3}", this, SU3c));
}
\end{verbatim}
The objects \texttt{U1Y}, \texttt{SU2w} and \texttt{SU3c} are all integers that
define the `line' of the group. The two strings for each group define the name
of the group coupling in the standard text output and the \LaTeX~output, 
respectively.
The groups can then be accessed at a later point using the standard text 
string (e.g. \texttt{getGroup("U1")}).

The next function that needs to be implemented is \texttt{createGaugeFields()}.
This function defines the fields that mediate the interactions. In the MSSM
these are the superpartners $(B^\mu,\tilde B)$, $(W^\mu_i, \tilde W_i)$ and
$(A^\mu_a, \tilde A_a)$. These are given by
\begin{verbatim}
void MSSM::createGaugeFields() {
  addField("B",new GaugeField("B","B",Utils::LorentzVector,
                              getGaugeGroup("U1")));
  addField("Bino",new GaugeField("Bino","\\tilde{B}",Utils::WeylSpinor,
                              getGaugeGroup("U1"),getGaugeField("B")));
  addField("W",new GaugeField("W","W",Utils::LorentzVector,
                              getGaugeGroup("SU2")));
  addField("Wino",new GaugeField("Wino","\\tilde{W}",Utils::WeylSpinor,
                              getGaugeGroup("SU2"),getGaugeField("W")));
  addField("gluon",new GaugeField("A","A",Utils::LorentzVector,
                              getGaugeGroup("SU3")));
  addField("gluino",new GaugeField("Aino","\\tilde{A}",Utils::WeylSpinor,
                              getGaugeGroup("SU3"),getGaugeField("gluon")));
}
\end{verbatim}

The \texttt{addField()} function takes a string as an index for the field and
a pointer to a field. This same function is used to add the 
\texttt{MatterField}s. Creating a \texttt{GaugeField} requires two strings.
Again the first is the text output string and the second is the \LaTeX~output.
These strings are followed by the spin of the particle. This is actually the
$(2s+1)(-1)^{2s}$ factor. This has been conveniently defined in the class
\texttt{Utils} for the scalar, fermion and vector spin types. The spin
is then followed by a pointer to the gauge group that the field is the mediator
of. The last argument, which is optional, is the SUSY partner of the field. We
can see above that the SUSY partner is set for the fermion but not the 
vector. This is because one field must be created first. Once created the 
partner can be set. Setting it in the creation of the fermion also sets it
in the vector, so this syntax sets both partners to each other with only one
reference to a partner.

After the \texttt{addGaugeFields} function is defined we implement the
\texttt{addMatterFields} function. This function sets all of the remaining
fields in the model. Again looking at our example from MSSM we can write the
function as 
\begin{verbatim}
void MSSM::createMatterFields() {
  numeric half(1,2);
  numeric sixth(1,6);
  numeric third(1,3);
  numeric twothirds(2,3);
  // Adding leptons/sleptons
  addField("leptonL",new MatterField("L","\\ell",Utils::WeylSpinor,famsize,
                                     getGaugeGroup("U1"),-half,
                                     getGaugeGroup("SU2"),1));
  addField("sleptonL",new MatterField("sL","\\tilde{\\ell}",
                                      Utils::Scalar,famsize,
                                      getGaugeGroup("U1"),-half,
                                      getGaugeGroup("SU2"),1,
                                      getMatterField("leptonL")));

  addField("leptonR",new MatterField("eR","e_R",Utils::WeylSpinor,famsize,
                                     getGaugeGroup("U1"),-1));
  addField("sleptonR",new MatterField("seR","\\tilde{e_R}",
                                      Utils::Scalar,famsize,
                                      getGaugeGroup("U1"),-1,
                                      getMatterField("leptonR")));

  // Adding quarks and squarks
  addField("quarkL",new MatterField("Q","Q",Utils::WeylSpinor,famsize,
                                    getGaugeGroup("U1"),sixth,
                                    getGaugeGroup("SU2"),1,
                                    getGaugeGroup("SU3"),1));
  addField("squarkL",new MatterField("sQ","\\tilde{Q}",
                                     Utils::Scalar,famsize,
                                     getGaugeGroup("U1"),sixth,
                                     getGaugeGroup("SU2"),1,
                                     getGaugeGroup("SU3"),1,
                                     getMatterField("quarkL")));

  addField("uR",new MatterField("uR","u_R",Utils::WeylSpinor,famsize,
                                getGaugeGroup("U1"),twothirds,
                                getGaugeGroup("SU3"),1));
  addField("suR",new MatterField("suR","\\tilde{u_R}",Utils::Scalar,famsize,
                                 getGaugeGroup("U1"),twothirds,
                                 getGaugeGroup("SU3"),1,
                                 getMatterField("uR")));

  addField("dR",new MatterField("dR","d_R",Utils::WeylSpinor,famsize,
                                getGaugeGroup("U1"),-third,
                                getGaugeGroup("SU3"),1));
  addField("sdR",new MatterField("sdR","\\tilde{d_R}",Utils::Scalar,famsize,
                                getGaugeGroup("U1"),-third,
                                getGaugeGroup("SU3"),1,
                                getMatterField("dR")));

  // Higgs fields 1 is upper, 2 is lower
  addField("H1", new MatterField("H1","{H^1}",Utils::Scalar,1,
                                 getGaugeGroup("U1"),-half,
                                 getGaugeGroup("SU2"),1));
  addField("sH1", new MatterField("sH1","\\tilde{H^1}",Utils::WeylSpinor,1,
                                  getGaugeGroup("U1"),-half,
                                  getGaugeGroup("SU2"),1,
                                  getMatterField("H1")));

  addField("H2", new MatterField("H2","{H^2}",Utils::Scalar,1,
                                 getGaugeGroup("U1"),half,
                                 getGaugeGroup("SU2"),1));
  addField("sH2", new MatterField("H2","\\tilde{H^2}",Utils::WeylSpinor,1,
                                  getGaugeGroup("U1"),half,
                                  getGaugeGroup("SU2"),1,
                                  getMatterField("H2")));
  // Now add higgs Vev
  Parameter upsilon1 = addParameter("HiggsVev1","\\upsilon_1",220.0,
                                    Parameter::vev);
  Parameter upsilon2 = addParameter("HiggsVev2","\\upsilon_2",220.0,
                                    Parameter::vev);
  addVev("HiggsVev1",getField("H1"), lst(getIndex("H1","SU2")==1),
         (upsilon1+Model::star));
  addVev("HiggsVev2",getField("H2"), lst(getIndex("H2","SU2")==2),
         (upsilon2+Model::star));
  addVevParameter(upsilon1);
  addVevParameter(upsilon2);
\end{verbatim}

The meaning of this code is easy to see. Again we see the last (optional) 
argument
in creating a \texttt{MatterField} is the superpartner. We can see now that
there are some new arguments to these fields though. There is a 
\texttt{famsize}
argument. This is an integer which defines how many generations the field has.
For example in MSSM we would set \texttt{famsize} to 3. Lastly we see that
after each gauge group is a number. This is the `charge' of the group. Most
groups just have a charge of 1. The $U(1)_Y$ does not, however. This has
fractional charges. The class \texttt{numeric} is one from {\small GiNaC} which
defines rational numbers, without rounding errors. These expressions will
be printed in \LaTeX~as they are (e.g. \texttt{half} = $\frac{1}{2}$).

In the last part of the code we see that we have created the VEVs. The
class \texttt{Parameter} is used in \EFF{} to represent all of the parameters
of the model. The \texttt{Parameter} class is printed out in symbolic form
during printing but is evaluated to its value during calls to \texttt{evalf()}.
The object \texttt{Model::star} is a placeholder which is where the relevant
field is placed. If we wanted to also add an imaginary VEV would could use
a similar syntax \texttt{imagVEV+Model::star} and in this instance 
\texttt{Model::star} is the imaginary part of the field.

As discussed in the text the parameters could be added in a different way. If 
we wanted to work with $\upsilon$ and $\tan \beta$ rather than $\upsilon_1$
and $\upsilon_2$ we could have used the code
\begin{verbatim}
Parameter upsilon = addParameter("HiggsVev","$\backslash\backslash$upsilon",246.0,
                                 Parameter::vev);
Parameter beta = addParameter("beta","$\backslash\backslash$beta",1.1,
                              Parameter::vev);
addVev("HiggsVev1",getField("H1"), lst(getIndex("H1","SU2")==1),
       (upsilon*sin(beta)+Model::star));
addVev("HiggsVev2",getField("H2"), lst(getIndex("H2","SU2")==2),
       (upsilon*cos(beta)+Model::star));
addVevParameter(upsilon);
addVevParameter(beta);
\end{verbatim}
but as discussed in the text, this type of setup will not properly define the
RGEs and the minimization routine may not succeed.

In the MSSM there are many terms which must be added from both the 
superpotential and from the soft breaking terms. We won't give them all here
but instead will present an example of how to add a term. We will look at how
to add the slepton mass term. This is given by
\begin{verbatim}
ex m0LL = addFamilyMatrix("m0LL", "{m^2_L}", famsize);
vector<idx*> sumIndex;
idx i = Utils::familyIndex(0,famsize);
idx j = Utils::familyIndex(1,famsize);
MatterField *sleptonL = getMatterField("sleptonL");
ex li = sleptonL->expression();
ex lj = Utils::conjugate(li);
lj = lj.subs(sleptonL->familyIndex()==j);
sumIndex = sleptonL->getIndices();
sumIndex.push_back(&j);
ex sleptonLTerm = Utils::real(Utils::sumIndices(m0LL*li*lj,sumIndex));
add(-sleptonLTerm);
\end{verbatim}

At first this looks quite horrific. But once the different parts are explained
it is obvious what everything means. The first term is used to define the
slepton mass term. Since there are three slepton generations the mass-squared
coupling must be defined as a $3\times 3$ matrix of parameters. The 
\texttt{addFamilyMatrix()} function provides this. This function creates a 
matrix whose elements are parameters. The parameters are created with the
indices appended to the name (e.g. the $m^2_{L11}$ element is given by
the parameter \texttt{m0LL11}). When the matrix is created it is created as
an identity matrix.

The next line defines a vector in which all the indices which are summed over
are placed. This is followed by defining two family indices. By default the
\texttt{FamilyMatrix} class uses the first two indices from the 
\texttt{Utils::familyIndex()} function. The \texttt{Utils} class defines some
universal symbols for different indices. A call to the \texttt{familyIndex}
function returns one of the symbols (given by the first integer) as an index
where the dimension of the index is given by the second argument (e.g. the 
number of families). As a note, there is a integer \texttt{Utils::max\_indices}
which defines how many symbolic indices are available.

The definition of the two family indices, \texttt{i} and \texttt{j}, is 
followed by a retrieval of the \texttt{MatterField} pointer to the slepton.
If we refer back to our definition of the slepton field we can see that the 
first string is ``slepton''. This is the tag that is used to retrieve the field
at a later time. There is also a function \texttt{getGaugeField} which behaves
in a similar manner, only it returns a \texttt{GaugeField} pointer. After
the retrieval of the slepton pointer we now get the expression of the slepton.
As \EFF{} seperates the real and imaginary parts of the scalars and fermions,
the expression is $\frac{1}{\sqrt{2}}({\cal R}(f) + i {\cal I}(f))$. Also the
function expression returns the fields with each index attached. For example a
left-handed quark would have its family index as well as the $SU(2)_L$ and
$SU(3)_c$ indices added. A vector also has a \texttt{LorentzIndex} attached.
The function \texttt{Utils::conjugate} then replaces all $i$'s with $-i$. The
default family index of a field is taken from 
\texttt{Utils::familyIndex(0,famsize)}. Since we want to contract the family
indices of the mass-squared coupling we must replace this index with
\texttt{Utils::familyIndex(1,famsize)} in the conjugate expression. This is
done via the \texttt{subs()} command. This command takes two forms. The first
is to use the \texttt{==} sign. This sets the argument on the left to be 
replaced by the argument on the right. This is useful when making one 
substitution. When many substitutions are desired we can instead use the
form \texttt{subs(lst,lst)}. The first \texttt{lst} is replaced by the second. 
A \texttt{lst} is a class from {\small GiNaC} and is just a sequence of 
expressions.

After the expression for $\phi$ and $\phi^*$ have been retrieved we want to
contract all the indices and create the expression $m^2_{Lij}\phi^*_j \phi_i$.
To do this we store all of the indices given in the slepton into the 
\texttt{sumIndices} vector we created earlier. We also realize that this vector
doesn't contain the extra family index in $m^2_L$, and now in \texttt{lj}, so 
we also add the index to the back of the list. Once the \texttt{sumIndices}
vector contains all the indices which we are summing over we can call the 
routine \texttt{Utils::sumIndices(ex,vector<idx*>)}. This routine iterates the
list of indices over all permutations and evaluates the ex at each set of
index values. The sum of all these permutations and evaluations is returned.
The last command on this line is the \texttt{Utils::real}. This 
returns all the parts of the 
expression that are not multiplied by $i$. In this example we shouldn't
need to call this command as $\phi^*\phi$ should be a real quantity. It is
also worth noting that instead of coding an expression as exp + c.c. it is 
easier and faster to write
\texttt{2*Utils::real()}. 

In the end we call the \texttt{add()} function. This adds the expression to the
Lagrangian. If we were developing terms of the superpotential we would instead
sum all the contributions from all the terms and return it. We don't want to
call the \texttt{add()} routine with the superpotential.

\section{Plotting Effective Potential and Running Couplings of MSSM}
Once we have developed the class \texttt{MSSM} with all the relevant groups,
fields and terms, we can work with this program. The following code is used
to create the model and initialize it
\begin{verbatim}
MSSM mssm;
Model::readCommandLine(argc,argv,&mssm);
mssm.couplings("mssm.gar");
mssm.initialize();
\end{verbatim}
The first line creates an \texttt{MSSM} object. The second line calls a static
function from the \texttt{Model} class which reads in standard command line
arguments and sets the relevant parameters of the model. The possible arguments
are
\begin{verbatim}
-p # - Sets the evaluation of the effective potential to tree-level (0) 
       or one-loop (1)
-t # - Sets the evaluation of the tadpoles to tree-level (0) or 
       one-loop (1)
-a # - sets the evaluation of the mass matrices to tree-level (0) or 
       one-loop (1)
-r   - Rederives the couplings from scratch
-s   - Saves the couplings to the file provided
-h   - Prints a help message
\end{verbatim}
The third line of code provides the model with a file to read the couplings in
from and save to, depending on what the program has been specified to do. The
last line initializes the model. This means generate (or read) the couplings,
find all the tree level mass matrices and generate the RGEs. At this stage the
model is able to be used.

At this point we will want to read our parameters in from a file. This is
done by the \texttt{loadParameters()} function. The code that would read the
parameters from the file \texttt{MSSM.dat} is
\begin{verbatim}
ifstream parin;
parin.open("MSSM.dat");
mssm.loadParameters(parin);
parin.close();
\end{verbatim}
This file has a simple structure. It is tab delimited where the first entry
is the name of the parameter (the text name given during its definition) and
the value of the parameter. There is also a parameter defined for every model.
This is \texttt{renormScale} and defines the current renormalization scale. 
It is defaulted to 91.2 GeV (the $Z^0$ mass) and when a parameter is read in
it is assumed to be entered at the scale currently given by 
\texttt{renormScale}. So if one wanted to input parameter
\texttt{a} at 91.2 GeV but parameter \texttt{b} at 200 GeV they would use
\begin{verbatim}
a	50.0
renormScale	200.0
b	20.0
\end{verbatim}

Once the parameters have been read in one may want to evolve the RGEs to a 
scale. At first the boundary conditions must be determined at one scale. This
is done by calling \texttt{mssm.rges().matchScales()}. This routine goes 
through the boundary conditions entered and uses the Newton-Raphson method to
find a suitable set of parameters that match all the boundary conditions 
given. Once that is done the parameters can be evolved to the scale desired
by calling \texttt{mssm.rges().evolve(scale)}. 

Once the parameters have been determined at the scale desired the mass spectrum
can be developed. This is done with a call to 
\texttt{mssm.masses().diagonalize()}. This will diagonalize all the mass
matrices at either tree-level or one-loop. There is an additional flag, 
\texttt{Model::CorrectMassStates} which determines whether the one-loop 
corrections are applied directly to the mass eigenstates derived from the
tree-level mass matrices, or if the one-loop corrections are applied based on
the interaction eigenstates. If one wanted to get the mass of a particle
they simply call \texttt{mssm.masses().massOf(field)} where field is the
expression with all the indices evaluated (e.g. instead of looking at the
top quark mass, one would have to look at the left-handed quark, with
$SU(2)_L$ index = 1, family index = 3 and the $SU(3)_c$ index set to 
any number between 1 and 3). In the SM, for example, one would fully expect
to find that the mass all of the colour indices of a particular quark are the
same. But it would be possible to develop a theory where this is not the case.

The last function that would be desired is to find the effective potential. 
This can be given as an analytic expression by a call to 
\texttt{mssm.treePotential()} or at one-loop order by a call to
\texttt{mssm.potential()} (with the correct value of 
\texttt{Model::\_potApprox} set). Using this we can use the code
\begin{verbatim}
ofstream potPlot;
potPlot.open("MSSM_Pot.dat");
double v1, v2;
ex pot = mssm.treePotential();
double v = 246.0;
double beta;

for(beta = -3.14; beta < 3.14; beta += 0.01) {
  mssm.getParameter("HiggsVev1") = v*cos(beta);
  mssm.getParameter("HiggsVev2") = v*sin(beta);
  potPlot << beta << "\t" << pot.evalf() << endl;
}
potPlot.close();
\end{verbatim}
to generate fig.~\ref{fig:MSSMpot}.

\section{Entering a New Group}
When studying models different symmetries are desired. These are given by
the group governing the interaction. \EFF{} provides a way in which a new
group can be defined. Inheriting the class \texttt{GaugeGroup} is how this
is done. This class requires the user to implement
\begin{verbatim}
constructor(string n, string cn, Model* m, char l);
ex structureConstant(GaugeIndex&, GaugeIndex&, GaugeIndex&) const;
ex generator(GaugeIndex&, MatterIndex&, MatterIndex&) const;
GaugeIndex gaugeIndex(int i=0) const;
MatterIndex matterIndex(int i=0) const;
bool isIndex() const;
ex Cr(bool) const;
ex C2(bool) const;
\end{verbatim}
The constructor defines the coupling (given by the text name \texttt{n}
and the \LaTeX~name \texttt{cn}). The \texttt{char l} is used to define the
line of the group. A constructor is generally of the form
\texttt{ClassName(n,cn,m,l) : GaugeGroup(n,cn,m,l)\{\}}. This just passes
all the handling of these arguments to the \texttt{GaugeGroup} superclass.
The functions \texttt{C2(bool)} and \texttt{Cr(bool)} are used to give the
evalation of $t_r^a t_r^a = C_2(r) \cdot {\bm 1}$ and 
$C(r) = \Tr[t_r^a t_r^a]$. If the boolean passed in is true then the adjoint
representation is used. If it is false then the fundamental representation is
used.

Next one must define the indices. This usually amounts to defining a set of
symbols for the adjoint and fundamental indices. The indices are then just
\texttt{idx} classes with the symbol and the correct dimension set. The
last remaining pieces are the structure constants and generators, given by
\texttt{structureConstant} and \texttt{generator}. These require the 
definition of a \texttt{tensor} class from {\small GiNaC}. An example of an 
$SU(3)$ structure constant is given here. The definition of the tensor is
\begin{verbatim}
class SU3Structure : public tensor {
  GINAC_DECLARE_REGISTERED_CLASS(SU3Structure,tensor)

 public:
  void print(const print_context &c, unsigned level = 0) const;
  ex eval_indexed(const basic &i) const;
};
\end{verbatim}
One can then define a helper function \texttt{SU3\_structure} which returns
the expression for the structure constant given the indices. This takes
the form
\begin{verbatim} 
inline ex SU3_structure(const ex &i1, const ex &i2, const ex &i3) {
  if(!is_a<GaugeIndex>(i1) || !is_a<GaugeIndex>(i2) ||
     !is_a<GaugeIndex>(i3))
    throw(std::invalid_argument(
          "Indices to SU(3) structure must be of type GaugeIndex"));

  if(ex_to<GaugeIndex>(i2).line() !=
     ex_to<GaugeIndex>(i3).line())
    throw(std::invalid_argument(
          "GaugeIndices must be of the same line and type"));

  return indexed(SU3Structure(), sy_anti(), i1, i2, i3);
}
\end{verbatim}

The structure constant can then be defined by
\begin{verbatim}
GINAC_IMPLEMENT_REGISTERED_CLASS(SU3Structure,tensor)
DEFAULT_CTORS(SU3Structure)
DEFAULT_ARCHIVING(SU3Structure)
DEFAULT_COMPARE(SU3Structure)
\end{verbatim}
and the \texttt{print} and \texttt{eval\_indexed} functions are
\begin{verbatim}
void SU3Structure::print(const print_context & c, unsigned level) const {
  if (is_a<print_tree>(c)) inherited::print(c, level);
  else {
    ostream& os = c.s;
    os << "f";
  }
}

ex SU3Structure::eval_indexed(const basic &i) const {
  GINAC_ASSERT(is_a<indexed>(i));
  GINAC_ASSERT(i.nops() == 4);
  GINAC_ASSERT(is_a<SU3Structure>(i.op(0)));
  GINAC_ASSERT(is_a<GaugeIndex>(i.op(1)));
  GINAC_ASSERT(is_a<GaugeIndex>(i.op(2)));
  GINAC_ASSERT(is_a<GaugeIndex>(i.op(3)));

  const GaugeIndex& g1 = ex_to<GaugeIndex>(i.op(1));
  const GaugeIndex& g2 = ex_to<GaugeIndex>(i.op(2));
  const GaugeIndex& g3 = ex_to<GaugeIndex>(i.op(3));

  // Numeric Evaluation
  if(static_cast<const indexed &>(i).all_index_values_are(info_flags::integer))
  {
    int i = ex_to<numeric>(g1.get_value()).to_int();
    int j = ex_to<numeric>(g2.get_value()).to_int();
    int k = ex_to<numeric>(g3.get_value()).to_int();
    int ip = i; int jp = j; int kp = k;
    if(ip>jp) Utils::swap(ip,jp);
    if(jp>kp) Utils::swap(jp,kp);
    if(ip>jp) Utils::swap(ip,jp);
    static ex half = numeric(1,2);
    double cyc = Utils::cyclicPermutation(i,j,k);
    if(ip==1) {
      if(jp==2 && kp==3) return cyc;
      else if(jp==4 && kp==7) return half*cyc;
      else if(jp==5 && kp==6) return -half*cyc;
    } else if(ip==2) {
      if(jp==4 && kp==6) return half*cyc;
      else if(jp==5 && kp==7) return half*cyc;
    } else if(ip==3) {
      if(jp==4 && kp==5) return half*cyc;
      else if(jp==6 && kp==7) return -half*cyc;
    } else if(ip==4 && jp==5 && kp==8) return sqrt(numeric(3,4))*cyc;
    else if(ip==6 && jp==7 && kp==8) return sqrt(numeric(3,4))*cyc;
    return 0;
  }
  // No further simplification
  return i.hold();
}
\end{verbatim}

This then completely defines the behaviour of the structure constants of the
group. A similar thing can be done for the generators.

\section{Defining a Diagram}
Defining a diagram is a useful way to do computations with \EFF{}. This is
also straightforward. One simply inherits the \texttt{Diagram} class
and implements the constructor 
\begin{verbatim}
MyNewDiagram(DElementVec &v, Model *m) : Diagram(v,m) {}
\end{verbatim}
and the \texttt{function()} function. This function is used to evaluate the
diagram at the current set of fields. The \texttt{DElementVec} is used to
specify a set of fields for each external leg and propagator of the diagram.
When the diagram is created the desired choice of fields for each part is
given. For example we want to compute the cross section for $\ee \to q \bar q$ 
at tree-level. We simply create some \texttt{DiagramElements} and put the 
appropriate fields in.
\begin{verbatim}
DElementVec eeqq(4);
exvector electrons, quarks;
electrons.push_back(real_left_electron);
electrons.push_back(imag_left_electron);
electrons.push_back(real_right_electron);
...
eeqq[0] = DiagramElement(electrons, true);
eeqq[1] = DiagramElement(electrons, true);
eeqq[2] = DiagramElement(quarks, true);
eeqq[3] = DiagramElement(quarks, true);
\end{verbatim}
This creates the \texttt{DElementVec} object with all the electrons and all the
quarks. This then gets given as input to the diagram \texttt{ee2qq}
\begin{verbatim}
ee2qq myee2qq(eeqq,my_model);
cout << "My matrix element is " << myee2qq.evaluate() << endl;
\end{verbatim}
The \texttt{evaluate} function goes over every permutation of the fields
given in the four \texttt{DiagramElement}s and computes a contribution to the
matrix element. This contribution is given by the \texttt{function} function.
This must be defined so that each combination of fields contributes the 
relevant part. This is done by determining the couplings of the set of fields
given. For example
\begin{verbatim}
ex coup123 = coupling(0,1,2);
ex coup234 = coupling(1,2,3);
ex coup134 = coupling(0,2,3);
ex coup1234 = coupling(0,1,2,3);
\end{verbatim}
The \texttt{true} statement in the definition of the \texttt{DiagramElement} 
indicates that the mass state couplings are to be used. If it were false the
interaction couplings would be used.

%\include{}
%\include{}
%\include{}

%\chapter{Abbreviations\label{abbreviations}}
%\input{thesis.abb}

\begin{spacing}{1.5}
\bibliography{thesis}
\bibliographystyle{JHEP-2}
\end{spacing}

\end{document}